&"template"					% si on utilise mylatexformat.ltx
\documentclass[a4paper,11pt]{book}
\usepackage[utf8]{inputenc}
\usepackage{utf8ext}
\usepackage{xspace}			   % espace intelligent dans les macros
\usepackage[english]{babel}
\usepackage[T1]{fontenc}
\usepackage{lmodern}

\usepackage{feynmp-auto}
\usepackage{setspace}

\usepackage[margin=28mm,includeheadfoot,bindingoffset=10mm]{geometry}
\usepackage{setspace}
\usepackage{graphicx,color} 
\usepackage{floatflt}
\addto\captionsfrenchb{}
\addto\captionsfrench{}
\addto\captionsfrenchb{}
\usepackage{amsmath, mathtools}
\usepackage{amssymb,amsfonts}	% tous les symboles mathématique de AMS
\usepackage{bm}				% lettre mathématiques grasses et 
\usepackage{bbm,upgreek}		% lettres math blackboard et grecques pour μm et pour β-decay
\usepackage{etoolbox}			% nombreuses fonctions avancées indispensables
\usepackage{calc}				% calcul infix des longueurs
\usepackage{eso-pic}			% pour placer des élémnst das le background lors du shipout
\usepackage{datetime}			% formatage des datees, accès au temps
\usepackage{slantsc}
\makeatletter
\patchcmd{\chaptermark}{\MakeUppercase}{\scshape\slshape}{}{}%
\patchcmd{\sectionmark}{\MakeUppercase}{\scshape\slshape}{}{}%
\makeatother
\renewcommand{\thechapter}{\Roman{chapter}} % num de chapitre: chiffres Romains
\numberwithin{equation}{section} 
\numberwithin{figure}{chapter}
\numberwithin{table}{chapter}
\usepackage{titlesec}                        %pour définir le format des titres
\titleformat{\chapter}[display]{\Huge\sffamily\bfseries}%
	{\chaptertitlename~\thechapter}{1ex}{}
\titleformat{\section}[hang]{\Large\sffamily\bfseries}%
	{\rlap{\thesection}}{2em}{}
\titleformat{\subsection}[hang]{\large\sffamily\bfseries}%
	{\rlap{\thesubsection}}{3em}{}
\usepackage[backref]{hyperref}
\hypersetup{linktocpage,colorlinks=true,linkcolor=[rgb]{0.5,0,0.6},citecolor=[rgb]{0,0,0.6},pdfdisplaydoctitle=true,bookmarksnumbered=true}
\csname endofdump\endcsname
\usepackage{coverpages}
\logos{logopsl2copie}{}{logoens2} %this is just an example
\author{Thimothée Thiery}
\vspace{1cm}
\thesis{THÈSE DE DOCTORAT\\ Université de recherche Paris Sciences et Lettres}
\specialite{}
\ecoledoct{}
\lab{Laboratoire de Physique Théorique\\ de l'École Normale Supérieure}
\title{Analytical Methods and Field Theory for Disordered Systems}
\date{5/09/2016 }
\jury{
 &\small{ M. Müller} &\small{ Rapporteur} \\
 &\small{A. Povolotsky} & \small{Rapporteur} \\
&\small{G. Biroli } & \small{Examinateur }   \\
&\small{F. Comets}  & \small{Examinateur }   \\
&\small{P. Le Doussal } & \small{Directeur de thèse}    \\
&\small{A. Rosso  }& \small{Examinateur}    \\
&\small{K. Wiese } & \small{Co-Directeur de thèse }   \\
}

\resume{Sed commodo posuere pede. Mauris ut est. Ut quis purus. Sed ac odio. Sed vehicula
hendrerit sem. Duis non odio. Morbi ut dui. Sed accumsan risus eget odio. In hac ha-
bitasse platea dictumst. Pellentesque non elit. Fusce sed justo eu urna porta tincidunt.
Mauris felis odio, sollicitudin sed, volutpat a, ornare ac, erat. Morbi quis dolor. Donec
pellentesque, erat ac sagittis semper, nunc dui lobortis purus, quis congue purus metus
ultricies tellus. Proin et quam. Class aptent taciti sociosqu ad litora torquent per conubia
nostra, per inceptos hymenaeos. Praesent sapien turpis, fermentum vel, eleifend faucibus,
vehicula eu, lacus.
Pellentesque habitant morbi tristique senectus et netus et malesuada fames ac turpis
egestas. Donec odio elit, dictum in, hendrerit sit amet, egestas sed, leo. Praesent feugiat
sapien aliquet odio. Integer vitae justo. Aliquam vestibulum fringilla lorem. Sed neque
lectus, consectetuer at, consectetuer sed, eleifend ac, lectus. Nulla facilisi. Pellentesque
eget lectus. Proin eu metus. Sed porttitor. In hac habitasse platea dictumst. Suspendisse
eu lectus. Ut mi mi, lacinia sit amet, placerat et, mollis vitae, dui. Sed ante tellus, tristique
ut, iaculis eu, malesuada ac, dui. Mauris nibh leo, facilisis non, adipiscing quis, ultrices
a, dui}
\motscles{Tintin, Milou, pétrole, pilosité}

\titleen{Dupondt's whool in the country of black gold}

\abstract{Lorem ipsum dolor sit amet, consectetuer adipiscing elit. Ut purus elit, vestibulum ut,
placerat ac, adipiscing vitae, felis. Curabitur dictum gravida mauris. Nam arcu libero,
nonummy eget, consectetuer id, vulputate a, magna. Donec vehicula augue eu neque.
Pellentesque habitant morbi tristique senectus et netus et malesuada fames ac turpis
egestas. Mauris ut leo. Cras viverra metus rhoncus sem. Nulla et lectus vestibulum urna
fringilla ultrices. Phasellus eu tellus sit amet tortor gravida placerat. Integer sapien est,
iaculis in, pretium quis, viverra ac, nunc. Praesent eget sem vel leo ultrices bibendum.
Aenean faucibus. Morbi dolor nulla, malesuada eu, pulvinar at, mollis ac, nulla. Curabitur
auctor semper nulla. Donec varius orci eget risus. Duis nibh mi, congue eu, accumsan
eleifend, sagittis quis, diam. Duis eget orci sit amet orci dignissim rutrum.
Nam dui ligula, fringilla a, euismod sodales, sollicitudin vel, wisi. Morbi auctor lorem
non justo. Nam lacus libero, pretium at, lobortis vitae, ultricies et, tellus. Donec aliquet,
tortor sed accumsan bibendum, erat ligula aliquet magna, vitae ornare odio metus a mi.
Morbi ac orci et nisl hendrerit mollis. Suspendisse ut massa. Cras nec ante. Pellentesque a
nulla. Cum sociis natoque penatibus et magnis dis parturient montes, nascetur ridiculus
mus. Aliquam tincidunt urna. Nulla ullamcorper vestibulum turpis. Pellentesque cursus
luctus mauris.}

\usepackage{amsthm}

\theoremstyle{plain}

\theoremstyle{definition}

\theoremstyle{remark}

\numberwithin{equation}{section}

\usepackage[normalem]{ ulem }
\usepackage{soul}

\newcommand{\bea}{\begin{eqnarray}}
\newcommand{\eea}{\end{eqnarray}}
\newcommand{\be}{\begin{equation}}
\newcommand{\ee}{\end{equation}}

\newcommand{\JN}{\mathbb{N}}
\newcommand{\JR}{\mathbb{R}}
\newcommand{\JZ}{\mathbb{Z}}
\newcommand{\nn}{\nonumber}
\newcommand{\ssp}{\hspace{3pt}}
\newcommand{\stab}{\hspace{15pt}}

\newcommand{\sP}{{\sf P}}
\newcommand{\sff}{{\sf f}}

\newcommand{\sU}{{\sf U}}
\newcommand{\sV}{{\sf V}}
\newcommand{\su}{{\sf u}}
\newcommand{\sv}{{\sf v}}

\newcommand{\cH}{{\cal H}}
\newcommand{\cD}{{\cal D}}
\newcommand{\sR}{{\sf R}}
\newcommand{\sh}{{\sf h}}
\renewcommand{\st}{{\sf t}}
\newcommand{\sx}{{\sf x}}

\newcommand{\cA}{{\cal A}}

\newcommand{\sE}{{\sf E}}

\newcommand{\sT}{{\sf T}}

\newcommand{\cE}{{\cal E}}

\begin{document}

%\includepdf[pages=-]{./couv2.pdf}

%%%%%%%%%%%%%%%%%% pages liminaires %%%%%%%%%%%%%%%%%
\frontmatter
%--------------------------------------------------

\chapter{Abstract}

This thesis presents several aspects of the physics of disordered elastic systems and of the analytical methods used for their study.
On one hand we will be interested in universal properties of avalanche processes in the statics and dynamics (at the depinning transition) of elastic interfaces of arbitrary dimension in disordered media at zero temperature. To study these questions we will use the functional renormalization group. After a review of these aspects we will more particularly present the results obtained during the thesis on (i) the spatial structure of avalanches and (ii) the correlations between avalanches.
On the other hand we will be interested in static properties of directed polymers in 1+1 dimension, and in particular in observables related to the KPZ universality class. In this context the study of exactly solvable models has recently led to important progress. After a review of these aspects we will be more particularly interested in exactly solvable models of directed polymer on the square lattice and present the results obtained during the thesis in this direction: (i) classification of Bethe ansatz exactly solvable models of directed polymer at finite temperature on the square lattice; (ii) KPZ universality for the Log-Gamma and Inverse-Beta models; (iii) KPZ universality and non-universality for the Beta model; (iv) stationary measures of the Inverse-Beta model and of related zero temperature models.

\tableofcontents

\chapter{Introduction, goal and outline of the manuscript}

The theoretical analysis of disordered systems is an outstanding challenge of modern statistical physics and probability theory that finds applications in (disorder is present in any experimental setup and {\it might} play a crucial role) and outside of physics (e.g. error-correcting codes and optimization problems). In such systems the interplay between thermal (and/or quantum) fluctuations, the disordered environment and interactions often creates a rich `glassy' phenomenology. In this thesis we focus on $d-$dimensional elastic interfaces in a $(d+1)-$dimensional disordered media, described by a single valued function $u:  x \in \JR^d \to u(x) \in \JR$ (the height of the interface). The latter are remarkable examples of disordered systems. On one hand they can be used to describe a variety of physical situations such as domain walls in disordered magnets, fractures fronts in brittle materials, contact lines of viscous fluids on rough substrates... On the other hand they are sufficiently simple to allow analytical approaches and tremendous theoretical progress. For such systems, while the elasticity tends to flatten the interface, thermal fluctuations and the disordered environment tend to roughen it. The energy landscape of the interface is fractioned into a multitude of metastable states which in some cases dramatically influence the static and dynamic properties of the interface.

\subsection*{Shocks and Avalanches} 

In the case of the statics, one can show that in `small' dimensions ($d\leq 4$ for interfaces with short-range elasticity), the temperature is irrelevant at large scale and the large scale properties of the interface are those of its ground state: the system is {\it pinned} by disorder (there are subtleties linked with the fact that the temperature is dangerously irrelevant). At least in some cases, this pinning phenomena is collective and one expects some {\it universality} and {\it scale invariance} to emerge: large scale properties of the interface are, up to some non universal constants, independent of the underlying disorder distribution (although there are several {\it universality classes} that one would like to classify). In particular the interface is {\it rough} and its static {\it roughness exponent}, $\zeta_s$ loosely defined by $u(x)-u(0) \sim x^{\zeta_s}$ is universal. As we will review this problem has been already well studied and a good understanding of these universal properties has been reached. A more recent question is to understand the properties of several successive metastable states of the interface. Confining the interface around some position $w$ and studying the evolution of the ground state of the interface as $w$ is varied, the ground state changes abruptly at a discrete sequence of positions $w_i$. These changes in the ground state $u(x) \to u(x) + S(x)$ define a sequence of {\it shocks} $S(x)$. As both $u(x)$ and $u(x) + S(x)$ are legitimate ground states of the interface displaying scaling and universality, one expects the shocks $S(x)$ to display scaling and universality inherited from the universal physics of disordered elastic interfaces.

\smallskip

A closely related phenomenon occurs in the zero temperature dynamics of the interface of sufficiently low dimensions when it is slowly driven with a mean velocity $\partial_t u(t,x) \sim v \to 0^+$ (at the {\it `depinning transition'}). The system reaches an out-of-equilibrium steady-state displaying scaling and universality with a depinning roughness exponent $u(t,x)- u(t,0) \sim x^{\zeta_d} $ that differs from the static one. In the steady-state, most of the time the interface is actually pinned by disorder in a metastable state, $\partial_t u(t,x) \sim 0$, and very rarely manages to cross an energy barrier. When it does the interface moves with a macroscopic velocity of order $O(1)$ during a finite time window $\Delta t = O(1)$ until it is pinned again in a new metastable state. The next jump occurs after a period of quiescence $T \sim 1/v \gg \Delta t$. The motion of the interface in between these metastable states is called an {\it avalanche}. The latter are very close cousins of the shocks between static ground states mentioned above, but are richer as they are a complex time-dependent phenomena for which more questions can be asked. Again these avalanches inherit the scale invariance and the universality of disordered elastic interfaces at the depinning transition.

\smallskip

 More generally avalanches occur in a wide range of complex systems, from snow avalanches to avalanches in the neural activity of the brain. While avalanches in some systems will fall in the universality class of the disordered elastic interface model, other may not. Important counter examples are e.g. avalanches at the yielding transition of amorphous materials (for which plastic deformations play an important role) or earthquakes (for which the presence of aftershocks, whose origin is still controversial, is certainly not captured by the simplest elastic interface model). In any case, characterizing and understanding the universality in avalanche processes, and in particular for the elastic interface model for which powerful analytical methods exist, is an important challenge. It indeed allows to understand and compare efficiently seemingly unrelated phenomena such as the fracture process of a brittle material or the jerky motion of a contact line between a viscous fluid and a rough substrate, and to assess the importance of various mechanisms in the dynamics. Important questions in the context of shocks and avalanches are the characterization of the distribution of the avalanche total size $S$ (the area swept by the interface), the duration $T$... As the avalanche processes mentioned above (at least for the elastic interface case) are scale-invariant processes, the latter are distributed with probability distribution functions (PDF) displaying (in a certain regime) a power-law behavior $P(S) \sim S^{-\tau_S}$, $P(T) \sim T^{-\tau_T}$. The critical exponents thereby defined are believed to be universal and related to the critical exponents of the depinning transition (for avalanches) or of the statics (for shocks). Scaling and universality is however not restricted to critical exponents and there exist {\it universal scaling functions}, allowing a refined characterization of universality classes. A perfect example of such universal scaling functions is the temporal shape of avalanches, which received much attention lately and was computed at and beyond mean-field. Going back to experiments, while on one hand the study of the temporal shape indeed showed universality between different avalanche processes, on the other hand it permitted to highlight the (non-universal) influence of Eddy currents for avalanches in Barkhausen noise experiments.

\subsection*{Out-of-equilibrium growth and the KPZ universality class}

 Another, seemingly unrelated phenomenon, is the out-of-equilibrium growth of an elastic interface $h(t,x) \in \JR$ driven by thermal fluctuations in the absence of quenched noise, typically thought of as separating a stable and an unstable phase of a thermodynamic system. The interface is rough $h(t,x) - h(t,0) \sim x^{\alpha}$ and exhibits non-trivial fluctuations and spatio-temporal patterns. For one dimensional interface $x \in \JR$ and local growth mechanisms with other reasonable assumptions, it is believed that a single universality class, the Kardar-Parisi-Zhang (KPZ) universality class, controls the large scale properties of growing interfaces. Remarkably, there is a very close connection with the statics of disordered elastic interfaces: the fluctuations of the free-energy $F(L,u)$ of a directed polymer ($d=1$ elastic interface case) of length $L$ at the temperature $T$ in a short-range random potential with end-points fixed as $u(0) = 0$, $u(L) = u$, defines a growing interface $h(t,x) := F(L=t,u=x)$ in the KPZ universality class. The KPZ universality class also encompasses models of interacting particles in one-dimension and has emerged over the years as a paradigmatic example of universality in out-of-equilibrium statistical physics. In this case the critical exponents are known exactly: the roughness exponent is $\alpha = 1/2$ and the height of the interface fluctuates widely on a scale of order $t^{1/3}$ (where here $t$ refers to the duration since the beginning of the growth).

\smallskip

  As for avalanches, universality goes however well beyond the sole critical exponents and the full distribution of the (rescaled) fluctuations of the interface are universal and, interestingly, depend only on global properties of the initial condition of the growing interface. As an example, for an interface growing from a flat initial condition, the fluctuations of the interface at a given point are distributed with the Tracy-Widom distribution for the largest eigenvalue of a random matrix in the Gaussian orthogonal ensemble, thus unveiling a remarkable connection between random matrix theory and the KPZ universality class. Observed in modern experiments, the emergence of such universal distributions related to extreme value statistics of random matrix theory is understood at the theoretical level through the analysis of {\it exactly solvable models} in the KPZ universality class, in particular models of directed polymers. While this property still lacks a `simple' explanation, the wide range of application of KPZ universality, of KPZ scaling and of the Tracy-Widom distribution, has the flavor of an extension of the central limit theorem to strongly correlated random variables. This has motivated in the past years a vast research effort aiming at the understanding of the KPZ fixed point, which thus still relies on the use of exactly solvable models.

 \subsection*{Analytical methods and the results obtained during the thesis}

While in a general setting I have tried throughout the thesis to improve the understanding and characterization of universal properties for models of disordered elastic interfaces in their strong disorder regime, my work can be divided following the subjects mentioned above.

 \medskip

\stab {\it On avalanches} \\
On one hand I have been interested in shocks and avalanche statistics for disordered elastic interfaces. To this aim I have used the functional renormalization group (FRG), a method already well developed. As any renormalization procedure, the latter directly aims at characterizing the large scale properties of the system through the identification of the appropriate fixed point. As I will review, the fixed points are perturbative in $\epsilon=4-d$ (for interfaces with short-range elasticity) and I have obtained results for avalanche statistics using this expansion at one-loop order, i.e. at first order beyond mean-field, i.e. at order $O(\epsilon)$.

 I have first focused on the spatial shape of avalanches. While on one hand the temporal shape of avalanches received a lot of attention, the spatial shape did not, surely because of the involved technical difficulties and the absence of an analytically tractable (and experimentally relevant) precise definition, in particular a centering procedure. I first obtained results at the mean-field level for the shape of peaked avalanches for model with short-range elasticity in $d=1$: there the shape becomes deterministic and given by a well defined spatial profile \cite{ThieryLeDoussalWiese2015}. Secondly, I focused on the mean shape of avalanches of fixed size centered on their seed for which I obtained results beyond mean-field, valid at order $O(\epsilon)$ \cite{ThieryLeDoussal2016a}. This `seed-centering' procedure introduced in this work appears as the most natural way, at least from the analytical point of view, to center spatial observables in the avalanche motion, and it could be used for other observables. In \cite{ThieryLeDoussal2016a} I perform simulations that show that the seed-centering can be successfully implemented in numerics, and in the future it would be interesting to confront these results with experiments for which the spatial shape is an accessible quantity, as is the case in some fractures experiments.

 In another project I investigated the correlations between successive avalanches and shocks. In general the question of correlations in avalanche processes has received a lot of attention, in particular in the context of earthquakes where these are linked to the notion of aftershocks, but in the elastic interface model they were always neglected and their sole existence was not put forward in the previous literature. While there are no correlations at the mean-field level where the avalanche process is a Lévy jump process \cite{ThieryLeDoussalWiese2015}, beyond mean-field I showed that there are always correlations. Furthermore these correlations are universal, of order $O(\epsilon)$, and controlled by the structure of the FRG fixed point \cite{ThieryLeDoussalWiese2016}. While these correlations do not correspond to the correlations observed in e.g. earthquake statistics, similar correlations probably exist in any system and understanding them is likely to be necessary to obtain a quantitative understanding of the correlations. In other systems well described by the elastic interface model, this work shows that in most cases (for interfaces of dimension below the upper critical dimension), there exist important correlations in the sequence of avalanches. Comparing these results with experiments would be very interesting.

\medskip

{\it On directed polymers} \\
On the other hand I have been interested in understanding the emergence of KPZ universality, or lack of thereof, in models of directed polymer on the square lattice. To this aim I have studied and discovered models with exact solvability properties, extending the already known exact solvability properties of the continuum directed polymer, the Bethe ansatz solvability and the exactly known stationary measure. In particular I have obtained: (i) Tracy-Widom GUE fluctuations for the point to point free-energy of the Log-Gamma polymer \cite{ThieryLeDoussal2014}; (ii) Tracy-Widom GUE fluctuations for the point to point free-energy of the Inverse-Beta polymer (a model I discovered during the thesis) and a classification of finite temperature Bethe ansatz exactly solvable models of directed polymer on the square lattice \cite{ThieryLeDoussal2015}; (iii) Tracy-Widom fluctuations for the large deviation function of a random walk on $\JZ$ in a time-dependent Beta distributed random environment, equivalent to the point to point free energy of the Beta polymer, and Gamma fluctuations in the diffusive regime of the random walk (suggesting a local breaking of KPZ universality due to an additional conservation law for the Beta polymer) \cite{ThieryLeDoussal2016b}; (iv) the stationary measures and mean quenched free-energy/optimal energy in the Inverse-Beta polymer and in the Bernoulli-Geometric polymer, an exactly solvable model of directed polymer on the square lattice at zero temperature dual to the Inverse-Beta polymer which I also discovered during the thesis \cite{Thiery2016}.

This `world' of exactly solvable models of directed polymer on the square lattice, in part unveiled by this thesis, now offers a set of models with different properties allowing to ask precise question about the KPZ fixed point and directed polymers in general. The Bethe ansatz approach to finite temperature models of directed polymer on the square lattice, developed in this thesis and at the same time by others, provides a new versatile tool which hopefully will permit to obtain a variety of interesting results for these models.

\subsection*{Goal and outline of the manuscript}

The goal of this manuscript is to provide a self-contained and pedagogical review of the subjects mentioned above, with an emphasis on theoretical techniques and aspects important to the understanding of the research papers \cite{ThieryLeDoussalWiese2015,ThieryLeDoussal2016a,ThieryLeDoussalWiese2016,ThieryLeDoussal2014,ThieryLeDoussal2015,ThieryLeDoussal2016b,Thiery2016} written during the thesis, whose main results will also be presented in the core of the manuscript.
\smallskip

In Chapter~\ref{chapI} I provide a broad introduction to disordered elastic systems, which will serve as a background for the understanding of the two main subjects studied during the thesis and presented thoroughly in the next chapters. In particular I re-obtain the static phase diagram of these systems, discuss the notion of strong disorder and the associated phenomenology, review early theoretical approaches and their caveats to motivate the use of the more sophisticated methods already mentioned, give a brief introduction to the more specialized subjects studied during the thesis and discuss some experimental evidences.

\smallskip

In Chapter~\ref{chapII} I focus on the avalanche processes of disordered elastic interfaces in the statics and in the dynamics (at the depinning transition) at zero temperature. I first introduce the notion of shocks and avalanches in $d=0$ toy models, and then generalize it to interfaces. I review the functional renormalization group approach to the statics and dynamics (at the depinning transition) of disordered elastic interfaces at zero temperature, with an emphasis on its applications to the computation of shocks and avalanches observables. I discuss the recent progresses made on the understanding of avalanche processes to motivate the main subjects studied during the thesis, the spatial structure of avalanches and the correlations in avalanche processes. The results obtained in the research papers \cite{ThieryLeDoussalWiese2015,ThieryLeDoussal2016a,ThieryLeDoussalWiese2016} are presented in the end of the chapter.

\smallskip

In Chapter~\ref{chapIII} I focus on the problem of the statics of a directed polymer at finite temperature in a random potential in dimension $1+1$. I recall the connection of this problem with the out-of-equilibrium growth of an interface in the KPZ universality class. I give an introduction to the KPZ universality class and review some recent remarkable progresses that were made (through the study of peculiar exactly solvable models) in the understanding of the KPZ universality class in $1+1$d. In particular I will present some known exact solvability properties of the continuum directed polymer -symmetries, stationary measure, Bethe ansatz solvability- that I tried in this thesis to generalize in discrete settings. I will then present the results obtained in the research papers \cite{ThieryLeDoussal2014,ThieryLeDoussal2015,ThieryLeDoussal2016b,Thiery2016}.

\chapter{Index of notations and abbreviations}

\begin{itemize}
\item{
use of subscripts: in this thesis the value at $x$ of an arbitrary function $f:x \in U \to f(x) \in V$ will often be denoted using a subscript: $f_x \equiv f(x)$. To avoid confusions, we will NEVER use subscripts to denote derivatives. \\
For $x \in \JR^d$, $\int_{x} \equiv \int_{x \in \JR^d} d^d x $ \\
For $q \in \JR^d$, $\int_q \equiv \int_{q \in \JR^d} \frac{d^d q }{(2 \pi)^d}$ \\
$\hat \delta^{(d)}(q) = (2 \pi)^d \delta^{(d)}(q)$ where $ \delta^{(d)}(q)$ is the usual Dirac delta distribution in $d$ dimension.\\
The Fourier transform of a function $f_x$ is denoted $f_q = \int_x e^{-i q \cdot x} f_x$. Thus $f_x = \int_q e^{i x \cdot q} f_q$. \\
$\overline{()}$ is the average over disorder.\\
$\sim$ denotes the equality in law between random variables \\
$\theta(x)$ is the Heaviside theta function \\
$\Gamma(x)$ is Euler's Gamma function \\
$A_d^{\gamma} = \frac{(2 \sqrt{\pi})^d}{2} \frac{\Gamma(\gamma)}{\Gamma(\gamma+1-d/2)}$ \\ 
$C^m_n = \frac{n!}{m! (n-m)!}$  \\
$(a)_n = \prod_{k=0}^{n-1} (a + k)$ \\
}
\end{itemize}

---------

\begin{itemize}
\item{
ABBM: Allesandro-Beatrice-Bertotti-Montorsi \\
BA: Bethe ansatz \\
BFM: Brownian-Force-Model \\
BW: Boltzmann Weight \\
CDF: Cumulative Probability Distribution \\
DES: Disordered Elastic System\\
DP: Directed-Polymer\\
DR: Dimensional Reduction \\
FP: Fixed Point \\
FPP: First Passage Percolation \\
FRG: Functional Renormalization Group \\
GOE/GUE/GSE: Gaussian Orthogonal/Unitary/Symplectic Ensemble of Random Matrix Theory. \\
gRSK: geometric/tropical RSK \\
GWN: Gaussian white noise  \\
IR: Infra-Red \\
KPZ: Kardar-Parisi-Zhang \\
KPZUC: KPZ universality class in $1+1$d \\
LL: Lieb-Liniger \\
LPP: Last Passage Percolation \\
LR: Long-Range\\
LT: Laplace Transform \\
MSHE: Multiplicative Stochastic-Heat-Equation \\
NF: Narayan-Fisher \\
PDF: Probability Distribution Function \\
RB: Random Bond \\
RF: Random Field\\
RG: Renormalization Group \\
RMT : Random Matrix Theory \\
RSK: Robinson–Schensted–Knuth \\
RV: Random Variable \\
RW: Random Walk \\
RWRE: Random Walk in a Random Environment \\
SHE: Stochastic-Heat-Equation \\
SR: short-range\\
TD-RWRE: Random Walk in a Time-Dependent Random Environment \\
TFP: Thermal Fixed Point \\
TW: Tracy-Widom \\
UV: Ultra-Violet \\
ZRP: Zero-Range-Process}
\end{itemize}

%%%%%%%%%%%%%%%%%% corps du texte %%%%%%%%%%%%%%%%%%%%
\mainmatter
%--------------------------------------------------

\chapter{Disordered elastic systems}\label{chapI}

The focus of this thesis is on disordered elastic systems (DES) and the goal of this section is to discuss some general properties of the latter that will underline all the manuscript. We will start by very general theoretical considerations on DES that are introduced in Sec.~\ref{secI1}, discuss the relevance of the disorder and draw the well-known static phase diagram in Sec.~\ref{Sec:StaticPhaseDiagram}. In Sec.~\ref{SecI3} we briefly introduce the more specialized topics that are the focus of Chapter~\ref{chapII} and \ref{chapIII}. Finally in Sec.~\ref{sec:Experiments} we discuss some experimental systems for which a theoretical approach using DES has been proposed. An alternative way to read this chapter is to start by the experimental observations of Sec.~\ref{sec:Experiments}. Here we have decided to first present the theoretical objects we will study in order to already try to be precise on the specific model with which one can attempt to understand a given experimental situation. The content of \ref{secI1} and \ref{Sec:StaticPhaseDiagram} is now standard and similar presentations can be found in \cite{Balents1996,ChauveThesis,GiamarchiKoltonRosso2006}

\section{The Hamiltonians} \label{secI1}

Although in this thesis we will also consider discrete systems, let us here focus on continuous systems introduce a general Hamiltonian for a DES of {\it internal dimension} $d \in \JN^*$ and {\it external dimension} $N \in \JN^*$. Let us first discuss the state space,

\subsection{The state space} \label{subsec:SecI1:StateSpace}

A state of the system is a real function
\be \label{Eq:SecI1:1}
u : x \in \JR^d \to  u(x) \equiv u_x  \in \JR^N  \ .
\ee
Where here we have introduced the subscript notation to indicate the dependence on the position in the internal space. The latter will be heavily used in the following. The space $\JR^d$ will be referred to as the {\it internal space}, whereas $\JR^N$ will be referred to as the {\it external space}. A state of the system can be embedded in the {\it total space} of dimension $d+N$ as the set of points $(x,u_x) \in \JR^{d+N}$. The case of {\it elastic interfaces} refers to the $N=1$ case. Indeed in this case the DES can be thought of as separating two phases of a thermodynamic system living in the total space. The case of {\it directed polymers } on the other hand refers to the $d=1$ case. Note that we already made a restricting hypothesis that will hold for all theoretical analysis present in this manuscript since more general DES not described by a single-valued function (\ref{Eq:SecI1:1}) could be considered. For the case of interfaces this is the hypothesis that there are no {\it overhangs}, while for the directed polymer case, this is precisely the hypothesis from which the word directed comes. More generally a point of the DES can thus be labeled by its internal coordinate $x$ and can only move in the external space.

\smallskip

In all the manuscript, whenever such a continuum description is used, we will assume the existence of two length-scales in the internal space: (i) a small-size cutoff $a$ below which the elastic and continuum description of the system breaks down; (ii) a large scale cutoff $L$ which represents the lateral extension of the system (the system is thus finite).  Boundary conditions will be discussed later.  Let us already say here that all the results obtained in this thesis actually concern the case of elastic {\it interfaces}, that is the special case $N=1$. The Hamiltonian of the system will be generally the sum of three contributions
\bea \label{Eq:secI1:Hamiltonian}
\cH_{V,w}[u] := \cH^{{\rm el}}[u] + \cH^{{\rm dis}}_{V}[u]+ \cH^{{\rm conf}}_w[u]   . \eea
As we detail below, $\cH^{{\rm el}}[u]$ is the {\it elastic Hamiltonian} of the system, $ \cH^{{\rm dis}}_{V}[u]$ is the {\it disorder Hamiltonian} (that depends on the realization of a random potential $V$) and $\cH^{{\rm conf}}_w[u]$ is a {\it confining Hamiltonian} that confines the system around an average position $w$. Here $\cH^{{\rm el}}[u]$ and $\cH^{{\rm dis}}_{V}[u]$ are the main players but the presence of a confining Hamiltonian will be important to define various quantities and to study avalanches. In this section we will keep $N$ arbitrary, mainly to emphasize the influence of the external dimension on the importance of the disorder on large scale properties. Let us now be more precise and define/give examples for each term that appears in (\ref{Eq:secI1:Hamiltonian}).\\

\subsection{ The elastic Hamiltonian} \label{subsec:SecI1:ElastHam}
Our typical choice for the elastic Hamiltonian will be the case of {\it short-range elasticity} that is modeled by the Hamiltonian
\bea
\cH^{{\rm el}}_{{\rm SR}}[u]  := \frac{c}{2} \int_{x} (\nabla_x u_x)^2   \ ,
\eea
where $c\geq 0$ is a constant. By rescaling the $x$ axis we will assume $c = 1$. We will also sometimes consider other types of elasticity. In the most general case the elastic Hamiltonian will be defined by, 
\bea \label{Eq:secI1:HelGen}
\cH^{{\rm el}}_g[u] :=\frac{1}{2} \int_{x,y}  g_{x,y}^{-1} u_x \cdot u_y \ ,
\eea
where we introduced the {\it elasticity kernel} $g_{x,y}^{-1}$
. We will suppose that the elasticity kernel is rotationally and translationally invariant (in internal space) $g_{x,y}^{-1} = g_{|x-y|}^{-1}$ and to be such that $ \cH^{{\rm el}}[u]$ defined above is a convex functional that attains its minimum for {\it flat} systems: $\cH^{{\rm el}}[cst] = 0$. We also suppose translational and rotational invariance in external space: $\cH^{{\rm el}}[u+cst]= \cH^{{\rm el}}[u]$. Note that the ground state of the elastic Hamiltonian is thus degenerate. Introducing the Fourier-transform of the elasticity kernel, $g_{q}^{-1} = \int_x e^{-i q x} g_{x,y}^{-1}$, the elastic Hamiltonian can be rewritten as
\bea
\cH^{{\rm el}}_g[u] =\frac{1}{2} \int_{q}  g_{q}^{-1} u_{-q} \cdot u_{q} \ .
\eea
And the case of short-range (SR) elasticity thus corresponds to the choice $g_q^{-1} = \sigma q^2$. In this introduction we will consider elastic kernels of the form $g_{q}^{-1} = |q|^{\gamma}$ and the important examples will be $\gamma=2$ (SR elasticity) and $\gamma=1$. The latter is known to be relevant in describing some systems with long-range (LR) elasticity as will be recalled in Sec.~\ref{sec:Experiments}. \\
 
 \subsection{The disorder Hamiltonian}  \label{subsec:SecI1:DisHam}
 The disorder Hamiltonians will be taken as the integral of a disorder potential $V : (x, u) \in \JR^d \times \JR^N \to V(x,u) \in \mathbb{R}$:
  \bea
  \cH^{{\rm dis}}_{V}[u] := \int_{x} V(x,u_x) \ .
  \eea
  Here $\{V(x,u) \}$ is a collection of random variables (RVs) that is drawn from a known probability distribution function (PDF). In Chapter~\ref{chapIII} we will define precisely the PDF of $\{V(x,u) \}$, as we will restrict our analysis to some specific distributions allowing exact treatments. We will however have in mind that some large scale properties should be {\it universal}, where here by universal we mean independent of the distribution of $V$ apart from some precise properties. In Chapter~\ref{chapII} on the other hand we will almost never specify the PDF of $\{V(x,u) \}$ as we will directly use methods that will make clear the universal character of our conclusions. Here let us only define the global properties of the PDF of $\{V(x,u) \}$ for which our results will hold.\\

\begin{figure}
\centerline{\includegraphics[width=6.0cm]{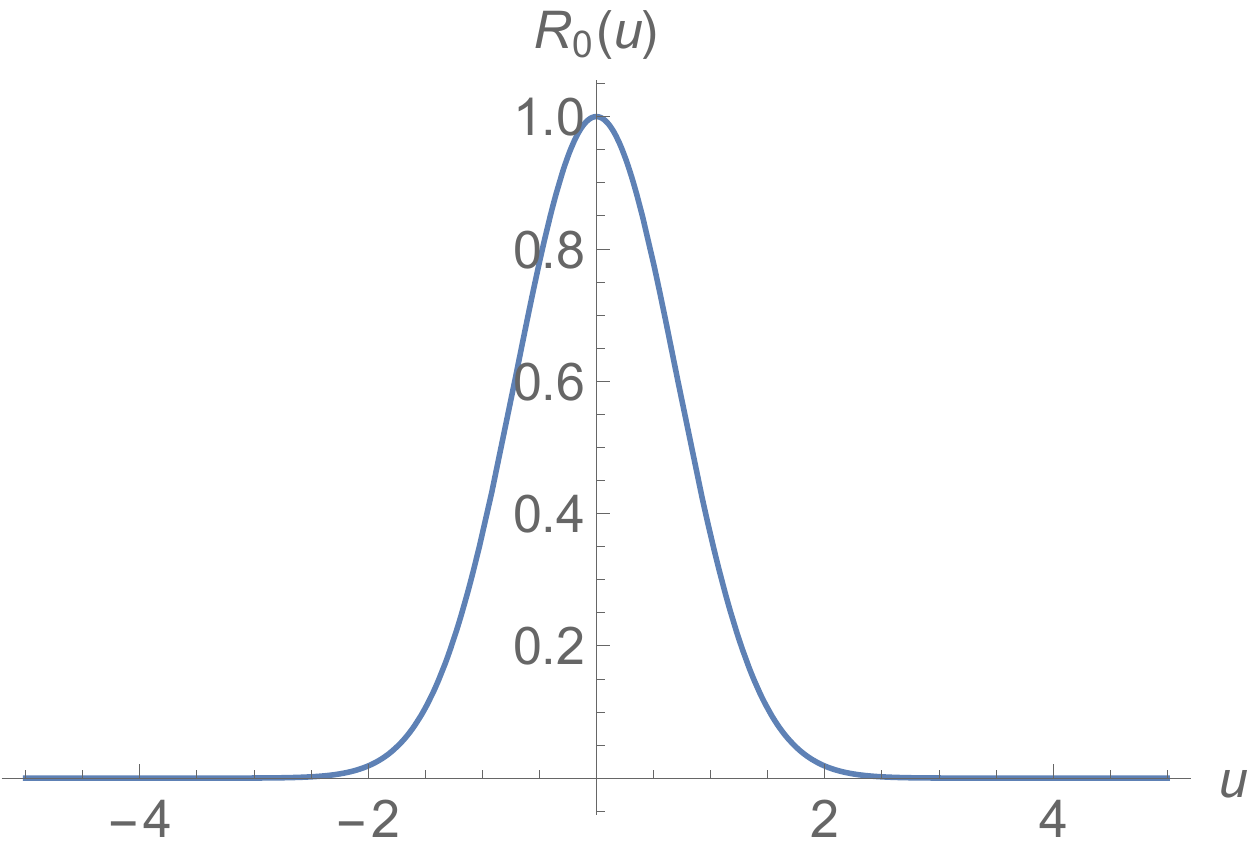} \includegraphics[width=6.0cm]{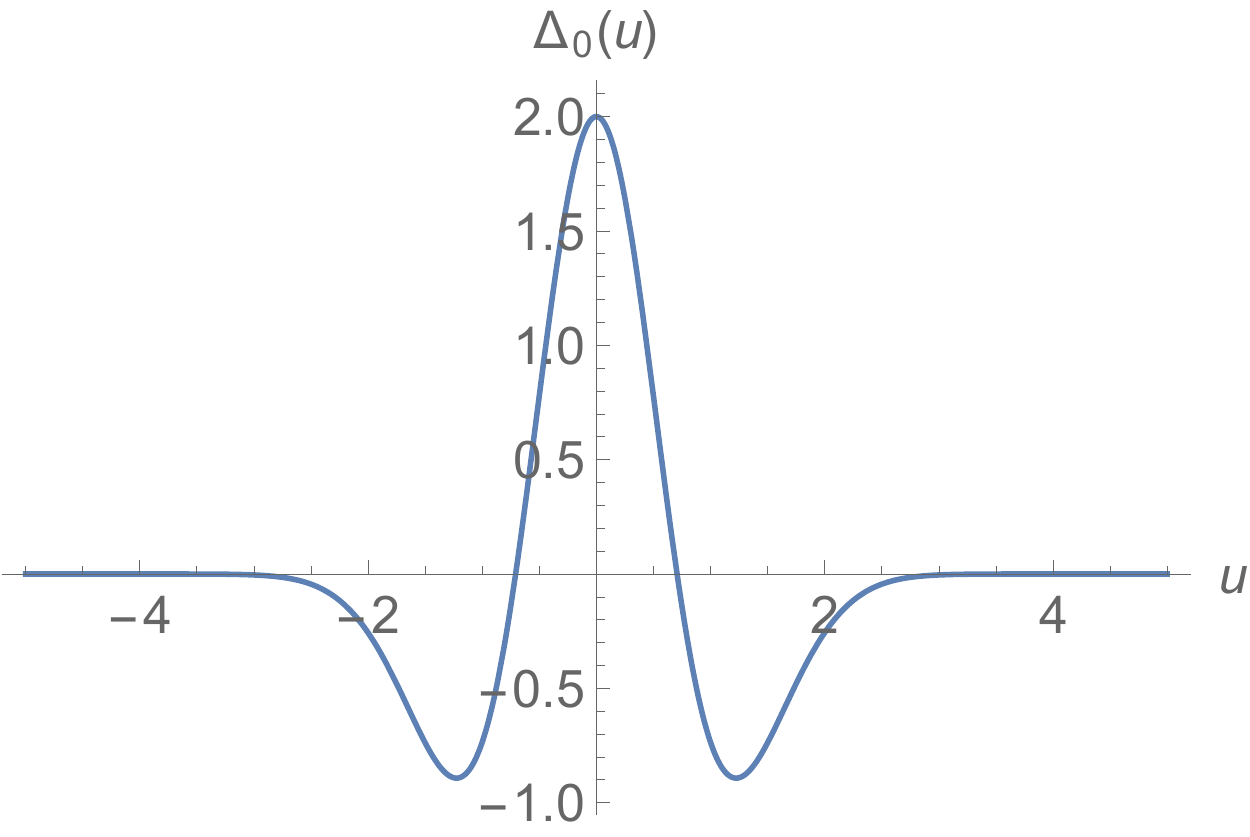} } 
\caption{Typical shape of the second cumulant of the disorder potential and of the disorder force for disorder of the random bond type with $N=1$ and correlation length $u_c \sim 2$.}
\label{fig:Intro:RB}
\end{figure}

\begin{figure}
\centerline{\includegraphics[width=6.0cm]{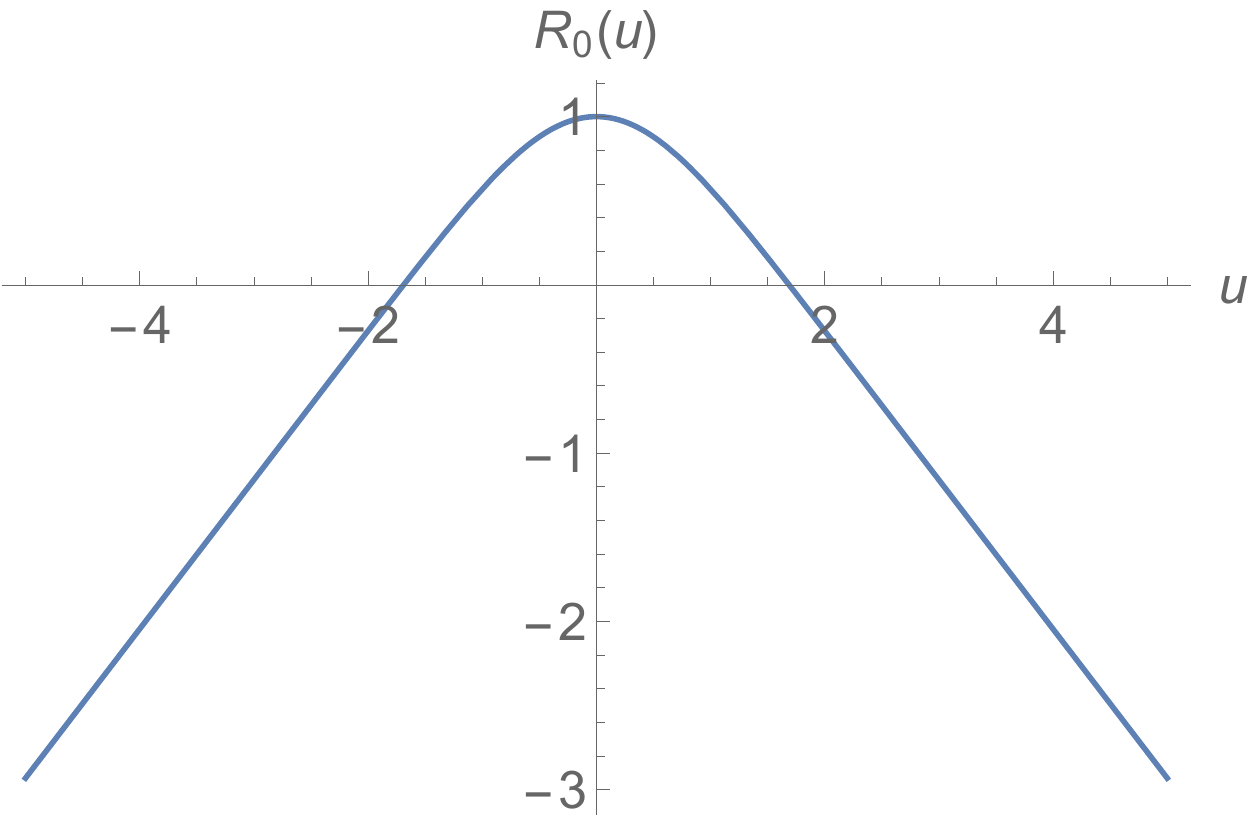}  \includegraphics[width=6.0cm]{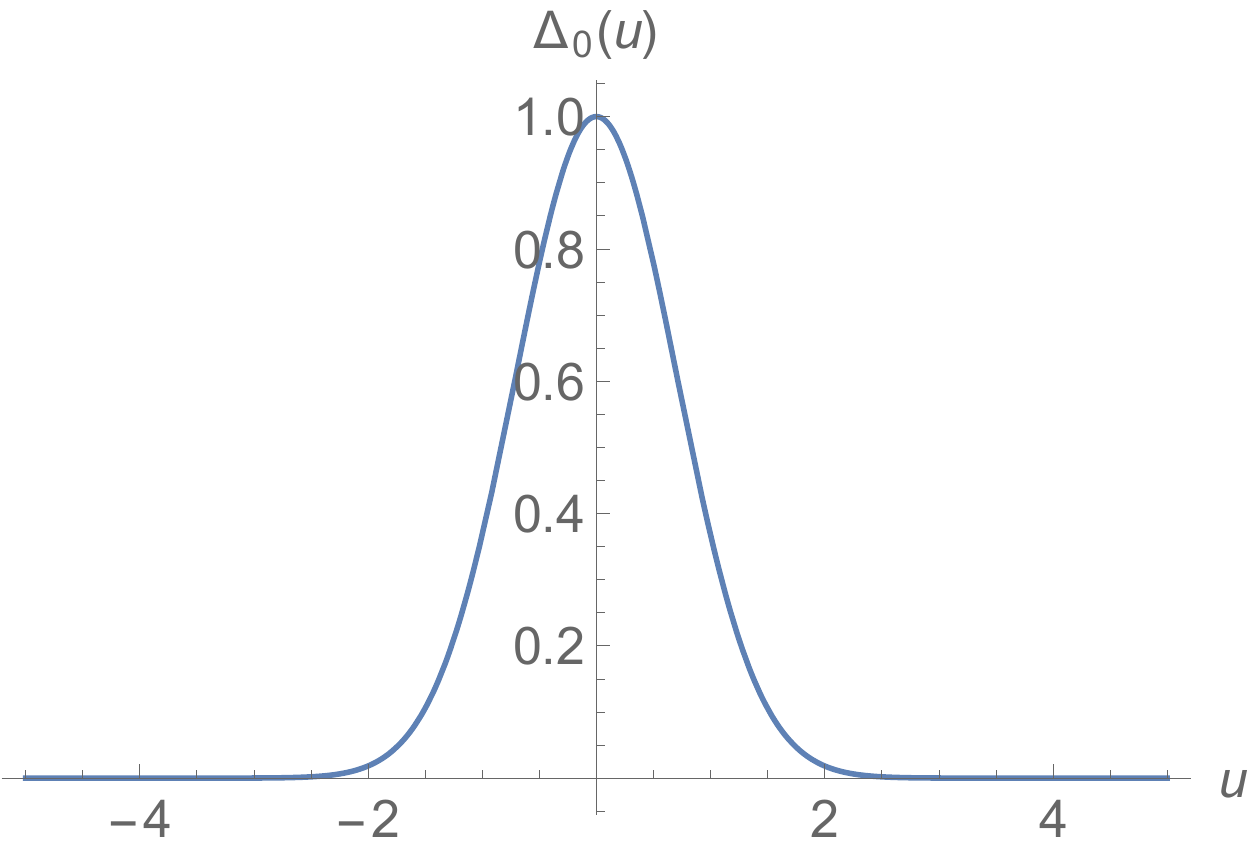} } 
\caption{Typical shape of the second cumulant of the disorder potential and of the disorder force for disorder of the random field type with $N=1$ and correlation length $u_c \sim 2$..}
\label{fig:Intro:RF}
\end{figure}

We will restrict our analysis to the case where the distribution of $V(x,u)$ at one point has no fat tails: all the positive moments $\overline{V(x,u)^n}$ (where from now on the overline $\overline{()}$ denotes the average over the random environment, i.e. over the distribution of  $\{V(x,u) \}$) are finite for $n \geq 0$: $\overline{V(x,u)^n} < + \infty$ (e.g. a Gaussian distributed disorder). We will suppose that $V(x,u)$ is homogeneously distributed. The symmetry in law $V(x+\Delta x , u + \Delta u ) \sim V(x , u)$ (where here and throughout this manuscript $\sim$ means `distributed in law as') will sometimes be referred to as the {\it statistical translational invariance of the disorder}. Concerning the {\it correlations} in the set of RVs $\{V(x,u) \}$, the two most important cases that we will consider, motivated by physical applications, are (i) disorder of the {\it random bond} type for which the correlations of the potential $V(x,u)$ are {\it short range} (SR); (ii) disorder of the {\it random field} type for which it is the {\it force} $F(x,u) := -\nabla_u V(x,u)$ acting on the system which has SR correlations. More precisely we will suppose
  \bea \label{Eq:secI1:Correlations}
  && \overline{V(x,u) V(x',u')}^c = \delta^{(d)}(x-x') R_0(u-u')   \ssp , \nn \\
  && \overline{F(x,u) F(x',u')}^c = \delta^{(d)}(x-x') \Delta_0(u-u')  \ssp , 
  \eea
  where $\Delta_0(u)= -\nabla_u^2R_0(u)$. Both are radial functions (i.e. even functions for $N=1$). In the random bond (RB) case $R_0(u)$ is decaying faster than any-power law at large $u$, i.e. typically as $R_0(u) \sim e^{-|u|/u_c}$ where $u_c$ is the {\it correlation length of the disorder}. The typical shape we have in mind for this case is shown in Fig.~\ref{fig:Intro:RB}, $R_0(u)$ is positive while $\Delta_0(u)$ is positive at small $|u|$ and negative at large $|u|$. If $R_0(u)$ is flat at $0$ and at infinity, we have $\int_u |u|^{d-1}\Delta_0(u) =0 $. The case of $0$ correlation length $u_c=0$ corresponds to $R_0(u) = \delta^{(N)}(u)$. In the random field (RF) case, $\Delta_0(u)$ has the same properties as $R_0(u)$ in the RB case with a correlation length $u_c$. In this case $R_0(u) \sim_{|u| \to \infty} - \sigma |u|$, see Fig.~\ref{fig:Intro:RF}. Finally we will also sometimes briefly consider periodic disorder: in this case $V_0(x,u + \Delta u) =V_0(x,u ) $ where $\Delta u$ is the period (this case is relevant e.g. in the context of charge density waves \cite{DSFisher1985}).\\


  {\it Remark:} The distinction between the internal and external space in the form of the correlations (\ref{Eq:secI1:Correlations}) might seem strange. The reason for this is that in our renormalization procedure, in Chapter~\ref{chapII}, we will try to describe the {\it effective disorder felt at large scale} by the interface. In doing so we will see that, e.g. starting from a RB disorder with $0$ correlation length $R_0(u) = \delta^{(N)}(u)$ the effective disorder at large scale will acquire a finite correlation length in the $u$ space. Conversely, starting with a non-zero correlation length in the internal space, the effective disorder at large scale will have a (decreasing to) $0$ correlation length in the internal space. For this reason we will already start our renormalization procedure with disorder having the same type of correlations than the effective disorder at large scale. Long-range correlations in internal space can break this picture and we will not study them, see e.g. \cite{FedorenkoLeDoussalWiese2006b}.
   
\subsection{ The confining Hamiltonian}   \label{subsec:SecI1:ConfHam}
Most of the questions one can ask about a disordered elastic system with Hamiltonian defined as in (\ref{Eq:secI1:Hamiltonian}) would be ill-defined without a confining term $ \cH^{{\rm conf}}_w[u] $ that confines the system around an average position $w \in \JR^N$. For example, due to the translational invariance in external space of $\cH^{{\rm el}}[u]$, the ground state of the Hamiltonian $\cH^{{\rm el}}[u] + \cH^{{\rm dis}}_{V}[u]$ for a `typical realization of a typical' disorder in an infinite space would be at infinity. There are various ways of regularizing the problem and in this thesis we will consider two possibilities. \\
  
In Chapter~\ref{chapII} we will confine the system around an average position $w \in \JR^N$ using a parabolic well as, for the case of a system with SR elasticity,
\bea \label{Eq:secI1:HMass}
\cH^{{\rm conf}}_w[u] := \frac{m^2}{2 } \int_x  (u_x -w)^2 \ , 
\eea
and the parameter $m>0$, which is the stiffness of the well, will be called the mass. In this case we have
\bea
\cH^{{\rm el}}_{{\rm SR}}[u]+\cH^{{\rm conf}}_w[u] = \frac{1}{2} \int_q (q^2+m^2)  (u_{-q} -w_{-q}) \cdot (u_{q} -w_q) \ .
\eea
Here we have noted $w_q = w \hat \delta^{(d)}(q)$. More generally for other types of elasticity we will consider confinement such that
\bea \label{Eq:secI1:HMass2}
\cH^{{\rm el}}_g[u]+\cH^{{\rm conf}}_w[u]  =  \frac{1}{2} \int_q \tilde{g}_q^{-1} (u_{-q} -w_{-q}) \cdot (u_{q} -w_{q} )
\eea
with 
\bea \label{Eq:SecI1:GammaElast}
\tilde{g}_q^{-1} = (\mu^2 + q^2)^\frac{\gamma}{2}
\eea
 and in these cases we will denote the mass as 
 \bea
 m = \sqrt{\tilde{g}_{q=0}^{-1}} = \mu^{\frac{\gamma}{2}} \ssp .
 \eea 
 Conversely we will note $\tilde{g}_{x,y}^{-1} = \int_q e^{iq \cdot (x - y)} \tilde{g}_{q}^{-1}$. In the remainder of the manuscript we will actually drop the tilde and except otherwise stated, the `elastic kernel' $g_q^{-1}$ will thus also contain the confining term as in (\ref{Eq:secI1:HMass2}). Note that the confining Hamiltonian introduces a length-scale $\ell_\mu := 1/\mu$ above which different parts of the interface are essentially elastically independent. Being interested in the regime where elasticity and disorder compete on equal footing, we will be interested in length-scales $|x-x'| \ll \ell_\mu$. We will thus consider the small $\mu$ limit. Taking also into account the scales $a$ and $L$ discussed earlier we will be interested in the regime $a \ll |x-x'| \ll \ell_\mu \ll L$. As such, boundary conditions will not play a role in this case and we will not discuss them. As we will recall below in Chapter~\ref{chapII}, adding such a confining term permits a convenient renormalization group approach to disordered elastic interfaces (the case $N=1$). There $\ell_{\mu}$ plays the role of an infra-red cutoff that smoothly cuts off fluctuations of the system at large length and $-\log \mu$ will play the role of time for the renormalization group flow. More precisely we will be interested in understanding the properties of the system as $w$ is varied. In particular, under some conditions that will be discussed, we will see that as $w$ is varied, the ground state of the interface changes discontinuously as a function of $w$. These abrupt changes define the notion of {\it shocks} or {\it static-avalanches}. Understanding universal properties of avalanche processes is one of the main goals of this thesis. In Chapter~\ref{chapII} the confining term (\ref{Eq:secI1:HMass}) will thus play a major role as moving $w$ will allow us to probe a certain sequence of metastable states of the interface. \\

In Chapter~\ref{chapIII} on the other hand we will consider the case of a {\it directed polymer in a two dimensional random medium}, that is the case $d=N=1$. There we will mostly consider the so-called point-to-point problem and suppose that both ends of the polymer at $x=0$ and $x=L$ are fixed as $u_0 = 0$ and $u_L = w$. This can be implemented by a confining Hamiltonian $\cH^{{\rm conf}}_w[u] = -E\left( \delta^{(N)}(u_0) + \delta^{(N)}(u_L - w)\right)$ with $E \to \infty$. There the focus will be on lattice models and on determining statistical properties of the DP at a finite temperature using exact methods. We will obtain finite $L$ results but we will mostly be interested in understanding some bulk properties for $L \to \infty$ which are believed to be universal. One of the reasons why we will use in this chapter a different confining Hamiltonian compared to Chapter~\ref{chapII} is that (\ref{Eq:secI1:HMass}) does not permit an exact solution.

\section{Static phase diagram and the strong disorder regime} \label{Sec:StaticPhaseDiagram}

In this section we consider the statistical mechanics of a disordered elastic system described  by an Hamiltonian of the form (\ref{Eq:secI1:Hamiltonian}) at a finite temperature $T$ and confined around $w=0$ as in (\ref{Eq:secI1:HMass2}). The goal of this section is to discuss qualitatively the relevance at large scale of the disorder and of the temperature and to draw the well known equilibrium phase diagram of DES. We will start by discussing the case of the pure system $V =0$ and describe the {\it thermal fixed point}.

\subsection{The pure system and the thermal fixed point}

We thus consider an elastic system $u: x \in \JR^d \to \JR^N$ described by the Hamiltonian, written in Fourier space,
\bea \label{Eq:secI21:Hpure}
{\cal H}_{{\rm pure}}[u] = \frac{1}{2} \int_q  (q^2 + \mu^2)^{\frac{\gamma}{2}} ~ u_{-q} \cdot u_q \ .
\eea
At a given temperature $T$, the thermal average of an observable $O$ of the $u$ field is defined by the path-integral
\bea
\langle O[u] \rangle :=  \frac{1}{Z[T]} \int {\cal D}[u]  O[u] e^{-\frac{1}{T} {\cal H}_{{\rm pure}}[u]}  \quad , \quad Z[T] =  \int {\cal D}[u] e^{-\frac{1}{T} {\cal H}_{{\rm pure}}[u]} \ .
\eea
The Hamiltonian (\ref{Eq:secI21:Hpure}) is quadratic and the theory is Gaussian: arbitrary integer moments of $u_x$ can be computed exactly using Wick's theorem. In particular the two-point function is obtained as
\bea
&&  \langle u_q \cdot u_{q'} \rangle  = \hat \delta(q+q') \frac{T}{(q^2 + \mu^2)^{\frac{\gamma}{2}}}    \nn \\
&& \langle u_x \cdot u_{x'} \rangle  =T \int_q \frac{e^{i q \cdot (x-x')}}{(q^2 + \mu^2)^{\frac{\gamma}{2}}} 
\eea
Hence we have
\bea \label{Eq:secI21:ThermalRugosityb}
\langle (u_{x} - u_{x'})^2 \rangle && = 2T \int_q \frac{1- e^{i q \cdot (x-x')}}{(q^2 + \mu^2)^{\frac{\gamma}{2}} } \nn \\
&& = 2 T |x-x'|^{2 \zeta_{{\rm Th}}}  F_d\left(  \frac{|x-x'|}{\ell_\mu} \right)
\eea
where we have introduced the {\it thermal roughness exponent}
\bea \label{Eq:secI21:ThermalRugosity}
\zeta_{{\rm Th}} := \frac{\gamma-d}{2} \ ,
\eea
as well as the scaling function $F_d(y) = \int_q \frac{1 - \cos(q_1)}{ (q^2 + y^2)^{\frac{\gamma}{2}}} $ where $q_1$ is the first coordinate of the $d$-dimensional vector $q$ and we recall $\ell_{\mu}=1/\mu$. At large distance $F_d(y)$ decays algebraically as $\frac{1}{|y|^{2 \zeta_{{\rm Th}}}}$: the different points along the interface become elastically independent and $\langle (u_{x} - u_{x'})^2 \rangle$ tends to a constant. In the regime we are interested in $|x-x'| \ll \ell_{\mu}$, $ F_d$ in (\ref{Eq:secI21:ThermalRugosityb}) is constant and can be forgotten and the mean square displacement $\langle (u_{x} - u_{x'})^2 \rangle$ displays a power-law behavior determined by the roughness exponent $\zeta_{{\rm Th}}$. Hence for $d \leq  \gamma$, $\zeta_{{\rm Th}} \geq 0$ and the interface is rough (it displays a logarithmic scaling for $d = \gamma$), while it is flat for $d > \gamma$ ( $\zeta_{{\rm Th}} < 0$). Hence, as expected, the effect of the temperature on the large scale fluctuations of a system of fixed internal dimension $d$ gets stronger as the range of elasticity is decreased.

\medskip

{\it The thermal fixed point} \\
Let us now discuss the notion of Thermal fixed point (TFP), first at the level of the Hamiltonian (\ref{Eq:secI21:Hpure}), and we consider the limit $\mu \to 0$ (which is the regime we are interested in). In this limit, the following scale transformation, equivalently written in Fourier or real space,
\bea \label{Eq:secI21:RescalingThermal}
&& x = b \tilde x \quad , \quad \tilde u_{\tilde x} = b^{-\zeta_{{\rm Th}}} u_{x=b \tilde x} \nn \\
&& q = b^{-1} \tilde q \quad  , \quad  \tilde u_{\tilde q} = b^{-d  -\zeta_{{\rm Th}}} u_{q = b^{-1} \tilde q} 
\eea
leaves the Hamiltonian ${\cal H}_{{\rm pure}}[u]$ invariant: it is a fixed point of the above scaling transformation. Note that physical observables are sensitive to the combination $\frac{1}{T} {\cal H}_{{\rm pure}}[u] $ (and not ${\cal H}_{{\rm pure}}[u]$ alone). In this theory, conserving the thermal averages of observables under the thermal rescaling (\ref{Eq:secI21:RescalingThermal}) does not impose a rescaling of the temperature: this is the thermal fixed point of the theory. The question of its stability can be simply considered by adding other terms in the Hamiltonian (\ref{Eq:secI21:Hpure}) and checking whether or not they decay to $0$ under the rescaling (\ref{Eq:secI21:RescalingThermal}) as $b \to \infty$ increases. Indeed finite $\tilde x$ corresponds to large $x$ if $b \gg 1$ and the theory in the variables $(\tilde x,\tilde u_{\tilde x})$ can be thought of as a coarse-grained effective theory describing the physics of the system at a larger scale. As an example, consider adding higher order derivative terms of the form, schematically, $\int_{x} (\nabla_x)^n u^p$ with $n\geq p$ and $p$ even (so that the elastic Hamiltonian is still invariant under translation and parity in the external space: ${\cal H}^{{\rm el}}[u] = {\cal H}^{{\rm el}}[u+cst] = {\cal H}^{{\rm el}}[-u] $)) in the Hamiltonian (\ref{Eq:secI21:Hpure}) for the case of short-range elasticity. Applying the scaling transformation we obtain
\bea
\int_{x} (\nabla_x)^n u^p \longrightarrow b^{d - n + p \frac{2-d}{2} } \int_{\tilde x} (\nabla_{\tilde x})^n (\tilde u_{\tilde x})^p \ .
\eea \\
Such a term always decays to zero under the coarse-graining procedure: the large scale physics will be described by the simple thermal fixed point we have considered. A similar analysis will allow us to discuss the relevance of the disorder at the TFP below but let us warn the reader here that this `renormalization procedure' will {\it not } be the one adopted in Chapter~\ref{chapII} to truly discuss the renormalization of the theory with disorder. The point of view we will adopt there will be to study the {\it effective action} of the field theory (which will be a replicated field theory in the case of the static problem). We will keep $\ell_{\mu}$ finite as a convenient infra-red cutoff, and compute the effective action in the limit $\mu \to 0$ and show that it takes a universal scaling form. Let us now briefly translate here the properties of the thermal fixed point in this language. The effective action of the theory at finite $\mu$ is defined as
\bea
&& e^{W_{\mu,T}[J]} := \int {\cal D}[u] e^{-\frac{1}{T} {\cal H}{{\rm pure}}[u] + \int_x J_x \cdot u_x}  \\ 
&& \Gamma_{\mu,T}[u] := -W_{\mu,T}[J] + \int_x J_x \cdot u_x \quad , \text{where } J \text{ is such that }  u_x = \frac{\delta  W_{\mu,T}[J] }{ \delta J_x} \ , \nn
\eea
where we have emphasized the dependence on $\mu$ and on the temperature of the effective action $\Gamma_{\mu,T}[u]$ and on the {\it generating function for connected diagrams} $W_{\mu,T}[J]$. For a Gaussian theory described by the quadratic Hamiltonian (\ref{Eq:secI21:Hpure}), it is trivial to compute these functionals and one obtains, up to an unimportant constant term, $\Gamma_{\mu,T}[u] = \frac{1}{T} {\cal H}_{{\rm pure}}[u]$: $\Gamma_{\mu,T}[u]$ can be computed for arbitrary $\mu$. In this language the scale invariance can be written by introducing the rescaled effective action
\bea
\tilde{ \Gamma}_{\mu}[\{ \tilde{u}_{\tilde x} \} ] = \Gamma_{\mu} [ \{ u_x =  \mu^{-\zeta_{{\rm Th}}} \tilde{u}_{\tilde{x}= \mu x} \} ] \ssp .
\eea
As $\mu \to 0$, for $\tilde x \leq 1 $ fixed of order $O(1)$, the field $\tilde{u}_{\tilde{x}}$ describes the large scale fluctuations in the scaling regime of the field $u_x$ in the original theory. At fixed $T$, $\tilde{u}_{\tilde{x}}$, the scale invariance of the thermal fixed point reads in this language
\bea \label{Eq:secI21:ThermalFRG}
- \mu \partial_{\mu} \tilde{ \Gamma}_{\mu }[\{ \tilde{u}_{\tilde x} \} ]  = 0 .
\eea
And this holds $\forall \mu$. In the theory with disorder the analysis will be much harder but we will in the end obtain an equation similar to (\ref{Eq:secI21:ThermalFRG}). We will only be able to obtain information on the effective action $\Gamma_{\mu}[u]$ in the limit $\mu \to \infty$ and in the limit $\mu \to 0$. In the latter limit, that is the one we will truly be interested in, the key point will notably be to identify the rescalings of the field $u$ and of the temperature $T$ such that a rescaled effective action $\Gamma_{\mu}[u]$ converges to a fixed point of a non-trivial FRG equation.

\subsection{Relevance and irrelevance of short-range disorder at the thermal fixed point}
Let us now consider the effect of adding a small Gaussian short-range disorder to the pure Hamiltonian (\ref{Eq:secI21:Hpure}):
\bea
 && \cH^{{\rm dis}}_{V}[u] := \int_{x} V(x,u_x) \nn \\
 && \cH[u] = {\cal H}_{{\rm pure}}[u]  + \cH^{{\rm dis}}_{V}[u] \ .
\eea
And we suppose for simplicity that $V(x,u)$ is Gaussian, with mean $0$ and a two-point correlation function
\bea \label{Eq:secI22:GaussianCorr}
\overline{ V(x,u) V(x',u') } = g \delta^{(d)}(x-x')  \delta^{(N)}(u-u')  \  ,
\eea
with $g \geq 0$ a parameter (a RB type disorder with $0$ correlation length). Under a general rescaling $x = b \tilde x$ and $\tilde{u}_{\tilde{x}} = b^{-\zeta} u_{x= b \tilde x}$, the disorder energy is rescaled as
\bea \label{Eq:secI22:RBRescaling}
\cH^{{\rm dis}}_{V}[u] = \int_{x} V(x,u_x) \quad \longrightarrow \quad  b^d \int_{\tilde x}  V(b \tilde x,b^{\zeta} \tilde{u}_{\tilde x}) \sim  b^{\frac{d - N \zeta}{2}} \int_{\tilde x} \tilde{V}(\tilde x , \tilde u_{\tilde x})     .
\eea
Here and throughout the rest of the manuscript $\sim$ means `distributed in law as' and $\tilde V (\tilde x , \tilde u)$ is a (new) centered Gaussian disorder with correlations $\overline{ \tilde{V}(\tilde x, \tilde u) \tilde V(\tilde x',\tilde u') } = g \delta^{(d)}(\tilde x-\tilde x')  \delta^{(N)}(\tilde u-\tilde u')$. Note that the rescaling (\ref{Eq:secI22:RBRescaling}) effectively assumes that the configuration of the field $u_x$ on which the rescaling is performed is independent of the disorder in order to use $\overline{ V(x,u_x) V(x',u_{x'}) } = g \delta^{(d)}(x-x')  \delta^{(N)}(u_x- u_{x'})$. This will be true here in the sense of the leading approximation for an expansion in $V$ since at leading order fluctuations of $u_x$ are controlled by the thermal fixed point and here $u_x$ can be thought of as a typical configuration of the DES at the TFP. Using $\zeta = \zeta_{{\rm Th}}$, one obtains that the disorder energy scales as $b^{ \frac{d - N (\gamma-d)/2}{2}}$. Hence small disorder is perturbatively irrelevant at the thermal fixed point if
\bea \label{Eq:secI22:ResN1}
N > \frac{2 d}{\gamma - d}   \   . 
\eea
The result (\ref{Eq:secI22:ResN1}) however only holds for $d < \gamma$ as we now explain. Indeed in larger dimension $d >  \gamma$, the thermal roughness exponent $\zeta_{{\rm Th}}$ is smaller than $0$. The system at the thermal fixed point is not rough but flat. The exponent $\zeta_{{\rm Th}}$ describes the algebraic speed at which the fluctuations $\langle (u_{x} - u_{x'})^2 \rangle$ converge to their asymptotic value. Using the rescaling $\tilde{u}_{\tilde{x}} = b^{-\zeta} u_x$ with $b \gg 1$ and $\zeta \leq 0$ means that distances of order $1$ in the coordinates $\tilde u $ correspond to infinitesimal distances in the coordinates $u$: for a realistic model this is dangerous since at small distances one expects the continuum description to break down. One also necessarily starts at some point to see the effect of the non-zero correlation length of the disorder $V$ which was assumed to be zero for simplicity here. For these reasons, for a realistic model of a disordered elastic system in a flat phase, comparing the effect of the elastic energy and of the disorder energy at large scale should rather be made by rescaling the lengths as $x = b \tilde{x} $ and $ u = \tilde{u}$. In such a rescaling the elastic energy is rescaled as ${\cal H}^{{\rm el}}[u] \to b^{d-\gamma} {\cal H}^{{\rm el}}[\tilde u]$ while the disorder energy is rescaled as ${\cal H}^{{\rm el}}[u] \to b^{\frac{d}{2}} {\cal H}^{{\rm el}}[\tilde u]$. Hence starting from a flat interface, i.e. e.g. in $d \geq \gamma$ for a model at the thermal fixed point or in arbitrary $d$ at zero temperature, weak disorder is relevant at large scale if
\bea
d   \leq  d_{{\rm uc}}:= 2  \gamma . 
\eea
Where we have introduced the {\it upper-critical dimension of the problem}. Let us now summarize our findings.

\subsection{Static phase diagram} \label{subsec:StaticPhaseDiagram}

\begin{enumerate}
{\item At finite temperature for $\gamma \leq d \leq 2 \gamma$ and for $d \leq \gamma$ and $N > \frac{2d}{\gamma-d}$ and at zero temperature for $d \leq 2 \gamma$, for arbitrary weak disorder, the elastic system is always rough at large scale with a roughness exponent larger than the thermal roughness exponent (the system pays more elastic energy than in the thermal phase to be able to visit regions of space with low values of the disorder potential). From the renormalization point of view, we will see that the effective action of the theory flows to a new fixed point at which scaling holds and in the scaling regime $|x-x'| \ll \ell_{\mu}$,
\bea  \label{Eq:SPD:Rug}
\overline{\langle (u_{x} - u_{x'})^2 \rangle} \sim |x-x'|^{2 \zeta_s}
\eea
where we defined the statics roughness exponent $\zeta_s \geq 0$. At the upper-critical dimension of the problem $d_{{\rm uc}} = 2 \gamma$, since the disorder is only marginally relevant at large scale, the roughness exponent $\zeta_s$ is expected to be $0$ and (\ref{Eq:SPD:Rug}) to be replaced by a logarithmic scaling. Here the fact that the large scale cutoff scale $\ell_{\mu}$ is equal to the one of the pure theory $\ell_{\mu}=1/\mu$ is a consequence of the so-called {\it Statistical-Tilt-Symmetry} (STS) of the problem as will be discussed later. At this fixed point, taken as a parameter of the effective action, the temperature of the systems is irrelevant and flows to $0$ when $\mu$ goes to $0$ as $\mu^{\theta}$. Asking that the combination $\frac{1}{T} \int_q (q^2 + \mu^2)^{\frac{\gamma}{2}} u_{-q} \cdot u_q$ that will enter into the effective action (again as a consequence of STS) of the problem converges to a well defined limit imposes
\bea
\theta = d - \gamma + 2 \zeta_s = 2 (\zeta_s - \zeta_{{\rm Th}} ) \geq 0 .
\eea
This regime will be called the {\it strong disorder regime} in the remainder of the manuscript. In this regime thus the temperature at large scale is irrelevant and the system optimizes its energy by balancing elasticity and disorder. We will also say that in this phase the system is {\it pinned} by disorder.}
{\item On the other hand, for $d \leq \gamma$ and $N >   \frac{2 d}{\gamma - d} $ (i) there exists a strong disorder fixed point at $T=0$ (ii) the thermal fixed-point is stable to weak-disorder. There are thus at least two phases. Starting from the strong disorder, zero temperature fixed point, it is believed that at least for small $N$, the strong disorder fixed point is stable to a perturbation by a small temperature. The question of whether or not there exists a finite critical value $N_{\rm uc} < \infty$ such that for $N >   N_{\rm uc} $ an arbitrary small temperature makes the system depart from the strong disorder fixed point and converge to the thermal fixed point is a difficult question which remains unanswered. This is true even in what might be the simplest case of SR elasticity ($\gamma=2$) in $d=1$, the so-called directed polymer problem, see Sec.~\ref{SecI3}. In this case, the problem is equivalent to the KPZ equation in dimension $N$ (see Sec.~\ref{SecI3}), and in this language $N_{\rm uc}$ is the unknown upper-critical dimension of the KPZ equation, which could be equal to $+ \infty$. In this phase, for a sufficiently large temperature / weak disorder, the large scale physics is described by the thermal fixed point, the system is rough but it is not pinned: its fluctuations are thermal.}
{\item Finally, for $d > 2 \gamma$ the system is always flat and the elasticity wins at large scale: the system is in its {\it ordered phase}.}
\end{enumerate}

\begin{figure}
\centerline{\includegraphics[width=9.0cm]{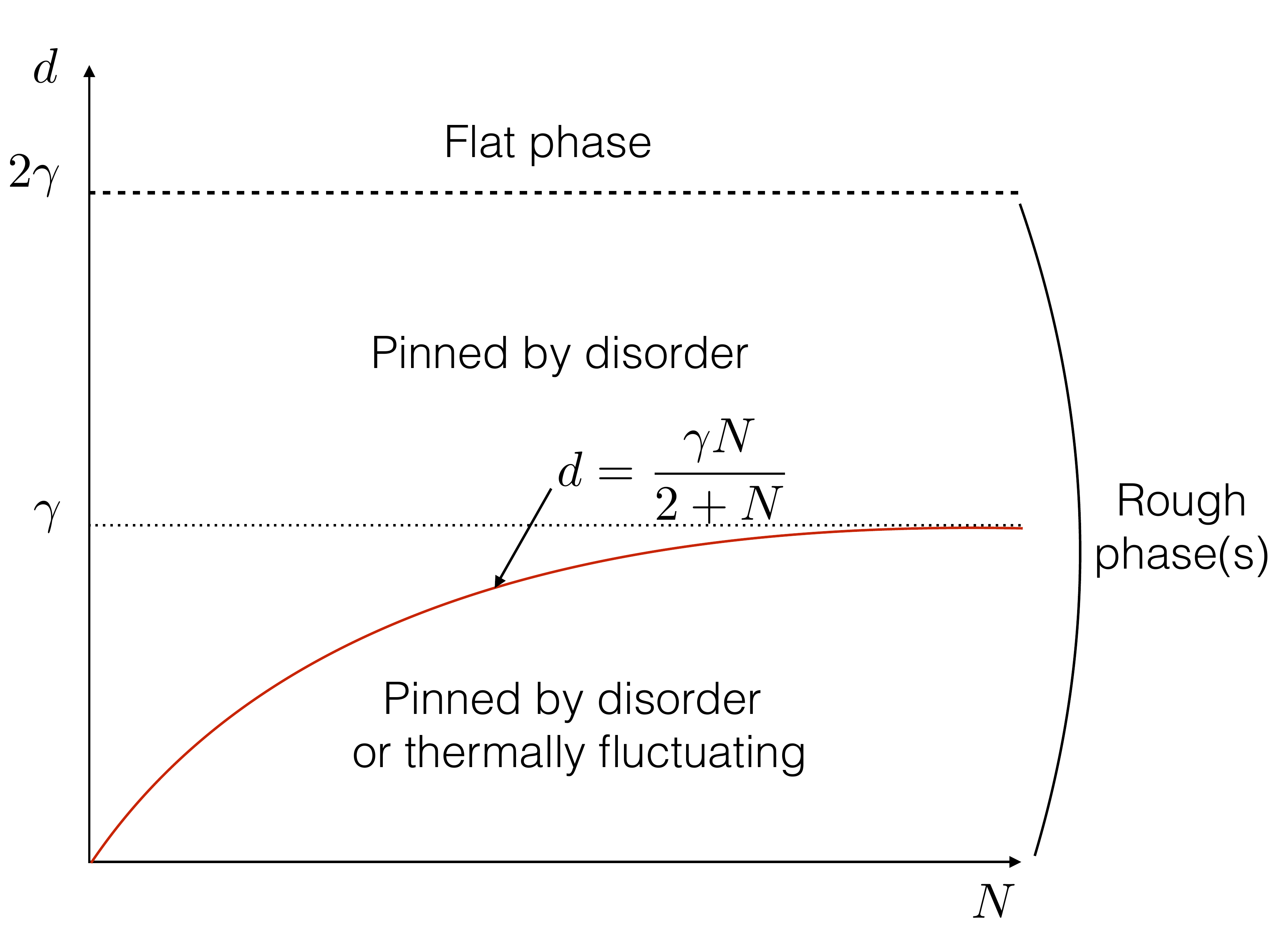}} 
\caption{Static phase diagram for a DES with elasticity kernel $g_q^{-1} = |q|^{\gamma}$ in a RB disorder potential at temperature $T$. For $d \geq 2 \gamma$ the system is always asymptotically flat. For $d \leq 2 \gamma$ and $ N < \frac{2d}{\gamma-d}$, i.e. $d > \frac{\gamma N}{2+N}$ the system is always pinned by disorder and is in a strong disorder / 0 temperature phase. For $d < \frac{\gamma N}{2+N}$ the system can a priori be either pinned by disorder (at the strong disorder FP), or thermally fluctuating (at the thermal FP), but is in any case always rough. The existence of a critical value $N_{{\rm uc}}$ above which the system is always in the thermal phase remains debated.}
\label{fig:StaticPhaseDiagram}
\end{figure}

From the above discussion we thus obtain the well-known phase diagram for the statics of disordered elastic systems as a function of the internal and external dimension $d$ and $N$ for RB disorder and elasticity of the type (\ref{Eq:secI1:HelGen}) with $g_q^{-1} = |q|^{\gamma}$ as presented in (\ref{fig:StaticPhaseDiagram}). This diagram can be more or less modified if one changes some assumptions that were made. For example\\
\\
(i) Disorder with long-range correlations: if the disorder has long-range correlations in either the external or internal space its influence is expected to increase and the large scale properties of the system can be different. Let us see qualitatively what changes for the case of random field disorder which will also be considered in this thesis (see Sec.~\ref{secI1}) with $0$ correlation length. In this case under rescaling the disorder energy behaves has, schematically,
\be \label{Eq:secI22:RFRescaling}
\int_{x} V(x,u_x) = \int_x \int_{u'} F(x,u'_x)   \longrightarrow   b^{\frac{d - N \zeta}{2} + \zeta} \int_{\tilde x} \int_{\tilde u '} \tilde{F}(\tilde x , \tilde u'_{\tilde x})   = b^{\frac{d - (N-2) \zeta}{2}} \int_{\tilde x} \tilde{V}(\tilde x , \tilde u_{\tilde x})   
\ee
Following the same path as before, one sees that such correlations do not modify the upper-critical dimension of the problem which is still $2 \gamma$, but increases the minimal value of $N$ above which the thermal phase is stable to weak disorder as $N_{\rm min} = 2  + \frac{2d}{\gamma- d} = \frac{2 \gamma}{\gamma-d}$. For a study of long-range correlations in internal space we refer to \cite{FedorenkoLeDoussalWiese2006b} and references therein. \\
\smallskip

(ii) Disorder with fat-tails: although it is not clearly visible in our derivation, disorder with fat-tails (i.e. for which there exists a finite $n_{{\rm max}}$ such that $\overline{V^{n}(x,u)} = + \infty$ for $n \geq n_{{\rm max}}$) can strongly modify the static phase diagram and corresponds to new universality classes. Naive intuition coming from the study of sums of independent random variables would suggest that $n_{{\rm max}} >  2$ would not change the phase diagram. In general not much is known about the behavior of disordered elastic systems in presence of fat-tail disorder. For the case of the directed polymer (DP) in a two-dimensional random media $N=d=1$ with SR elasticity $\gamma=2$, it is known from a Flory argument confirmed by numerical studies that $n_{\rm max} \leq 5$ actually suffices to change the behavior of the DP \cite{BiroliBouchaudPotters2007,GueudreLeDoussalBouchaudRosso2015}. The physical origin of this modification is that the strategy of optimization of energy of the DP changes in the presence of fat-tails disorder. While for a Gaussian disorder the DP optimizes its energy homogeneously along its internal direction, the optimization of energy becomes dominated by extreme value statistics in the presence of fat tails with $n_{\rm max} \leq 5$. The presence of a few sites in the energy landscape with very low values of the disorder potential then dominates the energy of the DP. We will not consider further in this thesis this complex question and restrict our analysis to $n_{{\rm max}} = + \infty$.

\medskip

The goal of this thesis is to understand properties of disordered elastic systems in the strong disorder regime of the phase diagram of Fig.~\ref{fig:StaticPhaseDiagram}. In this regime one expects (and would like to prove) that the system is rough with a non-zero roughness exponent $\zeta_s$ as in (\ref{Eq:SPD:Rug}). Furthermore one expects the roughness exponent $\zeta_s$ to be independent of the details of the distribution of $V(x,u)$: since the pinning of the DES is clearly a collective phenomenon (at least if the PDF of $V(x,u)$ has no fat tails) and $\zeta_s$ is a large scale property, one expects some universality to exist. As we will see in Chapter~\ref{chapII} using the Functional Renormalization Group, for disorder as discussed in Sec.~\ref{subsec:SecI1:DisHam} and in the case of interfaces $N=1$, there is indeed universality in the large scale properties of disordered elastic interfaces. The different universality classes are indexed by the choice of the range of elasticity (the experimentally most important being LR $\gamma=1$ and SR $\gamma=2$) and of the correlations of the disorder (RB, RF or random periodic).

\subsection{Early attempts at characterizing the strong disorder regime: Flory arguments and the Larkin model}

Let us now review two early theoretical approaches aiming at describing the strong disorder regime of Fig.~\ref{fig:StaticPhaseDiagram}. As usual we consider disordered elastic systems with an elastic kernel $g_{q}^{-1} \sim |q|^{\gamma}$.

\subsubsection{Flory scaling argument}

The Flory argument is a RG-type argument that consists in equating the scaling dimension of the elastic energy term and of the disorder term using simple rescaling of the different terms in the Hamiltonian as was performed in the last section to check the relevance of small disorder close to the thermal fixed point. In this senses it is similar to a dimensional analysis. We thus rescale $x = b \tilde x$ with $\tilde u_{\tilde x} = b^{-\zeta_s} u_{x = b \tilde x}$. The elastic Hamiltonian is rescaled as before as $\cH^{{\rm el}}[u] \to b^{d-\gamma + 2 \zeta_s} \cH^{{\rm el}}[\tilde{u}]$. For RB type disorder (see Sec.~\ref{subsec:SecI1:DisHam}) the rescaling of the disorder part of the Hamiltonian is as in (\ref{Eq:secI22:RBRescaling}) and we obtain
\bea \label{Eq:FloryRB}
d-\gamma + 2 \zeta_s = \frac{d - N \zeta_s}{2} \quad \Longrightarrow \quad \zeta_s = \frac{2 \gamma -d}{4 +N} \ssp.
\eea
In the RF type (see Sec.~\ref{subsec:SecI1:DisHam}) on the other hand we obtain, using (\ref{Eq:secI22:RFRescaling})
\bea \label{Eq:FloryRF}
d-\gamma + 2 \zeta_s = \frac{d - (N-2) \zeta_s}{2} \quad \Longrightarrow \quad \zeta_s = \frac{2 \gamma -d}{2+N} \ssp.
\eea
The flow in the Flory argument is the following. Here we are assuming that we are at a RG fixed point and that the effective disorder felt at large scale transforms as in (\ref{Eq:secI22:RBRescaling}) or (\ref{Eq:secI22:RFRescaling}). However, during such a coarse-graining procedure, there can be a non-trivial renormalization of the disorder coming from the optimization of the energy of the interface on small scales. This scaling argument should thus be taken with caution.

In particular we will see that (\ref{Eq:FloryRB}) disagrees with the {\it exact result} for directed polymers in $1+1d$ with short-range elasticity for which $\zeta_s=2/3$. Remarkably for the RF case at least for $N=1$, it is believed that (\ref{Eq:FloryRF}) is exact. This is due to the fact that, although the disorder is also corrected by the renormalization in this case and the hypothesis leading to (\ref{Eq:FloryRF}) are not correct, it is possible to show using the Functional Renormalization Group that the tail of $R_0(u) \sim - \sigma |u|$ is not corrected, and it is this uncorrected large distance behavior of $R_0(u)$ which dominates the optimization of the interface energy. For the RF case this remarkable result calls for a precise explanation showing that there are no corrections coming from the optimization on small scales, and for the important RB case the failure of the argument motivates the development of a true renormalization scheme.

\subsubsection{The Larkin model}

\ssp {\it Larkin model} \\
The Larkin model, introduced by Larkin in \cite{Larkin1970}, is a perturbative attempt at understanding the properties of the strong disorder fixed point by directly looking at the $T=0$ problem, i.e. by considering the ground state of the interface. It consists in linearizing the random potential $V(x,u)$ (taken e.g. of the RB type) around a given position and retaining only the first order term. If the system is confined around a position $w=0$ the potential is linearized as $V(x,u) = V(x,0) + \partial_u V(x,0) u  = V(x,0) - F(x) u$, where we introduced the force $F(x) :=  -\partial_u V(x,0)$. Note that by definition $F(x)$ does not depend on $u$. The latter is chosen centered, Gaussian with short-range correlations in the internal space and second moment $\overline{F(x) F(x')} = \delta^{(d)}(x-x') \Delta$ where $\Delta \geq 0$. Taking as usual an elastic Hamiltonian $\cH^{{\rm el}}[u] = \frac{1}{2} \int_q (q^2 + \mu^2)^{\frac{\gamma}{2}} u_{-q} u_q $, the ground state of the system satisfies
\bea \label{Eq:SecII2:Larkin1}
(q^2 + \mu^2)^{\frac{\gamma}{2}} u_q - F(q) =0 \ssp .
\eea
Hence, the position field is Gaussian, with correlations in Fourier space
\bea \label{Eq:SecII2:Larkin2}
\overline{u_{q'}  u_q} = \hat \delta^{(d)}(q+q') \frac{\Delta}{(q^2 + \mu^2)^{\gamma}} \ssp .
\eea
And in real space, for $d \leq d_{{\rm uc}}= 2 \gamma$ and for lengths $|x-x'| \ll 1/ \mu$, it is rough
\bea \label{Eq:SecII2:Larkin2b}
\overline{ (u_x - u_{x'})^2} \sim C_d \Delta |x-x'|^{2 \zeta_L}
\eea
 with $C_d= 2^{1-2 \gamma} \pi^{-d/2} \frac{|\Gamma(d/2-\gamma)|}{\Gamma(\gamma)}$ and the roughness exponent known as the Larkin exponent
\bea \label{Eq:SecII2:Larkin3}
\zeta_L := \frac{2 \gamma -d}{2} \ssp . 
\eea

{\it Dimensional reduction} \\
Again the Larkin exponent (\ref{Eq:SecII2:Larkin3}) is not correct. One example is that it does not reproduce the exact result already cited above $\zeta_s=2/3$ for the RB case with $N=d=1$ and $\gamma=2$. A natural question however is to understand whether or not one could extend and improve the previous calculation by taking into account higher order terms in the series expansion of the potential: $V(x,u) = \sum_{n =0}^{\infty} \frac{u^n}{n!} \partial_u^n V(x,u)|_{u=0}$. Solving the minimization problem (\ref{Eq:SecII2:Larkin1}) in an expansion in $u$ by adding higher order terms leads to a remarkable result: {\it to all orders in perturbation theory, it predicts (\ref{Eq:SecII2:Larkin2}) and the Larkin roughness exponent (\ref{Eq:SecII2:Larkin3}}). In the context of interfaces, this simplification of naive perturbation theory was first discussed in \cite{EfetovLarkin1977}. In the more general context of disordered systems it is known as the phenomenon of {\it dimensional reduction} which asserts that disorder averaged observables of a theory at $T=0$ are equivalent to thermal averages in the pure theory at finite temperature in dimension $d_{{\rm dr}} = d- \gamma$ (see \cite{ParisiSourlas1979} for a theoretical analysis of this property using supersymmetry and e.g. \cite{ChauveLeDoussal2001} for a diagrammatic approach). The thermal roughness exponent (\ref{Eq:secI21:ThermalRugosity}) $\zeta_{{\rm Th}} = \frac{\gamma-d}{2}$ is indeed equal to (\ref{Eq:SecII2:Larkin3}) using  $d \to \gamma-d$. Of course, if dimensional reduction was true everything would be rather simple. The problem here is that the Larkin analysis misses important non-perturbative effects and can only work at small scale as we now discuss. First, since the Larkin model is based on a perturbation theory with the disorder expanded around the flat interface configuration $u=0$, it is hard to believe that it is correct at large length scales for $d \leq 2 \gamma$ since it predicts a rough interface. More precisely, note that it effectively assumes that the force $F(x)$ does not depend on $u$ while a generic RB disorder has a finite correlation length $u_c< \infty$. For this reason, while on small scales $|x-x'| \leq L_c$ such that $ \Delta u$ is small $|\Delta u| \leq u_c$ the Larkin model can accurately describe the fluctuations of the system, it should certainly fail above. The length $L_c$ is known as the {\it Larkin length} and it can be estimated as, using (\ref{Eq:SecII2:Larkin2b}),
\bea \label{Eq:SecII2:Larkin4}
\sqrt{C_d \Delta} L_c^{\zeta_L} \sim  r \quad {\rm i.e.} \quad L_c \sim \left( \frac{u_c}{\sqrt{\Delta  C_d }} \right)^\frac{2}{2\gamma-d} \ssp .
\eea
In particular note that for $u_c=0$, $L_c=0$ and the Larkin model is nowhere consistent. What happens to the system above the Larkin length is that the elastic energy cost paid by the system to wander in the energy landscape on distances $\Delta u \gg u_c$ becomes manageable and the system starts to fully exploit the fact that there are a lot of minima. We will see in Chapter~\ref{chapII} that the Larkin length is linked to the notion of {\it shocks} and {\it avalanches}. Describing the optimization of energy on large scales is a complex problem that will be tackled using renormalization method in Chapter~\ref{chapII} and exact methods in Chapter~\ref{chapIII}.

\section{Various problems considered in this thesis} \label{SecI3}

In this section we consider a disordered elastic system described by the Hamiltonian (\ref{Eq:secI1:Hamiltonian}) and briefly introduce some questions that will be tackled during the thesis.

\subsection{ Shocks in the statics at zero temperature for elastic interfaces} \label{subsec:SecI3:Shocks}

In Chapter \ref{chapII} we will be interested in the statics at zero temperature for elastic interfaces, i.e. the $d \geq 1$ and $N=1$ problem. We will thus be interested in the ($V-$dependent) ground state $u_x^V(w)$ of the total Hamiltonian:
\bea
u_x^V(w) &&:= {\rm argmin}_{u_x : \JR^d \to \JR } \cH_{V,w}[u] \nn \\
&&= {\rm argmin}_{u_x : \JR^d \to \JR } \left( \cH^{{\rm el}}[u] + \cH^{{\rm dis}}_{V}[u]+ \cH^{{\rm conf}}_w[u]  \right) \ .
\eea
More explicitly we will be interested in the case of an elastic kernel $g_{q}^{-1}= \sqrt{q^2 + \mu^2}$ for an interface confined around a parabolic well at position $w$.
\bea
u_x^V(w) = {\rm argmin}_{u_x : \JR^d \to \JR } \left(  \frac{1}{2} \int_q g_q^{-1} (u_{-q} -w_{-q}) \cdot (u_{q} -w_{q} ) + \int_x V(x,u_x)\right) \ .
\eea
Following the previous section, the interesting case on which we will focus will be the low-dimension case $d \leq 2 \gamma$ (condition for the interface to be pinned by the disorder at large scale) and in the range of scales $a \ll |x-x'| \ll \ell_{\mu} \ll L$. In this range of scales one indeed expects scaling and universality to hold and we will be interested in understanding the process $u_x^V(w) $ as a function of $w$. Since for any $w$ the interface is pinned by disorder one expects that the evolution of $u_x^V(w)$ with $w$ contains jumps in between different metastable states of the disorder Hamiltonian. This will be made precise in Chapter \ref{chapII}, and we will see that these jumps, also called {\it shocks} will inherit the universality present in the physics of disordered elastic systems (yet to be precisely discussed).\\

{\it Irrelevance of the temperature?}
A natural question is to ask why we are only considering the zero temperature static problem in the questions outlined above. Indeed, we have `shown' in the previous section that the thermal fixed point of the interface is unstable and the large scale physics of the system is described by a strong disorder fixed point at zero temperature whenever $\frac{2 d}{\gamma - d} >1$, that is for $d \geq \gamma/3$. Since short-range elasticity is described by $\gamma =2$ and we will only be interested in the cases with a longer range of elasticity, especially $\gamma = 1$, for true interfaces of dimension $d \geq 1$, the thermal fixed point will always be unstable and the system is always expected to be in the pinned phase. Although it is true that for these problems the temperature does not play a role for large scale properties such as the roughness exponent, it does affect some small scales properties and in particular we will see that it smoothes the jump process described above. Therefore, though some of our results might also be relevant for the non-zero temperature case as we will discuss, we will focus on the zero-temperature problem.
\medskip

\subsection{ Avalanche dynamics at the depinning transition for elastic interfaces} \label{subsec:SecI3:Avalanches}

\subsubsection{Introduction to the depinning transition}

Another question we will be interested in is the dynamics of the interface at the depinning transition that we now introduce. The depinning transition is a dynamical phase transition that occurs in the over-damped dynamics (with viscosity coefficient $\eta$) of elastic interfaces with elastic kernel $g_{x,y}^{-1} = \int_q e^{iq \dot (x-y)} |q|^{\gamma}$, driven by a non-zero force $f$ in a random force field $F(x,u_x)$ with second moment as in (\ref{Eq:secI1:Correlations}), and we typically have in mind the case of a Gaussian force where $\Delta_0(u-u')$ is a short-range function with correlation length $u_c$. The equation of motion of the interface is 
\bea \label{Eq:SecII2:dep1bis}
\eta \partial_t u_{tx} = \int_{y} g_{x,y}^{-1} u_y + F(x,u_x) + f \ssp .
\eea
Note that this dynamics (which corresponds to type A in the classification of \cite{HohenbergHalperin1977}) is a somehow arbitrary choice on which we will focus. The presence of inertial or viscoelastic effects are not taken into account here and thus not all disordered elastic interfaces moving in nature can surely not be described by this dynamics. For some of them however, at least in some regime, this type of dynamics have been proposed as a relevant description (see Sec.~\ref{sec:Experiments}). For the case of SR elasticity $\gamma=2$ this equation is often referred to as the Quenched-Edwards-Wilkinson equation. Here the initial condition will be basically unimportant: we will be looking at the {\it out-of-equilibrium steady state} reached by the interface at $t\to \infty$. Indeed it can be proved that in our setting, starting from an initial condition such that all velocity along the interface are either positive or $0$ (i) they remain so for all time; (ii) up to a time translation the interface position field reaches a single well-defined steady state. The two last statements are often referred to as the Middleton theorem in the literature and were proved by Middleton in \cite{Middleton1992}. The first question in the depinning transition is to understand the {\it velocity-force characteristic} of the interface, that is
\bea  \label{Eq:SecII2:dep2}
v(f):= \lim_{t \to \infty} \frac{u_{tx}}{t} \ssp .
\eea
And here we are interested in the limit of an infinitely large interface $L \to \infty$. The basic physics of the depinning transition that we recall below is known since the work of Larkin in the framework of superconductors \cite{Larkin1970}, and was later developed in the interface context, see e.g. \cite{LeschhornTang1994} and references therein. The main observation is that for $f \leq f_c$, the interface does not move, $v(f)=0$ and for $f$ larger but close to $f_c$, the velocity force characteristic exhibits a power-law behavior with an exponent $\beta \geq 0$:
\bea \label{Eq:SecII2:dep3}
v(f) \sim (f-f_c)^{\beta} \quad {\rm for } \quad  f\geq f_c \ssp .
\eea
To estimate $f_c$, first note that when $f=0$, the interface is at rest. For $d \leq 2 \gamma$, the interface is rough while for $d \geq 2 \gamma$ is asymptotically flat. For a flat interface of internal length $L$, the typical disorder force acting on the interface scales as $F \sim \sqrt{\Delta(0)}L^{d/2}$, 
while the total driving force acting on the interface is $f L^d$. The latter always wins for $L \to \infty$
 and the interface starts to move: $f_c = 0$ for $d \geq 2 \gamma$, the disorder is irrelevant at large scale and $\beta =1$. For $d \leq 2\gamma$, the interface is rough when $f = 0$. On scales smaller than the Larkin length (\ref{Eq:SecII2:Larkin4}) $L \leq L_c$, the displacements of the interface are small, and one can estimate again the typical force acting on this portion of the interface as $F \sim \sqrt{\Delta(0)}L^{d/2}$. Such a small portion of the interface can stay pinned for sufficiently small $F$. Seeing the interface as a collection of $N= (L/L_c)^d$ domains of length $L_c$ pinned by fluctuations of the disorder, an estimate of the critical force (due to Larkin) is thus:
\bea \label{Eq:SecII2:dep4}
f_c \sim \sqrt{\Delta(0)}L_c^{-d/2} \ssp,
\eea
and an estimate of $L_c$ was given in $(\ref{Eq:SecII2:Larkin4})$. 

\subsubsection{The depinning transition as a continuous out of equilibrium phase transition}

\begin{figure}
\centerline{\includegraphics[width=9.0cm]{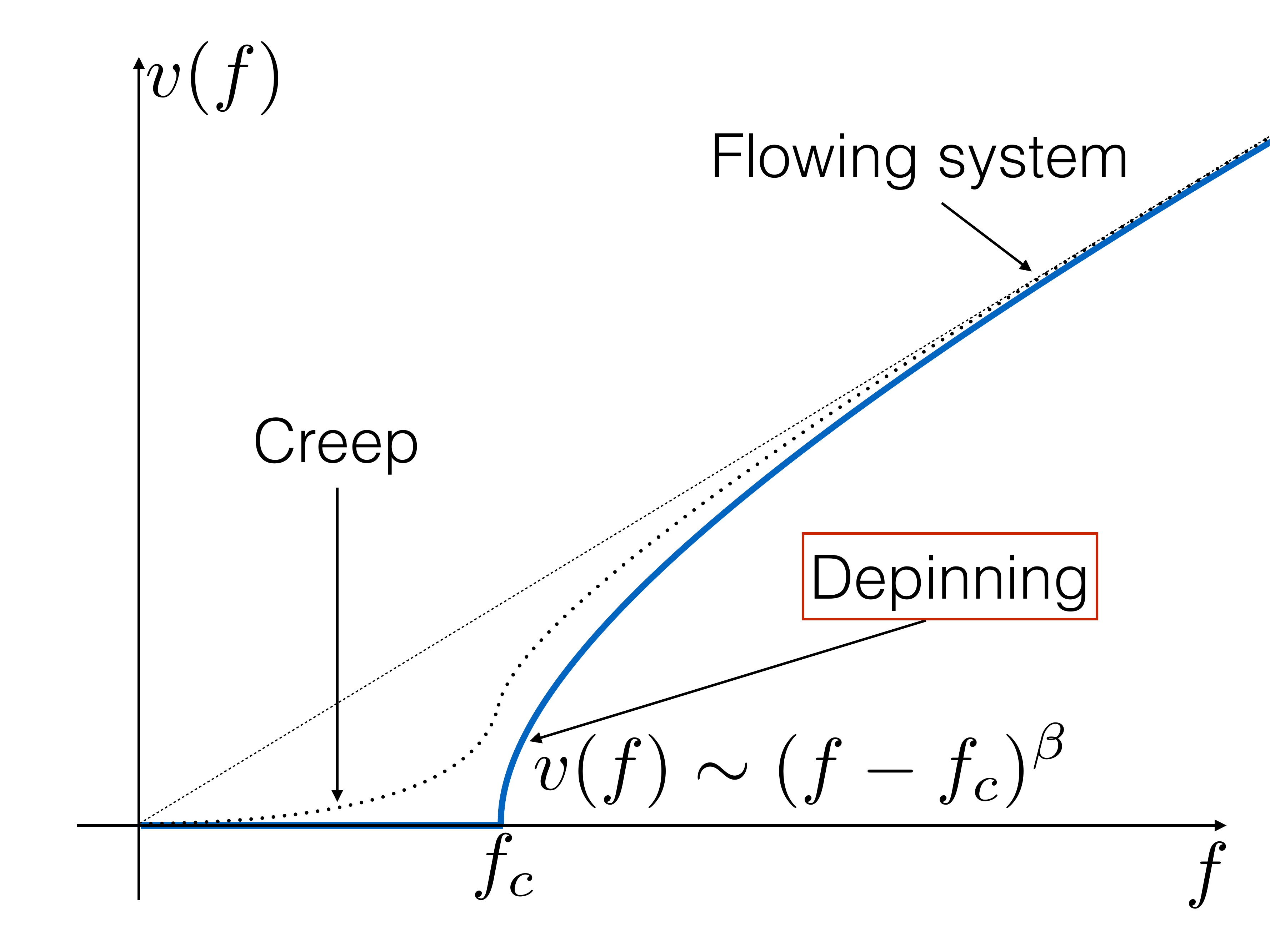}} 
\caption{Velocity-force characteristic of an infinite interface of dimension $d \leq 2 \gamma$ for the over-damped dynamics (\ref{Eq:SecII2:dep1bis}). Blue line: depinning velocity-force characteristic for the interface at $T=0$. Dotted black line: creep velocity-force characteristic for the interface at low temperature.}
\label{fig:Depinning}
\end{figure}

Following this simple analysis, the zero temperature dynamics of disordered elastic interfaces of dimension $d \leq 2 \gamma$ as described by (\ref{Eq:SecII2:dep1bis}) appears to exhibit a {\it dynamical phase transition} where the order parameter is the velocity of the interface and $f$ is the control parameter (see Fig.~\ref{fig:Depinning}). The description of this phase transition can be made in analogy with ordinary continuous phase transitions in equilibrium statistical mechanics, with the additional `complication' that there is also a time direction in the problem (see \cite{DSFisher1985} for the discussion of this point of view in the case of sliding charge density waves). The usual space scale invariance at the point of a continuous equilibrium phase transition becomes a space-time scale invariance in the steady state at the point of the dynamical phase transition which is basically summarized by saying that time scale as $t \sim x^z$ where $z$ is the {\it dynamics exponent} of the transition. In the steady-state, approaching the transition from above $f \to f_c^+$, there exists a growing correlation length $\xi \sim (f-f_c)^{-\nu}$ ($\nu \geq 0$ is another exponent) such that, for $|x-x'| \leq \xi$ and $|t-t'| \leq \xi^z$, the fluctuations of the position field satisfy the scaling form
\bea \label{Eq:SecII2:dep5}
\overline{ (u_{tx} - u_{t'x'})^2} \sim |x-x'|^{2 \zeta_d} {\cal G}\left( \frac{|t-t'|}{|x-x'|^z} \right)
\eea
where $\zeta_d$ is the {\it roughness exponent} and $ {\cal G}(y)$ is a scaling function that satisfies
\bea \label{Eq:SecII2:dep6}
{\cal G}(y) \to_{y \to 0} cst \quad, \quad {\cal G}(y) \sim_{y \to \infty} (y^{2 \zeta_d/z}) \ssp .  
\eea
The different critical exponents $\beta$, $\nu$, $\zeta_d$ and $z$ introduced above are, as will be shown later, not independent. More precisely the depinning transition can be described by only two independent critical exponents, which will be taken as $\zeta_d$ and $z$ in the following. $\zeta_d$ is analogous to the roughness exponent in the static problem $\zeta_s$ but is a priori different. As the depinning transition appears as a collective phenomenon, it is expected that these exponents have some universality. As we will recall in Chapter~\ref{chapII}, there are actually fewer universality classes at the depinning transition of the interface than in the corresponding static problem since it is now known that there is {\it a single universality class for short-range random forces}, corresponding to a random field universality class (although different from the corresponding universality class in the statics). In particular at depinning, the large scale properties in a random potential of the RF and RB type are similar. As the dimension becomes close to the upper-critical dimension $d_{{\rm uc}}=2 \gamma$, the disorder becomes irrelevant at large scale and the roughness exponent $\zeta_d$ must converge to $0$, while the value of the dynamic exponent converges to, as can simply be read off from (\ref{Eq:SecII2:dep1bis}), $z = \gamma$. As we know from the study of the static problem, for $d\leq 2 \gamma$ and without driving force, the system is pinned. As we will see, the non-trivial dynamics that occurs at the depinning transition is due to the fact that the interface will be most of the time pinned by disorder in a metastable state (as we will argue these are different from the static ground states). From time to time the interface will manage to cross the energy barrier and then moves with a velocity of order $1$ until it is pinned again by a new metastable state. Thus the interface dynamics at the depinning transition appears as an avalanche process. At large force the system is never pinned and flows with $v(f) \sim f/\eta$: disorder is washed out and only leads to small fluctuations around the deterministic behavior. The interesting regime to understand is thus clearly the avalanche process close to the depinning transition. The universal properties of this avalanche process will be, together with shocks between ground states presented before, at the core of Chapter \ref{chapII}.

\subsubsection{Creep and the temperature} 

The influence of the temperature on the depinning transition is much more subtle than on the large scale properties of the static ground state (although it can also be quite subtle). In the static problem, as we will see in Chapter~\ref{chapII}, non-zero temperature smoothes the shocks at small scales when the energy differences between successive minima become of the same order as the thermal energy. In particular in the static problem, the role of {\it energy barriers} between successive minima will be inexistent. On the contrary, the slow, non-trivial dynamics that is observed at the depinning transition of the interface is all about the interface being able to cross energy barriers (the fast motion observed after such a barrier has been crossed being an avalanche). Since a non-zero temperature allows the interface to cross an arbitrary large energy barrier, it has important effects on the dynamics, and the temperature is not irrelevant at large scale. In particular, {\it at non-zero $T$, for arbitrary force $f\geq 0$, the interface moves with a non-zero velocity} $v(f)$. This phenomenon is known as {\it creep}.  It was first described theoretically \cite{Nattermann1987,IoffeVinokur1987,NattermannShapirVilfan1990,FeigelmandGeshkenbeinLarkinVinokur1989}. Rather non-trivial assumptions and scaling arguments led to the creep law, valid for $ f \ll f_c$,
\bea \label{Eq:SecII2:creep1}
v(f) \sim e^{-  \frac{U_c}{T}  \left(\frac{f_c}{f} \right)^{\mu}} \quad , \quad \mu = \frac{D-2 + 2 \zeta_s}{2 - \zeta_s} \ssp ,
\eea
where $U_c$ is a system-dependent energy scale. Note in particular that (\ref{Eq:SecII2:creep1}) involves the static roughness exponent, while the creep is a non-equilibrium phenomenon. On the theoretical side the relation (\ref{Eq:SecII2:creep1}) was confirmed up to one loop accuracy using FRG \cite{ChauveGiamarchiLeDoussal1998,ChauveGiamarchiLeDoussal2000}. For the case of SR elasticity in $d=1$ in a random bond potential (for which the static exponent $\zeta_s=2/3$ is exactly known and thus $\mu=1/4$), it was also confirmed numerically in \cite{KoltonRossoGiamarchi2005}, and even experimentally using measurements on the dynamics of magnetic domain walls \cite{LemerleFerreChappertMathetGiamarchiLeDoussal1998}. Understanding more thoroughly the creep regime of an elastic interface is still a very active area of research \cite{AgoristasLecomte2016,Jagla2016} that we will not discuss in this thesis. Let us only note the recent numerical study \cite{FerreroFoiniGiamarchiKoltonRosso2016} that suggests that avalanches of the interface during the slow creep motion of the interface exhibit in different regimes scalings corresponding with either the scaling of static shocks, or the one of dynamic avalanches at the depinning transition studied in this thesis. From now on we will always restrict ourselves, for the dynamics, to the zero temperature case.

\subsection{ Static problem at finite temperature for directed polymers with SR elasticity and the KPZ universality class} \label{subsec:SecI3:KPZ}

\subsubsection{From the DP with SR elasticity to the KPZ equation}

Another focus of this thesis is the Kardar-Parisi-Zhang (KPZ) equation which is associated with the statics of a short-range elastic directed polymer ($d=1$ case) in dimension $N$ at finite temperature $T$ as we now recall. The partition sum for directed polymer (DP) with starting point $u_{0}=0$ and endpoint $u_{L}=u \in \JR^N$ can be written as a path integral
\bea \label{Eq:secIntroKPZ-1}
Z_L(u) = \int_{u(0)=0}^{u(L)=u} {\cal D}[u] e^{ - \frac{1}{2T} \int_{0}^{L} dx (\nabla_x u_x)^2  - \frac{1}{T} \int_0^L dx V(x,u_{x}) }  \ ,
\eea
and for now we suppose that the random potential is Gaussian with correlations
\bea  \label{Eq:secIntroKPZ-1ter}
\overline{V(x,u) V(x',u')}^c = \delta(x-x') R_0(|u-u'|)
\eea
with $R_0(u)$ a SR function (RB disorder). We will see in Chapter~\ref{chapIII} that (up to a subtlety on which we will comment later) that $Z_L(u)$ satisfies a stochastic partial differential equation (SPDE) known as the multiplicative stochastic heat equation (MSHE). In this equation $L$ plays the role of the time in the heat equation and we will thus make the change of variables
\bea
L \to t \quad , \quad u \to x \ssp .
\eea
In these variable $Z_L(u) \to Z_t(x)$ satisfies 
\bea \label{Eq:secI3:SHE}
\frac{\partial}{\partial t} Z_t(x)  = \left( \frac{T}{2} ( \nabla_x)^2 - \frac{1}{T} V(t,x) \right) Z_t(x)) \ .
\eea
And the initial condition is $Z_{t=0}(x) = \delta^{(N)}(x)$. Introducing the free-energy of the directed polymer through the change of variables $F_t(x) = - T \log Z_t(x) $, we obtain, for $V(t,x)$ a smooth disorder,
\bea
\partial_t F_t(x) = -\frac{1}{2} (\nabla_x F)^2 + \frac{T}{2} (\nabla_x)^2 F_t(x)  + V(t,x) .
\eea
Finally, making the change of variables $h(t,x) = -F_t(x)$, we obtain
\bea \label{Eq:secI3:KPZ}
\partial_t h(t,x) = \frac{1}{2} (\nabla_x h (t,x))^2 + \frac{T}{2} (\nabla_x)^2 h(t,x)  - V(t,x) \ssp .
\eea
For the case of $V(t,x)$ taken as a Gaussian white noise, this SPDE is known in the literature as the $N$-dimensional Kardar-Parisi-Zhang (KPZ) equation. Beware that $x$ now corresponds, in the elastic system language, to the $N$ dimensional coordinates in the external space, while $t$ corresponds to the one-dimensional coordinate that spans the internal space. The change of variables $Z_t(x) = e^{ T h(t,x)}$ that maps the MSHE to the KPZ equation is known in the literature as the Cole-Hopf transform.

\subsubsection{The KPZ equation as a model of out-of-equilibrium growth of interfaces}

Note that the KPZ equation (\ref{Eq:secI3:KPZ}) appears rather similar to the quenched Edwards-Wilkinson equation (i.e. (\ref{Eq:SecII2:dep1bis}) with SR elasticity $g_{x,y}^{-1} = - \delta^{(d)}(x-y) \nabla^2_y$) for the over-damped dynamics of a $d$ dimensional interface at zero temperature except for the few {\it important} differences that (i) it contains a non linear term $(\nabla_x h_{tx})^2$ (ii) the disorder is not quenched but rather depends on time and can be interpreted as a thermal disorder. Let us now interpret $h_{xt}$ as a $N$-dimensional interface and (\ref{Eq:secI3:KPZ}) has a SPDE for the dynamics of $h_{xt}$. Note that the non-linear term breaks the symmetry $h \to -h$ and makes the interface grow in the upward direction. In particular, even without driving term and at zero disorder $V= 0$, an interface described by the dynamics (\ref{Eq:secI3:KPZ}) starting from a non-flat initial condition would grow indefinitely: $\lim_{t \to \infty} h_{xt} = + \infty$. This should be compared with the Edwards-Wilkinson equation (\ref{Eq:SecII2:dep1bis}) with neither disorder nor driving. In this case the pure dynamics is rather simple: starting from an initial configuration $u_{x,t=0}$ the interface flattens through the effect of the elastic force. The interface dynamics described by (\ref{Eq:SecII2:dep1bis}) and (\ref{Eq:secI3:KPZ}) are thus radically different. While (\ref{Eq:SecII2:dep1bis}) can be thought of as the dynamics of an interface that separates two equivalent phases, (\ref{Eq:secI3:KPZ}) describes an {\it out-of-equilibrium} situation where one phase (the one below the interface) is favored compared to the other one. This is precisely for the purpose of describing such physical situations that (\ref{Eq:secI3:KPZ}) was first introduced in the seminal paper \cite{KPZ}. While in (\ref{Eq:SecII2:dep1bis}) all the complexity comes from the fact that the disorder is static, in (\ref{Eq:secI3:KPZ}) the complexity comes from the presence of the non-linear term that makes the problem an out-of-equilibrium problem. Indeed, without the non-linear term, solving (\ref{Eq:secI3:KPZ}) is trivial since the equation is linear in $h_{tx}$. 

\subsubsection{Introduction to KPZ universality in $1+1$d}

At least in $1+1$d (i.e. $N=1$) the equation (\ref{Eq:secI3:KPZ}) is believed to represent an important universality class of out-of-equilibrium {\it local} growth processes sharing the following properties \cite{Corwin2011Review}
\begin{enumerate}
\item{The interface is elastic, of the short-range type such as $(\nabla_x)^2 h$ in (\ref{Eq:secI3:KPZ}).}
\item{The growth rate at $x$ is non-linear in the local slope $(\nabla_x h)$ and thus favors one phase, as $(\nabla_x h)^2$ in (\ref{Eq:secI3:KPZ}).}
\item{The interface is subjected to thermal fluctuations, i.e. $V(t,x)$ in (\ref{Eq:secI3:KPZ}) has short-range correlations.}
\item{There is no quenched disorder in the system.}
\end{enumerate}
The importance of the KPZ universality class goes however well beyond growth processes (for recent review see \cite{Corwin2011Review,HalpinTakeuchi2015,QuastelSpohn2015}). One example is also that as we showed, it is equivalent to the static problem of a DP at a finite temperature $T$ (hence although the KPZ equation does describe an out-of-equilibrium situation, it is fair to say that it is a rather peculiar one). In the special case $N=1$ which will be the focus of Chapter~\ref{chapIII}, much is known about the problem and the associated universality. Let us recall here a few important features of this universality class (that will be proved in specific models in Chapter~\ref{chapIII}). Noting $v_{\infty}(\varphi) = \lim_{t \to \infty} \frac{1}{t} h(t,x = \varphi t)$\footnote{The ballistic scaling in this definition can be expected from the non-disordered case, e.g. starting from an initial condition $Z_{t=0}(x) = \delta(x)$ we obtain $Z_t(x) \sim  \frac{1}{\sqrt{t}} e^{- \frac{x^2}{2 T t}}$ and taking the log one obtains $h(t,x) \sim t \frac{x^2}{2 T^2 t^2} $.} (the non universal, deterministic asymptotic growth speed of the interface) for $h(t,x)$ an interface growing from an initial profile $h(t=0,x)=h_0(x)$ with a growth process in the $1+1$d KPZ universality class, we have 

\smallskip 

{\it Scale invariance and universality of critical exponents}\\
The centered profile
\bea \label{Eq:secI3:KPZU1}
\underline h(t,x) := h(t,x)-t v_{\infty}(x/t) \ssp,
\eea
has large time fluctuations such that for $t \gg 1$ and $\forall b,t,x$, we have the equality in law
\bea \label{Eq:secI3:KPZU2}
\underline h(t,x) \sim  b^{-\alpha} \underline h( b^z t,b x ) \ssp ,
\eea
with the {\it universal exponents}
\bea \label{Eq:secI3:KPZU3}
\alpha = 1/2 \quad , \quad z = 3/2 \ssp .
\eea
The critical exponent $\alpha$ is the roughness exponent of the interface. Note that it is equal to the roughness exponent of a Brownian motion. The critical exponent $z$ is the dynamic exponent. Note that going back to the DP language amounts to taking $x\to u$ and $t \to x$: in this language the roughness exponent of the DP $u \sim x^{\zeta_s}$ is thus $\zeta_s =1/z=2/3$ as already announced. Using (\ref{Eq:secI3:KPZU2}), one obtains, for $t \gg 1$ and $\forall x$
\bea \label{Eq:secI3:KPZU4}
\overline{(\underline h(t,x )-\underline h(0,0) )^2}  = |x|^{2 \alpha} {\cal G}_1 ( \frac{t}{|x|^z}) =  t^{2 \beta} {\cal G}_2 ( \frac{t}{|x|^z}) \ssp.
\eea
Where the choice of the writing is a matter of taste and ${\cal G}_1(y)$ are two scaling functions related by ${\cal G}_1(y)=y^{2\beta} {\cal G}_2(y)$, 
\bea \label{Eq:secI3:KPZU5}
\beta = \alpha/z = 1/3
\eea
is the {\it growth exponent}, often measured in numerics, ${\cal G}_1(y) \sim_{y \to 0^+} cst$ and ${\cal G}_1(y) \sim_{y \to \infty} y^{2\alpha/z}$. Note finally that, even if one takes $V(t,x)$ in (\ref{Eq:secI3:KPZ}) as a GWN, there is no simple way to see that the exponents (\ref{Eq:secI3:KPZU3}) are the true critical exponents of the KPZ universality class and that the equality in law (\ref{Eq:secI3:KPZU2}) holds: the KPZ equation is not invariant by rescaling (\ref{Eq:secI3:KPZU3}) and is certainly {\it not} the FP of its own universality class. This will be further discussed in Chapter~\ref{chapIII}.

\smallskip 

{\it Universality beyond critical exponents: universality of fluctuations}\\
The following convergence in law holds

\bea \label{Eq:secI3:KPZU6}
\lim_{t \to \infty} \frac{\underline{h}(t,0)}{t^{\frac{1}{3}}} =\lambda X,
\eea

where $\lambda$ is a non-universal constant and $X$ is a RV whose distribution is {\it universal and depends only on some global properties of the boundary conditions.} The classification of `sub-universality' classes corresponding to boundary conditions is probably still not complete (but almost, see \cite{Corwin2011Review}) but a few robust examples are: (i) starting from `droplet' initial condition, i.e. $h_0(x) = - w |x|$ with $w \to \infty$, leads to $X$ distributed with the Tracy-Widom GUE distribution, corresponding to the (rescaled) probability distribution function of the largest eigenvalue of a random matrix in the GUE ensemble \cite{TracyWidom1993}; (ii) starting from flat initial condition, i.e. $h_0(x) = 0$, leads to $X$ distributed with the Tracy-Widom GOE distribution, the distribution of the largest eigenvalue of a random matrix in the GOE ensemble \cite{TracyWidom1996}; (iii) starting from a stationary initial condition (see Chapter \ref{chapIII}) leads to $X$ distributed with the Baik-Rains distribution \cite{BaikRains2000}. Note that in the DP language (\ref{Eq:secI3:KPZU6}) means that the fluctuations of the DP free-energy scale with the length as $L^{1/3}$ and are distributed according to the same distributions. \\

Other remarkable known universal properties of the KPZ universality class in $1+1$d will be reviewed in Chapter \ref{chapIII}. Let us close this section by mentioning that the theoretical knowledge of this remarkable universality is due to the existence of various models in the $1+1$d KPZ universality class that possess {\it exact solvability properties.} This notably includes the usual continuum KPZ equation itself as we will recall but our focus in this thesis will be on {\it discrete exactly solvable models of directed polymers on the square lattice}.

\section{Experimental realizations} \label{sec:Experiments}

In this chapter we introduce a few of the physical systems for which a description by a disordered elastic interface has been proposed. They all have in common the fact that, at a mesoscopic scale in some regimes, they can be described by an interface with different elastic behaviors which look rough and exhibit complex fluctuations. They can however be regrouped in two different classes. In Sec.~\ref{subsec:ExpQuenched} we will give examples of interfaces pinned by quenched disorder, at or close to equilibrium (more precisely such that if no force acts on the system the interface is at rest). They will be described by various types of elasticities and disorder. In Sec.~\ref{subsec:ExpKPZ} on the other hand we will give examples of growing interfaces that are fundamentally out of equilibrium: they grow indefinitely. We will present interfaces whose large scale dynamics is believed to be captured by the standard KPZ universality class in $d=1+1$.

\subsection{Disordered elastic systems pinned in a quenched random environment} \label{subsec:ExpQuenched}

\subsubsection{Domain walls in magnetic systems and Barkhausen noise}

Considering a piece of ferromagnetic material of dimension $D \geq 2$ below its Curie temperature, it is known that a variety of static and dynamic properties of the material can be understood at a coarse grained level by describing only the {\it domain walls} (DW) between several domains of constant magnetization. If impurities are present in the material (without destroying the ferromagnetic order) and that the deformations of the DW are small (e.g. at sufficiently low temperature) it is possible to describe the domain wall by a simple elastic interface in a disordered medium without overhangs\cite{HuseHenley1985,Nattermann1987,JiRobbins1991,LemerleFerreChappertMathetGiamarchiLeDoussal1998} as in (\ref{Eq:SecI1:1}) with $d= D-1$ and $N=1$. Preparing the sample at low-temperature such that there are a few domain walls, the pinning of the domain walls by the disorder will make inhomogeneities in the magnetization persist. If the disorder is made of random magnetic impurities it will naturally be of the random bond type, but random field type disorder can also be studied in antiferromagnetic systems with random impurities under a constant field (see \cite{ChauveThesis} and references therein). The elasticity of the domain wall is naturally short-range ($\gamma=2$) as the energy cost of creating a domain wall is local and proportional to the area of the domain, but under certain conditions it is known that long-range elastic interactions can be relevant (see below). From this point a variety of situations can be investigated, in particular the static properties of the domain wall and its dynamics under an external magnetic field.
\medskip

In some experimental situations it is possible to directly visualize the domain walls and to investigate properties such as the roughness of the interface or its response to an external force. In particular in \cite{LemerleFerreChappertMathetGiamarchiLeDoussal1998} the authors investigated the so-called creep-regime of a domain wall in an effectively two-dimensional ferromagnetic material $(D=2)$ and investigated the so-called creep regime, that is the velocity-force characteristics of the domain-wall at very small applied force (=magnetic field). The authors obtained a remarkable confirmation of the so-called creep law discussed above (\ref{Eq:SecII2:creep1}), and also measured the roughness exponent of the domain-wall as $\zeta\simeq 0.69 \pm 0.07$, corresponding well to the theoretical exact value of the static roughness exponent of a directed polymer ($d=N=1$) in a RB disorder with SR elasticity $\zeta_s = 2/3$.

\medskip

The influence of the physics of domain walls has however also direct consequences on macroscopic properties of the sample. An important example is linked to the notion of {\it Barkhausen noise}. Applying a slowly increasing magnetic field to a magnetic sample, the magnetization increases following the hysteresis curve. The increase in magnetization is however non-smooth and proceeds by jumps. These can directly be measured (see \cite{Colaiori2008,DobrinevskiPhD}), and the first experimental report of the existence of this noisy signal is due to H. Barkhausen in \cite{Barkhausen1919}. Research on this process has led to distinguish two classes of magnets: (i) hard magnets, characterized by a `wide' hysteresis curve; (ii) soft magnets, characterized by `small' hysteresis curve. In the first class, the microscopic origin of the Barkhausen noise is attributed to the coherent reversal of domains of magnetizations. In the second class, the Barkhausen noise is attributed to the motion of domain walls which alternate periods where they are pinned for a long time by impurities, and period of fast motion where they jump from pinning configurations to pinning configurations. Plotting the magnetization as a function of time $M(t)$, the latter exhibits a so-called {\it avalanche dynamics} characterized by jumps $M(t + \Delta t) - M(t) = S$ interrupted by `long' ($\gg \Delta t$) periods of quiescence. Following the domain wall interpretation for the origin of these jumps, the size $S$ of the jumps of the magnetization are directly proportional to the volume (for samples in $D=3$) swept by the DW during its motion. It was found (see \cite{DurinZapperi2000,DurinZapperi2006b} and references therein) that the distribution of jumps $S$ and time $T$ are power law distributed in between two (widely separated) cutoffs:
\bea
P(S) \sim S^{-\tau_S} \quad,  \quad P(T) \sim T^{-\alpha} \ssp .
\eea 
As we will see in Sec.~\ref{subsec:avalanches}, these power-laws were argued to be related to the critical exponents of the interface at the depinning transition, and thus such measurements give access to properties of the interface (and vice versa). In Barkhausen experiments two universality classes for soft magnets in $D=3$ were found \cite{DurinZapperi2000,DurinZapperi2006b}: (i) polycrystalline materials for which the exponents depend on the driving rate and for which the exponent at slow driving are $\tau_S \simeq 3/2$. In this class it was argued that due to the presence of dipolar interactions, the elasticity of the domain-walls are effectively long-range with $\gamma=1$ \cite{CizeauZapperiDurinStanley1997}. Taking a look at the phase diagram in Sec.~\ref{subsec:StaticPhaseDiagram}, although here we are not in a static situation, we see that these systems have $d = 2 = 2 \gamma = d_{{\rm uc}}$: they sit right at the upper-critical-dimension of the problem. We will see in the following that the above exponents are indeed the mean-field exponents of avalanche motion. (ii) amorphous materials, for which the empiric exponents are $\tau_S \simeq 1.27$ and $\alpha \simeq 1.25$, independently of the driving rate, and for which the elasticity of the domain wall is short-range.

\subsubsection{Fractures fronts in brittle materials}

Another physical process for which the model of an elastic interface has been used is for the fracture of brittle materials (see \cite{Bonamy2009} for a review). Indeed it has been argued that for these systems the propagation of the crack front can be understood as the zero temperature over-damped dynamics of a line with long-range elasticity ($d= \gamma =1$) \cite{SchmittbuhlRouxVilotteMaaloy1995,RamanathanErtasFisher1997,BonamySantucciPonson2008,Ponson2008} in a disordered medium. The fracture proceeds again by avalanches, whose statistics can be experimentally measured by acoustic techniques, or in some experimental setup by direct visualization of the crack front \cite{SchmittbuhlMaloy1997,DeleplaceSchmittbuhlMaaloy1999,LaursonSantucciZapperi2010,SantucciGrobToussaint2010}. The experimentally obtained value of the roughness exponent was there reported as $\zeta \sim 0.35$ on large scales, while at small scales a value of $\zeta \sim 0.63$ was reported. In these systems, due to the long-range nature of the elasticity, an avalanche at one point of the interface generally triggers several avalanches at different points and when speaking about the distribution of the size of avalanches, one has to distinguish whether the size of single avalanches or of the cluster of avalanches is measured. In \cite{TallakstadToussaintSantucciSchmittbuhlMaaloy2011} the distribution of the size of single avalanches was reported to have a power-law exponent of $\tau_S^{{\rm ind}} \sim 1.56 \pm 0.04$.

\subsubsection{Some other related situations}
\stab {\it Contact lines of fluids on rough substrates}\\
It has been argued that the slow motion of the contact line of a fluid on a rough substrate could be well approximated by the motion of an elastic line with long-range elasticity at the depinning transition \cite{SchafferWongZen2000,MoulinetGuthmannRolley2002,MoulinetGuthmannRolley2004}. While some aspect of this dynamics agree (e.g. avalanches) well with the elastic interface theory, \cite{LeDoussalWieseMoulinetRolley2009}, the value of the experimentally measured roughness exponent $\zeta\sim0.5$ is still not understood, although it has been argued that it could be the sign of non-linear elastic terms \cite{LeDoussalWieseRaphaelGolestanian2004}.
\smallskip

{\it Earthquakes}\\
It has been argued that some features of earthquakes and geological faults could be captured by the model of an elastic interface in a disordered medium \cite{DSFisher1998,BenZionRice1993,BenZionRice1997,FisherDahmenRamanathanBenZion1997}. It is however a rather controversial issue and it is now clear that some important features of earthquakes, such as aftershocks and the Omori law \cite{Omori1894} are not contained in {\it the simplest} elastic interface model. We will come back to this specific issue in Sec.~\ref{subsec:SummaryAva}.
\smallskip

{\it Vortex lattices in superconductors}\\
Although it not a disordered elastic interface, let us mention here that features of the deformation of the vortex lattice in high-$T_c$ superconductors ($d=3$, $N=2$) are similar to those of disordered elastic interfaces. In particular it is known that the pinning of the lattice by the disorder plays an important role in high-$T_c$ superconductivity and that a similar depinning transition is observed. See \cite{GiamarchiLeDoussalBookYoung} for a review.
\smallskip

{\it Imbibition}\\
Let us finally mention here the problem of the invasion of a viscous fluid in a porous medium known as `imbibition' where in some regime the dynamics resembles the dynamics of an elastic interface and scale invariant avalanches are also observed \cite{PlanetSantucciOrtin2009}. Some aspects of the problem are however not captured by elastic interfaces (in particular in the experimentally much studied context of so-called forced flow imbibition, the conservation of the volume of the fluid imposes the mean velocity of the fluid at all time and thus generates a complex non-local dynamics along the front). We refer the reader to \cite{AlavaRostDube2004} for a review of this related subject.

\subsection{Out-of-equilibrium interface growth}  \label{subsec:ExpKPZ}

We now give a few examples of situations where the out-of-equilibrium growth of a $1$-dimensional interface was shown to display scale-invariant behavior in agreement with the $1+1$-d KPZ universality class. It should be stressed here that it is easier to find in the literature experimentally observed growth processes in $d=1+1$ for which the scaling behavior notably differs from the one of the KPZ universality class, see e.g. \cite{BarabisiStanleyBook,HalpinHealyZhang1995}. This obviously does not mean that the KPZ universality class does not exist in nature, but it is true that some of its conditions are not always easy to realize experimentally (e.g. absence of quenched noise). Below we mention three convincing experiments.

\subsubsection{Growth of bacterial colonies and cancerous cells}

Bacterial colonies growing on a Petri dish provide an experimental realization of a growing interface in $1+1d$. In \cite{WakitaItohMatsuyamaMatsushita} experiments on the growth of two types (B and D) of bacteria were performed. From microscopic observation it was observed that the microscopic growth mechanisms of the two types were quite different. While the type B bacteria formed long chain advancing simultaneously (thus inducing a non-local growth), for the type D bacterias the growth mechanisms were argued to be local. The found roughness exponent of the interface were found to be $\alpha_B \simeq 0.78 \pm 0.02$ and $\alpha_D \simeq 0.50 \pm 0.01$. The growth of type $D$ bacteria was therefore argued to provide an experimental example of a growing interface in the $1+1$d KPZ universality class. 

\smallskip

More recently in another biological context, the growth of cell colonies for cancerous and non cancerous cells on Petri dishes was investigated in \cite{HuergoPasqualeGonzalezBolzanArvia2012} with the aim of distinguishing both types of cells from their growth mechanisms. Although some distinguishing features were reported, both types of colonies were found to exhibit a KPZ type growth scaling with exponents measured as $\alpha\simeq 0.50 \pm 0.05 $, $\beta \simeq 0.32 \pm 0.04$ and $z \simeq 1.5 \pm 0.2$.

\subsubsection{Burning paper fronts}

Slowly burning sheets of paper also provide an example of interface growth in $1+1$d. In \cite{Myllys&al2001} this growth process was investigated for two different types of papers. The exponents were found in good agreement with the KPZ expected values: $\alpha \simeq 0.48 \pm 0.01$ and $\beta = 0.32 \pm 0.01$.

\subsubsection{Liquid crystal growths}

The most convincing experimental evidence of KPZ universality in growth process in $1+1$d comes from recent experiments on turbulent liquid crystal \cite{TakeuchiSano2010,TakeuchiSano2011,TakeuchiSano2012,Takeuchi2013,TakeuchiSano2014}. This experiment is very close in spirit to the original motivation for introducing the KPZ equation \cite{KPZ}: the interface is a true interface between two phases (called DSM1 and DSM2) of the same system. While the microscopic properties of each phase are rather complicated, at high electric field the DSM1 phase is unstable and the growth of a nucleus of the DSM2 phase in an initially prepared liquid crystal in the DSM1 phase exhibit fluctuation statistics in amazing agreement with the KPZ theory in $1+1$d. These highly reproducible experiments indeed allowed the authors to obtain the scaling exponents $\alpha \simeq 0.5 \pm 0.05$ and $\beta \simeq 0.336 \pm 0.001$, but also to exhibit strong evidence that the full rescaled fluctuations of the interface height at large time converges to the GUE and GOE distributions, depending on the shape of the original nucleus of stable phase. Traces of the Baik-Rains distribution in the stationary state were also obtained.

\chapter{Avalanches and shocks of disordered elastic interfaces} \label{chapII}

\section{Introduction} 

Avalanche-type dynamics occur in a large variety of complex systems: snow avalanches, earthquakes, fracture processes in disordered materials, fluctuations in the stocks market, Barkhausen noise in magnets, avalanches in the neural activity of the brain... In a general sense, a system is said to display avalanches if its response to a slow, smooth, external loading is discontinuous and proceeds via jumps. The most interesting situation to the statistical physicist is the case where these jumps span a wide range of space and time scales. If this occurs, then one might hope that the precise underlying dynamics of each system is mostly (except e.g. symmetries, etc) unimportant at large scales, i.e. that avalanche processes display some {\it universality}. In fact such systems do exist in nature, and the experimental and theoretical analysis of systems and models displaying avalanches has created a large research activity over the past decades \cite{SethnaDahmenMyers2001}. Some key conceptual frameworks on the theoretical side have been the analysis of avalanches in (i) cellular automaton models exhibiting {\it Self-Organized-Criticality} \cite{BakTangWiesenfeld1987}, as e.g. the Manna sandpile model \cite{Manna1991} and the Abelian sandpile model \cite{Dhar1999b}, see \cite{PruessnerBook} for a review; (ii) the random field Ising model \cite{DahmenSethna1996,TarjusBaczykTissier2013,TarjusTissier2016} and the mean-field spin glass SK model \cite{LeDoussalMullerWiese2010,LeDoussalMullerWiese2012}; (iii) models related to the concept of marginal stability \cite{MullerWyart2015}; (iv) disordered elastic systems. 

\medskip

In this part of the manuscript we will review some known results on avalanches (and the closely related notion of shocks) in disordered elastic interfaces. Understanding avalanche processes in this type of systems is an important issue. Indeed on one hand it is known that the model of an elastic interface in a disordered medium is relevant to describe a variety of physical situations (see Sec.~\ref{sec:Experiments}), and therefore on the theoretical side it is a perfect candidate to understand universality in some avalanche processes. On the other hand disordered elastic systems permit the use of a variety of existing analytical techniques. Theoretical progresses on this type of system are thus already possible and understanding them is a good starting point for other problems. For example it was recently argued that the Manna sandpile model is in the same universality class as the depinning of an interface in a short-range disordered medium \cite{LeDoussalWiese2014a}, or that the yielding transition in amorphous solids share some interesting properties with the depinning transition \cite{LinLernerRossoWyart2014} (although it is in a different universality class, in particular there is an additional independent critical exponent).

\medskip
The outline of this section is as follows: In Sec.~\ref{sec:ToyModelsOfShocksAndAva} we introduce the notion of shocks and avalanches in toy models of a particle on the real line ($d=0$ elastic interfaces). These notions are generalized to the case of interfaces in Sec.~\ref{sec:ShocksAndAvaForElasticInterfaces}. There we review the phenomenology associated with avalanches, in particular we discuss their scaling. In Sec.~\ref{sec:ShocksWithFRG} and Sec.~\ref{sec:AvaWithFRG} we review the functional renormalization group approach to the statics and to the depinning transition of disordered elastic interfaces, with an emphasis on its application to the study of shocks and avalanches. Finally in Sec.~\ref{sec:AvaResults}, we will review some important results on avalanches in disordered elastic interfaces, and summarize the results obtained during the thesis on this subject. These are presented more thoroughly in the original research papers \cite{ThieryLeDoussalWiese2015,ThieryLeDoussal2016a,ThieryLeDoussalWiese2016}.

\section[Avalanches for a particle]{Introduction: the avalanche process(es) of a particle on the real line} \label{sec:ToyModelsOfShocksAndAva}

In this section we begin our study of avalanche processes of elastic interfaces by studying toy models in $d=0$, i.e. a particle on the real line in a disordered medium. This analysis will prove relevant when discussing the avalanche processes of true elastic interfaces as a basis on which we will build some intuition on shocks and avalanches processes. In Sec.~\ref{subsec:ToyShock} we introduce the notion of shocks and in Sec.~\ref{subsec:ToyAva} we introduce the notion of avalanches. We will conclude by comparing these two notions in Sec.~\ref{subsec:ToyShockVsAva}

\subsection{Shocks between ground states for toy models of a particle without disorder} \label{subsec:ToyShock}

In this section we introduce the notion of shocks using toy models of a particle in a deterministic potential $V$ exhibiting several local minimas as would a true disorder potential. We study this simple case in order to maximize the clarity of the exposition. Exact results can also be obtained for models of shocks for a particle in a random potential: this includes the case of $V(u)$ taken with the correlations of a Brownian motion (the Sinai model \cite{Sinai1982}, a toy-model for the random field universality class of elastic interfaces) or with short-range correlations (that corresponds to the Kida problem in the context of Burgers's turbulence \cite{Kida1979} and to a toy-model for the random bond universality class for elastic interfaces). We refer the reader to \cite{LeDoussal2008} and references therein for exact results on these models, and we now begin our study of shocks.

\subsubsection{ Shocks for a particle in a cosine potential}
\begin{figure}
\centerline{\includegraphics[width=8.0cm]{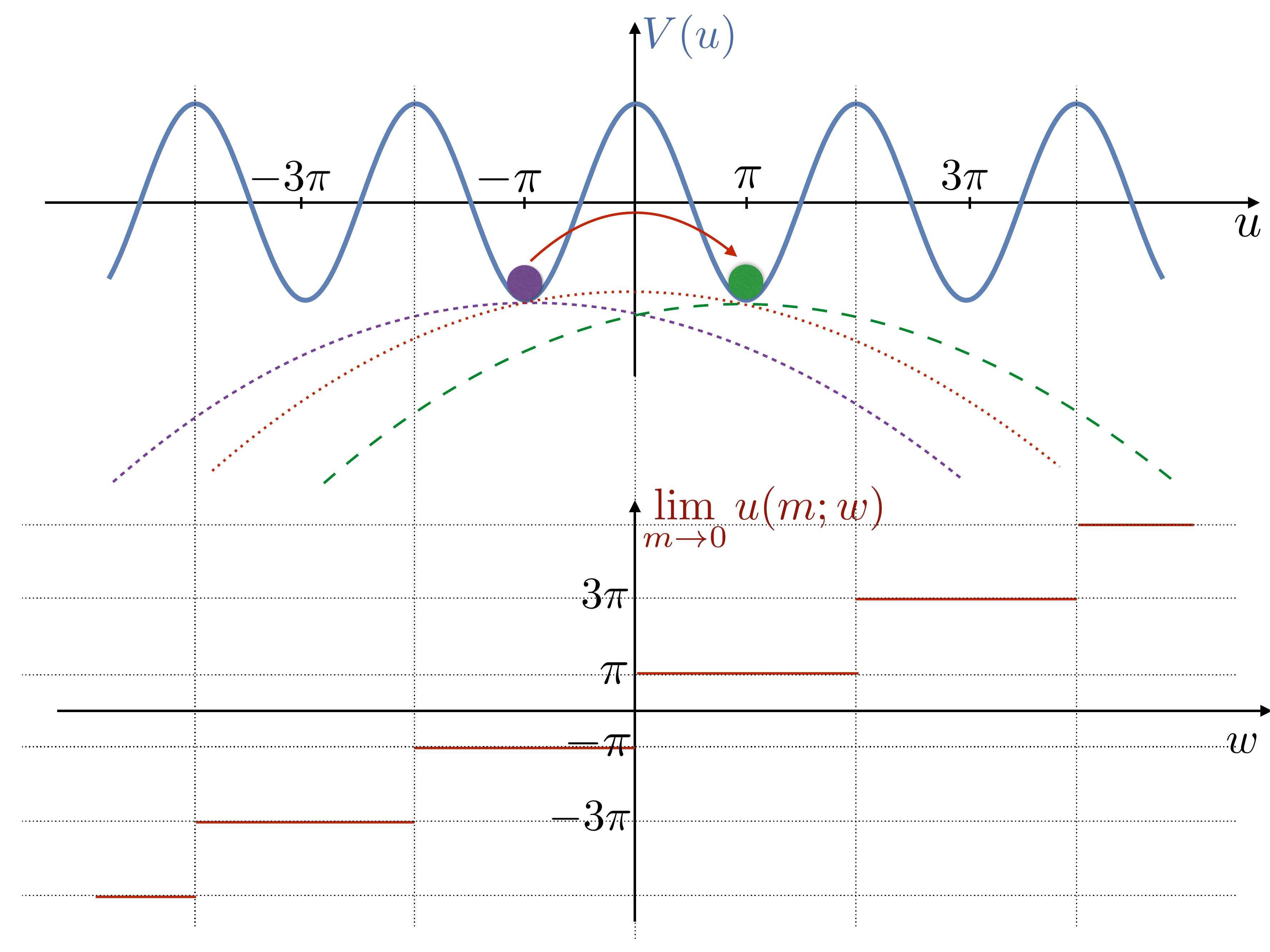} \includegraphics[width=8.0cm]{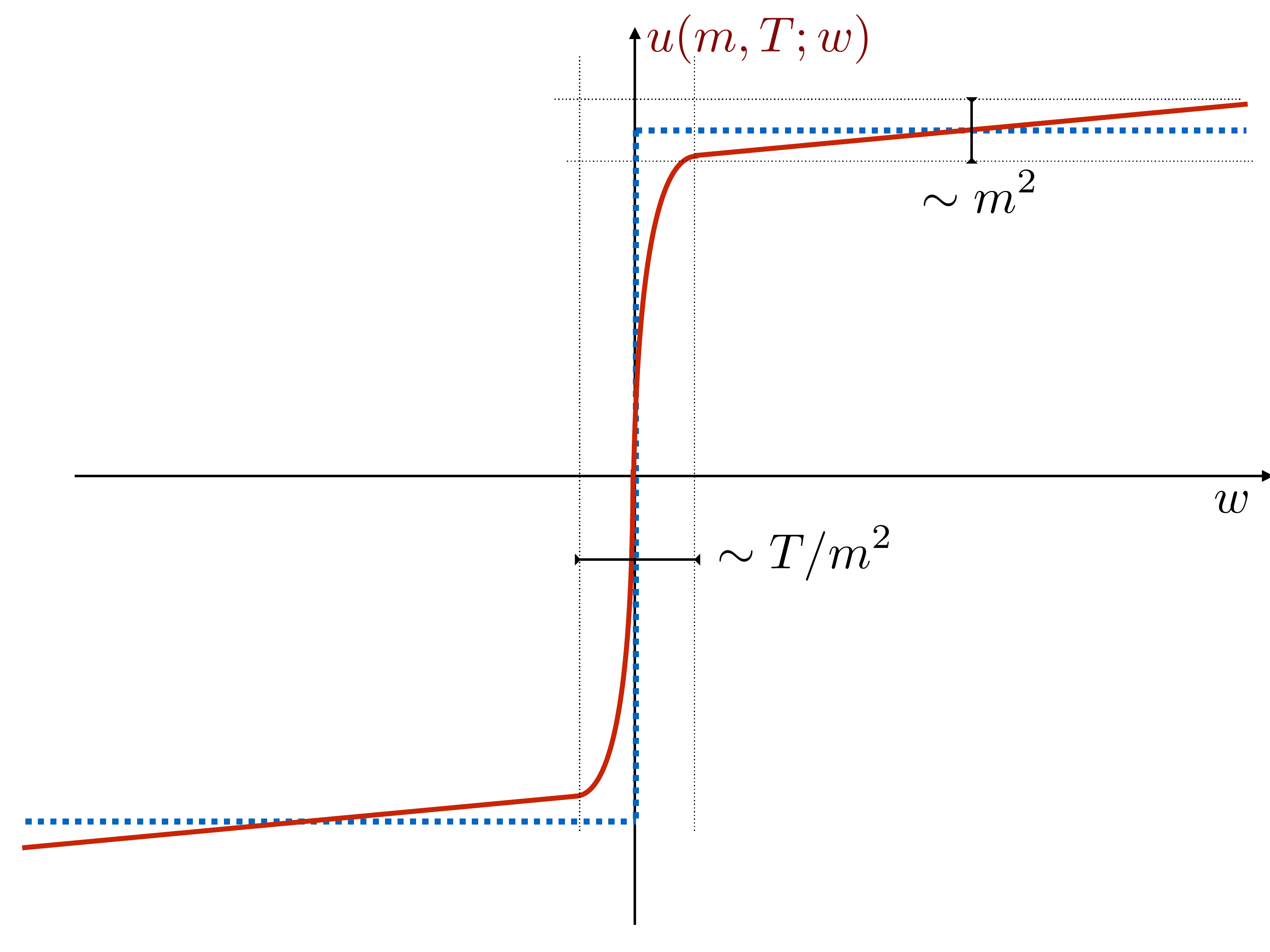}} 
\caption{Left: Shock process for a particle at zero temperature in a cosine potential. Left-up: The Maxwell construction allows to graphically find the position of the minimum of $V(u) + \frac{1}{2} m^2(u-w)^2$ for $m \neq 0$: the parabola $-\frac{1}{2} m^2(u-w)^2$ is raised until it intersects $V(u)$. The point of intersection corresponds to the position of the minimum. For $-2 \pi<w<0$ (resp. $0<w<2 \pi$), the minimum is at $-\pi$ (purple dot) (resp. $+ \pi$, green dot) (up to $O(m^2)$ corrections). At $w= 0$ (red dotted parabola) the minimum is degenerate. Left-bottom, the jump process obtained from the above picture in the limit $m \to 0$. Right: smoothening of a perfect shock for a particle in a double-well potential by non-zero $T$ and $m$ in the limit $T \ll m^2$ and $m \ll m_c$.}
\label{fig:ToyShocks}
\end{figure}

We consider the toy model of a particle on the real line $u \in \JR$ subject to a cosine potential $V(u) = \cos(u)$ and to a confining potential $\frac{1}{2} m^2(w-u)^2$. The total Hamiltonian of the particle is
\bea
\cH_w[u] :=  \cos(u) + \frac{1}{2} m^2(w-u)^2 \ .
\eea
The `disorder potential' $V(u)$ has an infinite number of exactly degenerate minima at $u_k = \pi + 2 \pi k$, $k\in \JZ$. For $m \neq 0$, the confining potential $ \frac{1}{2} m^2(w-u)^2 $ breaks this degeneracy, except at some special points $w_k$ (the points where $V(w)$ is maximum, that is when $w_k= 2 \pi k$, $k\in \JZ$) and the ground state
\bea
u(m;w) :={\rm argmin}_{u \in \JR} \cH_w[u] \ ,
\eea
is well defined except in this discrete set of points. Graphically the position of the ground state of the system can be obtained using the so-called `Maxwell construction': for a given $w$ with a non-degenerate ground state $u(m,w)$ with energy $E(m,w)$, by definition, $V(u) > E(m,w) -\frac{1}{2} m^2(u-w)$ $\forall u \neq u(m,w)$, with the equality at $u=u(m,w)$. Hence, $\forall C < E(m,w)$, the parabola $C-\frac{1}{2} m^2(u-w)$ does not intersect $V(u)$. Increasing $C$ from $-\infty$, the position of the ground state $u(m,w)$ is given by the abscissa of the first point of intersection of the parabola $C-\frac{1}{2} m^2(u-w)$ with $V(u)$ (see Fig.~\ref{fig:ToyShocks}). For large $m$, the ground state $u(m;w)$ closely follows $w$ and $u(m;w)$ is smooth as a function of $w$. For $m$ small enough, at the point $w=w_k$, the ground state of the system is degenerate between one point $u^+(m;w)$ that is close to $u_k$, and $u^{-}(m,w)$, that is close to $u_{k-1}$. The critical value where this degeneracy first occurs satisfies the equation
 $ \frac{1}{2} m_c^2 =  \frac{1}{2} \frac{d^2 V(u)}{du^2}|_{u=0} = \frac{1}{2}$, that is $m_c=1$. As $m \to 0$, it is trivial to see that $u^+(m;w)$ converges to $u_k$ and $u^-(m;w)$ converges to $u_{k-1}$. We obtain
 \bea
 \lim_{m \to 0} u(m;w) && = \sum_{k \in \JZ}  \theta(w-w_k) \theta(w_{k+1} - w) u_k \nn \\
 && =  u_{-1}+ \sum_{k=0}^{+\infty} \theta(w-w_k) S_k  \quad \text{for } w>0 \ .
 \eea
 Here $\theta$ denotes the Heaviside theta function, and in the second line we have introduced the size of the $k^{th}$ shock $S_k = u_k - u_{k-1}$. In this simple model these are of course all equal to $S_k = S = 2 \pi$ and $u_{-1} = -\pi$. The above formulae are ambiguous at the shock points $w=w_k$ since precisely at these points the ground state is degenerate, and thus $u(m;w)$ is ill defined. This is resolved by using e.g. the convention that $u(m,w)$ is left continuous. At small, but non-zero $m$, $m \ll m_c$, the shock process gets slightly modified, as $ u(m;w)$ is not exactly constant between $w_k$ and $w_{k+1}$. Small $m$ corrections are given by, for $w_k<w<w_{k+1}$,
 \bea
 && u(m; w) = u_k + m^2 \delta u_k(w)  + O(m^4) \quad , \quad  \delta u_k(w) := w-u_k \ .
 \eea
Let us now discuss the influence of the temperature in a toy model of a particle in a double-well potential.

\subsubsection{ Smoothening of the shocks by the temperature}
We now consider a particle on the real line $u \in \JR$ at equilibrium at a finite temperature $T>0$ in a double well potential $V(u) = -\frac{u^2}{2} + \frac{u^4}{4}$ and subject to a confining potential $\frac{1}{2} m^2(w-u)^2$. The Hamiltonian is
\bea
\cH_w[u] =  -\frac{u^2}{2} + \frac{u^4}{4} +\frac{1}{2} m^2(w-u)^2  .
\eea
And we consider the average position of the particle:
\bea \label{Eq:SecII1:SchokWithTemp}
u(m,T;w):= \frac{\int_{-\infty}^{\infty}  u e^{-\frac{1}{T} \cH_w[u]}}{\int_{-\infty}^{\infty}   e^{-\frac{1}{T} \cH_w[u]}}  .
\eea
First note that in the limit $m \to 0$, we obtain $\lim_{m\to 0} u(m,T;w) = 0$. On the other hand, taking first the zero temperature limit we obtain, for $m$ sufficiently small $m \leq m_{c} =1$,
\bea \label{Eq:SecII1:SchokWithTempShockLimit}
\lim_{T \to 0} u(m , T; w) = \theta(w) u^+(m) + \theta(-w) u^-(m)
\eea
with for $m \ll m_{c}$,
\be
u^+(m,w) = 1 +\frac{m^2}{3} (w-1)^3 + O(m^4)  \quad , \quad u^-(m,w) = -1 +\frac{m^2}{3} (w+1)^3 + O(m^4)  \ .
\ee
Let us now consider the limit $T$ small but non zero, with $m \leq m_c$ fixed. The integrals in (\ref{Eq:SecII1:SchokWithTemp}) are dominated by two saddle-points at $u^-(m,w)$ and $u^+(m,w)$. One easily obtains, noting $\Delta E(m,w) := \cH[u^{+}(m,w)]-\cH[u^{-}(m,w)]$ the difference of energy between the right minimum and the left minimum,
\be
u(m,T;w) = \frac{{\rm sign}( u^+(m,w)) e^{-\frac{1}{T} \Delta E(m,w)  }   +{\rm sign}( u^-(m,w)) }{ \frac{1}{|u^+(m,w)|} e^{-\frac{1}{T} \Delta E(m,w) } + \frac{1}{|u^-(m,w)|}} + O(1/\sqrt{T})
\ee
Taking now the small $m$ limit on this expression we obtain
\be
u(m,T;w) \simeq  \tanh\left( \frac{m^2 w + O(m^4)}{T}  \right)  + O(1/\sqrt{T})  \ssp .
\ee
In particular, one retrieves for large $|w|$ or for small temperature the shock limit (\ref{Eq:SecII1:SchokWithTempShockLimit}). For non-zero $T$, the shock is smoothened on a scale
\bea
w_T \sim \frac{T}{m^2}    \ssp    .
\eea
Read differently, the shock is smoothened on small scales when the energy difference between the two minima $\sim m^2 w$ is smaller than the thermal energy $\sim T$. Note that the height of the barrier of potential between the two minima does not play a role in this problem.

\subsection{Avalanches in the dynamics of a particle on the real line} \label{subsec:ToyAva}

In this section we now discuss the notion of avalanches in the zero temperature over-damped dynamics of a particle. We first consider the case of a particle in an abstract force landscape $F(u)$, and then recall some features of the very instructive exact solution of the Alessandro-Beatrice-Bertotti-Montorsi (ABBM) model where the force $F(u)$ is a Brownian motion. We refer the reader to \cite{LeDoussalWiese2008a} for the study of other cases.
  
\subsubsection{The avalanche process of a particle in a smooth force landscape $F(u)$.}

Let us thus consider the over-damped dynamics of a particle in a force landscape $F(u)$. Here we typically have in mind the case $F(u) = a(u) sin( b(u) u)$ with $a(u)$ and $b(u)>1$ some smooth bounded functions: the force landscape has a lot of minimas and does not wander too far away from zero (see Fig.~\ref{fig:ShocksVsAva}). The temperature is $0$ and similarly to the static problem, we consider the over-damped dynamics of a particle driven by a parabolic well at a constant velocity $v>0$:
\bea \label{Eq:SecII2:ToyAva}
\eta \partial_t u_t = m^2( v t- u_t ) + F(u_t) \ssp .
\eea
Let us suppose that at $t=0$ the particle is at rest and sits in a stable minima of its energy landscape: $u_{t=0}=u_0$ with
\bea
m^2 u_0 = F(u_0) \quad ,  \quad  m^2 - F'(u_0) >0  \ssp .
\eea
Now, note that since $F(u)$ is continuous, it is clear from (\ref{Eq:SecII2:ToyAva}) that $u_t$ is ${\cal C}^1$ and 
\bea
\forall t \geq 0 \quad, \quad \partial_t u_t \geq 0 \ .
\eea
This allows to make the change of variable $u(v,m;w) = u_{t=w/v}$. Plugging it into (\ref{Eq:SecII2:ToyAva}) one obtains
\bea
\eta v \partial_w u(v,m;w)  = m^2(w - u(v,m;w)) + F(u(v,m;w)) \ . 
\eea
Hence taking the limit $v \to 0$, the {\it quasi-static process}
\bea
u(m;w):= \lim_{v \to 0} u(v,m;w) =  \lim_{v \to 0} u_{t = w/v}  \ssp ,
\eea
satisfies
\bea \label{Eq:SecII2:ToyAvaQS}
0=m^2(w-u(m;w) ) + F(u(m;w))  \ssp .
\eea
Of course, $\forall v >0$, the function $ u(v,m;w)$ is ${\cal C}^1$ as a function of $w$. However, the limit $v \to 0$ can be singular, and indeed as in the case of shocks in the statics, if $m$ is sufficiently small $m < m_c$ (the latter being now given by $m_c^2 = {\rm max}_{u} F'(u)$), the quasi-static process $u(m;w)$ can exhibit discontinuities as a function of $w$. The classic construction of $u(m;w)$ is given in red in Fig.~\ref{fig:ShocksVsAva}: starting from $u(m;0)=u_0$, it is obtained by following whenever possible the root of (\ref{Eq:SecII2:ToyAvaQS}) (hence $u(m;w)$ is left continuous, $\forall w >0$, $u(m;w^-)=\lim_{\delta w \to 0^+} u(m;w - \delta w)=u(m;w)$), or when not possible at some $w_k$ with $k \in \JN$, $u(m;w_k^+)=\lim_{\delta w \to 0} u(m;w_k + \delta w)$ is given by the smallest root of (\ref{Eq:SecII2:ToyAvaQS}) that is larger than $u(m;w_k^-)$. At these points of discontinuity the quasi-static process makes a jump $S_k =u(m;w_k^+)-u(m;w_k^-)$  that we call an avalanche. Following similar arguments as before for the case of shocks in the statics, it is clear that in the limit $m \to 0$, this process becomes a pure jump process
\bea  \label{Eq:SecII2:ToyAvaJump}
u^{{\rm jump}}(w) := \lim_{m \to 0}  m^{-\zeta} u(m; m^{-\zeta} w)= u_0 +\sum_{k=0}^{\infty} \theta(w-w_k) S_k \ssp    .
\eea
Here we have introduced the {\it roughness exponent} $\zeta \geq 0$. The latter accounts for the fact that in general, e.g. for the case of a random process $F(u)$, the jumps become rarer and bigger as $ m \to 0$. In the case of shocks in the periodic potential, we had $\zeta = 0$ precisely because the potential was periodic and in the limit $m\to 0$ only one shock occurred per period of the potential, with its size being equal to the period. Scaling $u$ and $w$ by $\zeta$ in (\ref{Eq:SecII2:ToyAvaJump}) allows us to obtain a non-trivial jump process in the limit. The value of $\zeta$ depends on the precise properties of $F(u)$ and is therefore non-universal. Why this exponent is called $\zeta$, i.e. how it is related to the roughness exponent for interfaces, will become clearer in the next section. In the following we will set $\zeta = 0$ for simplicity but the discussion can be repeated with $\zeta \geq 0$. \\

Let us now come back to the time process $u_t$ and draw some conclusions on the dynamics of the particle. Between jumps, $w_k <w<w_{k+1} $, $\partial_w u^{{\rm jump}}(w) = 0$. Inverting the derivative and the limits with $t = w/v$ we obtain,
\bea
0=\partial_w u^{{\rm jump}}(w) = \lim_{m \to 0}  \lim_{v \to 0}  \frac{1}{v} \partial_t u_t =0 \ssp .
\eea
Hence between jumps, in the successive limit $v \to 0$ and (= then) $m \to 0$ the velocity of the particle is {\it not} of order $v$. Rather it is $o(v)$. Let us now `zoom in' around the $k^{th}$ jump and look at a window around $t_k$, i.e. $t_k<t  < w_k/v +\Delta t$ with by definition $\Delta t = o(1/v)$ since this smooth process happens during a time scale that is not captured by the jump process (\ref{Eq:SecII2:ToyAvaJump}). Since this process happens on such a short-time scale, we can forget during the jump that the well keeps moving and approximate $vt  \sim w_k + o(1)$ and the dynamics during the jump is given by
\bea
\eta \partial_t u_t = m^2(w_k -u_t) + F(u_t) + o(1)
\eea
At $t=0$ the right hand-side is approximately equal to $ m^2(w_k -u(m;w_k)) + F(u(m;w_k))  = 0$, but here this equilibrium point is unstable since a jump occurs. After a short transient time of order $\tau_m = \eta/m^2$, the right hand-side is soon of order $1$ and in the limit of small $m$, it is dominated by the force $F(u)$. Hence during the jump the velocity of the particle is of order $1$ and the jump occurs on a time scale of order $1$. This discussion highlights a characteristic feature of the temporal dynamics of the avalanche process we are interested in. Although it is clear that, since $F(u)$ is bounded, the mean velocity (here mean refers to the average over space) of the particle is equal to $v$ (i.e. $\lim_{t \to \infty} \frac{u_t - u_0}{t} = v$), the dynamics is {\it intermittent}. Most of the time the particle is actually pinned by disorder and its velocity is $o(v)$ (if one again uses a notion of probability by taking a random time $t$ this occurs with a probability of order $1$), and from time to time the particles is in an avalanche and its velocity is of order $1$ (this occurring with probability $O(v)$). This is quite different for a smooth motion (obtained e.g. either by taking $F \to 0$ or $v \to \infty$) for which one expects to observe the velocity of the particle to be of order $v$ with probability $1$. These considerations will become clearer in the ABBM model.

\subsubsection{ The ABBM model.} \label{subsec:ABBM}

Let us now study our first true random process and consider the ABBM model. A possible definition of the latter is the stochastic process
\bea \label{Eq:SecII2:ToyAvaABBM}
\eta \partial_t u_t = m^2(vt - u_t) + F(u_t) \ssp ,
\eea
with the initial condition $u_{t=0} = 0$ and the force $F(u)$ is a one-sided (i.e. $F(0)=0$) Brownian-motion (BM) with correlations
\bea
\overline{(F(u') - F(u))^2} = 2 \sigma |u-u'| \ssp .
\eea
This is one definition of the so-called Alessandro-Beatrice-Bertotti-Montorsi (ABBM) model. It was first introduced as a phenomenological model to describe Barkhausen noise \cite{ABBMTh,ABBMEx}. In this context $u_t$ denotes the measured magnetization of the disordered magnetic sample under the applied magnetic field $\sim vt$. It was later argued, first on phenomenological grounds \cite{ZapperiCizeauDurinStanley1998}, then from first principles using FRG \cite{LeDoussalWiese2012a} that it correctly describes the avalanche of the center of mass of realistic interfaces at the depinning transition at the mean-field level. The model presents the peculiarity of being in some sense exactly solvable and a lot of exact quantities can be computed (see \cite{Colaiori2008,DobrinevskiPhD} for a review). We will not recall them all here, but only focus on analyzing the exact results for the stationary velocity distribution and for the avalanche size distribution to highlight in a more concrete model some of the considerations of the previous section.
\medskip

 To simplify the discussion first note that, using the scale invariance of the BM, (\ref{Eq:SecII2:ToyAvaQS}) can almost be entirely rescaled so that all parameters are equal to $1$. Indeed, introducing $S_m :=\sigma/m^4$ and $\tau_m:= \eta/m^2$, rescaling $u_t = S_m \tilde{u}_{\tilde{t}} $, $t = \tau_m \tilde{t} $, noting $\tilde{v} = v/v_m$ with $v_m = S_m/\tau_m$, one obtains in these dimensionless units
\bea \label{Eq:SecII2:ToyAvaABBM2} 
\partial_t u_t = v t - u_t  +F(u_t)
\eea
where here we have dropped the tildes and now $F(u)$ is a one-sided Brownian-Motion (BM) with correlations $\overline{(F(u') - F(u))^2} = 2  |u-u'|$. 

\medskip

{\it Stationary velocity distribution}\\
The stationary distribution of the velocity of the particle was already obtained in the original paper of ABBM \cite{ABBMTh}. At long time $t \gg \tau_m$, the probability distribution function (PDF) of the velocity of the particle $\dot{u}_t :=\partial_t u_t$ is stationary and equal to 
\bea \label{Eq:SecII1:ABBM:Pv}
P(v;\dot u)= \frac{1}{\Gamma(v)}( \dot{u})^{-1+ v} e^{ - \dot{u}} \theta(\dot u)  \ssp ,
\eea
where $\Gamma(v)$ is the Euler's Gamma function. As expected the mean-velocity of the particle is equal to $v$: $ \int_{0}^{\infty} \dot u P(v;\dot u) = v$. The behavior of the PDF (\ref{Eq:SecII1:ABBM:Pv}) is however completely different depending on whether $v >v_m=1$ or $v<1$. For $v>1$, $P(v;\dot u = 0)=0$: in the stationary state the particle is never pinned by the potential. In this phase the PDF $P(v , \dot u)$ is maximum for $\dot u = v-1$ and both the most probable velocity and the mean velocity are of order $O(v)$: the motion of the particle is more or less smooth and the particle mostly follows the imposed driving. On the other hand when $v <1$, $P(v;\dot u = 0)= + \infty$ and there is an accumulation of the probability at $0$: the particle is almost always at rest. In the limit $v \to0$, in the sense of distribution, $\lim_{v \to 0} P(v; \dot u)  = \delta(\dot u) $. At $\dot{u} = O(1)$ fixed on the other hand
\bea
P(v; \dot u)  =  v \hat \rho(\dot u) + O(v^2) \quad , \quad  \hat \rho( \dot u )  :=\frac{1}{\dot u } e^{- \dot u} \ssp  . 
\eea
While $ \hat \rho(\dot u)$ is not normalizable, i.e. $\hat \rho_0 = \int  \hat \rho(\dot u ) = +\infty$, it controls all the moments of $P(v; \dot u)$. To see this, it is useful to consider the Laplace transform of $P(v; \dot u)$, defined as, for $\lambda <1$,
\bea \label{Eq:SecII1:ABBM:Poisson1}
\int_{0}^{\infty} d \dot u e^{\lambda \dot{u}}   P(v; \dot u) && = e^{-v \log(1- \lambda)} = e^{ v \int_{0}^{\infty} d \dot u( e^{\lambda \dot u} -1)  \hat \rho(\dot u ) } \nn \\
&& = 1 +  v \int_{0}^{\infty} d \dot u( e^{\lambda \dot u} -1)  \hat \rho(\dot u ) \\
&&  + \frac{v^2}{2}\int_{0}^{\infty} \int_{0}^{\infty} d \dot u_1  d \dot u_2 ( e^{\lambda \dot u_1} -1)( e^{\lambda \dot u_2} -1)  \hat \rho(\dot u_1 )  \hat \rho(\dot u_2 ) + \dots  \nn
\eea
Now, if $\hat \rho_0$ defined above was finite, we could define a normalized probability distribution $P_{{\rm ava}}(\dot u) := \frac{  \hat \rho(v)}{\hat \rho_0}$ and the above equality could be rewritten
\be \label{Eq:SecII1:ABBM:Poisson2}
\int_{0}^{\infty} d \dot u e^{\lambda \dot{u}} P(v; \dot u)  =  \sum_{m=0}^{\infty} \frac{(\hat \rho_0 v)^m}{m!} e^{- \hat \rho_0 v} \int d \dot u_1 \cdots d \dot u_m e^{\lambda(\dot u_1+ \cdots + \dot u_m) } P_{{\rm ava}}(\dot u_1) \cdots  P_{{\rm ava}}(\dot u_m)  \  .
\ee

Proving explicitly that (\ref{Eq:SecII1:ABBM:Poisson1}) can be resummed into (\ref{Eq:SecII1:ABBM:Poisson2}) if $\hat \rho_0$ is finite is not complicated. A similar equality is shown in Appendix D of \cite{ThieryLeDoussalWiese2015}. The formula (\ref{Eq:SecII1:ABBM:Poisson2}) can equivalently be rewritten
\bea \label{Eq:SecII1:ABBM:Poisson3}
P(v; \dot u) = \sum_{m=0}^{\infty} \frac{(\hat \rho_0 v)^m}{m!} e^{- \hat \rho_0 v} \underbrace{(P_{{\rm ava}} \star \cdots \star P_{{\rm ava}})}_{m  {\rm \ times \ self-convolution \ of \ } P_{{\rm ava}} }(\dot u)
\eea
The interpretation of this formula is as follows. At each time, the velocity $\dot u$ of the particle is the sum of $m$ velocities that are independently drawn from $P_{{\rm ava}}(\dot u)$. $P_{{\rm ava}}(\dot u)$ is thought of as the PDF of the velocity during an avalanche, and $m$ is the number of active avalanches at an arbitrary time $t$. The latter is drawn from a Poisson distribution with intensity $\hat \rho_0 v$. Thus in the ABBM model, avalanches are independent from one another. This is characteristic of a Lévy jump process. One of the question tackled in this thesis is to understand if this independence property is also true for models of interfaces in a short-range random force landscape (we will see that it is not!). Taking the limit $v \to 0$ of (\ref{Eq:SecII1:ABBM:Poisson3}) we obtain
\bea \label{Eq:SecII1:ABBM:Poisson4}
P(v; \dot u) = (1- \hat \rho_0 v) \delta(\dot u) + \hat \rho_0 v P_{{\rm ava}}(\dot u) +O(v^2) \ssp .
\eea
This is the picture of avalanche motion that was described in the last section. At small $v$ the particle does not move at all with a probability close to $1$, and sometimes move (with a probability of order $v$) at a velocity of order $1$. For the ABBM model, we note that $\hat \rho_0$ is infinite and the previous interpretation is a bit tedious. This is due to the scale invariance of the BM even at small scales: the particle is never truly pinned in a typical realization of the BM and it always makes microscopic jumps, formula (\ref{Eq:SecII1:ABBM:Poisson4}) does not truly hold. For a realistic model with a smooth disorder at small scale (\ref{Eq:SecII1:ABBM:Poisson4}) will truly hold. Even in the ABBM model however, this interpretation holds at the level of the moments as seen using (\ref{Eq:SecII1:ABBM:Poisson2}). In particular
\bea
\int_{0}^{\infty} d\dot u (\dot u)^n P(v;\dot u) && =  \frac{\Gamma(v+n)}{\Gamma(v)} = v\times (v+1) \times \dots (v+n-1) \nn \\
&& = v (n-1)!  +O(v^2) \nn \\
&& = v  \int d\dot u  \rho(\dot u) \dot u +O(v^2) \ssp .
\eea
While for $v \gg 1$, the $n^{th}$ integer moment is of order $v^n$, characteristic of a smooth motion at the typical velocity $v$, for $v \ll 1$ all integer moments are of order $v$: an avalanche occurs with probability of order $v$ but if it does, the velocity inside the avalanche is of order $1$.\\

{\it Avalanche sizes}

The most convenient way to define the PDF of avalanche sizes in the ABBM model is to consider a driving during a finite duration $T_{{\rm d}}$ (i.e. $w(t) = v \theta(T_{{\rm d}}- t) \theta(t)$. Defining the avalanches size as the total motion of the particle $S = u(t= +\infty ) -u(t=0) = \int_{0}^{\infty} \dot{u}_t $, the PDF of avalanche sizes $P(S)$ was computed in \cite{DobrinevskiLeDoussalWiese2011b}. The result is
\bea \label{Eq:SecII1:ABBM:Size1}
P(S) = \frac{v T_{{\rm d}}}{2 \sqrt{\pi} S^{3/2}} e^{- \frac{(S-v T_{{\rm d}})^2}{4 S}} \ssp .
\eea
Hence, for $v T_d \ll 1$ (small driving) the PDF of avalanche size exhibits a power-law behavior $ S^{-3/2}$ (this exponent was first identified in \cite{BertottiDurinMagni1994}) in between two cutoff scales, a small scale cutoff $ \sim \frac{(v T_{{\rm d}})^2}{S_m}$ and a large scale cutoff $\sim S_m$ (here $S_m = \sigma/m^4 = 1$) in dimensionless units. Since the limit $m \to 0$ can obviously be taken on the expression (\ref{Eq:SecII1:ABBM:Size1}), this shows that the proper rescaling of $u$ that we discussed earlier in the general case here corresponds to $u \to  m^{-\zeta} u $ with $\zeta =4$ for the ABBM model. The limit $v T_d \to 0$ at $S$ fixed defines the {\it avalanche size density in the ABBM model}:
\bea \label{Eq:SecII1:ABBM:Size2}
P(S) = v T_{{\rm d}} \rho(S) + O((v T_{{\rm d}} )^2) \quad , \quad \rho(S) := \frac{1}{2 \sqrt{\pi} S^{3/2}} e^{- S/4} \ssp .
\eea
As for the case of the stationary velocity distribution, the density $\rho(S)$ is not normalizable due to a divergence at small $S$. One could apply a similar treatment as we did before for the stationary velocity distribution and show that $P(S)$ can be rewritten in a certain sense as an infinite series of self-convolutions of $\rho(S)$ with itself, a property that defines a Lévy jump process (see also Appendix D of \cite{ThieryLeDoussalWiese2015}). It is also possible to precisely relate $\rho(S)$ to the avalanches observed in the quasi-static dynamics (this will be shown in Sec.~\ref{subsec:FRGDepAva}), and show that the motion of the particle in the ABBM model between two points where the velocity is zero are distributed according to (\ref{Eq:SecII1:ABBM:Size2}). Again the accumulation of avalanches of small sizes is due to the scale invariance of the BM. Note finally that the exponent $3/2$ can simply be understood as follows. Assume that at a time $t=0$ an avalanche has started and the velocity of the particle is $v_0$ and its initial position is $u_0$. Taking the limit $v \to 0$ in (\ref{Eq:SecII2:ToyAvaABBM}) and differentiating with respect to $u$ we obtain
\bea
\frac{d \dot{u}_t}{du} = -1 + \xi(u)   \ .
\eea
Where now the velocity $\dot{u}_t = \partial_t u_t$ is seen as a function of the position of the particle $u$ and $\xi(u)$ is a GWN. Hence the velocity of the particle performs a Brownian motion in `time' $u$ with a unit negative drift $-1$. Hence the next point $u =u_0+ S$ where $\dot{u}_t$ is $0$ corresponds to the next passage time to the origin of a Brownian motion with a unit negative drift, which is indeed power-law distributed with an exponent $3/2$ (see e.g. \cite{gardiner2004handbook}), and the unit drift provides an exponential cutoff as in (\ref{Eq:SecII1:ABBM:Size2}). The exponent $\tau = 3/2$ plays an important role in avalanche statistics (recall that it is observed in some Barkhausen noise experiments \cite{DurinZapperi2000}) and it is interesting to understand its value as a consequence of this well-known property of the BM.

\subsection{Shock process versus avalanche process for a particle} \label{subsec:ToyShockVsAva}

Let us conclude this section by comparing the shock and avalanche processes defined before for a particle in a smooth, bounded potential $V(u)$ that has a lot of minimas. The shock process was defined by the minimization of the total energy of the particle
\bea \label{Eq:SecII1:ShockVsAva1}
u^{{\rm shock}}(m;w) := {\rm argmin}_{u \in \JR}  \left( V(u) + \frac{1}{2} m^2(u-w)^2 \right) \ .
\eea
And the latter becomes a true jump process in the limit $m \to 0$ as discussed before. In all this section we will keep $m$ small but finite. Assuming that $V(u)$ is differentiable, the shock process verifies $\forall w \in \JR$,
\bea \label{Eq:SecII1:ShockVsAva2}
0=m^2(w-u^{{\rm shock}}(m;w)) +F(u^{{\rm shock}}(m;w) )
\eea
and it is by definition the root of this equation with the smallest energy. On the other hand in the dynamical case, the quasi-static process was defined again as a root of the same equation (see (\ref{Eq:SecII2:ToyAvaQS}))
\bea \label{Eq:SecII1:ShockVsAva3}
0=m^2(w-u^{{\rm q.s.}}(m;w)) +F(u^{{\rm q.s.}}(m;w) )
\eea 
with a specific rule that we recall. Starting from an arbitrary root of (\ref{Eq:SecII1:ShockVsAva3}) $u^{{\rm q.s.}}(m;w)$ is always increasing (since the driving was positive $v = 0^+$), is continuous whenever possible (i.e. it follows a given root when the root exist), and whenever the root it follows ceases to exist, it jumps to the smallest of bigger roots. Let us now note, $\forall w \in \JR$, the set of $n(w)$ roots of the equation $0=m^2(w-u) +F(u)$ as $(u_1(w) , \cdots , u_{n(w)} (w))$ with $u_i(w) < u_{i+1}(w)$.  Of course $n(w)$ a priori varies as a function of $w$. Let us suppose that at a given $w=w_0$, $u^{{\rm q.s.}}(m;w_0)$ is the $i_0^{th}$ root of the equation (\ref{Eq:SecII1:ShockVsAva3}): $u^{{\rm ava}}(m;w) = u_{i_0}(w)$. Note that as $w$ increases, from the rules specified before, $\forall w \geq w_0$, $u^{{\rm q.s.}}(m;w)$ is the $i^{th}(w)$ root of (\ref{Eq:SecII1:ShockVsAva3}): $ u^{{\rm q.s.}}(m;w) = u_{i(w)}(w)$ and it is clear that $i(w)$ cannot increase if $F(u)$ is continuous. On the contrary it decreases as we now explain. Since $F(u)$ is bounded, the first root of the equation $u_1(w)$ ceases to exist at some finite $w = w_1 \geq w_0$. In the sequence of roots of the equation $(u_1(w) , \cdots , u_{n(w)} (w))$, at $w=w_1$, $u_1(w)$ is then replaced by $u_2(w)$ (the second smallest root becomes the smallest root as the smallest root ceases to exist). Hence when this occurs, either $i(w)$ decreases by one unit at $w_1$ (since the sequence of roots is shifted to the left at $w_1$), or just before $w_1$, $ u^{{\rm q.s.}}(m;w)$ is already the smallest root and continues being the smallest root at $w_1^+$. This shows that after a finite driving $\Delta W$ (which however diverges as $m\to 0$), the quasi-static process {\it follows the smallest root of the equation} (\ref{Eq:SecII1:ShockVsAva3}). Hence if one starts the dynamics at $w_0 = -\infty$, the quasi-static process driven to the right $v = 0^+$, noted $u^{{\rm q.s.}}_+(m;w)$ always follows the smallest root of (\ref{Eq:SecII1:ShockVsAva3}). Similarly, the quasi-static process starting at $+ \infty$ and driven to the left with $v=0^-$ always follows the largest root of (\ref{Eq:SecII1:ShockVsAva3}). Hence there are two canonical quasi-static processes and the shock process. They all follow a sequence of roots of the same equation and
\be \label{Eq:SecII1:ShockVsAva4}
u^{{\rm q.s.}}_+(m;w) = u_{i_1(w)}(w) \quad , \quad u^{{\rm q.s.}}_-(m;w) = u_{i_{n(w)}(w)}(w)  \quad , \quad u^{{\rm shock}}(m;w) = u_{i_{{\rm shock}}(w)}(w)  \ssp .
\ee
where $\forall w$, $u_{i_{{\rm shock}}(w)}(w)$ is the root with the smallest energy. In general there is no symmetry between these different jump processes that would allow to get one from a simple translation/reflection of another. While $u^{{\rm shock}}(m;w)$ always follows the ground state of the system, in general $u^{{\rm q.s.}}_+(m;w)$ visits a sequence of {\it metastable states}. This sequence of states is sometimes referred to in the literature, especially in the case of interfaces, as the Middleton states \cite{Middleton1992}. An interesting consequence/characterization of this is related to the irreversibility of the quasi-static process.\\

{\it Dissipation of energy and hysteresis in the avalanche process}

\begin{figure}
\centerline{\includegraphics[width=9.0cm]{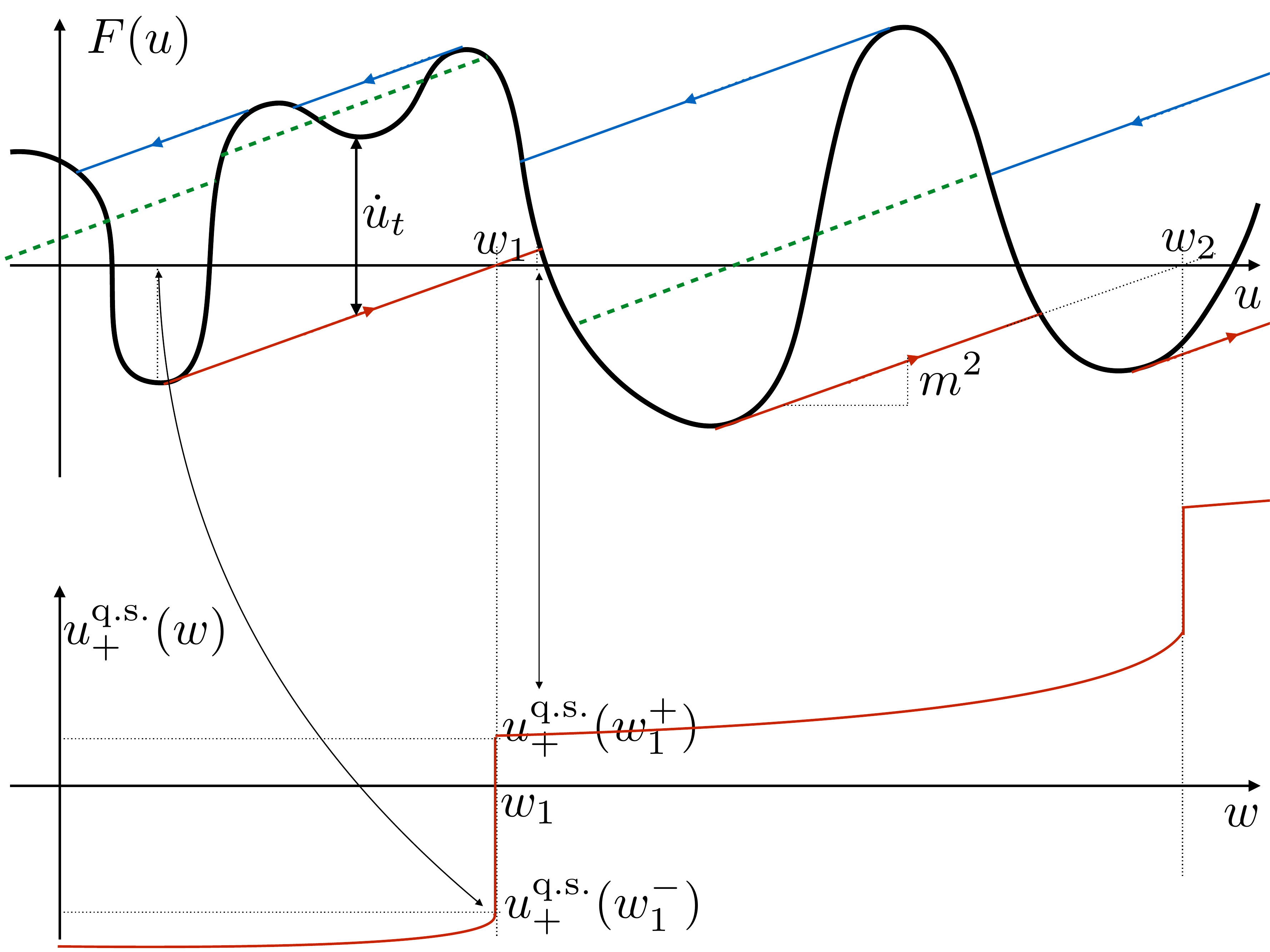} } 
\caption{ Top: The different jump processes  (here smoothened by a non zero $m$) of a particle in a force landscape $F(u)$ (black line). The forward quasi-static process $u^{{\rm q.s.}}_+(m;w)$ follows the smallest root of the equation $m^2 (u-w) = F(u)$ (red construction above). When the root ceases to exist, $u^{{\rm q.s.}}_+(m;w)$ jumps to the smallest one on the right (such as the shock at $w_1$ above). During the jumps of the forward quasi-static process, the particle moves with a velocity equal to $m^2(w_k - u_t) + F(u_t)$, that is the difference of height between the red line and $F(u)$. During such a jump, the particle thus dissipates an energy equal to the area between the red line and $F(u)$. In the backward quasi-static process (blue construction above), the particle follows a different sequence of metastable states. In the static shock-process (green-dashed construction above), the particle jumps between ground states at the same-energy. Hence, during a jump in the static shock process, the algebraic area enclosed in between the green-dashed line and $F(u)$ is 0. Dragging slowly the particle from $-\infty$ to $+\infty$ and back, the dissipated energy (`the area of the hysteresis loop') is equal to the sum of the areas enclosed in between the blue and black line and red and black line. Bottom: Forward quasi-static process deduced from the top picture. Here two avalanches at $w_1$ and $w_2$ are visible.}
\label{fig:ShocksVsAva}
\end{figure}

Let us first remark that, at the point of a shock in the static ground state of the particle, the total energy of the particle is conserved: $ \cH_{w_k}[u^{{\rm shock}}(m;w_k^+)] =  \cH_{w_k}[u^{{\rm shock}}(m;w_k^-)]$. This is true since shocks between ground states precisely occur at the position where the latter is degenerate. On the other hand in the dynamics, one should not forget that between the beginning and the end of an avalanche $u^{{\rm q.s.}}_+(m;w_k) = u_0$ and $u^{{\rm q.s.}}_+(m;w_k^+) = u_0 + S$, $S>0$, the dynamics of the particle actually (\ref{Eq:SecII2:ToyAvaABBM2})  plays a role. During the shock $vt = w_k$ and the dynamics is
\bea \label{Eq:SecII1:ShockVsAva5}
\eta \partial_t u_t = m^2(w_k - u_t) + F(u_t) \ .
\eea
Multiplying by $\partial_t u_t$ and integrating between the beginning $t_0$ and the end $t_f$ of the shock one obtains
\bea \label{Eq:SecII1:ShockVsAva6}
\eta \int_{t_0}^{t_f} (\partial_t u_t )^2 = \cH_{w_k}[u_0]-\cH_{w_k}[u_0+S] >0    \ssp .
\eea
Note that the left-hand side has no reasons to vanish in the limit $m \to 0$. This shows that there is a dissipation of energy in the avalanche process (graphically it can be represented as an area as in Fig.~\ref{fig:ShocksVsAva}). In a protocol where one drags slowly the particle from $-\infty$ to $+\infty$ and back, the particle first follows the forward quasi-static process $u^{{\rm q.s.}}_+(m;w)$, and then the backward quasi-static process $u^{{\rm q.s.}}_-(m;w)$. These are different, see Fig.~\ref{fig:ShocksVsAva}, and the system exhibits {\it hysteresis} (we refer to \cite{DobrinevskiPhD} for a study of avalanches on the hysteresis loop of the ABBM model).

\section[Avalanches for an interface]{Shocks and Avalanches of elastic interfaces}\label{sec:ShocksAndAvaForElasticInterfaces}

In this section we discuss the generalization of the notion of shocks and avalanches introduced in the previous section for models of particle to $d$-dimensional interfaces. We begin with the case of shocks in Sec.~\ref{subsec:shocks}, and study the case of avalanches in Sec.~\ref{subsec:avalanches}.

\subsection{Shocks for an interface} \label{subsec:shocks}

\subsubsection{Introduction}

Let us begin with the notion of shocks for an interface. We consider a $d$-dimensional elastic interface $u: x \in \JR^d \to u_x \in \JR$ with elasticity of range $\gamma$ pinned by an harmonic well at the position $w \in \JR$ and subject to a `nice' random potential $V(x,u_x)$ with short-range correlations (by nice we mean as discussed in Sec.~\ref{subsec:SecI1:DisHam}). The Hamiltonian is thus
\bea \label{Eq:SecII2:GSInitHam}
\cH_{V,w}[u]:= \frac{1}{2} \int_{x,y} g_{x,y}^{-1} (u_x- w) (u_y -w) + \int_x V(x,u_x) \ssp ,
\eea
where we recall $g_{x,y}^{-1}= \int_{q} e^{i q (x-y)} \sqrt{q^2 + \mu^2}$ with $\mu >0$. Here and throughout the rest of the manuscript $\mu$ is thought of as small (i.e. $\ell_{\mu}=1/\mu$ is very large compared to all eventual microscopic scales of the models, as e.g. the Larkin length (\ref{Eq:SecII2:Larkin4})), although $\ell_{\mu}$ is kept small compared to $L$ (in order not to feel the boundary conditions). As seen in Sec.~\ref{Sec:StaticPhaseDiagram}, for $d < 2 \gamma$, the ground state of the interface at fixed $w$,
\bea \label{Eq:SecII2:GSdef}
u_x(w) := {\rm argmin}_{u_x :  \JR^d \to \JR} \cH_{V,w}[u] \ssp ,
\eea
is rough with a non zero roughness exponent $\zeta_s > 0$:
\bea \label{Eq:SecII2:ScalingGS}
\overline{ (u_x(w) -u_{x'}(w))^2} \sim |x-x'|^{2 \zeta_s} \quad , \quad \text{for} \ssp |x-x'| \leq \ell_{\mu} = 1/\mu \ssp .
\eea
Here again the fact that the large scale cutoff scale $\ell_{\mu}$ is equal to the one of the pure theory $\ell_{\mu}=1/\mu$ is a consequence of the Statistical-Tilt-Symmetry (STS) of the problem, of which we postpone the discussion to Sec.~\ref{subsec:FRGStatic}. What happens to the ground state as a function of $w$? Although it cannot be proved in all generality, as in the $d=0$ problem of a particle discussed in the last section, one expects the ground state of the interface to exhibit jumps at an ensemble of discrete locations $\{w_i , i \in \JZ \}$ (whose density increases with the system size as $L^d$) as $w$ is changed (at these precise points the ground state (\ref{Eq:SecII2:GSdef}) is ill-defined since it is actually degenerate). This was confirmed numerically in e.g. \cite{MiddletonLeDoussalWiese2006,LeDoussalMiddletonWiese2008}. Hence we write, $\forall i \in \JZ$,
\bea
u_x(w_i^+)= u_x(w_i^-) + S^{(i)}_x  \ssp .
\eea
 Here $S^{(i)}_x$ is the {\it local size at $x$} of the $i^{th}$ avalanche. Using a stability argument it is trivial to show that $u_x(w)$ is strictly increasing as a function of $w$ and hence $S^{(i)}_x \geq 0$. The {\it sequence of shocks} $(w_i , S^{(i)}_x)$ is random and characterizing its statistical properties is one of the main problems studied in this thesis. In the study of shocks for particles on the real line of the previous section, we saw that shocks only occurred for a small enough confinement. Using FRG we will see in Sec.~\ref{sec:ShocksWithFRG} that a similar property holds for interfaces, and that shocks only occur for $\mu \leq \mu_c $ where $\mu_c$ is linked to the Larkin length (see Sec.~\ref{subsec:StaticPhaseDiagram}) as $\mu_c:=1/L_c$. Similarly, for non-zero $\mu$, as for the $d=0$ case, one expects the motion of the interface to also contain some smooth part. We will discuss later the appropriate scaling limit that one should take in order for $u_x(w)$ to be a pure jump process. Before we continue let us write here
\medskip

{\it A word of caution}

Let us stress that most of what we will say about shocks and avalanches for disordered elastic interfaces relies on a large number of unproven assumptions. The notion of shocks and avalanches can be seen as a phenomenological picture and it is rather hard to see it emerge from the theory `from first principles' (except e.g. in $d=0$ as in the previous section). It is mostly built from studies of toy models and numerical simulations. However as we will see, it is a very useful phenomenological picture as it allows to efficiently interpret -and is consistent with- the output of the calculations of the Functional Renormalization Group. In this sense, the existence of shocks can be seen as a hypothesis, or an ansatz, that will be plugged into the theory, and the consistency of the ansatz will always have to be verified.

\subsubsection{Shock observables and scaling}

 A few important shock observables are (i) the {\it lateral extension of the shocks} $\ell^{(i)}$; (ii) its total size $S^{(i)}$ and (iii) its local size at $x$, $S^{(i)}_x$.\\
(i) The lateral extension $\ell^{(i)}$ denotes the diameter of the domain $x \in \JR^d$ where $S^{(i)}_x$ is non-zero. Note that in general it is actually not obvious that the latter is not infinite (this was recently shown in a specific model, the Brownian Force Model with SR elasticity in $d=1$, defined below, in \cite{DelormeLeDoussalWiese2016}), but here we will assume that it is so, and this eventually implies that there is some small cutoff scale $\delta u$ such that by $S^{(i)}_x = 0$ we mean $S^{(i)}_x \leq \delta u$. \\
(ii) The total size of the shock is defined as $S^{(i)}:=\int_x S^{(i)}_x$. Let us now introduce the {\it density of avalanche sizes} $\rho(S)$, defined as 
\bea \label{Eq:SecII2:DefrhoS}
\rho(S):= \overline{ \sum_i \delta(S-S^{(i)}) \delta(w-w_i)} \ssp  .
\eea
The latter does not depend on $w$ for a statistically translationally invariant disorder. $\rho_0 = \int_{0}^{\infty} dS \rho(S)$ is the mean number of shocks per unit of $w$, and 
\bea \label{Eq:SecII2:DefPS}
P(S):=\frac{1}{\rho_0} \rho(S)
\eea
 is the normalized probability distribution function (PDF) of avalanche sizes. In the following we will denote $\langle \rangle_{\rho}$ and $\langle \rangle_{P}$ the average with respect to $\rho$ and $P$.

\medskip

Let us now discuss some properties of the avalanche observable previously defined.

\medskip
{\it Scaling of avalanches} \\
Since two points $x$ and $x'$ with $x- x' \geq \ell_\mu$ are essentially statistically independent, one expects not to observe avalanches with an extension much larger than $\ell_{\mu}$: $\ell_{\mu}$ is a large scale cutoff for the avalanches lateral extension probability distribution function (PDF) $P(\ell)$. On the other hand in the regime of lengths smaller than $\ell_{\mu}$, one expects scale invariance to hold. That is, reintroducing an eventual short scale cutoff $\ell_0$, {\it no} length scale should have any influence for $\ell_0 \ll |x-x'| \ll \ell_{\mu}$). This leads us to the scaling hypothesis that in the scaling regime $P(x \ell)/P(\ell)$ does not depend on $\ell$, i.e. it is a function of $x$, $f(x) \leq 0$. It is easily seen that $f(x_1x_2) = f(x_1) f(x_2)$ and is $f$ is continuous, it well known that $f(x)$ must be a power-law. Hence we expect, in the scaling regime that
\bea \label{Eq:SecII2:Pofl}
\ell_0 \ll \ell \ll \ell_{\mu} \Longrightarrow  P(\ell) \sim \frac{1}{\ell^{\tau_\ell}} \ssp .
\eea
i.e. $P(\ell)$ is a power-law in the scaling regime characterized by an exponent $\tau_\ell$. Since both ground states $u_x(w_i^+)$ and $u_x(w_i^-)$ are statistically equivalent and scale as in (\ref{Eq:SecII2:ScalingGS}), one also expects the local shape $S_x $ to scale as $|x-x_0|^{\zeta_s}$ where $x_0$ denotes a point on the border of the avalanche and the scaling is expected to hold for $ \ell_0 \ll |x-x_0| \ll \ell$. Hence the local size well inside the avalanche $S_x \sim \ell^{ \zeta_s} $ must be power-law distributed as, in the scaling regime $\ell_0^{\zeta_s} \ll S_x \ll \ell_{\mu}^{\zeta_s}$,
\bea \label{Eq:SecII2:PofSx}
P(S_x) \sim \frac{1}{S_x^{\tau_{S}^{{\rm loc}}} } \quad , \quad \tau_{S}^{{\rm loc}} := 1 + \frac{\tau_{\ell} - 1}{\zeta_s} \ssp  .
\eea
Similarly, one expects the total size to scale as $S \sim \ell^{d+\zeta_s}$ and to be distributed as, in the scaling regime
\bea \label{Eq:SecII2:ScalingRegimeS}
S_0 \sim \ell_0^{-d-\zeta} \ll S \ll S_\mu \sim \mu^{-d-\zeta}
\eea
 one expects
\bea \label{Eq:SecII2:PofS}
P(S) \sim \frac{1}{S^{\tau_{S}}} \quad , \quad \tau_{S} := 1 + \frac{\tau_{\ell} - 1}{d + \zeta_s} \ssp  .
\eea
Hence, as $\mu \to 0$ avalanches become larger and larger. It is thus expected that the appropriate scaling to obtain a non-trivial pure-jump process is to take
\bea \label{Eq:SecII2:Rescalingu}
\tilde{u}_{\tilde x} (\tilde w) && := \lim_{\mu \to 0} \mu^{\zeta_s} u_{x=\mu^{-1} \tilde x}(  w= \mu^{-\zeta_s} \tilde w ) \nn \\
&& = cst  + \sum_{i \in \JZ} \theta(\tilde w - \tilde w_i ) \tilde{S}^{(i)}_x \ssp .
\eea
 This a posteriori justifies the notation $\zeta$ used in (\ref{Eq:SecII2:ToyAvaJump}). In these units the large scale cutoff on the total size of avalanches $\tilde S = \int_x \tilde{S}_x$ is now of order $1$, while the low scale cutoff is now of order $\mu^{d + \zeta}$. In the following we will temporarily ignore the rescaling (\ref{Eq:SecII2:Rescalingu}) and do as if $u_x(w)$ itself was performing a jump process:
\bea \label{Eq:SecII2:uJump}
u_x(w) = cst  + \sum_{i \in \JZ} \theta(w -  w_i ) S^{(i)}_x \ssp .
\eea
Knowing that in doing so we are ignoring some smooth parts (typically of order $O(\mu^{\gamma})$, see (\ref{Eq:SecII2:GSInitHam})) that disappear in the scaling limit (\ref{Eq:SecII2:Rescalingu}). Using the fact that $\overline{u_x(w)}=w$ and using the definition (\ref{Eq:SecII2:DefrhoS}), we obtain the first moment of the avalanche total size density:
\bea
\langle S \rangle_{\rho} = L^d \ssp .
\eea

As was shown in (\ref{Eq:SecII2:Pofl}), (\ref{Eq:SecII2:PofSx}) and (\ref{Eq:SecII2:PofS}), the different power-law exponents of avalanche observable distributions are not independent. Rather they are linked to one another by scaling relations involving the roughness exponent $\zeta_s$. As for the roughness exponent $\zeta_s$, an important question is to understand whether are not the power-law exponents defined above are universal or not, and how many universality classes they are. A conjecture by Narayan and Fisher (NF), originally proposed in \cite{NarayanDSFisher1993a} in the context of avalanches at depinning, actually states that the exponents can all be obtained from the sole knowledge of $\zeta_s$. If the latter is true, it means that there is a single critical exponent governing both the static ground state of an elastic interface in a disorder media (that can be measured using a snapshot of the interface), and the power law exponents (that describe the complex shock process of switches between the ground states of the interface). This also implies that there are exactly as many universality classes of shocks as there are universality classes for the statics of disordered elastic interfaces. Let us derive here the NF conjecture following e.g. \cite{LeDoussalWiese2008c} (see also \cite{DobrinevskiLeDoussalWiese2014a,DobrinevskiPhD} for a closely related approach).

\subsubsection{The NF conjecture}

The NF conjecture permits to obtain the power-law exponent of the avalanche total size distribution $\tau_S$ self consistently when $ 1 < \tau_S <2$. It is based on several hypotheses. The first is that the avalanche size density obeys the following scaling form, already motivated by the previous discussion,
\bea \label{Eq:SecII2:NF1}
\rho(S) = L^{\alpha} \mu^{\rho} \frac{1}{S^{\tau_S}} f_{{\rm cut}}\left(\frac{S}{S_\mu} \right)  g_{{\rm cut}}\left(\frac{S}{S_0} \right) \ssp .
\eea
Where (i) $f_{{\rm cut}}(x)$ is a large scale cutoff function such that $f_{{\rm cut}}(x) \approx 1$ for $x \leq 1$ and $f_{{\rm cut}}(x)$ quickly decays to $0$ for $x \geq 1$ ; (ii) $g_{{\rm cut}}(x)$ is a small scale cutoff function such that $g_{{\rm cut}}(x) \approx 1$ for $x \geq 1$ and $g_{{\rm cut}}(x)$ quickly decays to $0$ for $x \leq 1$. The presence of the large scale cutoff function was already justified before. The presence of a small scale cutoff function is necessary for $\tau_S>1$ since in this case, the mean number of avalanches per unit $w$ diverges as $S_0 \to 0$: it is dominated by the small scale cutoff as
\bea \label{Eq:SecII2:NF2}
\rho_0 = \int_{0}^{\infty} dS \rho(S) \sim L^{\alpha} \mu^{\rho} S_0^{1- \tau_S} \ssp .
\eea
The low scale cutoff $S_0$ is a priori purely of microscopic origin (or set to an arbitrary value in a simulation) and does not scale with either $\mu$ or $L$. For $\tau<2$, the first moment of the avalanche size density is dominated by the large scale cutoff and
\bea \label{Eq:SecII2:NF3}
\langle S \rangle_{\rho} = L^d \sim L^{\alpha} \mu^{\rho} S_{\mu}^{2 - \tau_S} \sim L^{\alpha} \mu^{ \rho -(2-\tau_S)(d + \zeta_s) } \ssp  .
\eea
Hence we obtain from this relation
\bea \label{Eq:SecII2:NF4}
\alpha = d \quad , \quad  \tau_S = 2- \frac{\rho}{d + \zeta_s} \ssp .
\eea
The true difficulty is therefore to obtain $\rho$. The NF conjecture is that {\it the density of avalanches per unit of applied force stays constant as $\mu \to 0$}. $\rho_0$ in (\ref{Eq:SecII2:NF2}) must therefore be proportional to the applied force which scales as $\mu^{\gamma}$, and the NF conjecture thus states
\bea \label{Eq:SecII2:NF5}
\tau_S = 2 - \frac{\gamma}{d + \zeta_s} \ssp .
\eea
Although its derivation assumed here $\tau_S>1$ (if $\tau_S <1$, $\rho_0$ is convergent at small $S_0$ but dominated by the large scale cutoff as $\rho_0 \sim \mu^{\rho} S_\mu^{1-\tau_S})$, i.e. $\gamma/(d+ \zeta_s) < 1$, it validity might actually be more general as suggested e.g. in $\cite{FerreroFoiniGiamarchiKoltonRosso2016}$. The NF conjecture was confirmed up to one-loop accuracy by FRG calculations in \cite{LeDoussalWiese2008c,LeDoussalWiese2011b,LeDoussalWiese2012a}. Note that for models of interfaces at their upper-critical dimension $d = d_{\rm uc} = 2 \gamma$, since the fluctuations of the interface are there expected to show a logarithmic scaling (hence $\zeta_s=0$ at $d= d_{\rm uc}$) we obtain $\tau_S = 3/2$. That is, we obtain the same exponent as the avalanche size exponent in the ABBM model (see Sec.~\ref{subsec:ToyAva}). We will see later that this is not a coincidence.

\subsection{Avalanches for an interface} \label{subsec:avalanches}

In this section we now discuss the notion of avalanches for a d-dimensional interface at the depinning transition.

\subsubsection{Alternative approach to the depinning transition and avalanches}
To discuss the notion of avalanches at the depinning transition we will actually not discuss the dynamics described by the equation of motion (\ref{Eq:SecII2:dep1bis}) (elastic interface driven by a force $f$), but rather, in analogy with the static problem and as in the $d=0$ models of Sec.~\ref{subsec:ToyAva}, study the interface dynamics when driven by a harmonic well:
\bea \label{Eq:SecII2:dep2-1}
&& \eta \partial_t u_{tx} = \int_{y} g_{x,y}^{-1}(u_{ty} - w(t) ) + F(x,u_{tx}) \nn \\
&& w(t) = v t \ssp .
\eea
Here $v >0$, the hypotheses on the random force are as in Sec.~\ref{subsec:SecI1:DisHam} and as usual $g_{x,y}^{-1} = \int_q (q+\mu^2)^{\frac{\gamma}{2}} e^{iq (x-y)}$ with $\mu \geq 0$. Here again, taking an initial condition such that the velocities of the interface are positive at $t=0$, they remain so for all time and at late times the interface position field converges to a well-defined steady state (Middleton theorem \cite{Middleton1992}). Here the driving $w(t) = vt$ imposes the mean velocity of the interface, in the steady-state,
\bea \label{Eq:SecII2:dep2-2}
\overline{\partial_t u_{tx}} =v \ssp .
\eea
Since we are actually interested in describing the dynamics at the depinning transition, we will be interested in the limit $v \to 0^+$. In this limit, the interface fluctuations are expected to display a scale invariant behavior identical to (\ref{Eq:SecII2:dep5}) (only the exponents are the same, the scaling function can be different). As for our study of the static problem, this scale invariant behavior is expected to occur in the range of scales $|x-x'| \leq \ell_\mu$, $|t-t'| \leq \ell_{\mu}^z$ with $\ell_{\mu}^{-1} = 1/\mu$. As in the static problem, the fact that the large scale cutoff $\ell_{\mu}$ is equal to the one of the pure theory $\ell_{\mu}=1/\mu$ is a consequence of the Statistical-Tilt-Symmetry (STS) of the problem, of which we postpone the discussion to Sec.~\ref{subsec:FRGDep}. To study the depinning transition, we will consider the quasi-static process, as we did for the $d=0$ model of a particle (see Sec.~\ref{subsec:ToyAva}),
\bea \label{Eq:SecII2:dep2-3}
u_{x}(w) := \lim_{v \to 0^+} u_{t=w/v , x} \ssp .
\eea
The latter satisfies, $\forall w \in \JR$,
\bea \label{Eq:SecII2:dep2-4}
0 = \int_{y} g_{x,y}^{-1}(u_y(w) - w) + F(x,u_x(w)) \ssp .
\eea
Note that the static ground state of the interface studied in Sec.~\ref{subsec:shocks} is also a solution of this equation, and is by definition the one of minimum energy. As for the $d=0$ models, in the steady-state, the correct solution of (\ref{Eq:SecII2:dep2-4}) is the leftmost one (see Sec.~\ref{subsec:ToyShockVsAva}) and a priori differs from the ground state. This sequence of Middleton states can have roughness different from the one of the ground state and a result of FRG is that they are indeed different. As in the statics we expect that for $\mu \leq \mu_c$ (associated with the Larkin length as $\mu_c :=1/L_c$, see (\ref{Eq:SecII2:Larkin4})), the quasi-static process $u_x(w)$ is non-analytic and displays avalanches at discrete locations $w_k$, $k \in \JZ$. In the appropriate scaling limit $u \sim \mu^{-\zeta_d}$, $x \sim \mu^{-1}$ and $w \sim \mu^{-\zeta_d}$, it is expected to become a pure jump process:
\bea \label{Eq:SecII2:dep2-5}
u_x(w) = cst + \sum_{k \in \JZ} \theta(w-w_k) S^{(k)}_x  \ssp .
\eea
During an avalanche, the interface dynamics plays an important role and is described by (\ref{Eq:SecII2:dep2-1}) with $vt \to w_k^+$. This is a clear difference with the statics, which is in particular responsible for the fact that energy is dissipated during the quasi-static process (see Sec.~\ref{subsec:ToyShockVsAva} for the complete discussion in $d=0$). Note that since the sequence of metastable states visited by the interface in this forward quasi-static process is a priori different from the sequence of ground states in the static problem (in particular these states have different roughness exponent), the sequence of avalanches is also different. As for the static problem, we however expect similar scaling and universality to occur in avalanches in the interface dynamics, with different exponents however. Interesting observables associated with the $k^{th}$ avalanches are (i) as in the static case: $w_k$ the location of the avalanche, $\ell^{(k)}$ the lateral extension of the avalanche, $S^{(k)}_x$ the local size of the avalanche at $x$, $S^{(k)} = \int_x S^{(k)}_x$ the total size of the avalanche; (ii) observables that only exist in the dynamics: $T^{(k)}$ the duration of the avalanches, $x^{(k)}_0$ the first point that becomes unstable at the beginning of the avalanche (the `seed' of the avalanche, note that this is another important difference with avalanches in the statics where this notion does not make sense), $v^{(k)}(t,x)= \partial_{t'} u_{t= w_k/v+t x}$ the velocity field inside the avalanche, $v^{(k)}$ the mean velocity inside the avalanche, $E^{(k)} = \eta \int_{tx}( \partial_t u_{tx})^2$ the energy dissipated during the avalanche. For these observables, scaling now notably imposes similar relations as in the static case and some new relations, associated with new observables and that involve the dynamic exponent $z$:
\bea \label{Eq:SecII2:dep2-6}
S_x \sim \ell^{\zeta_d} \quad , \quad S \sim \ell^{d +\zeta_d} \quad , \quad T \sim \ell^z \quad ,  \quad v \sim \ell^{\zeta_d - z} \ssp . 
\eea 
These observables are expected to be distributed with PDF with power-law behavior as in the static case. From (\ref{Eq:SecII2:dep2-6}) it is clear that the different power-law exponents are not independent. It is in this context that the Narayan-Fisher conjecture was first introduced \cite{NarayanDSFisher1993a}. It can be `shown' using the exact same arguments as in Sec.~\ref{subsec:shocks} and reads
\bea
\tau_S  = 2 - \frac{\gamma}{d + \zeta_d} \ssp .
\eea

\subsubsection{A side remark/a word of caution: avalanche power-law exponents depend on the driving }

We have up to now discussed avalanches for an elastic interface driven by a parabolic well: the driving is soft ($\mu \to 0$) and homogeneous on the system. This type of driving is known to be relevant in various experimental situations and in particular to reproduce the driving by a force right at the depinning transition, but other driving can be considered and can be relevant in other situations. Another driving that has already been considered is the case of avalanches for an interface such that the position of part of the interface is imposed to be $w$ (see e.g. \cite{AragonKoltonDoussalWieseJagla2016,PaczuskiBoettcher1996}). Here we consider the situation where an interface of internal length $L$, dimension $d$, elastic Hamiltonian (\ref{Eq:secI1:HelGen}) with $g_{q}^{-1} = |q|^{\gamma}$, in a random potential $V(x, u)$, is free to move in $\mathbb{R}$, except on a subspace $\mathbb{E}_{d_{{\rm dr}}} \in \JR^d $ of dimension $d_{{\rm dr}}$ where its position is imposed to be $w(t)=vt$. The extreme case $d_{{\rm dr}} = 0$ corresponds to an interface driven at a single point. Noting $x=(x_1 , \cdots , x_d) \in \JR^d$ the $d-$dimensional coordinates of $x$ and taking for concreteness $\mathbb{E}_{d_{{\rm dr}}} = \{ x=(x_1 , \cdots , x_{d_{{\rm dr}}}  , 0 , \cdots , 0)  \in \JR^d , (x_1 , \cdots , x_{d_{{\rm dr}}}) \in \JR^{d_{{\rm dr}}} )$ we thus study the problem
\bea
&& \eta \partial_t u_{tx} = \int_{y} g_{x,y}^{-1} u_y + F(x,u_x) \nn \\
&& u_{tx} = w(t) \quad {\rm for  } \ssp  x  \in   \mathbb{E}_{d_{{\rm dr}}}
\eea
for a very slow driving $w(t) = v t$ and $v \simeq 0^+$, and we study the forward quasi-static process $u_x(w) = \lim_{v \to 0^+} u_{t = w/v , x}$ Obviously for $x\in \mathbb{E}_{d_{{\rm dr}}}$, $u_{x}(w) = w$ is a smooth function of $w$, but for $x \notin   \mathbb{E}_{d_{{\rm dr}}}$ one still expects to observe scale invariance and some non-analytic behavior with avalanches at discrete positions $w_i$: $u_x(w_i^+)= u_x(w_i^-) + S^{(i)}_x$. Since the roughness exponent $\zeta_d$ is a bulk property of the system, it is expected that the total size of these shocks still satisfies the relation $S \sim \ell^{d + \zeta_d}$ with the same roughness exponent as before, where $\ell$ is the linear extension of the shocks in the space perpendicular to the driven space $\mathbb{E}_{d_{{\rm dr}}}^{\perp}$. Here the only large scale cutoff for the avalanche linear extension is $L$ and the large scale cutoff for the avalanche total sizes is $S_L \sim L^{d + \zeta_d}$. Defining again the avalanche size distribution as $\rho(S) := \overline{ \sum_i \delta(S-S^{(i)}) \delta(w-w_i)} $, we thus expect that it displays a power-law behavior in between two cutoff scales $S_0$ and $S_L$, as
\bea
\rho(S) = L^{\alpha} \frac{1}{S^{\tau_S}} f_{{\rm cut}}\left(\frac{S}{S_L} \right)  g_{{\rm cut}}\left(\frac{S}{S_0} \right),
\eea
where $f_{{\rm cut}}$ and $g_{{\rm cut}}$ are two scaling functions such that $f_{{\rm cut}}(x) \approx 1$ for $x \leq 1$, $f_{{\rm cut}}(x)$ quickly decays to $0$ for $x \geq 1$, $g_{{\rm cut}}(x) \approx 1$ for $x \geq 1$ and $g_{{\rm cut}}(x)$ quickly decays to zero for $x \leq 1$. Let us now apply a reasoning similar to the one used in the derivation of the NF conjecture (\ref{Eq:SecII2:NF5}). If $1<\tau_S < 2$ the mean-density of avalanches per unit $w$ is dominated by the small scale cutoff as
\bea
\rho_0 = \int_{0}^{\infty} \rho(S) dS \sim L^{\alpha } S_0^{1 - \tau_S} \ssp .
\eea
The first moment of the avalanche size distribution must still be $\langle S \rangle_{\rho} = L^d$ and is dominated by the large scale cutoff as
\bea \label{Eq:SecII2:NF2.02}
\langle S \rangle_{\rho} = L^d \sim L^{\alpha } S_L^{2- \tau_S} \sim L^{\alpha + (2 - \tau_S)(d + \zeta_s)  } \ssp .
\eea
In this setting, it is natural to think that the mean number per unit of $w$ of avalanches scales as $L^{d_{{\rm dr}}}$, implying $\alpha = d_{{\rm dr}}$. Indeed, here avalanches can only be triggered by the depinning of one of the points in the vicinity of one of the driven points $\mathbb{E}_{d_{{\rm dr}}}^{\perp}$ and the number of avalanches is thus expected to be proportional to the number of driven points. Using (\ref{Eq:SecII2:NF2.02}) we thus obtain a generalization of the Narayan-Fisher conjecture:
\bea
\tau_S = 2 - \frac{d- d_{{\rm dr}} }{d + \zeta_d} \ssp .
\eea
This relation is in agreement with the result \cite{AragonKoltonDoussalWieseJagla2016,PaczuskiBoettcher1996} for the case $d=1$ and $d_{{\rm dr}}=0$. We are unaware whether or not this general conjecture already appeared in the literature.

\section[FRG approach to shocks]{The functional renormalization group treatment of shocks in disordered elastic systems: a short review} \label{sec:ShocksWithFRG}
In this section we now review the use of the functional renormalization group to calculate shock observables. We will begin by recalling the important results of FRG for the statics of $d$-dimensional interfaces in Sec.~\ref{subsec:FRGStatic}, and in Sec.~\ref{subsec:FRGStaticSchocks} we will show how to apply FRG to the study of shocks.

\subsection{The functional renormalization group for the statics of disordered elastic interfaces} \label{subsec:FRGStatic}

\subsubsection{Introduction}

We thus consider the static problem of determining the statistical properties of the ground state of a disordered elastic interface of internal dimension $d$ in a quenched random potential $V(x,u)$
\bea \label{Eq:SecII3:FRGStat1}
u_x(w)&& := {\rm argmin}_{u : \JR^d \to \JR} \cH_{V,w}[u] \\
&& = {\rm argmin}_{u : \JR^d \to \JR} \left( \frac{1}{2} \int_{x,y} g_{x,y}^{-1}(u_x-w)(u_y-w) + \int_x V(x,u_x) \right) \ssp . \nn 
\eea
where as usual $g_{x,y}^{-1} = \int_{q} e^{iq (x-y)} (q^2 + \mu^2)^{\frac{\gamma}{2}}$. The disorder potential is chosen centered, e.g. Gaussian, with second cumulant\footnote{The fact that we simply assume here that the correlations of the potential in the internal space $x$ are described by a $\delta$ distribution, while the correlations in the external space $u$ are given by a function $R_0(u)$ even for short-range disorder notably comes from the fact that this structure is stable under renormalization. That is, starting from a bare disorder with more complex short-range correlations in internal space, at large scale it will look just as if we started from (\ref{Eq:SecII3:FRGStat2}). Conversely, starting from a bare disorder with `trivial' short-range correlations $R_0(u-u') = \sigma \delta(u-u')$, the renormalized disorder at large scale will still be short-range in $u$ space, but with a finite correlation length.}
\bea \label{Eq:SecII3:FRGStat2}
\overline{V(x,u) V(x' ,u')} =\delta^{(d)}(x-x') R_0(u-u') \ssp , 
\eea
where $R_0$ is the {\it bare disorder cumulant} that will be chosen either as associated with disorder of the random bond type, the random field type, or eventually periodic disorder $R_0(u+\Delta u ) = R_0(u)$ that can be relevant to some applications (see Sec.~\ref{subsec:SecI1:DisHam} for definitions).  In low dimensions $d \leq 2\gamma$, we know from Sec.~\ref{subsec:StaticPhaseDiagram} that the disorder is relevant at large scale, that the ground state is rough and exhibits scaling in the range of scales $|x-x'| \ll \ell_\mu:=1/\mu$, and we will be interested in the $\mu \to 0^+$ limit. The application of renormalization group ideas to disordered elastic systems has a long and rich history that we will not thoroughly review here. The phenomenon of dimensional reduction recalled in Sec.~\ref{subsec:StaticPhaseDiagram} warns us that naive perturbation theory in the disorder badly fails and one has to find a way to do better. The way out proposed by the FRG is as follows. Let us first introduce here the replicated action of the theory.

 A convenient way to perform disorder averages is to consider the replicated action of the theory (see  \cite{LeDoussal2008} for some background on replicas) for the statics at temperature $T$. Replicating the field $u_x \to u_x^a$, $a= 1 , \cdots , n$, the action is
\bea \label{Eq:SecII3:FRGStat3}
S[w;\{u_x^a \}]:= \frac{1}{2T} \sum_{a=1}^{n} \int_{x,y} g_{x,y}^{-1} (u_x^a-w) (u_y^a-w)  - \frac{1}{2T^2} \sum_{a,b=1}^{n} \int_{x} R_0(u_x^a - u_x^b) + \cdots \ssp .
\eea
here the dots indicate the eventual presence of higher cumulants of the random potential that will be generated by the renormalization procedure anyway. Physical, disorder averaged observables of the ground state (\ref{Eq:SecII3:FRGStat1}) of the Hamiltonian are obtained by taking the limit of a path integral formula
\bea \label{Eq:SecII3:FRGStat4}
\overline{ O[\{u_x(w)\}]} = \lim_{n,T \to 0} \int \prod_{a=1}^{n} O[\{u_x^1\}] \cD[u^a] e^{-S[w;\{u_x^a \}]} \ssp .
\eea

{\it Fisher's breakthrough and the development of FRG}\\
Performing the naive perturbation theory of an observable (\ref{Eq:SecII3:FRGStat4}) using an analytic $R_0(u)$ again gives the dimensional reduction result with the Larkin roughness exponent $\zeta_L = \frac{2\gamma-d}{2}$. What is even more surprising with this result is the following: expanding the even and analytic function $R_0(u)$ in (\ref{Eq:SecII3:FRGStat3}) as  $R_0(u) = \sum_{n=0}^{+ \infty} \frac{1}{2n!}C_{2n} u^{2n} $ and performing the rescaling valid at the Larkin fixed point
\bea \label{Eq:SecII3:FRGStat4c}
x = \mu^{-1} \tilde{x} \quad , \quad u_x^a = \mu^{-\zeta_L} \tilde{u}_{\tilde x = \mu x} \quad,  \quad T = \mu^{-d + \gamma - 2 \zeta_L} \ssp ,
\eea
the disordered part of the action (\ref{Eq:SecII3:FRGStat3}) is rescaled as
\bea \label{Eq:SecII3:FRGStat5c}
&& \frac{1}{2\tilde T^2} \int_{\tilde x} \sum_{a,b=1}^n \sum_{n=0}^{\infty} \mu^{\alpha_n }  \frac{1}{2n!}C_{2n} (\tilde u_{\tilde x}^a-\tilde u_{\tilde x}^b )^{2n} \nn \\
&& \alpha_n = -d -2 n \zeta_L -2 (-d + \gamma - 2 \zeta_L) = (2\gamma -d)(1-n) \ssp .
\eea
Here for $d < 2 \gamma$, $C_0$ thus flows to $0$ and is unimportant at large scale at the Larkin FP, $C_1 = R''(0)$ is left invariant as expected, but all the higher cumulants $C_{2n}$ with $n >1$ do flow and are relevant at the Larkin FP. This should lead the system to flow away from the Larkin FP. However, as observed in \cite{EfetovLarkin1977}, the contribution of all higher cumulants simplify in the calculation of observables (the dimensional reduction (DR) property). This was attributed to an underlying supersymmetric property of the FT \cite{ParisiSourlas1979}, or equivalently in a diagrammatic language, to the `mounting property' of diagrams associated with the field theory (\ref{Eq:SecII3:FRGStat3}) \cite{ChauveLeDoussal2001}. The Larkin result is however as we know incorrect (see the discussion in Sec.~\ref{subsec:StaticPhaseDiagram}). Escaping the Larkin FP using a RG procedure calls for (i) a functional RG to take into account the fact that all cumulants of the potential become simultaneously relevant for $d < 2\gamma$; (ii) a RG scheme that somehow escapes DR. The solution first noted by Fisher in \cite{DSFisher1986} is as follows. Using a one-loop Wilson's shell RG on the replicated action (\ref{Eq:SecII3:FRGStat3}) with $\mu =0$ but with a UV cutoff $\Lambda>0$ that is sent to $\Lambda_l = \Lambda e^{-l}$ to renormalize the complete function $R_0(u)$ the bare cumulant $R_0(u)$ is renormalized into a function $R_l(u)$ (it corresponds to the cumulant of a renormalized disorder seen at large scale by the manifold, see below). Following the RG flow, remarkably, the function $R_l(u)$ becomes {\it non-analytic at a finite scale} $l=l_c< \infty$: the function $\Delta_l(u) = - R_l''(u)$ exhibits a cusp around $0$: $\Delta_l(u) - \Delta_l(0) \sim \Delta_l'(0^+) |u| + O(u^2)$. The non-analyticity forbids the expansion (\ref{Eq:SecII3:FRGStat5c}) and escapes DR. Several fixed point (FP) functions $R(u)$ were found corresponding\footnote{Other FP functions exist and correspond to disorder with long range correlations $R_0(u) \sim_{u \to \infty} u^{\alpha}$.} to the three classes listed above, i.e. random bond, random field and random periodic. The found fixed point functions are of order $O(\epsilon)$ with $\epsilon = d_{{\rm uc}}- d= 2 \gamma - d$, hence allowing to compute perturbatively any observable in an expansion in the non-analytic renormalized disorder second cumulant $R_l'(u)$. It was later argued in \cite{BalentsBouchaudMezard1996} that the non-analyticity in $R_l(u)$ is related to the presence of shocks in the ground state (this will be clear below). Many developments followed this work and here we name a few: (i) refinement of the result to two-loops using perturbative RG \cite{ChauveLeDoussalWiese2000a,LeDoussalWieseChauve2003}; (ii) development of exact RG approaches \cite{ChauveLeDoussal2001,SchehrDoussal2003}; (iii) clarification of the role of the temperature \cite{BalentsLeDoussal2004,LeDoussal2008}. References for the application of FRG to the depinning transition and to avalanches will be discussed in Sec.~\ref{subsec:FRGDep}. In the following we will use the most modern approach to FRG as presented e.g. in \cite{LeDoussal2008} and only state the results. We refer the reader to \cite{Wiese2005,WieseLeDoussal2006} for what are probably the most pedagogical introductions to FRG.
 
\subsubsection{Definition of the different functionals and the statistical-tilt-symmetry}
 
Our preferred approach to FRG, as presented \cite{LeDoussal2008} to which we refer the reader for more details, is to study the flow of the effective action of the replicated theory as the strength of the confining well is varied from $\mu \to \infty$ to $ \mu \to 0$. In the limit $\mu \to \infty$ the fluctuations are frozen and the effective action is basically the bare action of the theory, while in the limit $\mu \to 0$ that we want to study, the effective action takes a universal scaling form. This type of approach is common in non-perturbative RG (see \cite{Delamotte2012} for a review) but here our final results will be perturbative. Although this presentation can be quite cumbersome the first time it is probably the clearest way to understand the validity and the interpretation of the main results of the FRG.  \\

{\it The renormalized disorder functional}\\
Let us first define, for each realization of the disorder $V$, the {\it renormalized disorder at the scale $\mu$ for a well centered at $w_x$, $\hat V_{\mu}[\{w_x\}]$} as
\bea \label{Eq:SecII3:FRGStat5}
e^{-\frac{1}{T} \hat V_{\mu}[\{w_x \}]} = \int \cD[u] e^{- \frac{1}{T} \left(   \frac{1}{2} \int_{x,y} g_{x,y}^{-1}(u_x-w_x)(u_y-w_y) + \int_x V(x,u_x)   \right) }
\eea
Hence here we are considering the usual theory with a well position that is now inhomogeneous in space. The renormalized disorder $\hat V_{\mu}[\{w_x\}]$ is a functional of the well position $w_x$. It converges in the limit $T \to 0$ to the energy of the ground state of the Hamiltonian in the well $w$. In the limit $\mu \to \infty$, $u_x=w_x$ and $V_{\mu}[\{w_x\}] = \int_x V(x,w_x)$. More generally the renormalized disorder combines the effects of the elasticity, the thermal fluctuations and the disorder and we think of it as a renormalized disorder seen by the interface on a scale $\ell_{\mu}=1/\mu$. \\

{\it The $W$ functional}\\
On the other hand, let us consider, in the theory with $w_x =0$, the {\it generating functional for connected correlations in the replicated field theory}:
\bea\label{Eq:SecII3:FRGStat6}
e^{W_\mu[\{ j_x^a \}] } :=  \int \prod_{a=1}^{n} \cD[u^a] e^{-S[0,\{u_x^a \}] + \sum_a \int_x j_x^a u_x^a } 
\eea
This functional is a standard object considered in field theory and is the sum of all connected diagrams. Writing $\langle \rangle_{S[0,.]}$ the average with respect to the replicated action with $w=0$ (\ref{Eq:SecII3:FRGStat3}), the connected correlations $G_{x_1, \cdots , x_n}^{a_1, \cdots a_n} := \langle u_{x_1}^{a_1} \cdots u_{x_n}^{a_n} \rangle_{S[0,.]}$ appear in the polynomial expansion of $W_{\mu}$ as
\be \label{Eq:SecII3:FRGStat7}
W_{\mu} [\{ j_x^a \}] = W_\mu[0] + \frac{1}{2} \sum_{ab} \int_{x,y} G_{x,y}^{a,b} j_x^a j_y^b + \frac{1}{4!} \sum_{abcd} \int_{x,y,z,t} G_{x,y,z,t}^{a,b,c,d} j_x^a j_y^b j_z^c j_t^d +O(j^6) \ssp .
\ee
And connected correlations of the replicated field are obtained by applying functional derivatives to $W_{\mu}$. $W_{\mu}$ is an even functional of $j_x$ by statistical parity invariance of the disorder. Here we assumed that it is analytic $\forall T>0$.\\

{\it The STS symmetry and the form of $G_{x,y}^{a,b}$} \\
The Statistical-Tilt-Symmetry originates from the statistical translational invariance of the disorder. The latter implies, for an arbitrary function $\phi_x$ (constant in replica space)
\bea \label{Eq:SecII3:FRGStat8}
S[0 , \{ u_x^a  + \phi_x\}] = S[0 , \{ u_x^a \}] + \frac{1}{T} \int_{x,y} g_{x,y} u_x^a \phi_y +  \frac{1}{2T} \sum_{a,b}\int_{x,y} g_{x,y} \phi_x \phi_y \ssp .
\eea
This implies the identity for $W_{\mu} [\{ j_x^a \}]$, using
 $j_x =\frac{1}{T} \int_y g_{x,y}^{-1} \phi_y $,
\bea \label{Eq:SecII3:FRGStat9}
W_{\mu} [\{ j_x^a + j_x \} ] = W_{\mu} [\{ j_x^a  \}] +T \sum_a \int_{x,y} g_{x,y}^{-1} j_x^a j_y + n \frac{T}{2} \int_{x,y} g_{x,y}^{-1} j_x j_y \ssp .
\eea
Taking a derivative with respect to $j_x$ at $j_x = 0$ we obtain
\bea \label{Eq:SecII3:FRGStat9bb}
\sum_a  \frac{\delta}{\delta j_x^a}W_{\mu} [\{ j_x^a\}] = T \sum_a \int_{y} g_{x,y} j_y^a 
\eea

This being valid $\forall n \in \JN$ and for any sources $j^a_x$, this implies an infinite series of identities for the `coefficients' of the series expansion of $W_{\mu} [\{ j_x^a\}]$. In particular, it implies for the quadratic part
\bea \label{Eq:SecII3:FRGStat10}
\sum_{b} G_{x,y}^{ab} = T g_{x,y} \ssp ,
\eea
This is the same result as the one that would be obtained in the pure theory: {\it the sum of the connected correlations $\sum_{b} \langle u_x^a u_y ^b \rangle_{S[0, .]} =T g_{x,y}$ is not modified by the disorder.} This indicates that connected correlations decay as $e^{—|x-y|/\ell_{\mu}}$ with $\ell_\mu:=1/\mu$. The parameter $\mu$ is not modified by the renormalization. This fact was already heavily used before. Other relations extracted from (\ref{Eq:SecII3:FRGStat9bb}) are $\sum_{a_{2 n }} G_{x_1 , \cdots , x_{2n}}^{a_1, \cdots, a_{2n}} =0$\\

{\it Relating the $W_{\mu}$ functional and the renormalized disorder $V_\mu$.}\\
It is an elementary calculation to show that the $W_{\mu}$ functional is related to the renormalized disorder functional defined in (\ref{Eq:SecII3:FRGStat5}) as
\bea\label{Eq:SecII3:FRGStat7b}
e^{W_\mu[\{ j_x^a = \frac{1}{T} \int_{y} g_{x,y}^{-1} w_x^a \} ] }  e^{-\frac{1}{2T} \sum_a \int_{x,y} g_{x,y}^{-1} w_x w_y } = \overline{ \prod_{a=1}^n e^{ - \frac{1}{T} \hat V_{\mu}[\{w_x^a \}] }} \ssp .
\eea
This important relation (its consequences will be shown below) was first shown in \cite{LeDoussal2006b,LeDoussal2008}. Expanding in cumulants the right hand side of (\ref{Eq:SecII3:FRGStat7}), and in replica sums the $W_\mu$ functional, one sees that {\it the expansion in cumulants of the renormalized disorder functional exactly gives $W_\mu[\{ j_x^a = \frac{1}{T} \int_{y} g_{x,y}^{-1} w_x^a \} ]$} as, noting $\tilde{W}_{\mu} [\{w_x^a\}] :=W_\mu[\{ j_x^a = \frac{1}{T} \int_{y} g_{x,y}^{-1} w_x^a \} ]$
\bea \label{Eq:SecII3:FRGStat8b}
\tilde{W}_{\mu} [\{w_x^a\}] && = \tilde{W}_{\mu} [0] + \frac{1}{2T} \sum_{ab} \int_{x,y} g_{x,y}^{-1} w_x^a w_y^b  + \frac{1}{2T^2}\sum_{ab} \hat R[\{w_x^{a,b}\}]   + \nn \\
&& +  \sum_{m \geq 3} \frac{1}{n! T^m} \sum_{a_1, \cdots , a_m} \hat S^{(m)}[ \{ w_{x_1}^{a_1} \} , \cdots , \{ w_{x_m}^{a_m} \}] \ssp .
\eea
where we have introduced the notation $w_x^{a,b} = w_x^a - w_x^b$ and
\bea \label{Eq:SecII3:FRGStat9b}
&& \tilde{W}_{\mu} [0] = - \frac{n}{T} \overline{\hat V_{\mu}[\{w_x^a \}]} \\
&& \hat R[\{w_x^{a,b}\}] := \overline{\hat V_{\mu}[\{w_x^a \}] \hat V_{\mu}[\{w_x^c \}]}^c  \\
&&  \hat S^{(m)}[ \{ w_{x_1}^{a_1} \} , \cdots , \{ w_{x_m}^{a_m} \}]  := (-1)^m \overline{\hat V_{\mu}[\{w_x^1 \}] \cdots \hat V_{\mu}[\{w_x^m \}]}^c 
\eea
Here STS implies that $\tilde{W}_{\mu} [0] $ does not depend on $\{w_x^a \}$, $\hat R[\{w_x^{a,b}\}]$ only depends on the difference $w_x^a -w_x^b$, and similarly the higher order cumulants of the renormalized disorder satisfy $\hat S^{(m)}[ \{ w_{x_1}^{a_1} + w_x \} , \cdots , \{ w_{x_m}^{a_m}  +w_x \}] = \hat S^{(m)}[ \{ w_{x_1}^{a_1} \} , \cdots , \{ w_{x_m}^{a_m} \}]$. Note that this expansion in cumulants is not trivially related to the expansion in $j_x^a$ performed in (\ref{Eq:SecII3:FRGStat7}) (e.g. $\hat R[\{w_x^{a,b}\}]$ itself has an expansion in $w$). \\

{\it The effective action}\\
The last functional to introduce before we give the important results of the FRG is {\it the effective action functional} $\Gamma_{\mu}[\{u_x^a\}]$. As usual in Field-Theory it is defined as the Legendre transform of the $W_{\mu}$ functional:
\bea \label{Eq:SecII3:FRGStat10b}
&& \Gamma_{\mu}[\{u_x^a\}] = - W_{\mu}[\{j_x^a\}] + \sum_a \int_x j_x^a u_x^a  = - \tilde{W}_{\mu}[\{w_x^a\}] + \frac{1}{T} \int_{x,y} g_{x,y}^{-1} u_x^a w_y^a \nn \\ 
&& u_x^a = \frac{\delta W_{\mu} [\{j_y^b\}]}{\delta u_x^a} = T \int_z g_{x,z} \frac{\partial \tilde{W}_\mu [\{w_y^b\}]}{ \delta w_x^a} \ssp .
\eea
In terms of diagrams it corresponds to the sum of $1$-particle irreducible diagrams generated by the action $S[0;\{u_x^a\}]$. The physical, disordered averaged observables are contained in $ W_{\mu}$ which can be obtained from the effective action $\Gamma_{\mu}$ by inverting the Legendre transform (the latter is actually an involution). This was performed in \cite{LeDoussal2008}. In particular, it is shown that $\Gamma_{\mu}[\{u_x^a\}]$ admits an expansion as
\bea \label{Eq:SecII3:FRGStat11}
\Gamma_{\mu}[\{u_x^a\}]= && \Gamma_{\mu} [0] + \frac{1}{2T} \sum_{ab} \int_{x,y} g_{x,y}^{-1} u_x^a u_y^b  - \frac{1}{2T^2}\sum_{ab} R[\{u_x^{a,b}\}]   + \nn \\
&& -  \sum_{m \geq 3} \frac{1}{n! T^m} \sum_{a_1, \cdots , a_m} S^{(m)}[ \{ u_{x_1}^{a_1} \} , \cdots , \{ u_{x_m}^{a_m} \}]  \ssp .
\eea
And inverting the Legendre transform gives `nice' relations between the cumulants of the renormalized disorder and the `Gamma cumulants'. In particular we have $\Gamma_{\mu}[0]=-\tilde{W}_{\mu} [0]$ and what will turn out to be the most important relation, the equality between functionals
\bea \label{Eq:SecII3:FRGStat12}
R[\{w_x\}] = \hat  R[\{w_x\}] \ssp .
\eea

\subsubsection{Exact RG approach: The Morris-Wetterich equation and the scaling hypothesis}
Let us first formulate the FRG result using an exact RG formalism and later connect it to perturbative approaches. To formulate the exact RG equation, it is convenient to define a slightly different effective action functional 
\bea \label{Eq:SecII3:FRGStat13}
\hat \Gamma_{\mu}[\{u_x^a\}]=\Gamma_{\mu}[\{u_x^a\}] -\frac{1}{2T} \sum_{ab} \int_{x,y} g_{x,y}^{-1} u_x^a u_y^b  \ssp .
\eea
The latter behaves well even for $\mu \to \infty$. Indeed, for $\mu \to \infty$, it is clear that the functional $W_\mu$ in (\ref{Eq:SecII3:FRGStat6}) can be evaluated using a saddle-point calculation around $u_{x}^a = w_x^a$. Inserting the result into the Legendre transform (\ref{Eq:SecII3:FRGStat10}) shows that $\lim_{\mu \to \infty} \hat \Gamma_{\mu}[\{u_x^a\}] $ is exactly the expansion in cumulants of the bare disorder $V(x,u)$:
\bea \label{Eq:SecII3:FRGStat14}
\lim_{\mu \to \infty} \hat \Gamma_{\mu}[\{u_x^a\}] = - \frac{1}{2T^2} \sum_{a,b=1}^{n} \int_{x} R_0(u_x^a - u_x^b) + \cdots \ssp,
\eea
where again we have added dots to signify the presence of higher order cumulants if the initial disorder is non-Gaussian\footnote{Note here that something bad can happen if the bare disorder has fat tails since then the expansion in cumulants is ill-defined.}. As a function of $\mu$, the effective action functional satisfies the Morris-Wetterich equation \cite{Wetterich1993,Morris1994,SchehrDoussal2003,LeDoussal2008}
\bea \label{Eq:SecII3:FRGStat15}
 -\mu \partial_{\mu} \hat \Gamma_{\mu}[ \{u_x^a\}] =  \beta[\hat \Gamma_{\mu}] [ \{u_x^a\}] \ssp ,
\eea
where the functional $\beta$ function is 
\be \label{Eq:SecII3:FRGStat16}
\beta[\hat \Gamma_{\mu}] [ \{u_x^a\}] = - \frac{1}{2} \sum_{a} \int_{x,y,z} \mu (\partial_\mu g_{x,y}) g_{y,z}^{-1} \left(G^{-1} \right)^{a,a}_{z,x}  \quad , \quad  G^{a,b}_{x,y} = 1 - T \int_{z} g_{x,z} \frac{\delta^2  \hat \Gamma_{\mu}}{\delta u_z^a u_y^b} \ssp .
\ee
At our level of rigor (\ref{Eq:SecII3:FRGStat14}) is an (awfully complicated) well posed problem: we have an initial condition at $\mu \to \infty$\footnote{If a small scale cutoff $a$ is assumed then the initial condition (\ref{Eq:SecII3:FRGStat14}) holds at $\mu = 1/a$.} for a differential equation that we want to solve. Actually we do not want to follow completely the RG flow from $\mu \to \infty$ to $\mu = 0^+$, but rather, although it is not obvious, show that close to $\mu =0$, an appropriately rescaled version of $ \hat \Gamma_{\mu}[ \{u_x^a\}]$ tends to a fixed point functional. More precisely, reintroducing explicitly $T$ as a parameter and rescaling,
\be \label{Eq:SecII3:FRGStat16b}
x = \mu^{-1} \tilde{x} \quad,  \quad u_x =  \mu^{-\zeta_s} \tilde{u}_{\tilde x}  \quad , \quad T \sim \mu^{- \theta} \tilde{T} 
\ee
the scaling hypothesis can be phrased in an unambiguous way as follows. We require that the effective action in the `tilde' variables, that describe the large scale physics of the original theory,
\bea \label{Eq:SecII3:FRGStat17b}
\tilde{\hat \Gamma}_{\mu}[\tilde T ; \{ \tilde{u}_{\tilde x} \} ] = \Gamma_{\mu} [ T=\mu^{-\theta} \Tilde{T} ,  \{ u_x = \mu^{\zeta_s} \tilde{u}_{\tilde{x} = \mu x} \} ]
\eea
converges, as $\mu \to 0$, to a constant, well-defined action:
\bea \label{Eq:SecII3:FRGStat18b}
\lim_{\mu \to 0} \tilde{\hat \Gamma}_{\mu}[\tilde T ; \{  \tilde{u}_{\tilde x} \} ] =\tilde{\hat \Gamma}^*[\tilde T ; \{ \tilde{u}_{\tilde x} \} ] \ssp .
\eea
Inserting the scaling form (\ref{Eq:SecII3:FRGStat17b}) into the RG equation (\ref{Eq:SecII3:FRGStat15}), $\tilde{\hat \Gamma}_{\mu}[\tilde T ; \{ \tilde{u}_{\tilde x} \} ]$ itself satisfies a Morris-Wetterich equation with a rescaled $\beta$ function:
\bea \label{Eq:SecII3:FRGStat19b}
- \mu \partial_\mu \tilde{\hat \Gamma}_{\mu}[\tilde T ; \{ \tilde{u}_{\tilde x} \} ] = \tilde{\beta}[\tilde{\hat \Gamma}_{\mu} ][ \{ \tilde{u}_{\tilde x} \} ] \ssp .
\eea
And since $-\mu \partial_{\mu} = - \partial_{\log(\mu)}$ and $\log(\mu) \to_{\mu \to 0} - \infty$, (\ref{Eq:SecII3:FRGStat18b}) actually implies that the limiting functional is a fixed point of the rescaled Beta function
\bea \label{Eq:SecII3:FRGStat20b}
\tilde{\beta}[\tilde{\hat \Gamma}^* ] = 0 \ssp .
\eea 
The hypothesis (\ref{Eq:SecII3:FRGStat18b}) is thus rather strong and allows us not to follow completely the RG flow (since we are thus only interested in the fixed point), while defining the Beta function in (\ref{Eq:SecII3:FRGStat15}) as generating the flow associated with $-\mu \partial_\mu$ (an not e.g. $\partial_{\mu}$) is not a random choice. Obviously these are all strong hypotheses (that will not be proven). Before we describe the solution of this problem at $T = 0$, let us note that since the effective action contains the term $\frac{1}{2T} \sum_{ab} \int_{x,y} g_{x,y}^{-1} u_x^a u_y^b$ where $g_{x,y}^{-1}$ is the bare propagator that is not corrected by the renormalization (STS), the exponent $\theta$ must be given by $d- \gamma + 2 \zeta_s$ and there is only one unknown critical exponent here.

\subsubsection{Solving the Morris-Wetterich equation at $T=0$ in $d=d_{{\rm uc}}- \epsilon$: the multi-local expansion}

The main result of the FRG approach to disordered elastic interface is: there exists a solution of (\ref{Eq:SecII3:FRGStat20b}) such that, in the $T \to 0$ limit, it converges to a fixed point functional $\tilde{\hat \Gamma}^*$ that admits a perturbative expansion in $\epsilon = 2 \gamma -d$. In this limit the action is non-analytic around $u_x^a =0$. This non-analyticity is smoothed at $T \neq 0$ on a small scale called the {\it thermal boundary layer}, $u_x^a \sim \tilde{T} \sim \mu^{\theta} T$. Considering carefully the $T \to 0$ by taking this smoothing into account allows to obtain the $\beta$ function directly at  $T=0$ \cite{LeDoussal2008}. In the end the structure of the solution is as follows. The functional $R[\{u_x\}]$ can be separated into its local $R(u)$ and non-local part $\tilde{R}[u]$ as
\bea \label{Eq:SecII3:FRGStat17}
R[\{u_x\}] = \int_x R(u_x)  + \tilde{R}[\{u_x\}] \nn \\
\eea
where the decomposition is unambiguously defined by the fact that $\tilde{R}[\{u_x\}]$ is $0$ for a constant field $u_x = u$ for which $R[\{u_x = u\}] = L^d R(u)$. At the fixed point, the function $R(u)$ is $O(\epsilon)$, the non-local part $\tilde{R}[\{u_x\}]$ is $O(\epsilon^2)$ and the higher-order cumulants are $S^{(m)}[ \{ u_{x_1}^{a_1} \} , \cdots , \{ u_{x_m}^{a_m} \}]$ are $O(\epsilon^m)$. They can be computed by plugging the decomposition (\ref{Eq:SecII3:FRGStat11}) into (\ref{Eq:SecII3:FRGStat15}) and assuming that they scale with $\epsilon$ as written above. Using this decomposition, the differential equation for the local part can be closed, in principle, up to an arbitrary order in $\epsilon$. We now show this differential equation up to order $O(\epsilon^2)$ as obtained in  \cite{ChauveLeDoussalWiese2000a}. Let us first introduce the loop integrals
\bea \label{Eq:SecII3:FRGStat18}
&& I_1= \mu^{d- 2 \gamma} \tilde{I}_1 \quad, \quad \tilde{I}_1=\int_q \frac{1}{(q^2 + 1)^{\gamma}}  \\
&& I_A= \mu^{2 d - 4 \gamma} \tilde{I}_A \quad,  \quad \tilde{I}_A = \int_{q_1, q_2} \frac{1}{(q_1^2 +1)^{\frac{\gamma}{2}} (q_2^2 + 1)^{\frac{\gamma}{2}} ((q_1+q_2)^2 + 1)^{\gamma}} + O(\epsilon) \nn
\eea
Note that the combination $\epsilon \tilde{I}_1$ stays finite as $\epsilon \to 0$. We will often use
\bea
A_d^{\gamma} && := \frac{1}{\epsilon \tilde{I}_1} = \frac{(2 \sqrt{\pi})^d}{2} \frac{\Gamma(\gamma)}{\Gamma(\gamma+1-d/2)} \\
&& = 2^{-1+2\gamma} \pi^{\gamma} \Gamma(\gamma) \ssp .
\eea
Rescaling 
\bea
R(u) =  A_d^{\gamma}  \mu^{\epsilon-4 \zeta_s} \tilde{R}(\mu^\zeta_s u)
\eea
one obtains \cite{ChauveLeDoussalWiese2000a,LeDoussal2008}
\bea \label{Eq:SecII3:FRGStat19}
- \mu \partial_{\mu} \tilde{R}(u) && = \underbrace{(\epsilon - 4 \zeta_s) \tilde{R}(u) + \zeta u \tilde{R}'(u)}_{{\rm rescaling }}  + \underbrace{\left( \frac{1}{2} \tilde{R}''(u)^2 - \tilde{R}''(0)\tilde{R}''(u) \right)}_{1-{\rm loop}} \nn \\
&& + \underbrace{\frac{1}{2} X \left( ( \tilde{R}''(u) - \tilde{R}''(0)) \tilde{R}'''(u)^2 \right)- \frac{\lambda_s}{2} X  (\tilde{R}'''(0^+)^2 ) \tilde{R}''(u) }_{2-{\rm loops}} \nn \\ 
&& + O( \tilde{R}^4 ) 
\eea
where $\lambda_s = 1$ and $X = \frac{2 \epsilon (2 \tilde{I}_A - \tilde{I}_1^2)}{(\epsilon \tilde{I}_1 )^2} $, i.e. $X = 1 + O(\epsilon)$ for $\gamma=2$ (short-range elasticity) and $X= 4 \ln(2) + O(\epsilon)$ for $\gamma=1$. We are looking for a solution of the equation $- \mu \partial_{\mu} \tilde{R}(u)  = 0$. Since we are expanding around $d = d_{{\rm uc}}$ where the disorder is only marginally relevant, $\zeta_s$ is expected to be $O(\epsilon)$ and thus the solution of $- \mu \partial_{\mu} \tilde{R}(u)  = 0$ in (\ref{Eq:SecII3:FRGStat19}) is also, as announced, $O(\epsilon)$. The value of the exponent $\zeta_s = \epsilon \zeta_1 + \epsilon^2 \zeta_2^2 + O(\epsilon^3) $ has to be adjusted so that a solution of (\ref{Eq:SecII3:FRGStat19}) with the desired properties hold. Note that if $\tilde{R}^*(u)$ is a fixed point of (\ref{Eq:SecII3:FRGStat19}), $\tilde{R}(u) = \frac{1}{\kappa^4} \tilde{R}^*(\kappa u)$ is also a fixed point and thus there are several {\it families} of fixed points. Thus when talking about a FP one has to specify one scale. Universal quantities can nevertheless be constructed as e.g. $\tilde{R}''''(0)$, $\tilde{R}(0)/(\tilde{R}''(0))^2$... (see \cite{ChauveLeDoussalWiese2000a}). A standard choice is to fix the value at $0$ as  $\tilde{R}^*(0) = \epsilon$. It was found that
\begin{itemize}
\item{A single value of $\zeta_{1}$, $\zeta_2$ leads to a fixed point function in the random bond universality class with $\tilde{R}^*(u)$ quickly decaying to $0$. The latter was obtained using a shooting method in (\cite{ChauveLeDoussalWiese2000a}) as (the $O(\epsilon)$ result is coherent with the previous result from \cite{DSFisher1986}) $\zeta_1 = 0.20829806(3)$ and $\zeta_2= 0.006858(1)$ for SR elasticity ($\gamma=2$). }
\item{A single value of $\zeta_s$ leads to a FP in the random field universality class, i.e. with $\tilde{R}^*(u) \sim_{|u| \to \infty} |u|$ and $\tilde{R}^{*\prime \prime} (u)$ decaying quickly to $0$. It is given by $\zeta_s= \epsilon/3 + O(\epsilon^2)$ (independently of $\gamma$). This result was actually argued to hold to any order \cite{ChauveLeDoussalWiese2000a} in $\epsilon$, in agreement with the simple Flory argument of Sec.~\ref{subsec:StaticPhaseDiagram}.}
\item{For periodic disorder $\tilde{R}^*(u)  = \tilde{R}^*(u+1)$, the value of $\zeta$ is necessarily $0$. Interestingly in this case it was found in \cite{ChauveLeDoussalWiese2000a} that  $\tilde{R}^*(u) = f(\epsilon) u(1-u)$ with $f(\epsilon) = O(\epsilon)$ a function. This form was also conjectured to hold to all order.}
\end{itemize}

\begin{figure}
\centerline{\includegraphics[width=6.0cm]{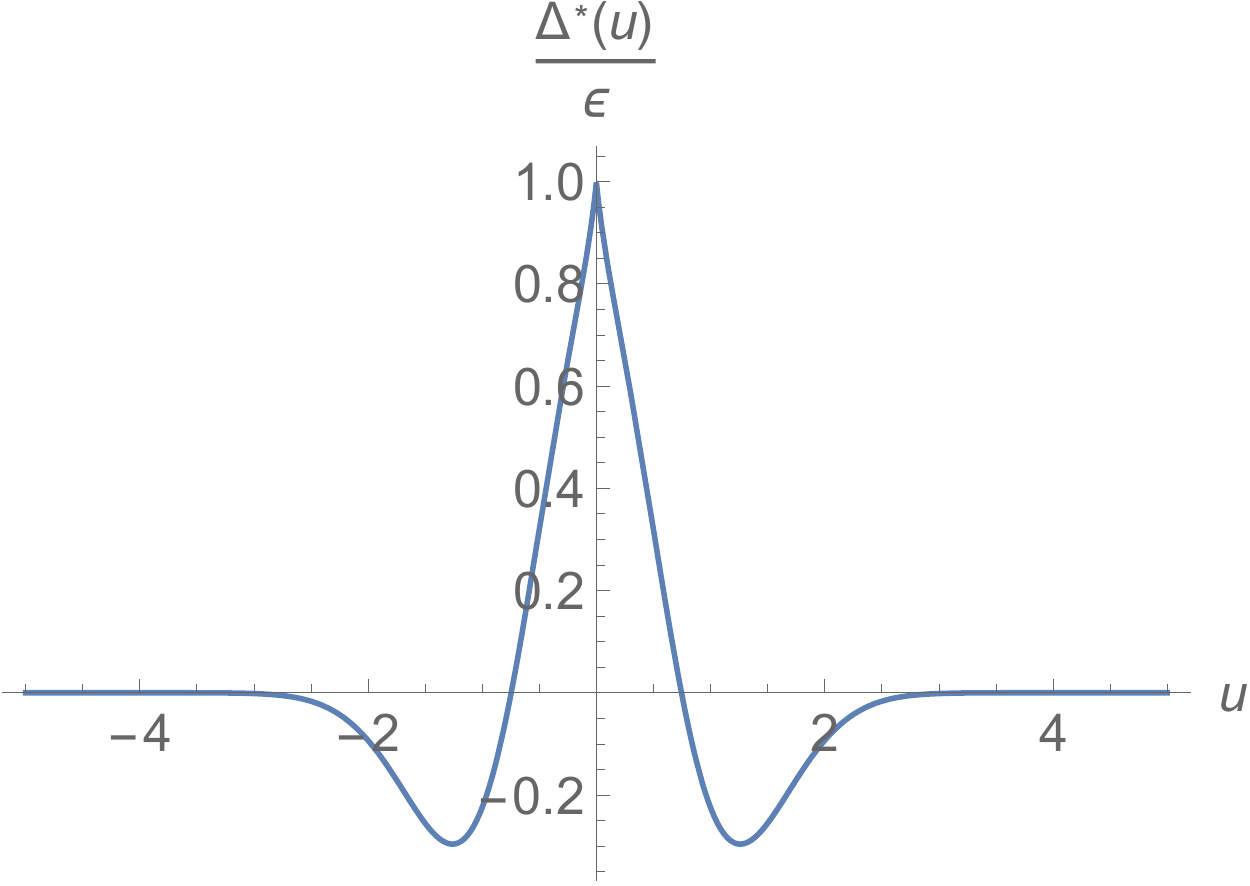} \includegraphics[width=6.0cm]{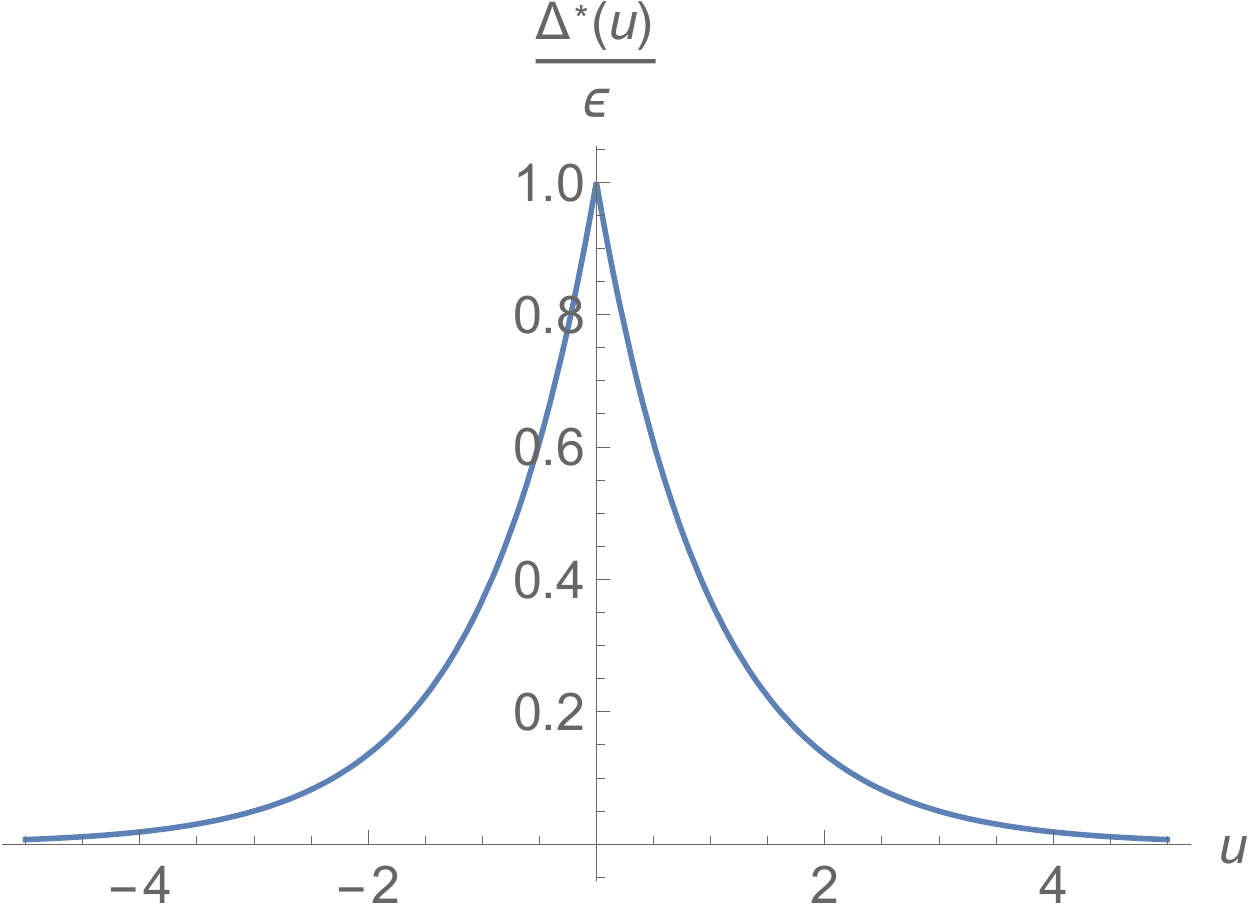}} 
\caption{Cartoon of the shape of FRG fixed point functions $\Delta^*(u)=-R^{*\prime \prime}(u)$ for the RB universality class (left) and the RF universality class (right).}
\label{fig:FRGFixedP}
\end{figure}

These different fixed points were argued to be stable in \cite{DSFisher1986,ChauveLeDoussalWiese2000a}. The typical shape of the RF and RB FP functions $\Delta^*(u)= -R^{* \prime \prime}(u)$, here normalized as $\Delta^*(0) = \epsilon$, are plotted in Fig.~\ref{fig:FRGFixedP}. Other generic long-range fixed points with $\tilde{R}(u)\sim_{u \to \infty} u^{2(1-\alpha)}$ with $\zeta_s^{\alpha} = \frac{\epsilon}{2(1+\alpha)}$ (holding presumably to all order) were also found and argued to be stable as long as they lead to a roughness exponent larger than the RB roughness exponent \cite{ChauveLeDoussalWiese2000a} (otherwise the system flows to the RB FP). Let us conclude this section with a few remarks:
\begin{enumerate}
\item{The first line of (\ref{Eq:SecII3:FRGStat19}) (i.e. the rescaling and the one-loop part) were already obtained by Fisher in \cite{DSFisher1986} using a Wilson's RG scheme and can be easily obtained using a standard perturbative RG from the beginning.}
\item{It is instructive to study the flow of $\tilde{R}''(0)$ and $\tilde{R}''''(0)$: to one-loop these close as $-\mu \partial_{\mu}\tilde{R}''(0) = (\epsilon - 2 \zeta_s) \tilde{R}''(0) + \tilde{R}'''(0)^2$ and $-\mu \partial_{\mu}\tilde{R}''''(0) =  \epsilon \tilde{R}''''(0) + 3 \tilde{R}''''(0)^2 + 4 \tilde{R}'''(0)\tilde{R}'''''(0)$. Starting from an initially smooth disorder at some scale $\mu_0$, $R_{\mu_{0}}(u)$, at the beginning of the flow the function stays analytic: $\tilde{R}'''(0)= \tilde{R}'''''(0)=0$. At the beginning of the flow $\tilde{R}''(0)$ thus does not flow, and if it were so for all time one would find $\zeta_s = \epsilon/2$ i.e. the dimensional reduction result. However, $\tilde{R}''''(0)$ becomes infinite, and thus the function non-analytic around $0$, in a finite renormalization time. More precisely, for $\mu > \mu_c = \mu_0 \left( \frac{3\tilde{R}''''_0(0) }{3 \tilde{R}''''_0(0) + \epsilon} \right)^{\frac{1}{\epsilon}}$, one obtains $\tilde{R}''''(0)=\frac{\tilde{R}''''_0(0) \epsilon  \mu _0^{\epsilon }}{(3 \tilde{R}''''_0(0)+\epsilon ) \mu ^{\epsilon }-3 \tilde{R}''''_0(0) \mu _0^{\epsilon }}$ and $\tilde{R}$ becomes non-analytic around $0$ for $\mu \leq \mu_c$: the function $\tilde{\Delta}(u) = - \tilde{R}''(u)$ acquires a linear cusp around $0$: $\tilde{\Delta}(u) - \tilde{\Delta}(0) \simeq \tilde{\Delta}'(0^+)  |u| + O(u^2) $. The occurrence of this cusp will be related to shocks below. Note that once the cusp appears, $\tilde{R}''(0) $ starts to flow and the system escapes dimensional reduction. Note that the cusp appears when the system probes length scales larger than $L_c' := 1/\mu_c = \frac{1}{\mu_0} \left(1  +\frac{\epsilon}{3 \tilde{R}''''_0(0)} \right)^{\frac{1}{\epsilon}}$ . Using the simple estimate $\tilde{R}''''_0(0) \sim \frac{\Delta(0)}{u_c^2}$, one shows that $L_c'$ essentially reproduces the Larkin length $L_c$ computed in (\ref{Eq:SecII2:Larkin4}). Thus, following the renormalization group flow, the system actually believes that it is flowing to the Larkin fixed point up to a scale corresponding to the Larkin length where the cusp, associated with metastability in the system (see below), appears. The fact that no `supercusp' appears later in the flow, i.e. all derivative of $\tilde{R}$ at $0$ up to the fourth one remain finite, was discussed in \cite{ChauveLeDoussalWiese2000a}.}
\item{The second line of (\ref{Eq:SecII3:FRGStat19}) was first obtained in \cite{ChauveLeDoussalWiese2000a} where the authors used standard (although functional) perturbative RG to directly attempt to renormalize the field theory at $T=0$. The problem with this approach is that one then encounters so-called `anomalous terms' involving $\tilde{R}'''(0)$ and one has to decide whether it is $\tilde{R}'''(0^+)$ or $\tilde{R}'''(0^-)$. The proper way to do so is to regularize the non-analytic behavior by studying the zero temperature limit of the renormalized theory at finite temperature as outlined above and reviewed in \cite{LeDoussal2008}. In \cite{ChauveLeDoussalWiese2000a} the authors nevertheless obtained (\ref{Eq:SecII3:FRGStat19}) with a value of $\lambda$ a priori not specified by the perturbative method, which was imposed to be one to respect the `potentiality' of the problem, i.e. the existence of FP function of the RB type (see \cite{ChauveLeDoussalWiese2000a}).}
\item{The FRG equation (\ref{Eq:SecII3:FRGStat19}) is both UV {\it and} IR universal. It is UV universal in the sense that, by construction, it does not depend on the microscopic details of the models. The more subtle property is to prove that it does not depend on the chosen IR cutoff scheme \cite{ChauveLeDoussalWiese2000a}, that was here chosen as a massive scheme.}
\end{enumerate}

\subsection{Applying the functional renormalization group to shocks} \label{subsec:FRGStaticSchocks}
In this section we now discuss the application of FRG methods presented in Sec.~\ref{subsec:FRGStatic} to the study of the shock statistics of the interface presented in Sec.~\ref{subsec:shocks}. We begin by linking the non-analyticity of the fixed point effective action of FRG with the occurrence of shocks. We then show how FRG can be use to compute shocks observables on the example of the density of total size of shocks.

\subsubsection{The cusp and the shocks} 

As already remarked we first note that as $T \to 0$, the renormalized disorder potential for a constant well $\hat V_{\mu}(w) :=  \hat V_\mu[\{w_x = w\}]$ defined in (\ref{Eq:SecII3:FRGStat5}) converges to (assuming no degeneracy of the ground state, which is true with probability $1$ except at some discrete positions)
\bea \label{Eq:SecII3:FRGStatShock1}
\hat V_\mu(w) = \frac{1}{2} \int g_{x,y}^{-1}(u_x(w)- w)(u_y(w) - w)  + \int_x V(x,u_x(w)) \ssp,
\eea
where as usual $u_x(w)$ denotes the ground state of the interface as defined in (\ref{Eq:SecII3:FRGStat1}). Using the saddle-point structure in (\ref{Eq:SecII3:FRGStat1}), one shows that the {\it renormalized force at the scale $\mu$}, defined by
\bea \label{Eq:SecII3:FRGStatShock2}
\hat F_\mu (w) := - \partial_w \hat V_\mu(w)
\eea
admits the expression at zero temperature
\be \label{Eq:SecII3:FRGStatShock3}
\hat F_\mu (w) = \int_{x,y} g_{x,y}^{-1} (w-u_x(w)) = \int_x f_x(w) \quad {\rm with } \quad f_x(w) = m^2(w - u_x(w)) \ssp .
\ee
where the second equality is true for elastic kernel of the form $g_{x,y}^{-1} = \int_{q} e^{iq(x-y)} (q^2 + \mu^2)^{\frac{\gamma}{2}}$ and we recall that $m = \mu^{\frac{\gamma}{2}}$. Hence, using (\ref{Eq:SecII3:FRGStat9b}), (\ref{Eq:SecII3:FRGStat12}) and the definition of the local part (\ref{Eq:SecII3:FRGStat17}), one sees that the {\it local part of the second cumulant of the renormalized force at the scale $\mu$}, defined by,
\bea \label{Eq:SecII3:FRGStatShock4}
\Delta(u):= -R''(u)
\eea
is linked to an observable of the ground state as
\bea \label{Eq:SecII3:FRGStatShock5}
 \Delta(w-w')= \frac{1}{L^d} \partial_w \partial_{w'} \overline{\hat V_\mu(w) \hat V_\mu(w') } = \frac{m^4}{L^d} \int_x \int_y \overline{ (w-u_x(w))(w-u_y(w)}^c \ssp .
\eea
Or equivalently, in terms of the position of the {\it center of mass of the interface} defined by
\bea \label{Eq:SecII3:FRGStatShock6}
u(w):= \frac{1}{L^d} \int_x u_x(w) ,
\eea
we have
\bea \label{Eq:SecII3:FRGStatShock7}
\Delta(w-w') = L^d m^4 \overline{ (u(w)-w) (u(w')-w')} \ssp .
\eea
This relation, first shown in \cite{LeDoussal2006b,LeDoussal2008}, has deep consequences. On its left hand side it involves the second cumulant of the renormalized disorder that naturally appears in the effective action of the replicated theory. As a function of $m = \mu^{\frac{\gamma}{2}}$, the rescaled cumulant 
\bea \label{Eq:SecII3:FRGStatShock8}
\tilde{\Delta}(w):=  (A_d^{\gamma})^{-1} \mu^{2 \zeta_s -\epsilon} \Delta(\mu^{-\zeta_s} w)
\eea
obeys a RG equation that is the second derivative of (\ref{Eq:SecII3:FRGStat19}). In the limit $\mu \to 0$ it converges to a fixed point function, depending on the universality class of the initial bare disorder. This fixed point function is non-analytic and exhibits a cusp around $0$ (see Fig.~\ref{fig:FRGFixedP}) and for small $\mu$ we thus have $\tilde{\Delta}(w) - \tilde{\Delta}(0) \simeq \tilde{\Delta}'(0^+) |w| + O(w^2)$. On the right hand side on the other hand it involves a simple observable linked to the ground state of the interface. This relation thus provides a protocol to {\it  measure the FRG function $\Delta(u)$}. The latter was first implemented in numerics with an excellent agreement with the theory \cite{MiddletonLeDoussalWiese2006}, and later also in experiments \cite{LeDoussalWieseMoulinetRolley2009} (for the related case of the depinning). On the other hand, since the left hand side of (\ref{Eq:SecII3:FRGStatShock7}) is non-analytic beyond the Larkin scale $\mu_c \sim L_c^{-1}$ this {\it shows} that the right-hand side is {\it also non-analytic beyond the Larkin scale}. More precisely, using $\overline{u(w)}=w$, (\ref{Eq:SecII3:FRGStatShock7}) can be rewritten as, introducing $\hat u(w) = u(w)-w$,
\bea \label{Eq:SecII3:FRGStatShock9}
\overline{( \hat u(w) - \hat u(0) )^2} =-2 \frac{\Delta(w)}{L^d m^4} \ssp .
\eea
For a smooth motion, the left-hand side of (\ref{Eq:SecII3:FRGStatShock9}) is $O(w^2)$ for small $w$. Obviously this could just mean that with some probability $\hat u(w) \sim \sqrt{w}$ close to $w = 0$. The natural interpretation from the study of $d=0$ models, numerics (and physical intuition) is, however, that with a small probability proportional to $w$, the center of mass of the interface makes a jump of size $S/L^d$, distributed with a PDF $P(S)$ and (\ref{Eq:SecII3:FRGStatShock9}) is rewritten
\bea \label{Eq:SecII3:FRGStatShock10}
\rho_0 w \frac{\langle S^2 \rangle_P}{L^{2d}} + O(w^2) = - \frac{2 \Delta'(0^+) w}{L^d m^4} + O(w^2) \ssp ,
\eea
where $\langle \rangle_P$ denotes the average with respect to $P$ and $\rho_0$ already introduce in Sec.~\ref{subsec:shocks}, is the density of shocks per unit length. Equivalently, using the definitions of the density $\rho(S)$ and PDF $P(S)= \rho(S)/\rho_0$ of avalanche total size introduced in Sec.~\ref{subsec:shocks}, see (\ref{Eq:SecII2:DefrhoS}) and (\ref{Eq:SecII2:DefPS}), as well as the relation shown there $\langle S \rangle_{\rho}  = \rho_0 \langle S \rangle_P = L^d$, (\ref{Eq:SecII3:FRGStatShock10}) is rewritten
\bea \label{Eq:SecII3:FRGStatShock11}
S_m:=\frac{\langle S^2 \rangle_{\rho} }{2 \langle S \rangle_{\rho}} =\frac{\langle S^2 \rangle_{\rho} }{2 \langle S \rangle_{\rho}}= \frac{\sigma}{m^4} \quad , \quad \sigma = - \Delta'(0^+) \geq 0 \ssp .
\eea
Note that the relation (\ref{Eq:SecII3:FRGStatShock11}) is exact here since $\Delta(w)$ is exactly given by (\ref{Eq:SecII3:FRGStatShock7})\footnote{Up to a non universal scale it can also be computed in an $\epsilon$ expansion using FRG but that is not what we are doing here.}. It can also be easily obtained by {\it assuming that the shock decomposition}
\bea \label{Eq:SecII3:FRGStatShock12}
u_x(w) = cst + \sum_i S^{(i)}_{x} \theta(w-w_i) \ssp ,
\eea
holds at small $\mu$. The basic idea of applying FRG to shocks is to {\it interpret} the short-scale singularities that appear in the FRG flow of the effective action as consequences of the presence of shocks as written in (\ref{Eq:SecII3:FRGStatShock12}). Here (\ref{Eq:SecII3:FRGStatShock11}) actually does not tell much about the shock statistics since $\Delta'(0^+)$ contains one non-universal scale $\kappa$ as discussed in Sec.~\ref{subsec:FRGStatic}. Here we have thus linked this non universal scale to a precise (non-universal) observable in (\ref{Eq:SecII3:FRGStatShock11}). On the other hand the result of the FRG is that all higher order cumulants of the renormalized disorder potential and the full effective action can be obtained using the structure of the $\epsilon$ expansion as functions of $\Delta(u)$. In the end this will imply that all higher order moments of the shock total size density $\langle S^n \rangle_{\rho}$ (and other shocks observables) can be expressed using the $\epsilon$ expansion in terms of only one non-universal scale, that we can choose as $S_m$. The true input of FRG in the study of shock statistics basically works in three steps:\\
(i) Assume the shock decomposition (\ref{Eq:SecII3:FRGStatShock12}) and relate a given shock observable to a disorder averaged observable of $u_x(w) $ (actually we will need disorder average observables of the ground state for different well position $w$ in the same environment, see below).
\\ (ii) Compute the disorder averaged observable in the limit $\mu \to 0$ using the results of the FRG. \\
(iii) Draw the consequences for the shock observable.

\medskip

\subsubsection{ Shocks and the $\epsilon$ expansion: the case of the one-shock total size distribution}
Let us now briefly recall how the above program goes for the one-shock total size density $\rho(S)$. The following is based on \cite{LeDoussalWiese2008c,LeDoussalWiese2011b}.
Let us first study the scaling of $S_m$ defined in (\ref{Eq:SecII3:FRGStatShock11}) with $\mu$ and $\epsilon$. Note that since for small $m$, $\Delta(w)$ takes the scaling form $\Delta(w) = A_d^{\gamma}\mu^{\epsilon-2 \zeta_s} \tilde{\Delta}( \mu^{\zeta_s} w)$ with $\tilde{\Delta}(w) = O(\epsilon)$ a function close to one of the fixed points of the FRG equation, we have, defining $\tilde{\sigma} = \tilde{\Delta}'(0^+)$,
\bea \label{Eq:SecII3:FRGStatShock13}
S_m= \frac{\langle S^2 \rangle_{\rho} }{2 \langle S \rangle_{\rho}} = A_d^{\gamma} \frac{\mu^{\epsilon-\zeta_s}}{m^4} \tilde{\sigma} = A_d^{\gamma} \mu^{- d - \zeta_s} \tilde{\sigma} \ssp ,
\eea
and we recall $m = \mu^{\gamma/2}$. First, note that $S_m$ diverges as $\mu \to 0$ and that the scaling (\ref{Eq:SecII3:FRGStatShock13}) is consistent with the scaling hypothesis 
\bea \label{Eq:SecII3:FRGStatShock14}
\rho(S) = L^{\alpha} \mu^{\rho} \frac{1}{S^{\tau_S}} f_{{\rm cut}}\left(\frac{S}{S_m} \right)  g_{{\rm cut}}\left(\frac{S}{S_0} \right) \ssp .
\eea
with $\tau_S \leq 2$ that was made in the derivation of the NF conjecture in Sec.~\ref{subsec:shocks} (there $S_m$ was denoted $S_\mu$ in (\ref{Eq:SecII2:NF1})). Indeed, the second moment $\langle S \rangle_{\rho}$ diverges as $S_m \to \infty$ and is controlled by the massive cutoff.\\
On the other hand, note that $S_m = O(\epsilon)$. Therefore it is natural to assume, and the end result will be consistent with this assumption, that all moments of the avalanche size distribution scale as, for $n\geq 2\in \JN $, $ \langle S^n \rangle_{\rho} \sim S_m^{n-1} \sim \epsilon^{n-1} \mu^{-(n-1)(d + \zeta_s) }$. The proper object which is expected to have a well defined (and universal) $\mu \to 0 , \epsilon \to 0$ limit is thus the density of shocks total size in units of $S_m$:
\bea \label{Eq:SecII3:FRGStatShock15}
\tilde{\rho}(\tilde{S}):=  S_m \rho(S_m \tilde{S}) \ssp .
\eea
which is $O(1)$. The latter is normalized as $\int_{\tilde{S} >0}   \tilde{S} \tilde{\rho}(\tilde{S} ) d \tilde{S} = L^d$ and $\int_{\tilde{S} >0}  \tilde{S}^2 \tilde{\rho}(\tilde{S} ) d \tilde{S}  = 2 L^d$. Let us introduce (almost) its Laplace transform (dropping from now on the tilde)
\bea  \label{Eq:SecII3:FRGStatShock16}
Z(\lambda):= L^{-d}\int_{S>0} \left( e^{\lambda S } -1 \right) \rho(S) dS \ssp .
\eea
Here the $-1$ in the definition of $Z(\lambda)$ is to ensure that (\ref{Eq:SecII3:FRGStatShock16}) is finite even when $\tau_S >1$ (which will be true, at least close to $\epsilon =0$) and the small scale cutoff $S_0$ in (\ref{Eq:SecII3:FRGStatShock14}) is sent to $0$. The $L^{-d}$ ensures that it is finite as $L \to \infty$. Altogether, $Z(\lambda) = \lambda + 2 \lambda^2 + O(\lambda^3)$, and note that the coefficient in front of $\lambda$ and $\lambda^2$ are exact (i.e. they are consequences of our definition). The real input of FRG is to provide the $\epsilon$ expansion of the coefficients in front of the higher order terms in $\lambda^n$, $n \geq 3$.

\smallskip

(i) The first step is now to relate $Z(\lambda)$ to a disorder averaged observable of the position field, assuming that the shock decomposition (\ref{Eq:SecII3:FRGStatShock12}) holds. This was done in \cite{LeDoussalWiese2008c,LeDoussalWiese2011b} and the result is
\bea \label{Eq:SecII3:FRGStatShock17}
Z(\lambda) = \partial_{\delta} G_{\delta}(\lambda)|_{\delta=0^+} \quad , \quad G_{\delta}(\lambda) :=L^{-d}  \overline{e^{\frac{\lambda L^d}{S_m} (\hat u(w+ \delta) - \hat u(w) )}}
\eea
Here $\hat u(w+ \delta) = \frac{1}{L^d} \int (u_x(w) -w)$. Let us give a quick justification that (\ref{Eq:SecII3:FRGStatShock17}) holds: if there is no shock between $w$ and $w+ \delta$, $L^{nd}(\hat u(w+ \delta) - \hat u(w) )^n$ is of order $O(\delta^n)$ and such events do not contribute to (\ref{Eq:SecII3:FRGStatShock17}). The order $O(\delta)$ in $L^{nd}\overline{(\hat u(w+ \delta) - \hat u(w) )^n}$ is thus dominated by the probability $\rho_0 \delta$ that one shock occurred and $\overline{(\hat u(w+ \delta) - \hat u(w) )^n} \sim \rho_0 \delta \langle S^n \rangle_P + O(\delta^2)= \delta \langle S^n \rangle_{\rho}+ O(\delta^2) $. (Here we neglect possible contributions coming from the simultaneous occurrence of more than one shock at $w^+$, this hypothesis is sometimes referred to as the fact that the shocks are dilute, and the probability that two shocks occurred is thus $O(\delta^2)$). This justifies (\ref{Eq:SecII3:FRGStatShock17}) at each order in an expansion in $\lambda$.

\smallskip

(ii) In the second step, we now want to compute the right hand-side of  (\ref{Eq:SecII3:FRGStatShock17}) using FRG. To do this we need to consider the replicated action for several copies of the same disordered elastic system. We therefore consider the theory for $r=2$ position fields $u_{x}^i$ coupled to different parabolic wells centered at positions $w_i$ in the same disordered environment ($w_1= w$ and $w_2= w +\delta$). The Hamiltonian of the problem is
\be \label{Eq:SecII3:FRGStatShock18}
{\cal H}[\{u\}, \{w \}] = \sum_{i=1}^{r} {\cal H}_{\rm el}[u^i, w^i] + \sum_{i=1}^r \int_x V(u_x^i , x) \ .
\ee 
This leads to a replicated action of the form
\be \label{Eq:SecII3:FRGStatShock19}
S[u] =  \frac{1}{2 T} \sum_{a,i} \int_{xx'}  g_{xx'}^{-1}(u_{ax}^i - w_i)(u_{ax'}^i-w_i)   - \frac{1}{2T^2}  \sum_{a,i; b ,j} \int_x R_{0}(u_{ax}^i - u_{bx}^j) + \cdots 
\ee
where $a$ is a replica index and $R_0$ is, as in (\ref{Eq:SecII3:FRGStat2}), the bare cumulant of the disorder $V(x,u)$. The results from FRG of the previous section for the $r=1$ case can be generalized\footnote{Actually it is exactly the same theory since (\ref{Eq:SecII3:FRGStatShock19}) is equivalent to taking all $w_i = 0$ but coupling the action to a source that depends on the replica index as done in Sec.~\ref{subsec:FRGStatic}, see (\ref{Eq:SecII3:FRGStat7b}).} to this new problem \cite{LeDoussalWiese2008c,LeDoussalWiese2011b}. In particular one shows that the effective action of the theory is given by, in the limit $\mu \to 0$,
\be \label{Eq:SecII3:FRGStatShock20}
\Gamma[u] =  \frac{1}{2 T} \sum_{a,i} \int_{xx'}  g_{xx'}^{-1}(u_{ax}^i - w_i)(u_{ax'}^i-w_i)    - \frac{1}{2T^2}  \sum_{a,i; b ,j} \int_x R(u_{ax}^i - u_{bx}^j) + O(\epsilon^2) \ssp .
\ee
Here $R(u) = O(\epsilon)$ is the same renormalized disorder correlator as already introduced in the previous section, while the neglected terms are higher-order terms in $\epsilon$ that can be expressed as loop integrals with higher powers of $R$. Using the effective action (\ref{Eq:SecII3:FRGStatShock20}) is often referred to as the improved tree theory. Here one can a priori use either the renormalized disorder correlator $R(u)$ as computed to order $O(\epsilon)$ from the FRG, or the true renormalized disorder correlator $R(u)$, since (\ref{Eq:SecII3:FRGStatShock20}) is then still true up to $O(\epsilon^2)$. The observable $G_{\delta}(\lambda)$ is then computed by singling out the first replica as
\bea \label{Eq:SecII3:FRGStatShock21}
G_{\delta}(\lambda) =L^{-d} \lim_{T,n \to 0}  \int {\cal D}[u] e^{\frac{\lambda}{S_m} \int_x \left(  u_{1x}^2(w_2) - u_{1x}^1(w_1) -w_2 +w_1  \right)  -  S[u]}
\eea
This path integral is evaluated using a saddle-point calculation on the effective action (\ref{Eq:SecII3:FRGStatShock20}). The saddle-point equations are solved in the limit $T,n \to 0$ (see \cite{LeDoussalWiese2011b}).

\smallskip

(iii) In the end one obtains the result, valid for $\lambda <1/4$,
\bea \label{Eq:SecII3:FRGStatShock22}
&& Z(\lambda ) =  \lambda + Z(\lambda)^2 + O(\epsilon) \nn \\
&& Z(\lambda)= \frac{1}{2} \sqrt{1 - 4 \lambda}  + O(\epsilon) \ssp .
\eea
Which corresponds to a tree result for $\rho(S)$:
\bea \label{Eq:SecII3:FRGStatShock23}
\rho(S) = \frac{L^d}{2 \sqrt{\pi} S^{3/2} } e^{-S/4} + O(\epsilon) \ssp .
\eea 
In terms of the avalanche size distribution exponent we have thus
\bea \label{Eq:SecII3:FRGStatShock24}
\tau_S = \frac{3}{2} + O(\epsilon) = 2 - \frac{\gamma}{d + \zeta_s} + O(\epsilon) \ssp .
\eea
i.e. the result agrees well with the NF conjecture to order $O(1)$ since $\zeta_s=O(\epsilon)$ and $d = 2\gamma - \epsilon$. The results (\ref{Eq:SecII3:FRGStatShock22}), (\ref{Eq:SecII3:FRGStatShock23}), (\ref{Eq:SecII3:FRGStatShock24}) have the status of mean-field theory results for the shock total size density. We will see that these are equivalent to the mean-field theory results obtained in the dynamics and there we will describe more precisely what this mean-field theory corresponds to. Before we turn to the analysis of avalanches at depinning let us make a few comments on (\ref{Eq:SecII3:FRGStatShock23})
\begin{itemize} 
\item{Dimensions in (\ref{Eq:SecII3:FRGStatShock23}) are reintroduced using (\ref{Eq:SecII3:FRGStatShock15}), in particular note that, as announced, the dimensionful result for $\rho(S)$ contains only one non-universal scale that was here chosen as $S_m$ (defined in (\ref{Eq:SecII3:FRGStatShock13})).}
\item{The exponent $\tau_S= 3/2$ appears completely universal, i.e. independent of both the UV (hidden in $\sigma$) and IR (the massive cutoff) details of the models.}
\item{The form of the distribution (\ref{Eq:SecII3:FRGStatShock23}) (i.e. the exponential cutoff) is UV independent but non-universal in the sense that it depends on the chosen IR regularizing scheme, here a massive cutoff with $\ell_{\mu}=1/\mu  \ll L$. We note that such massive cutoffs have been argued to be relevant in some experimental setups, as e.g. for fluid contact line experiments \cite{LeDoussalWieseMoulinetRolley2009}.}
\item{The above calculations have been extended using FRG to one-loop accuracy in \cite{LeDoussalWiese2008c,LeDoussalWiese2011b}. One of the result shown there is that the avalanche size exponent $\tau_S$ does agree with the NF conjecture. To this date this is the most precise calculation and we thus have $\tau_S = 2 - \frac{\gamma}{d + \zeta_s} + O(\epsilon^2)$.}
\end{itemize}

\section[FRG approach to avalanches]{The functional renormalization group treatment of avalanches in disordered elastic systems: a short review} \label{sec:AvaWithFRG}
In this section we now review the use of the functional renormalization group to calculate avalanche observables. We will begin by recalling the important results of FRG for the dynamics of $d$-dimensional interfaces at the depinning transition in Sec.~\ref{subsec:FRGDep}, and in Sec.~\ref{subsec:FRGDepAva} we will show how to apply FRG to the study of avalanches.

\subsection{The functional renormalization approach to the depinning transition} \label{subsec:FRGDep}

\subsubsection{Introduction}

In this section we now review important FRG results on the depinning transition. We will try to make as many parallels as possible with the FRG theory for the static problem and therefore give less details in this section. As for the static problem at $T=0$, a naive perturbative approach to the depinning transition gives the trivial dimensional reduction result. The FRG approach escapes this phenomenon through a renormalization procedure which involves a non-analytic action. As for the statics, this was first found by a one-loop perturbative analysis using Wilson's RG \cite{NarayanDSFisher1992b,NattermannStepanowTangLeschhorn1992,LeschhornNattermannStepanowTang1997,NarayanDSFisher1993a}. This analysis surprisingly led to results identical to the one-loop results in the static problem, while the two problems are clearly different (in particular the random bond universality class is unstable at depinning, see below). The FRG analysis was extended to two-loops in \cite{LeDoussalWieseChauve2002}, and there the differences with the statics appeared. An interesting extension to the creep regime was also made in \cite{ChauveGiamarchiLeDoussal1998}, though some questions remain open. Pedagogical reviews of FRG for the depinning transition can be found in \cite{ChauveThesis,DobrinevskiPhD}. Let us now introduce the main objects and state the results. We study the problem
\bea \label{Eq:SecII3:FRGDep1}
&& \eta_0 \partial_t u_{tx} = \int_{y} g_{x,y}^{-1}(w(t)-u_{ty}) + F(x,u_{tx}) \nn \\
&& w(t) = v t \ssp .
\eea
As usual $g_{x,y}^{-1} = \int_q \sqrt{q^2 + \mu^2} e^{iq (x-y)}$ and the random force is chosen centered, Gaussian with a second cumulant
\bea \label{Eq:SecII3:FRGDep2}
\overline{F(x,u) F(x',u')} = \delta^{(d)}(x-x') \Delta_0(u-u') \ssp .
\eea
We are interested to study (\ref{Eq:SecII3:FRGDep1}) in the successive limits: (i) $t \to \infty$ (more precisely we want to describe correlations in the steady state reached by an interface described by (\ref{Eq:SecII3:FRGDep1}), the latter being uniquely defined up to trivial time-translations \cite{Middleton1992} and thus the initial condition of  (\ref{Eq:SecII3:FRGDep1}) will be unimportant); (ii) $v \to 0^+$ (depinning regime); (iii) $\mu \to 0$ (where observables are expected to reach a universal scaling limit and avalanches to occur). Disorder-averaged observables of the position field are computed using the Martin-Siggia-Rose (MSR) formalism (see \cite{MSR,Janssen1976} for historical references and \cite{Janssen1992,TauberBook2014} for reviews).
\bea \label{Eq:SecII3:FRGDep3}
&& \overline{ O[\{u_{tx}\}]} := \int \cD u \cD \hat u  ~ O[\{u_{tx}\}] ~e^{- S[u , \hat u] }  \ssp , \nn \\
&&  S[u, \hat u]:= S_0[u , \hat u] +S_{{\rm dis}} [u , \hat u]  +S_{{\rm dri}}[u, \hat u]    \ssp . 
\eea
Where here $\hat u_{t,x} \in i \JR$ is the MSR response field and we have split the action between the quadratic part $S_0$, the disorder part $S_{{\rm dis}}$ and the driving part $S_{{\rm dri}}$ as:
\bea \label{Eq:SecII3:FRGDep4}
&& S_0[u , \hat u] :=  \int_{tx} \hat u_{tx} \left( \eta_0 \partial_t u_{tx}  + \int_{y} g_{x,y}^{-1} u_{ty} \right)  \ssp , \nn \\
&& S_{{\rm dis}} [u , \hat u] := -\frac{1}{2} \int_{t,t',x} \hat u_{tx} \hat u_{t'x}  \Delta_0(u_{tx} - u_{t'x}) + \cdots \ssp , \nn \\
&& S_{{\rm dri}}[u, \hat u] = -\int_x \int_y  \hat u_{tx} g_{x,y}^{-1} w(t)   = -m^2 \int_x \hat u_{tx} vt \ssp .
\eea  
Here we have implicitly adopted the Ito convention to interpret  (\ref{Eq:SecII3:FRGDep1}) and the dots in $ S_{{\rm dis}} [u , \hat u]$ indicates eventual higher order cumulants of the bare disorder force if the latter is non Gaussian. In the following we will denote by $\langle \rangle_S$ the average with respect to the MSR action (\ref{Eq:SecII3:FRGDep3}). The latter is identical to the disorder average $\overline{()}$ in the steady state of (\ref{Eq:SecII3:FRGDep1}) for observables of the position field $u_{tx}$, but in the following we will also consider averages involving the MSR field $\hat u_{tx}$.

\subsubsection{Definition of the different functionals and the statistical tilt symmetry}
As in the statics, we consider the generating function for connected correlations:
\bea
e^{W_\mu[\{j_{tx}\} ,\{ \hat j_{t,x} \} ] }:= \langle e^{\int_{t,x} j_{t,x} u_{tx} + \int_{t,x} \hat{j}_{t,x} \hat u_{tx}} \rangle_S \ssp,
\eea
and its Legendre transform, the effective action of the theory,
\bea
\Gamma_\mu [u , \hat u] := -W[j , \hat{j}] + \int_{t,x} j_{t,x} u_{tx} + \int_{t,x} \hat{j}_{t,x} \ssp,
\eea
where on the right hand side the sources $j $ and $\hat j$ are given in terms of the fields $u$ and $\hat u$ by inverting
\bea
u_{t,x} = \frac{\delta W}{ \delta j_{t,x}} \quad , \quad  \hat{u}_{tx} = \frac{\delta W}{ \delta \hat{j}_{t,x}} \ssp .
\eea
Let us begin with \\

{\it The Statistical Tilt Symmetry in the dynamics and the relation statics/dynamics}

Note that the total action $S[u, \hat u]$ has the symmetry, for any {\it time-independent} function $\phi_x$, $S[u+\phi_x, \hat u]=S[u, \hat u] + \int_{txy} \hat u_{tx} g_{x,y}^{-1} \phi_y$. Hence, for any observable $O[u, \hat u]$, we obtain the STS relation,
\bea \label{Eq:SecII3:FRGDep5}
\langle  O[u - \phi , \hat u ] \rangle_S = \int \cD u \cD \hat u ~ O[\{u_{tx}\}] ~e^{- S[u + \phi , \hat u] } = \langle O[u  , \hat u ] e^{—\int_{txy} \hat u_{tx} g_{x,y}^{-1} \phi_y} \rangle_S \ssp .
\eea
In a differential form the latter is rewritten, applying $\frac{\delta}{\delta \phi_x}|_{\phi_x =0}$ to the last identity,
\bea \label{Eq:SecII3:FRGDep6}
\int_{t} \langle \frac{\delta }{ \delta u_{tx}} O[u , \hat  u] \rangle_S  = \langle O[u, \hat u]  \int_{t,y}  g_{x,y}^{-1} \hat u_{ty} \rangle_S \ssp ,
\eea
where we have explicitly used the symmetry $g_{x,y}^{-1}=g_{y,x}^{-1}$. For the observable $O[u , \hat  u] = u_{t_f , x_f}$, we thus obtain
\bea \label{Eq:SecII3:FRGDep7}
\delta^{(d)}(x-x_f) = \int_{t y} g_{x,y}^{-1} \langle \hat u_{t,y}  u_{t_f, x_f}\rangle_{S} \ssp .
\eea
Introducing the response function
\bea \label{Eq:SecII3:FRGDep8}
\sR(x_f - x_i , t_f-t_i) := \langle \hat u_{t_i,x_i}  u_{t_f,x_f}\rangle_S \ssp ,
\eea
(recall here that from causality and the Ito convention that $\sR(* , t)=0$ for $t \leq 0$) we thus obtain that
\bea \label{Eq:SecII3:FRGDep9}
\int_{t ,y} g_{x,y}^{-1} \sR(x_f-y , t_f-t) = \delta^{(d)}(x-x_f) \ssp .
\eea
I.e. inverting the elastic kernel and going into Fourier space we obtain
\bea \label{Eq:SecII3:FRGDep10}
\sR(q,\omega = 0 ) = \frac{1}{(q^2 + \mu^2)^{\frac{\gamma}{2}} } \quad , \quad \sR(q , \omega ) := \int_{q,\omega} e^{- iq x - i \omega t } R(x,t) \ssp .
\eea
On the other hand, note that in the theory without disorder, the bare response function $\sR_0(x_f - x_i , t_f-t_i) := \langle \hat u_{t_i,x_i}  u_{t_f,x_f}\rangle_{S_0}$ is easily obtained as
\bea \label{Eq:SecII3:FRGDep11}
\sR_0(q,\omega) =  \frac{1}{ i \eta_0 \omega + (q^2 + \mu^2)^{\frac{\gamma}{2}} } \ssp . 
\eea 
Hence the zero frequency part of the response function $\sR(q, \omega)$ is not modified by the disorder. Note that STS does not say anything on the non-zero frequency part. Indeed the latter will be modified and the viscosity coefficient $\eta$ receives a correction from the disorder. Before we proceed, let us emphasize here an idea that can be seen in much of the calculation performed in the dynamics: there is an analogy between the time in the dynamic theory and the replica index in the statics. This can be seen by comparing the calculations performed above with the one in the discussion of STS in the static theory in Sec.~\ref{subsec:FRGStatic}. This analogy can be pushed quite far \cite{LeDoussalWieseChauve2002}. Note that in the dynamics, when two position fields are far from each other in time, they are effectively independent but see the same disorder, just as would two replicated fields with different replica index in the statics. More generally there will be close links between observables at $0$ frequency in the dynamics and observables in the static theory.\medskip

\subsubsection{Main Results of FRG in the dynamics}

The result of FRG for depinning is that the effective action of the theory $\Gamma_{\mu}[u,\hat u]$ takes, in the limit $v \to 0^+$ first and $\mu \to 0$ then, a scaling form as follows. Rescaling 
\bea \label{Eq:SecII3:FRGDep12}
&& x = \mu^{-1} \tilde{x} \quad,  \quad  t = \mu^{-z} \tilde{t} \quad,\quad   \nn \\
&& u_{t,x} =  \mu^{-\zeta_d} \tilde{u}_{\tilde t, \tilde x}  \quad , \quad  \hat{u}_{t,x} = \mu^{d+z - \gamma + \zeta_d }\tilde{\hat{u}}_{\tilde t , \tilde x} \ssp .
\eea
The rescaled effective action
\be \label{Eq:SecII3:FRGDep13}
\tilde{ \Gamma}_{\mu}[\{\tilde{u}_{\tilde t, \tilde{x}}\} , \{\tilde{\hat u}_{\tilde t, \tilde{x}}\}  ] = \Gamma_{\mu} [ \{u_{t,x}=  \mu^{-\zeta_d} \tilde{u}_{\tilde{t}= \mu^{z} t , \tilde{x}= \mu x} \},\{\hat{u}_{t,x}=  \mu^{d+z-\gamma+\zeta_d} \tilde{\hat{u}}_{\tilde{t}= \mu^{z} t , \tilde{x}= \mu x} \}]
\ee
converges, as $\mu \to 0$, to a fixed point of a rescaled functional Beta function
\bea \label{Eq:SecII3:FRGDep14}
\lim_{\mu \to 0}\tilde{ \Gamma}_{\mu}[\{\tilde{u}_{\tilde t, \tilde{x}}\} , \{\tilde{\hat u}_{\tilde t, \tilde{x}}\}  ] =\tilde{ \Gamma}^*[\{\tilde{u}_{\tilde t, \tilde{x}}\} , \{\tilde{\hat u}_{\tilde t, \tilde{x}}\}  ]  \quad,  \quad \tilde{\beta}[\tilde{ \Gamma}^*]= 0 \ssp .
\eea
In this limit the effective action can be computed in an expansion in $\epsilon = 2 \gamma- d$. As for the statics the critical exponents $\zeta$ and $z$ are adjusted so that a fixed point indeed exists. The latter has the same form as the initial action of the theory, except that\\
(i) one must replace $\eta$ by a renormalized value
\bea \label{Eq:SecII3:FRGDep15}
\eta \to  \mu^{\gamma-z} \tilde{\eta}_\mu 
\eea
where $\tilde{\eta}_{\mu}$ flows with the RG and converges to a constant $\eta^*:=\lim_{\mu \to 0} \eta_{\mu}$ (which depends on the flow, i.e. it is non-universal). The Beta function associated with the flow of $\eta$ was computed up to order $O(\epsilon^2)$ in \cite{LeDoussalWieseChauve2002}. \\
(ii) The disorder part of the effective action is changed to
\bea \label{Eq:SecII3:FRGDep16}
\Gamma_{{\rm dis}}[u,\hat u] := -\frac{1}{2} \int_{t,t',x} \hat u_{t,x} \hat u_{t',x} \Delta(u_{tx} - u_{t',x}) +O(\epsilon^2) \ssp ,
\eea
where, as in the statics, the $O(\epsilon^2)$ contains higher order cumulants (and the non-local part of the second order cumulant) of the renormalized pinning force. The latter admits an expansion in $\Delta$, that can be computed using a standard loop expansion and handling carefully the non-analyticities (see below). Close to the fixed point, $\Delta(u)$ takes a scaling form
\bea \label{Eq:SecII3:FRGDep17}
\Delta(u) = A_d^{\gamma} \mu^{\epsilon - 2 \zeta_d} \tilde{\Delta}(\mu^{\zeta_d} u)
\eea
where $\tilde{\Delta}(u) = O(\epsilon)$ is closed to a FP of the following FRG flow equation (computed up to two-loops in \cite{LeDoussalWieseChauve2002}):
\bea \label{Eq:SecII3:FRGDep18}
- \mu \partial_{\mu} \tilde{\Delta}(u) && = \underbrace{(\epsilon - 2 \zeta_d) \tilde{\Delta}(u) + \zeta_d u \tilde{\Delta}'(u)}_{{\rm rescaling }}  - \underbrace{ \frac{1}{2} [(\tilde{\Delta}(u) - \tilde{\Delta}(u))^2 ]''  }_{1-{\rm loop}} \nn \\
&& + \underbrace{\frac{1}{2} X [ ( \Delta(u) - \Delta(0))\Delta'(u)^2 ]''- \frac{\lambda_d}{2} X  (\tilde{\Delta}'(0^+)^2 ) \tilde{\Delta}''(u) }_{2-{\rm loops}} \nn \\ 
&& + O( \tilde{\Delta}^4 ) 
\eea
where $\lambda_d = -1$. As in the statics, if $\tilde{\Delta}^*(u)$ is a fixed point, $\frac{1}{\kappa^2}\tilde{\Delta}^*(\kappa u)$ is also a fixed point and thus there are several families of fixed points. The fixed point toward which the system flows depends on the full flow (in particular it depends on the starting point, i.e. microscopic parameters). Remarkably, to one-loop order, the flow equation (\ref{Eq:SecII3:FRGDep18}) is identical to the one of the statics. Namely, taking two derivatives of (\ref{Eq:SecII3:FRGStat19}) and defining $\tilde{\Delta}(u)= - \tilde{R}''(u)$, one obtains exactly (\ref{Eq:SecII3:FRGDep18}), up to the change $\lambda_s \to \lambda_d$ (we remind $\lambda_s=1$). Hence, up to two loop order, the FRG equation for the depinning and the statics only differs by the coefficient in front of the `anomalous terms' involving the $\tilde{R}'''(0^+) = - \tilde{\Delta}'(0^+)$. This change has drastic consequences:
\begin{itemize}
\item{The random bond universality class of the statics, the universality class for short-range correlated disorder, is unstable in the dynamics and flows to the random field universality class (see below). This is a signature of irreversibility in the depinning process: since the interface always moves forward in the dynamics, it does not know whether it is dragged in a random force landscape that is the derivative of a potential or not. This difference was confirmed numerically in \cite{RossoLeDoussalWiese2006a}.}
\item{The random field universality class thus appears as the unique universality class for the depinning with non-periodic short-range disorder. Moreover this random field universality class is different from the static one. The roughness exponent was shown to be different with $\zeta_d \simeq  \frac{\epsilon}{3}(1 + 0.143313 \epsilon) + O(\epsilon^3)$ for SR elasticity ($z \simeq 2-\frac{2}{9} \epsilon-0.0432087\epsilon^2+O(\epsilon^3)$) and $\zeta_d \simeq  \frac{\epsilon}{3}(1 + 0.39735 \epsilon) + O(\epsilon^3)$ ($z = 1- \frac{2}{9} \epsilon -0.1132997\epsilon^2 + O(\epsilon^3)$) for LR elasticity. The fixed point function $\tilde{\Delta}^*(u)$ is also different from the static one, as confirmed numerically in \cite{RossoLeDoussalWiese2006a}.}
\item{The random periodic universality class is also different and the fixed-point function is non-potential. See \cite{LeDoussalWieseChauve2002} for more details.}
\end{itemize}
This shows that it is crucial, when performing a perturbative calculation in the statics or in the dynamics in terms of the non-analytic, renormalized disorder, to evaluate carefully the possible non-analytical terms that arise because of the short-distance singularities of the fixed point functions. In the statics as we explained, the correct evaluation of such anomalous terms require to take very carefully the $T \to 0$ limit. In the dynamics the situation is somehow simpler and the short-distance singularities are regularized by the non-zero velocity: since the renormalized disorder correlator always comes with $\Delta(u_{xt} -u_{x t'})$ and $u_{xt}\geq u_{xt'}$ if $t \geq t'$, there should never be ambiguities regarding the side of the cusp that appears. In particular, the higher cumulants of the renormalized pinning force (the $O(\epsilon^2)$ in (\ref{Eq:SecII3:FRGDep16})) can be computed using a standard loop expansion (at least up to two-loop \cite{LeDoussalWieseChauve2002}) directly at $T=0$ and $v = 0^+$.\\

{\it Other exponents at the depinning transition} \\
In our field theory approach to depinning, the correlation length $\xi$ mentioned in the introduction to the depinning transition in Sec.~\ref{subsec:SecI3:Avalanches} is equal to $\ell_{\mu}:=1/\mu$ since $\mu$ is not corrected by the renormalization (STS). On the other hand, an appropriate definition of the critical force is the mean force exerted by the well on the interface as $v \to 0^+$ in the steady state:
\bea \label{Eq:SecII3:FRGDep19}
f_c(\mu) := \lim_{v \to 0^+}\mu^{\gamma} \overline{vt - u_{tx}} \ssp .
\eea
Hence, at non-zero velocity, slightly above the depinning transition, we expect that $f-f_c(\mu) = \mu^{\gamma} \overline{vt - u_{tx}} - \lim_{v \to 0}\mu^{\gamma} \overline{vt - u_{tx}}  $ to scale as $\mu^{\gamma - \zeta_d}$. Hence the exponent $\nu$ defined by $\xi \sim (f-f_c)^{-\nu}$ is thus 
\bea \label{Eq:SecII3:FRGDep20}
\nu = \frac{1}{ \gamma - \zeta_d} \ssp .
\eea
On the other hand the exponent $\beta$ defined by $v = \overline{ \partial_{t} u_{tx}} \sim(f-f_c)^{\beta}$ must scale as $ \mu^{z - \zeta_d} = \mu^{\beta( \gamma- \zeta_d) } $ and hence
\bea \label{Eq:SecII3:FRGDep21}
\beta = \frac{z- \zeta_d}{\gamma- \zeta_d} \ssp .
\eea

\subsection{Applying the functional renormalization group to avalanches} \label{subsec:FRGDepAva}

\subsubsection{Introduction}

Let us now extend our analysis of Sec.~\ref{subsec:avalanches} of shocks in the statics to the case of avalanches in the dynamics. We first note the relation, shown in $\cite{LeDoussalWiese2008a}$, that extends the relation (\ref{Eq:SecII3:FRGStatShock7}) of the statics:
\bea  \label{Eq:SecII3:FRGDepAva1}
\Delta(w-w') = L^d m^4 \overline{(u(w) -w) (u(w') -w') }^c \ssp ,
\eea
where $\Delta(w-w')$ is the renormalized second cumulant of the pinning force at depinning and $u(w)$ is the center of mass of the interface in the forward quasi-static process
\bea \label{Eq:SecII3:FRGDepAva2}
u(w):= \frac{1}{L^d} \int_x u_x(w) \quad , \quad u_x(w) := \lim_{v \to 0^+} u_{t=w/v , x} \ssp .
\eea
Note that it slightly differs from (\ref{Eq:SecII3:FRGStatShock7}) by the fact that the average in (\ref{Eq:SecII3:FRGDepAva1}) is a connected average. While in the statics we have $\overline{u(w)} = w$, in the dynamics $\overline{u(w)} \neq  w$ and the difference $m^2(w - \overline{u(w)})$ is the critical force of the system. Apart from this the interpretation and consequences of the formula (\ref{Eq:SecII3:FRGDepAva1}) are mostly equivalent to those of (\ref{Eq:SecII3:FRGStatShock7}) and we refer the reader to the discussion after (\ref{Eq:SecII3:FRGStatShock7}). In particular as $\mu \to 0$, beyond the Larkin length, the left-hand-side of (\ref{Eq:SecII3:FRGDepAva1}) takes a universal, non-analytic scaling form (\ref{Eq:SecII3:FRGDep17}). This non-analyticity is interpreted as the occurrence of avalanches in the forward quasi-static process and we assume that in the scaling limit $u_x(x) \sim \mu^{-\zeta_d}$,
\bea  \label{Eq:SecII3:FRGDepAva3}
u_x(w) = cst+ \sum_i S^{(i)}_{x} \theta(w-w_i) \ssp .
\eea
As in the statics, the density for the total size of the shocks, $S^{(i)} = \int_x S^{(i)}_x$, is defined as
\bea  \label{Eq:SecII3:FRGDepAva4}
\rho(S) := \overline{ \sum_{i} \delta(w-w_i) \delta(S-S^{(i)}) } \ssp ,
\eea
and does not depend on $w$ in the steady-state. Without repeating the same discussion as in the statics we obtain the two important exact relations
\bea   \label{Eq:SecII3:FRGDepAva5}
\langle S \rangle_{\rho} = L^d \quad , \quad S_m:= \frac{\langle S^2 \rangle_{\rho} }{2 \langle S \rangle_{\rho}}  = \frac{\sigma}{m^4}  =A_d^{\gamma} \mu^{- d - \zeta_d} \tilde{\sigma} \ssp ,
\eea
where $\sigma=-\Delta'(0^+)$ and $\tilde{\sigma} = - \tilde{\Delta}'(0^+)$ are $O(\epsilon)$. Let us now see how to relate more generally an observable associated with the avalanche motion of the interface to an observable of the MSR field theory. Before we do so let us note that it is important to have in mind the discussion of avalanches in the $d=0$ case given in Sec.~\ref{subsec:ToyAva}.

\subsubsection{The MSR response field as the generator of avalanche motion}

Let us now consider the theory for the interface velocity field $\dot{u}_{tx} = \partial_t \dot{u}_{tx}$. It is obtained by taking a derivative of (\ref{Eq:SecII3:FRGDep1})
\bea   \label{Eq:SecII3:FRGDepAva6b}
&& \eta_0 \partial_t \dot u_{tx} = \int_{y} g_{x,y}^{-1}(v-\dot{u}_{ty}) + \partial_t F(x, u_{tx}) 
\eea
We consider, for an arbitrary source $\lambda_{tx}$, the generating functional of the velocity field in the steady state of (\ref{Eq:SecII3:FRGDepAva6b}) 
\bea \label{Eq:SecII3:FRGDepAva7b}
G[\lambda_{tx}]:= \overline{ e^{ \int_{tx} \lambda_{tx} \dot{u}_{tx} } } \ssp . 
\eea
Let us suppose that $\lambda_{tx}$ is non-zero only in a time window $t \in [0,T]$. The latter is taken to be large compared to the typical time scale of the avalanche motion (so that every avalanche that occurs in this time window terminates), and small compared to the waiting time in between successive avalanche (that is of order $O(1/v)$). This will be automatically ensured later by taking the limit $v \to 0$ first and $T \to \infty$ afterwards. Let us now denote: (i) $p_{t_i}[\dot{u}_{tx}]$ the probability distribution functional of the velocity field of the interface knowing that an avalanche started somewhere along the interface at time $t_i$, normalized as $\int \cD[\dot u]p_{t_i}[\dot{u}_{tx}]=1$; (ii) $\rho_0$ the mean number of avalanches per unit of driving. From the general picture of avalanche motion built on $d=0$ toy models in Sec.~\ref{subsec:ToyAva}, we know (i.e. that is how we would like to interpret the FRG results) that in the limit $v \to 0^+$, $G[\lambda_{tx}]$ admits the expansion
\bea \label{Eq:SecII3:FRGDepAva8b}
G[\lambda_{tx}] = (1 - \rho_0 v T) +  \rho_0 v \int_{t_i=0}^T \int \cD[\dot u] e^{\int_{tx}\lambda_{tx}\dot{u}_{tx}}p_{t_i}[\dot{u}_{tx}] +O(v^2) \ssp .
\eea 
Equivalently we can write, introducing the {\it density functional of the velocity field inside an avalanche starting at time $t_i$}: $\rho_{t_i}[\dot{u}_{tx}] = \rho_0 p_{t_i}[\dot{u}_{tx}] $, 
\bea \label{Eq:SecII3:FRGDepAva9b}
G[\lambda_{tx}]-1 = v \int_{t_i=0}^T \int \cD[\dot u] \left( e^{\int_{tx}\lambda_{tx}\dot{u}_{tx}}-1\right) \rho_{t_i}[\dot{u}_{tx}] +O(v^2) \ssp .
\eea
On the other hand, $G[\lambda_{tx}]$ can be computed using the MSR field theory associated with the velocity theory. The latter is a simple adaptation of (\ref{Eq:SecII3:FRGDep3})-(\ref{Eq:SecII3:FRGDep4}) and we obtain
\bea \label{Eq:SecII3:FRGDepAva10b}
&& G[\lambda_{xt}]=\int D[\tilde u]D[ \dot{u}] e^{ \int_{xt} \lambda_{xt} \dot{u}_{xt}  - S[\dot u , \tilde  u] +m^2 v \int_{tx} \tilde u_{tx}} \ssp ,
\eea
where $S[\dot u , \tilde  u]:= S_0[\dot u , \tilde  u] +S_{{\rm dis}} [\dot u , \tilde  u]  $ with
\bea \label{Eq:SecII3:FRGDepAva11b}
&& S_0[\dot u , \tilde  u] :=  \int_{tx} \tilde u_{tx} \left( \eta_0 \partial_t \dot u_{tx}  + \int_{y} g_{x,y}^{-1} \dot u_{ty} \right)  \ssp , \nn \\
&& S_{{\rm dis}} [\dot u , \tilde  u] := -\frac{1}{2} \int_{t,t',x} \tilde u_{tx} \tilde u_{t'x}  \partial_t \partial_{t'} \Delta_0(u_{tx} - u_{t'x})  \ssp .
\eea

Taking now the expansion in $v$ of (\ref{Eq:SecII3:FRGDepAva10b}) we obtain
\bea \label{Eq:SecII3:FRGDepAva12b}
G[\lambda_{xt}]- 1 = m^2 v \langle\int_{tx}\tilde{u}_{tx} e^{ \int_{xt} \lambda_{xt} \dot{u}_{xt}} \rangle_{S} + O(v^2) \ssp .
\eea
Hence, comparing (\ref{Eq:SecII3:FRGDepAva12b}) with (\ref{Eq:SecII3:FRGDepAva10b}) we identify
\bea \label{Eq:SecII3:FRGDepAva13b}
Z[\lambda]:=  \int_{t_i=0}^T \int \cD[\dot u] \left( e^{\int_{tx}\lambda_{tx}\dot{u}_{tx}} -1 \right)\rho_{t_i}[\dot{u}_{tx}] = m^2  \int_{tx} \langle \tilde{u}_{tx}  e^{ \int_{xt} \lambda_{xt} \dot{u}_{xt}} \rangle_{S} \ssp .
\eea
Note now that from a diagrammatic point of view, all diagrams that contribute to $\langle \tilde{u}_{tx} e^{ \int_{xt} \lambda_{xt} \dot{u}_{xt}}\rangle_{S}$ correspond to `histories of the interface motion' such that the first non-zero velocity of the interface is at a time larger or equal to $t$ and at a position $x$. It is thus natural to identify 
\bea \label{Eq:SecII3:FRGDepAva14b}
Z_{t_i,x_i}[\lambda]:= \int \cD[\dot u] \left( e^{\int_{tx}\lambda_{tx}\dot{u}_{tx}} -1 \right)\rho_{t_i,x_i}[\dot{u}_{tx}] = m^2 \langle \tilde{u}_{t_i x_i}  e^{ \int_{xt} \lambda_{xt} \dot{u}_{tx}} \rangle_{S} \ssp ,
\eea
 where $\rho_{t_i ,x_i}[\dot{u}_{tx}]$ is the density for the velocity field inside {\it avalanches that are triggered at time $t_i$ at the position $x_i$}. This formula first appeared in \cite{ThieryLeDoussal2016a}. Although, as it is usual when dealing with avalanches, one could debate some heuristic steps that were taken in its derivation, it is also coherent with the more controlled setting of avalanches following a kick in the force in a non-stationary setting in a solvable model, and we refer to \cite{ThieryLeDoussal2016a} for a more complete discussion of (\ref{Eq:SecII3:FRGDepAva14b}). Let us now discuss how to use FRG to obtain a simplified action that allows us to compute the $\epsilon$ expansion of the right-hand side of (\ref{Eq:SecII3:FRGDepAva14b}) in the limit $\mu \to 0$, and thus obtain the $\epsilon$ expansion of a generic observable associated with the avalanche motion.

\subsubsection{The simplified action for the motion inside avalanches in the velocity theory} 

One way to use FRG to compute avalanche observables at depinning is to follow the same route as presented for the statics in Sec.~\ref{subsec:FRGStaticSchocks}. This route is presented in detail in \cite{LeDoussalWiese2012a} and shown to be equivalent, to one-loop accuracy and for observables associated with a single avalanche such as (\ref{Eq:SecII3:FRGDepAva14b}), to a simplified theory. We now present this simplified theory following \cite{LeDoussalWiese2012a} and \cite{DobrinevskiPhD}. The essential steps will be: \\
(i) Express the bare disorder force-force cumulant $\Delta_0(u)$ and viscosity coefficient $\eta_0$ of the theory by formally inverting the one-loop expressions for $\Delta(u)$ and $\eta$ obtained from one-loop perturbative FRG.\\
(ii) Express the MSR action for the {\it velocity theory} in terms of the renormalized disorder $\Delta(u)$ and viscosity $\eta$.\\
(iii) Take into account the fact that the scale of avalanches, $S_m$, is $O(\epsilon)$ to obtain a simplified action valid to describe the velocity field inside a single avalanche.\\
Let us start with the first step. The one-loop expression for $\Delta(u)$ in terms of $\Delta_0(u)$ is \cite{LeDoussalWieseChauve2002}
\bea  \label{Eq:SecII3:FRGDepAva9}
\Delta(u) = \Delta_0(u) - I_1 (\Delta_0(u) - \Delta_0(0) ) \Delta_0'(u) \ssp .
\eea
It can formally be inverted\footnote{The formal expansions and inversions of series performed here are controlled under the assumption that $\Delta$ and $\Delta_0$ are small and of the same order.} as
\bea  \label{Eq:SecII3:FRGDepAva10}
\Delta_0(u) = \Delta(u) + I_1  (\Delta(u) - \Delta(0) ) \Delta'(u)  \ssp .
\eea
In the same way \cite{LeDoussalWieseChauve2002}
\bea  \label{Eq:SecII3:FRGDepAva11}
\eta = \eta_0(1 - \Delta_0''(0^+) I_1 )\quad , \quad \eta_0= \eta(1 + \Delta''(0^+) I_1 ) \ssp .
\eea
Of course these expressions and the inversions that were made are completely formal: while the bare disorder correlator appearing on the left of (\ref{Eq:SecII3:FRGDepAva10}) can be a smooth function of $u$, we know that the renormalized disorder on the right of (\ref{Eq:SecII3:FRGDepAva10}) is non analytic for $\mu \leq \mu_c$. As usual with perturbative RG, this type of manipulations and one-loop expression (\ref{Eq:SecII3:FRGDepAva9}) are only appropriate to obtain the RG flow of the parameters close to a perturbative FP (see e.g. \cite{Delamotte:2002vw}). More precisely, taking the derivative $-\mu \partial_\mu$ of (\ref{Eq:SecII3:FRGDepAva9}) at fixed $\Delta_0(u)$ and then replacing on the right-hand side $\Delta_0(u) $ by the expression (\ref{Eq:SecII3:FRGDepAva10}) leads to the correct one-loop Beta function (\ref{Eq:SecII3:FRGDep18}). This inversion allows however to obtain self-consistently a bare action that will lead, using a one-loop perturbative calculation of an observable, to the correct result up to order $O(\epsilon)$. Hence to obtain our avalanche observables (\ref{Eq:SecII3:FRGDepAva14b}) we will perform one-loop perturbative calculations with an action similar to (\ref{Eq:SecII3:FRGDepAva11b}) with the replacement $\eta_0 \to \eta$, $\Delta_0 \to \Delta$ and reminding ourselves in the end that counter-terms involving the loop-integral $I_1$ as in (\ref{Eq:SecII3:FRGDepAva10}), (\ref{Eq:SecII3:FRGDepAva11})  must be added. \\

The above considerations are the steps (i) and (ii) in the announced program. Let us now finally perform the last step, namely use that $S_m$, the upper-cutoff of the avalanche size distribution, is $O(\epsilon)$. We thus need to rescale (\ref{Eq:SecII3:FRGDepAva14b}) in a way that allows us to study the limit $\mu \to 0$ and $\epsilon \to 0$. We will do the rescaling in two steps for clarity. We thus consider
\bea \label{Eq:SecII3:FRGDepAva15b}
\tilde{Z}_{t_i,x_i}[\tilde{\lambda}]:= Z_{t_i,x_i}[\lambda=\tilde{\lambda}/S_m] = m^2 \langle \tilde{u}_{t_i x_i} e^{ \int_{xt} \frac{\tilde{\lambda}_{xt}}{S_m} \dot{u}_{xt}} \rangle_{S} 
\eea

Let us first take care of the scaling with $\mu$: we must adapt the scaling of (\ref{Eq:SecII3:FRGDep12}) to the fields of the velocity theory. We thus take,  
\bea  \label{Eq:SecII3:FRGDepAva13}
&& x = \mu^{-1} \bar x \quad,  \quad  t = \mu^{-z} \bar t \quad,\quad   \nn \\
&& \dot{u}_{t,x} =  \mu^{z-\zeta_d} \dot{\bar u}_{\bar t, \bar x}  \quad , \quad  \tilde{u}_{t,x} = \mu^{d - \gamma + \zeta_d }\bar{\tilde{u}}_{\bar t , \bar x} \ssp ,
\eea
using here $\bar ()$ for rescaled quantity, and where the dots mean either $\partial_t$ or $\partial_{\bar t}$, depending on the field to which they are applied. The elastic kernel is rescaled as $g_{\mu^{-1} \bar x ,\mu^{-1} \bar y}^{-1} = \mu^{d + \gamma} \bar g_{\bar x , \bar y}^{-1}$ with $\bar g_{\bar x , \bar y}^{-1} =\int_q e^{iq \mu^{-1} (x-y)} (q^2 +1)^{\frac{\gamma}{2}} $ and we remind ourselves that in the limit $\mu \to 0$ the rescaled renormalized disorder and friction are close to one of the fixed points of the FRG flow:
\bea  \label{Eq:SecII3:FRGDepAva14}
\eta \sim \mu^{\gamma- z} \tilde{\eta}^* \quad , \quad \Delta(u) \sim A_d^{\gamma} \mu^{\epsilon - 2 \zeta_d} \tilde{\Delta}^*(\mu^{\zeta_d} u) \ssp .
\eea
Using this rescaling, our `model' avalanche observable (\ref{Eq:SecII3:FRGDepAva14b}) is expressed as, using that $S_m = A_d^{\gamma} \mu^{- d - \zeta_d} \tilde{\sigma}^*$,
\bea  \label{Eq:SecII3:FRGDepAva15}
\tilde{Z}_{t_i,x_i}[\tilde{\lambda}] =  \mu^{ d + \zeta_d} \int \cD \dot{\bar u} \cD \bar{\tilde u} ~ \bar{\tilde u}_{\bar{t}_i \bar{x}_i} ~e^{   (A_d^{\gamma})^{-1} \int_{\bar t, \bar x} \tilde{\lambda}_{tx} \frac{\dot{\bar u}_{tx}}{\tilde{\sigma}^*}   - \bar S[\bar{\dot u} , \bar{\tilde u }]}  \ssp ,
\eea
where the action $\bar S[\bar{\dot u} , \bar{\tilde u }]$ is now as in (\ref{Eq:SecII3:FRGDepAva10b}) with $\eta \to \tilde{\eta}^* $, $\tilde{\Delta} \to \tilde{\Delta}^*$ and the dependence on $\mu$ has completely disappeared (apart from the prefactor). Finally, we must now `zoom' in on the $O(\epsilon)$ scale of the avalanche motion that is controlled by $\tilde{\sigma}^* = - \tilde{\Delta}^{*\prime}(0^+) = O(\epsilon)$. To this aim we thus rescale
\bea  \label{Eq:SecII3:FRGDepAva16}
\dot{\bar u}_{\bar t, \bar x} = A_d^{\gamma} \tilde{\sigma}^{*} \dot{\underline{u}}_{\bar t, \bar x} \quad , \quad \bar{\tilde{u}}_{\bar t, \bar x} = (A_d^{\gamma} \tilde{\sigma}^{*})^{-1} \underline{\tilde{u}}_{\bar t, \bar x} \ssp .
\eea
This final rescaling allows us to make an expansion of the renormalized disorder correlator around $u =0$: for $\underline{u}$ of order $1$ we have
\bea  \label{Eq:SecII3:FRGDepAva17}
\tilde{\Delta}^*(A_d^{\gamma} \tilde{\sigma}^{*} \underline{u}) =\tilde{\Delta}^*(0)- A_d^{\gamma} (\tilde{\sigma}^{*})^2  |\underline{u}| + \frac{1}{2} \tilde{\Delta}^{*\prime \prime}(0)\left(A_d^{\gamma} \tilde{\sigma}^{*} \right)^2 \underline{u}^2 + O(\epsilon^4) \ssp .
\eea
This rescaling leaves the quadratic part of the action invariant but changes the disorder part as, following the different changes of variables,
\bea  \label{Eq:SecII3:FRGDepAva18}
S_{{\rm dis}} [\dot u , \tilde  u] && = -\frac{1}{2} \int_{t,t',x} \tilde u_{tx} \tilde u_{t'x}  \partial_t \partial_{t'} \Delta(u_{tx} - u_{t'x})  \nn \\
&& =  -\frac{A_d^{\gamma}}{2 }  \int_{\bar t,\bar t',\bar x} \bar{\tilde u}_{\bar t \bar x} \bar{\tilde u}_{\bar t' \bar x} \partial_{\bar t} \partial_{\bar t'} \tilde{\Delta}^*(\bar u_{\bar t \bar x} - \bar u_{\bar t' \bar x})  \nn \\
&& =  -\frac{1}{2 A_d^{\gamma} (\tilde{\sigma}^*)^2 }  \int_{\bar t,\bar t',\bar x} \underline{\tilde u}_{\bar t \bar x} \underline{\tilde u}_{\bar t' \bar x} \partial_{\bar t} \partial_{\bar t'} \tilde{\Delta}^*( A_d^{\gamma} \tilde{\sigma}^{*} (\underline u_{\bar t \bar x} - \underline u_{\bar t' \bar x}))    \\
S_{{\rm dis}} [\dot u , \tilde  u] && = -\int_{\bar t  , \bar x }(\underline{\tilde u}_{\bar t \bar x})^2 ~\dot{\underline{u}}_{\bar t, \bar x} + \frac{ A_d^{\gamma} \tilde{\Delta}^{*\prime \prime}(0)}{2 }  \int_{\bar t ,\bar t' , \bar x }  \underline{\tilde u}_{\bar t \bar x} \underline{\tilde u}_{\bar t' \bar x}  \underline{\dot u}_{\bar t \bar x} \underline{\dot u}_{\bar t' \bar x}  + O(\epsilon^2) \nn  \ssp .
\eea
We refer the reader to \cite{LeDoussalWiese2012a} for more details on the simplification in the last line that notably uses $\underline{\dot u}_{\bar t \bar x} \leq \underline{\dot u}_{\bar t' \bar x}$ for $\bar t \leq \bar t'$. Hence our model observable  (\ref{Eq:SecII3:FRGDepAva15}) can be calculated when $\mu$ is close to $0$ using the $\epsilon$ expansion as
\bea  \label{Eq:SecII3:FRGDepAva19}
\tilde{Z}_{t_i,x_i}[\tilde{\lambda}] = \frac{1}{S_m} \int \cD \dot{\underline u} \cD \bar{\underline u} ~ \bar{\tilde u}_{\underline{t}_i \underline{x}_i}~e^{  \int_{\bar t, \bar x} \tilde{\lambda}_{tx} \dot{\underline u}_{tx}   - \underline S[\underline{\dot u} , \underline{\tilde u }]} ,
\eea
where the action $\underline S[\underline{\dot u} , \underline{\tilde u }]$ is similar to (\ref{Eq:SecII3:FRGDepAva10b}) with $\eta \to \tilde{\eta}^*$, $\mu \to 1$ and the disorder part is as in (\ref{Eq:SecII3:FRGDepAva18}). The observable can be computed to order $O(\epsilon)$ using one loop perturbative RG. Possible divergences appearing in the calculation are canceled by counter-terms associated to the renormalization of $\eta$ and $\Delta$ (\ref{Eq:SecII3:FRGDepAva10}) and (\ref{Eq:SecII3:FRGDepAva11})\footnote{A subtlety linked to calculations using the simplified action (\ref{Eq:SecII3:FRGDepAva18}) is that one has also to take care of a formal renormalization of $\mu$ forbidden by STS in the full theory. We refer the reader to \cite{DobrinevskiPhD,LeDoussalWiese2012a} for more details on this issue.}. In the following we conclude our introduction to the analysis of avalanches using FRG by focusing on the mean-field theory that is obtained by retaining only the terms of order $O(1)$ in (\ref{Eq:SecII3:FRGDepAva18}), i.e. we set $\tilde{\Delta}^{*\prime \prime}(0) \to 0$.

\subsubsection{The mean-field theory: the Brownian Force Model}

Let us now discuss in more details avalanche observables in the dynamics of elastic interfaces to lowest order in $\epsilon = 2 \gamma -d$. As discussed in the previous section we thus only need to consider an interface whose {\it velocity field} dynamics inside an avalanche is described by the action
\bea   \label{Eq:SecII3:SecBFM1}
&& \overline{ O[\{\dot u_{tx}\}]} := \int \cD \dot u \cD \tilde u  ~ O[\{\dot u_{tx}\}] ~e^{- S[\dot u , \tilde  u] }  \ssp , \nn \\
&&  S[\dot u , \tilde  u]:= S_0[\dot u , \tilde  u] +S_{{\rm dis}} [\dot u , \tilde  u]  +S_{{\rm dri}}[\dot u , \tilde  u]   \nn \\
&& S_0[\dot u , \tilde  u] :=  \int_{tx} \tilde u_{tx} \left( \eta \partial_t \dot u_{tx}  + \int_{y} g_{x,y}^{-1} \dot u_{ty} \right)  \ssp , \nn \\
&& S_{{\rm dis}} [\dot u , \tilde  u] := -\sigma\int_{tx}\tilde{u}_{tx}^2 \dot{u}_{tx}\ssp , \nn \\
&& S_{{\rm dri}}[\dot u , \tilde  u] := -m^2 v \int_{tx}  \tilde u_{tx} \ssp .
\eea  
Here we have reintroduced a possible driving velocity $v \geq 0$, the different units and the renormalized parameters $\eta$ and $\sigma$. Let us first comment on the nature of this theory.\\

{\it The Brownian Force Model} \\
It is a simple exercise \cite{LeDoussalWiese2012a} to show that the MSR action (\ref{Eq:SecII3:SecBFM1}) is equivalent to the following stochastic equation for the velocity field of the interface:
\bea \label{Eq:SecII3:SecBFM2}
\eta \partial_t \dot{u}_{tx} = \int_{y} g_{x,y}^{-1} (v - \dot{u}_{tx}) + \sqrt{2 \sigma \dot{u}_{tx} } \xi_{tx} \ssp, 
\eea
where $\xi_{xt}$ is a centered and normalized Gaussian white noise (GWN): 
\bea \label{Eq:SecII3:SecBFM3}
\overline{ \xi_{tx} \xi_{t',x'} } = \delta^{(d)}(x-x') \delta(t-t') \ssp .
\eea
In turn, the equation (\ref{Eq:SecII3:SecBFM2}) appears as the time derivative of an equation for the position field of the interface as
\bea \label{Eq:SecII3:SecBFM4}
\eta \partial_t u_{tx} = \int_{y} g_{x,y}^{-1} (vt  - u_{tx}) + F(x,u_{xt}) \ssp, 
\eea
where for each $x$, $F(x,u)$ is a Brownian motion (BM) in $u$ independent of the others and with increments
\bea \label{Eq:SecII3:SecBFM5}
\overline{ (F(x,u) - F(x,u'))^2} := 2 \sigma |u-u'| \ssp .
\eea
This theory was called the Brownian Force Model (BFM) in \cite{LeDoussalWiese2011b,LeDoussalWiese2012a,DobrinevskiLeDoussalWiese2011b}. Note that the emergence of a BM should not be surprising: we obtained the BFM as the mean-field theory for the depinning of interfaces in short-range disorder by linearizing the correlator of the renormalized pinning force around the cusp because avalanches are small, that is of order $O(\epsilon)$. For an arbitrary pinning force $\tilde{F}(x,u)$ that is stationary with $\overline{\tilde{F}(x,u)\tilde{F}(x',u')}^c = \delta^{(d)}(x-x') \Delta(u-u')$ and has a cusp, we have
\be \label{Eq:SecII3:SecBFM6}
\overline{ (\tilde{F}(x,u) -\tilde{ F}(x,u'))^2} = 2 (\Delta(0) - \Delta(u-u')) \simeq - 2 \Delta'(0^+) |u-u'| + O(|u-u'|^2) \ssp .
\ee
hence we retrieve generally the BFM with $\sigma = - \Delta'(0^+)$ through such considerations. A subtle issue here is, however, that the BM is not a stationary process. While in (\ref{Eq:SecII3:SecBFM2}) we did not define precisely the initial condition since it is implicit that we are looking at the stationary process for the velocity field $\dot{u}_{tx}$ (which exists), the process in (\ref{Eq:SecII3:SecBFM4}) has generally no stationary state. The definition (\ref{Eq:SecII3:SecBFM4}) thus requires some precisions. One way is to make the Brownian motion $F(x,u)$ stationary in $u$ by considering Brownian bridges in a large box \cite{LeDoussalWiese2012a} $F(x,0) = F(x,W)=0$ with $W \gg 1$ and looking at the process in the middle of the box in a width of order $1$: $u \to W/2 + u$ with $u =O(1)$. The correlations between the different BM constructed in this way are then to leading order $\overline{F(x,u) F(x',u')}^c = \delta^{(d)}(x-x') (\Delta(0) - \sigma|u-u'|)$. This is, however, a bit artificial and in this setup $\Delta(0)$ is huge, $\Delta(0) \sim W$. From a more pragmatic point of view one can consider one-sided Brownian motion with the initial condition that the interface is at rest at $t=0$:
\bea  \label{Eq:SecII3:SecBFM7}
F(x,0) =0 \quad , \quad \dot{u}_{tx} =u_{tx}=0 \ssp ,
\eea
In this setup one can now consider an arbitrary (non-stationary) driving $w(t)$ with the equation of motion 
\bea \label{Eq:SecII3:SecBFM7b}
\eta \partial_t u_{tx} = \int_{y} g_{x,y}^{-1} (w(t) - u_{tx}) + F(x,u_{xt}) \ssp, 
\eea
and $w(t=0)=0$. If $w(t)$ is always increasing, $\dot{w}(t)\geq 0$, the interface dynamics in the velocity theory is described by the same MSR action as in (\ref{Eq:SecII3:SecBFM1}) with $S_{{\rm dri}}[\dot u , \tilde  u] = m^2 \int_{tx} \tilde u_{tx} \dot{w}(t)$. 
\medskip

{\it The BFM as a FRG fixed point in any $d$ and the scaling exponents of the BFM}\\
As was first remarked in \cite{LeDoussalWiese2011b}, and can be checked by differentiating the static FRG equation for $\tilde{R}(u)$ (\ref{Eq:SecII3:FRGStat19}) three times with respect to $u$ or the dynamic FRG equation (\ref{Eq:SecII3:FRGDep18}) once with respect to $u$ to obtain in both cases a FRG equation for $\tilde{\Delta}'(u)$, the model defined by
\bea \label{Eq:SecII3:SecBFM7c}
\tilde{\Delta}'(u) = \frac{d}{du}\left( - \tilde{\sigma}  |u| \right) =- \tilde{\sigma} {\rm sign}(u)
\eea
is a fixed point of the FRG equation in any $d$ with the exponent $\zeta_s= \zeta_d = \epsilon$ and the dynamic exponent $z = \gamma$. Furthermore this fixed point was argued in \cite{LeDoussalWiese2011b} to be stable and to be an exact fixed point for an arbitrary number of loops. In \cite{DobrinevskiLeDoussalWiese2011b} it was even shown that it is a fixed point of the (more complicated and not shown in this manuscript) FRG equation for the dynamics at non-zero velocity. Note that inserting the scaling (\ref{Eq:SecII3:FRGDepAva13}) in (\ref{Eq:SecII3:SecBFM1}) it is easily seen that the critical exponents $z = \gamma$ and $\zeta= \epsilon $ does lead to a $\mu$ independent action (this is linked to the exact scale invariance of the BM).

\medskip

{\it Back to the ABBM model}\\
Let us now look at the dynamics of the center of mass in the BFM model. Defining $u_t =\frac{1}{L^d}\int_{x}u_{tx}$, we obtain from (\ref{Eq:SecII3:SecBFM2})
\bea \label{Eq:SecII3:SecBFM8}
\eta \partial_t\dot{u}_t = m^2(v- \dot{u}_t) + \sqrt{2 \sigma_L} \xi_t
\eea
where we have used the identity in law $\frac{1}{L^d}\int_{x}\sqrt{\dot{u}_{t,x}} \xi_{tx} = \sqrt{\sigma_L \dot{u}_t} \xi_t $  with $\sigma_L = \sigma/L^d$ and $\xi_t$ a unit centered GWN $\overline{\xi_t \xi_{t'}}^c = \delta(t-t')$. The equation (\ref{Eq:SecII3:SecBFM8}) is equivalent to the time-derivative of the equation of motion of a particle in the ABBM model already considered in Sec.~\ref{subsec:ToyAva}\ref{subsec:ABBM}. Hence {\it the mean-field theory for the motion of the center of mass of an elastic interface inside an avalanche at the depinning transition is the ABBM model}. In particular, the mean-field value for the power-law exponent $\tau_S$ is, as in the shocks case, $3/2$. Note that this value is also consistent with the NF conjecture applied to the BFM in any $d$ since $\tau_S = 2 - \gamma/(d + \zeta)=3/2$ as $\zeta= \epsilon = 2 \gamma -d$. Let us remind here the reader that this mean-field exponent is linked to the first return time to the origin of the one-dimensional BM as shown in Sec.~\ref{subsec:ToyAva}. The mean-field nature of the ABBM model was already argued on phenomenological grounds in \cite{ZapperiCizeauDurinStanley1998,Colaiori2008}. Here it has been derived from first principles using FRG but more importantly it is now clear how to go beyond the predictions of the ABBM model. Namely, the BFM, first introduced in \cite{LeDoussalWiese2011a,LeDoussalWiese2012a,DobrinevskiLeDoussalWiese2011b}, provides the proper mean-field theory to describe {\it spatial correlations} in the avalanche process. Finally, FRG also permits to go beyond mean-field and to compute corrections in an $\epsilon$ expansion. Before we close this chapter, let us finally recall here that the BFM has an important exact solvability property and show an application of this property.

\subsubsection{ Exact solvability of the BFM and the avalanche size distribution} 
We now discuss a remarkable solvability property of the BFM. We consider a non-stationary case with an inhomogeneous driving $w_x(t)$ described by the equation of motion
\bea \label{Eq:SecII3:SecBFM9}
\eta \partial_t u_{tx} = \int_{y} g_{x,y}^{-1} (w_x(t) - u_{tx}) + F(x,u_{xt}) \ssp, 
\eea
where as before the BM is one sided $F(x,0) = 0$ and at $t=0$ the interface is at rest $u_{tx} =0$ and $w_x(t=0)=0$. We suppose that the driving is always increasing $\dot{w}_x(t)\geq 0$ during a finite amount of time and note the total displacement $w_x = \int_0^{\infty} \dot{w}_x(t) dt$. Our goal is to compute the generating function for the velocity field for an arbitrary source $\lambda_{xt}$:
\bea \label{Eq:SecII3:SecBFM10}
G(\lambda_{xt}) && := \overline{ e^{\int_{t \geq 0 , x} \lambda_{xt} \dot{u}_{xt}} }  \ssp  \\
&&= \int \cD[\dot u] \cD[\tilde{u} ] e^{- \int_{tx} \tilde u_{tx} \left( \eta \partial_t \dot u_{tx}  + \int_{y} g_{x,y}^{-1} \dot u_{ty}  \right) + \sigma\int_{tx}\tilde{u}_{xt}^2 \dot{u}_{tx} +m^2 \int_{tx} \tilde{u}_{tx} \dot{w}_{x}(t) } e^{ \int_{t \geq 0 , x} \lambda_{tx} \dot{u}_{xt}}  \nn
\eea
Here we have rewritten the average over disorder using the MSR action. As first remarked in \cite{LeDoussalWiese2011a}, a remarkable simplification of the BFM is that the action for the velocity theory is {\it linear in} $\dot{u}_{tx}$. The path integral over $\dot u$ thus simply leads to a functional Dirac delta distribution. One then easily obtains
\bea \label{Eq:SecII3:SecBFM11}
G(\lambda_{xt})= e^{m^2 \int_{tx} \tilde{u}^{\lambda}_{tx} \dot{w}_x(t) },
\eea
where $\tilde{u}^{\lambda}_{tx}$ is the solution of the `instanton' equation:
\bea \label{Eq:SecII3:SecBFM12}
\eta \partial_t \tilde{u}_{tx}- \int_{y} g_{x,y}^{-1} \tilde{u}_{ty} +\sigma \tilde{u}_{tx}^2 + \lambda_{tx}=0
\eea
with the condition $\tilde{u}_{tx}=0$ for $t \geq t_{{\rm max}} = {\rm min} \{ t \in \JR , \lambda_{xt'} =0 ~ \forall t' \geq t \}$. A remarkable feature of (\ref{Eq:SecII3:SecBFM12}) is that that it does not depend on the driving: the dependence of the observable on the driving only appears in (\ref{Eq:SecII3:SecBFM11}). Before we show a simple application of this formula let us mention that the solution (\ref{Eq:SecII3:SecBFM11}-\ref{Eq:SecII3:SecBFM12}) can also be obtained without using the MSR formalism, see \cite{DobrinevskiLeDoussalWiese2011b}.
\medskip

{\it Distribution of avalanche total size in the BFM} \\
As an application of (\ref{Eq:SecII3:SecBFM11}-\ref{Eq:SecII3:SecBFM12}) consider the calculation of the PDF $P(S)$ of the total displacement of the interface $S = \int_{t \geq 0 , x} \dot{u}_{tx}$ for a homogeneous driving $w_x= w$. The Laplace transform of $P(S)$
\bea \label{Eq:SecII3:SecBFM13}
G(\lambda ) := \int_{S=0}^{\infty} P(S)e^{\lambda S} dS \ssp ,
\eea
is obtained using (\ref{Eq:SecII3:SecBFM11}-\ref{Eq:SecII3:SecBFM12}) with $\lambda_{xt} = \lambda$. The solution of \ref{Eq:SecII3:SecBFM12} is time and space independent and reads
\bea \label{Eq:SecII3:SecBFM14}
\tilde{u}_{tx}^\lambda = \frac{m^2}{\sigma} Z(\lambda S_m)  \quad , \quad Z(\lambda) = \frac{1}{2}\left( 1- \sqrt{1-4 \lambda} \right) \quad , \quad S_m := \sigma/m^4 \ssp .
\eea
Note that the function $Z(\lambda)$ already appeared in the mean-field calculation of the shocks total size density in (\ref{Eq:SecII3:FRGStatShock22}). Here we thus obtain
\bea \label{Eq:SecII3:SecBFM15}
G(\lambda)= e^{ \frac{L^d w}{S_m}Z(\lambda S_m) } \ssp .
\eea
Performing the Inverse-Laplace transform of (\ref{Eq:SecII3:SecBFM15}) as in \cite{LeDoussalWiese2012a}, we obtain the result given in the discussion of the ABBM model (\ref{Eq:SecII1:ABBM:Size1}) (there it was given in dimensionless units $S_m=1$ and with $v T_d = L^d w$):
\bea \label{Eq:SecII3:SecBFM16}
P(S)= \frac{L^d w}{2 \sqrt{\pi} \sqrt{S_m} S^{3/2}} e^{- \frac{(S- L^d w)^2}{4 S S_m}} 
\eea
In order to see the link between this quantity and avalanches in the quasi-static steady state of the interface, consider now the density of avalanche total size $\rho_{x=0}(S)$ triggered at an arbitrary time $t$ at position $x=0$ in the quasi-static steady state (for the velocity theory) of the BFM. Its `Laplace transform' is obtained using (\ref{Eq:SecII3:SecBFM14}) with $\lambda_{xt} = \lambda$. We can again use the instanton equation to evaluate the path-integral over $\dot u_{tx}$ and we obtain
\bea \label{Eq:SecII3:SecBFM17}
\int_{S>0} \left( e^{\lambda S} -1 \right) \rho_{x=0}(S) = m^2 \langle  \tilde{u}_{t,x=0} \rangle_S =  m^2 \tilde{u}_{t,x=0}^{\lambda} =  \frac{1}{S_m} Z( \lambda S_m) \ssp .
\eea
Inverting (see e.g. \cite{LeDoussalWiese2012a}), we obtain
\bea \label{Eq:SecII3:SecBFM18}
\rho_{x=0}(S) = \frac{1}{2 \sqrt{\pi} \sqrt{S_m} S^{3/2}} e^{- \frac{S}{4 S_m}} \ssp .
\eea
And we note the equality
\bea \label{Eq:SecII3:SecBFM19}
\rho_{x=0}(S)=\frac{1}{L^d}\frac{\partial P(S)}{\partial w}|_{w=0} \ssp ,
\eea
that shows the link between avalanches defined in the quasi-static steady state or in the non-stationary setting. Here the factor $ \frac{1}{L^d}$ accounts for the fact that avalanches contributing to $P(S)$ can be triggered with equal probability at any point of the interface. Using the fact that the BFM is a Lévy jump process it is also possible to `invert' (\ref{Eq:SecII3:SecBFM19}) and obtain $P(S)$ in terms of $\rho_{x=0}(S)$. This is a similar calculation as the one given in Sec.~\ref{subsec:ToyAva} for the stationary velocity distribution of the ABBM model and it is detailed in \cite{ThieryLeDoussalWiese2015}. Let us conclude this section by remarking that the density of avalanche total size in the BFM (\ref{Eq:SecII3:SecBFM18}) is the same as the one for shocks in the statics (\ref{Eq:SecII3:FRGStatShock23}) at the level of mean-field theory. This is not surprising since the differences between depinning and statics only appear at two-loop order in FRG.

\section[Summary of the thesis]{Summary of (and more context around) the results obtained during the thesis} \label{sec:AvaResults}

In this section we present the main results obtained on shocks and avalanches during this thesis. We begin in Sec.~\ref{subsec:SummaryAva} with a quick summary of the previous section and also present the actual research context around the obtained results. The next sections present the main results obtained in \cite{ThieryLeDoussalWiese2015,ThieryLeDoussal2016a,ThieryLeDoussalWiese2016}.

\subsection{Introduction} \label{subsec:SummaryAva}

\stab {\it Summary of the previous sections}\\
In the last section we have introduced the notion of shocks and avalanches in disordered elastic systems. We have shown how the Functional Renormalization Group can be efficiently used to calculate universal properties of these jump processes: the latter inherit the universality and the scale invariance of the FRG fixed points and gives a natural interpretation to the non-analytic nature of these FPs. For non periodic, short-range disorder we have shown that there are a priori two universality classes for shocks, random bond and random field disorder, and a unique universality class for avalanches that corresponds to random field disorder, which is however different from the Random Filed universality class for shocks. Close to the upper-critical-dimension, which depends on the range of the elasticity of the interface as $d_{{\rm uc}} = 2 \gamma$, we have identified the relevant mean-field theory to describe the motion inside avalanches at the depinning transition as the BFM model. The center of mass dynamics in the BFM model was shown to be equivalent to the ABBM model. Based on these constructions, we can now ask various questions about the universality in avalanche processes using mean-field approaches, but also beyond mean-field in a controlled $\epsilon = d_{{\rm uc}} - d$ expansion of observables using the structure of the FRG FPs. Let us now review some known results and introduce the subjects which will be the focus of Sec.~\ref{subsec:PresBFM}, Sec.~\ref{subsec:PresShape} and Sec.~\ref{subsec:PresCorrel}.

\smallskip

{\it Critical exponents}\\
The first focus of the community has been on the determination of the exponents characterizing the power-law distribution of quantities such as the extension, duration or total size of the avalanches. Since those are linked to one another by scaling relations involving the critical exponents of the statics (for shocks) or of the depinning transition (for avalanches), an important question was to understand whether or not the exponents can be entirely deduced from the exponents of the statics and depinning transition. The NF conjecture that was presented earlier, first proposed for avalanches at depinning \cite{NarayanDSFisher1993a} and  later generalized to the case of shocks \cite{LeDoussalMiddletonWiese2008,LeDoussalWiese2008c}, provides a precise affirmative answer to this question. Since it is however based on unproven assumptions, it is still important to obtain an independent derivation of these exponents. At the mean-field level the exponent $\tau_S= 3/2$, first derived in \cite{BertottiDurinMagni1994}, agrees with the NF conjecture. More recently, the NF conjecture was shown to hold up to one-loop both for shocks and avalanches in \cite{LeDoussalWiese2008c,LeDoussalWiese2011b,LeDoussalWiese2012a}.

\smallskip

{\it Universal distribution}\\
Besides critical exponents, it is interesting to obtain the full PDF of avalanche observables. Although these in general depend on the IR cutoff of the theory (i) cutoffs such as the massive scheme discussed in this thesis have proved relevant in the description of some experimental setups \cite{LeDoussalWieseMoulinetRolley2009}; (ii) the scaling with the cutoff of the different avalanche observables distributions on the IR cutting length is also expected to be universal, e.g. here $S_\mu \sim \mu^{- d - \zeta}$; (iii) this implies the universal scaling behavior of various avalanche observables since the (sufficiently high order) moments of avalanche observables distributions are dominated by their cutoff. For the ABBM model the PDF of avalanche size and duration were obtained in \cite{LeDoussalWiese2008a,LeDoussalWiese2011a,LeBlancAnghelutaDahmenGoldenfeld2013,DobrinevskiLeDoussalWiese2011b}. Still in the ABBM model, the distribution of the maximum velocity inside an avalanche was also computed in \cite{LeBlancAnghelutaDahmenGoldenfeld2013,LeBlancAnghelutaDahmenGoldenfeld2012} where the authors also obtained the dependence of the exponents on the velocity. The joint distribution of size and duration was obtained in \cite{DobrinevskiLeDoussalWiese2011b}. The distribution of the extension of avalanches was computed in the BFM with SR elasticity in $d=1$ in \cite{DelormeLeDoussalWiese2016}. Results beyond-mean field for the avalanche total and local size distribution were obtained at one loop-order in \cite{LeDoussalWiese2008c,LeDoussalWiese2011b} (shocks) and \cite{LeDoussalWiese2012a} (avalanches). These notably predicted the characteristic `bump' that is observed in numerics \cite{RossoLeDoussalWiese2009a} in the avalanche size distribution close to the large-scale cutoff.

\smallskip

{\it Universal scaling functions}\\
Recently universality in avalanche processes has been pushed one step further and a lot of attention was devoted to the study of {\it universal scaling functions.} Indeed, the full velocity field of the interface $\dot{u}(t,x)$ inside an avalanche (where $t$ refers to the time since the beginning of the avalanche and $x$ is the $d$-dimensional internal coordinate according to some  centering procedure) is expected to be universal and scale invariant. For example, using the scaling (\ref{Eq:SecII2:dep2-6}) and a sum rule, for avalanches of fixed duration $T$ inside the scaling regime $T_0 \ll T  \ll T_m$, one expects to have the equality in law
\bea \label{Eq:SecII-V-Intro1}
\dot{u}(t,x) \sim  T^{\zeta_d/z-1 } {\sf v}^{{\rm fixed \ duration}}(t/T , x/T^{1/z}) \ssp ,
\eea 
where the rescaled spatio-temporal process ${\sf v}^{{\rm fixed \ duration}}(t, x)$ is a well defined $T-$independent stochastic process. Alternatively, for avalanches of fixed total size $S$ in the scaling regime, one expects
\bea \label{Eq:SecII-V-Intro2}
\dot{u}(t,x) \sim  S^{1- (z+d)/(\zeta_d + d)} {\sf v}^{{\rm fixed \ size}}(t/S^{z/(d+\zeta_d)} , x/S^{1/(d+\zeta_d)}) \ssp ,
\eea
where the rescaled spatio-temporal process ${\sf v}^{{\rm fixed \ size}}(t, x)$ is a well defined $S-$independent stochastic process. Of course it is one thing to write (\ref{Eq:SecII-V-Intro1}) or (\ref{Eq:SecII-V-Intro2}) but it is another to prove it and to characterize in some way these rescaled stochastic processes. In recent years a lot of attention has been devoted to the study of the {\it mean temporal shape of avalanches at fixed duration or size}.
\bea
&& {\cal F}_{{\rm temporal  \ shape}}^{{\rm fixed  \ duration}} (t) := \overline{\int_x  {\sf u}^{{\rm fixed \ duration}}(t , x) }  \ssp , \nn \\
&& {\cal F}_{{\rm temporal  \ shape}}^{{\rm fixed  \ size}} (t) := \overline{\int_x  {\sf u}^{{\rm fixed \ size}}(t , x) }  \ssp .
\eea
At the mean-field level in the ABBM model, the closely related mean temporal shape at fixed size as a function of the interface position was first computed in \cite{BaldassarriColaioriCastellano2003,ColaioriZapperiDurin2004}. The mean temporal shape at fixed duration was computed in \cite{PapanikolaouBohnSommerDurinZapperiSethna2011}, with the remarkably simple result ${\cal F}_{{\rm temporal  \ shape}}^{{\rm fixed  \ duration}} (t) \sim t(1-t)$. Interestingly, these remarkable observables are also well suited to investigate non-universal effects in avalanche processes which are also interesting for practical applications. In particular, it was known experimentally that the mean temporal shape at fixed duration of Barkhausen pulses present an asymmetry and are skewed to the left, a fact which was attributed to the slow relaxation of Eddy currents affecting the domain wall dynamics \cite{ZapperiCastellanoColaioriDurin2005,Colaiori2008}. Years later these effects were introduced in a modified ABBM model, the temporal shape was again computed analytically and presented the asymmetry observed experimentally \cite{DobrinevskiLeDoussalWiese2013}. Finally, results beyond mean-field (at one-loop) were recently obtained for the average temporal shape at fixed duration and size in \cite{DobrinevskiPhD,DobrinevskiLeDoussalWiese2014a}. The very recent comparison with experiments confirmed the increase of precision brought by one-loop corrections \cite{DurinBohnCorreaSommerDoussalWiese2016}.

\smallskip

{\it The spatial shape of avalanches}\\
On the other hand the equally interesting {\it spatial shape of avalanches} was left aside from theoretical studies until recently. This was a rather disappointing state of affairs since the latter could also be measured in some modern experimental setups on e.g. fracture processes. In \cite{ThieryLeDoussalWiese2015} (presented in Sec.~\ref{subsec:PresBFM}, as will be detailed in the next section,  we made the first progress in this direction and showed that at the mean-field level, i.e. in the BFM, the shape of avalanches in $d=1$ becomes {\it deterministic} in the limit of peaked avalanches $S/\ell^{d+\zeta} \gg 1$ and we identified the limiting shape. Fluctuations around this deterministic profile were also studied in an expansion in $\ell^{d+\zeta}/S$. Comparison with numerical simulations showed a good to perfect agreement. In \cite{ThieryLeDoussal2016a} (presented in Sec.~\ref{subsec:PresShape}) we went further in analyzing the spatial shape of avalanches. We first obtained {\it the mean velocity field} inside avalanches of fixed total size at the mean-field level. This observable contains both the mean temporal shape at fixed size previously studied and a new result, the mean {\it spatial shape at fixed total size}. Going beyond mean field we were able to compute the one-loop corrections to the mean spatial shape at fixed total size. Comparison with numerical simulations showed a good agreement.

\smallskip

{\it What about correlations?}\\
Up to now all the observables that were mentioned concern what one may call `one-shock/avalanche statistics'. However, an interface in a given disordered medium generally experiences a {\it sequence of shocks/avalanches} $\{w_i ,S^{(i)}_x \}$. The observables mentioned before do not fully characterize the properties of this sequence since there can be correlations between different shocks/avalanches. In the context of earthquakes the study of these correlations has been a major focus of the field. From the phenomenological point of view the main result is the Omori law that characterizes the number of aftershocks after a main shock \cite{Omori1894}. Several mechanisms have been advanced to explain these strong correlations, all involving an additional dynamical variable \cite{BurridgeKnopoff1967,JaglaKolton2009}. For elastic interfaces, in an attempt to explain the Omori law from simple mechanisms, correlations between avalanches were until recently only studied as a result of such additional degrees of freedom in the interface dynamics, such as relaxation processes \cite{JaglaLandesRosso2014,Jagla2014} or memory effects \cite{DobrinevskiLeDoussalWiese2013}. Overall there is a belief that these correlations are not captured by the simple interface model. While this is certainly true, it is still clearly of interest to understand first the correlations in interface models (as the example of the temporal shape of Barkhausen avalanches teach us, it is important to first thoroughly understand what is universal to understand what is not!). These universal correlations were remarkably left aside from all theoretical studies until recently. There was even a belief that avalanches were uncorrelated in elastic interfaces model. While this is certainly true for the ABBM and the BFM model, as we very precisely show in \cite{ThieryLeDoussalWiese2015} (presented in Sec.~\ref{subsec:PresBFM}), correlations always exist in non mean-field models. Using FRG we showed in \cite{ThieryLeDoussalWiese2016} (presented in Sec.~\ref{subsec:PresCorrel}) that correlations between shocks in disordered elastic interfaces are universal, of order $\epsilon$ and are controlled by the renormalized disorder correlator $\Delta(u)$. While as we showed, the one-shock statistics is only sensitive to the behavior of $\Delta(u)$ around the cusp, correlations between static shocks feel the full shape of $\Delta(u)$. We obtained quantitative results on the correlations that notably permits and original and unambiguous distinction between the RB and RF universality class.

\smallskip

Before we begin a more detailed presentation of the results obtained during the thesis, let us mention here that the list of problems introduced above is of course very incomplete. Other interesting important open problems of the field will be mentioned in the conclusion.

\subsection{Presentation of the main results of \cite{ThieryLeDoussalWiese2015} } \label{subsec:PresBFM}

\stab {\it An exact formula for the local size of avalanches following an arbitrary driving in the BFM on an arbitrary graph}

In \cite{ThieryLeDoussalWiese2015} we consider the BFM on an arbitrary graph and consider the avalanches following a stepped driving (defined below). That is we consider the equation of motion
\begin{equation}\label{overdampedposition}
\eta \partial_t u_{it} =  \sum_{j=1}^N c_{ij} u_{jt} - m^2( u_{it}-w_{it} ) + F_i(u_{it})
\end{equation}
where (i) $i=1,\cdots,  N \in \JN$ label the points of the graph; (ii) $u_{it}$ is the position of the $i^{th}$ point at time $t$; (iii) the points are linked to one another by a time-independent elasticity matrix $c_{ij}$ such that $\sum_{j}c_{ij}=0$ and $c_{ij} \geq 0$ for $i \neq j$; (iv) the random forces $ F_i(u)$ are a collection of independent one-sided BM with $\overline{[F_i(u)-F_i(u')]^2} =  2 \sigma |u-u'| $ and $\sigma \geq 0$; (v) the interface is at rest at time $0$ and $u_{it=0} = F_i(0)=w_{it=0} =0$; (vi) for $t\geq 0$ the driving verifies $\dot{w}_{it} \geq 0$ and $w_i:=w_{i,t=+\infty}  < \infty$.

Under these conditions, we obtain an exact formula for the joint distribution of avalanche local size defined as $S_i := u_{i,t=\infty}$ (which is smaller than $\infty$ with probability 1) for an arbitrary driving $\vec w = (w_1 , \cdots, w_N)$. We obtain (in dimensionless units $w_i \to w_i/S_m$ and $S_i \to S_i/S_m$ with $S_m = \sigma/m^4$)

\bea \label{EqRes:Nlayers1}
&& P_{\vec w}( \vec S) 
= \left(\frac{1}{2\sqrt{\pi}}\right)^{\!\!N} \left(\prod_{i=1}^N  S_i \right) ^{\!\!-\frac{1}{2}} \exp { \left(-\frac{1}{4} \sum_{i=1}^N \frac{( w_i-\sum_{j=1}^N  C_{ij}  S_j )^2}{ S_i} \right)} \det\left(  M_{ij}  \right)_{N \times N} \nn  \\
&&  M_{ij} = C_{ij} + \delta_{ij} \frac{ w_i- \sum_{k=1}^N  C_{ik}S_k}{ S_i} \quad , \quad C_{ij} = \delta_{ij} -\frac{1}{m^2} c_{ij}  \ .
\eea
The formula is obtained both (i) using an `instanton' method similar as (\ref{Eq:SecII3:SecBFM10}-\ref{Eq:SecII3:SecBFM12}) to obtain formally the Laplace Transform (LT) of (\ref{EqRes:Nlayers1}), the LT is then formally inverted using heuristic calculations involving Grassmann variables; (ii) an exact proof by deriving the Kolmogorov backward equation satisfied by (\ref{EqRes:Nlayers1}) in the case where the driving is $\dot{w}_{it} = w_i \delta(t)$. 

{\it Exact formula for the densities and the BFM as a Lévy-Jump process} \\
Based on the formula (\ref{EqRes:Nlayers1}) and on the form of its Laplace Transform, we show that the BFM is an {infinitely divisible process. Namely $\forall k \in \JN$, $\forall \vec w_1 , \cdots , \vec w_k \in (\JR_+)^N$ such that $\vec w = \sum_{i=1}^k \vec{w}_i$, we have
\bea \label{EqRes:Nlayers2}
P_{\vec w} (\vec S) = \left(P_{\vec w_1} \star P_{\vec w_2} \star \cdots \star  P_{\vec w_k}   \right) (\vec S) \ssp .
\eea
This rigorously shows that the BFM on an arbitrary graph is a Lévy jump process (see e.g. \cite{gardiner2004handbook}). The motion of the interface is a succession of jumps independently generated in time and in space by the densities $\rho_i(\vec S) = \frac{\partial P_{\vec w}(\vec S)}{\partial w_i}|_{\vec w = 0}$ which, for each site $i$, corresponds to the density of avalanches triggered at the $i^{th}$ site (which depends non-trivially on $i$ on an arbitrary graph). Based on (\ref{EqRes:Nlayers1}) we obtain an exact formula for these densities and are able to take the $k \to \infty$ limit of (\ref{EqRes:Nlayers2}) as
\be \label{EqRes:Nlayers3}
\int d^N \vec S  e^{\vec \lambda \cdot \vec S } P_{\vec w}(\vec S) = \sum_{n=0}^{\infty} \sum_{(i_1 ,\dots , i_n )}\frac{w_{i_1} \dots  w_{i_n}}{n!}  \prod_{l=1}^n  \int  d^N \vec s_{i_{l}}( e^{ \lambda \vec s_{i_l} } -1)  \rho_{i_1} (\vec s_{i_1}) \dots \rho_{i_n} (\vec s_{i_n})  \ssp 
\ee
In (\ref{EqRes:Nlayers3}) each term $w_{i_1} \dots  w_{i_n}$ corresponds to events where the total motion $\vec S$ of the interface was generated by $n$ elementary avalanches triggered by the density $\rho_i$ at the site $i_1 , \cdots , i_n$ and $\vec S = \vec s_{i_1} + \cdots +  \vec s_{i_n}$ (note that there can be several avalanche triggered from the same seed, i.e. the terms $i_k = i_{k'}$ are contained in the above summation).

\medskip

The formulae (\ref{EqRes:Nlayers1}) and (\ref{EqRes:Nlayers3}) are rather remarkable as they contain in principle all the information on the spatial structure of avalanches in a completely general version of the BFM model. The possibility to obtain such a formula is linked to the non-trivial exact solvability property of the BFM. Although it was previously known that the avalanches in the BFM are independent, this independence property was never described as precisely as in (\ref{EqRes:Nlayers3}). It is however fair to say that extracting more information (e.g. a marginal probability) from these formulae is quite hard. In \cite{ThieryLeDoussalWiese2015} we are able to make progress in the fully connected model for arbitrary $N$. The study of the large $N$ limit from these formulae is rather instructive and we encourage readers to read the $5^{th}$ section of \cite{ThieryLeDoussalWiese2015}. Here we present only one result extracted from (\ref{EqRes:Nlayers1}) namely the deterministic shape taken by peaked avalanches in the continuum BFM model with SR elasticity in $d=1$.

\medskip

{\it The shape of peaked avalanches in the BFM with SR elasticity in $d=1$}

Taking the continuum limit of the above formulae, we obtain a formula for the density of avalanches in the BFM with SR elasticity in $d=1$ on a line of length $L$ with periodic boundary conditions as (in dimensionless unit, see \cite{ThieryLeDoussalWiese2015} for more details and Sec.~\ref{subsec:FRGDepAva} for the definition of the BFM in the continuum)
\begin{equation}\label{EqRes:Nlayers4}
\rho[S_x] \sim  \frac{ (\int_0^L   d x S_x) \int_{0}^L \frac{1}{S_x^2} }{(\prod_x S_x)^{\frac{1}{2}}} \exp{ \left(-  \int_{0}^L d x \frac{(S_x - \nabla^2 S_x )^2}{4 S_x} \right)} \ssp .
\end{equation}
This allows us to obtain observable such as the mean shape of avalanches using a path integral on the shape of avalanches $S_x$ with statistical weight $\rho[S_x]$. We show that for avalanches of extension $\ell$ and total size $S$, in the limit of peaked avalanches $S/\ell^4 \gg 1$, the {\it centered, reduced} shape $s(x) = \frac{1}{S \ell} S_{(x-x_0)/l} $, defined such that the support of the reduced shape is $x \in [-1/2 , 1/2]$ and $\int_{-1/2}^{1/2}s(x)dx=1$, becomes deterministic $s(x) = s_0(x)$ (given by a saddle-point of the path integral associated with (\ref{EqRes:Nlayers4})) and solves

\be \label{EqRes:Nlayers5}
{\cal A} \, \phi(x) =   \phi ^{(4)}(x)+\frac{5 \phi
   '(x)^4}{\phi (x)^3}-\frac{10 \phi '(x)^2 \phi ''(x)}{\phi (x)^2}  .
\ee
where $s_0(x) = \phi_0(x)^2$ with $\phi_0(x)$ the solution of (\ref{EqRes:Nlayers5}) with the saddle-point parameter ${\cal A}_0$ ensuring $\int_{-1/2}^{1/2} \phi_0(x)^2 = 1$. (\ref{EqRes:Nlayers5}) was solved numerically with a high precision. The solution is predicted to decay close to the boundary as $s_0(x) \sim_{x \to 1/2^-}(1/2- x)^4$. This is confronted with numerical simulations of the BFM (see \cite{ThieryLeDoussalWiese2015} for more details) with a very good agreement, as shown in Fig.~\ref{fig:PresNlayers}. In \cite{ThieryLeDoussalWiese2015} we also investigate the $\sqrt{\ell^4/S}$ corrections to this deterministic behavior and obtain, based on the optimal shape $s_0(x)$, the tails of the PDF of aspect ratios $S/\ell^4$. These additional results are successfully confronted with numerical simulations (see \cite{ThieryLeDoussalWiese2015}.

\begin{figure}
\centerline{\includegraphics[width=8.0cm]{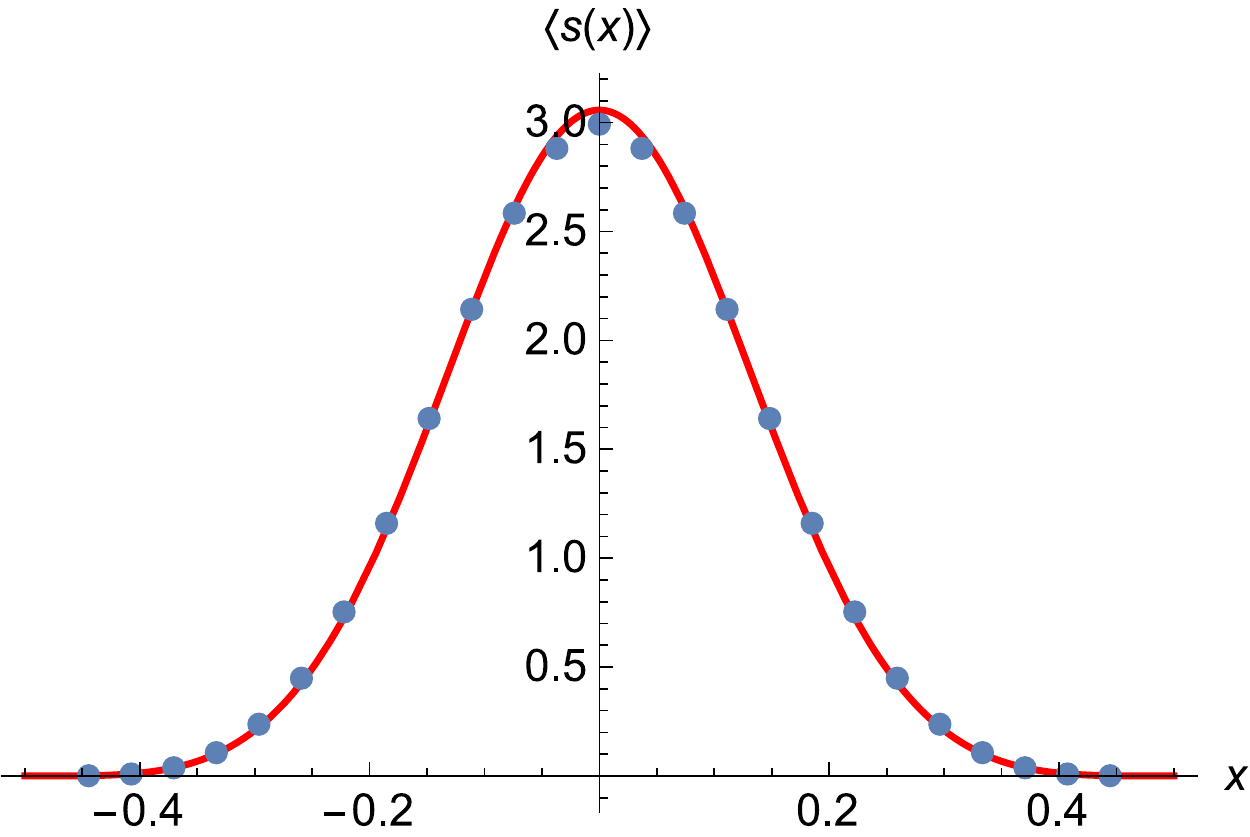} ~~ \includegraphics[width=8.0cm]{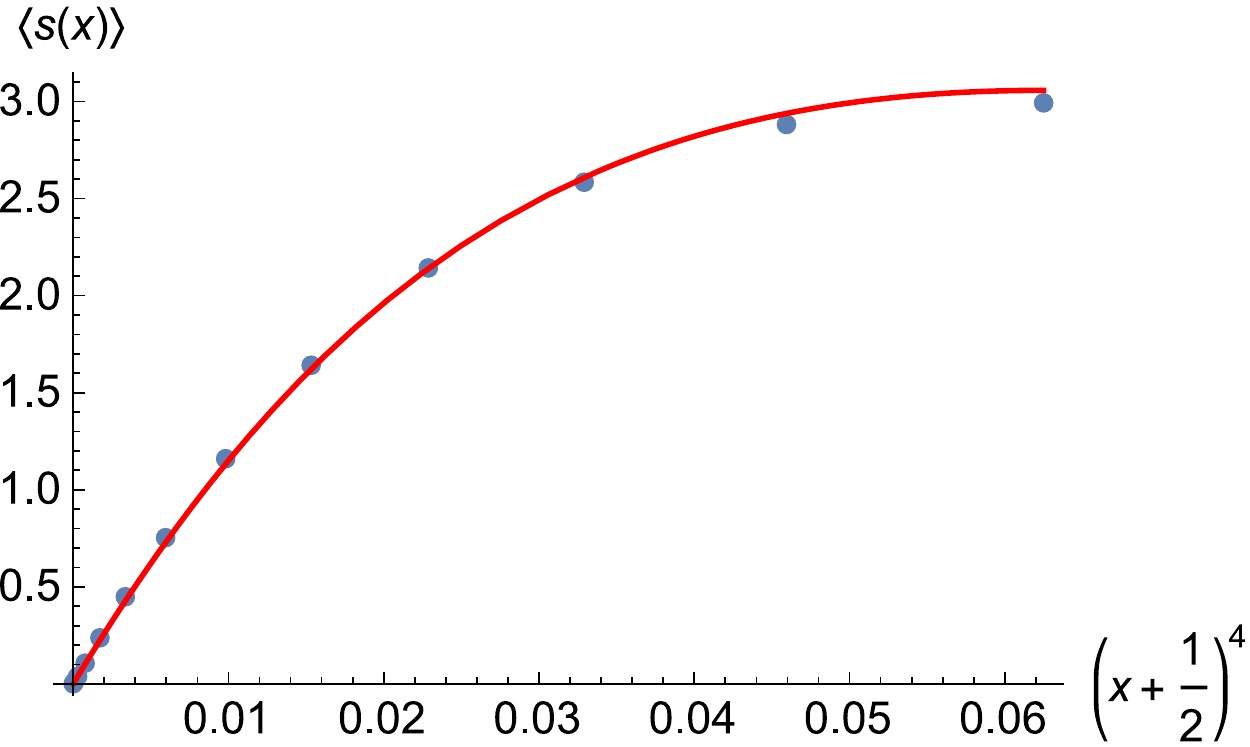} }
\caption{Left: Mean shape obtained by averaging over the $1000$ avalanches with the largest $S/\ell^4$ in the simulations of the BFM of \cite{ThieryLeDoussalWiese2015} (blue dots, compared to the optimal shape $s_{0}(x)$  solution of (\ref{EqRes:Nlayers5}) (red line). Right: test of the predicted behavior $s_0(x)\sim (x+1/2)^4$ close to the boundaries.  Figures taken from \cite{ThieryLeDoussalWiese2015}.}
\label{fig:PresNlayers}
\end{figure}

\subsection{Presentation of the main results of \cite{ThieryLeDoussal2016a}} \label{subsec:PresShape}

\stab {\it The seed-centering}

In \cite{ThieryLeDoussal2016a} we pursued the analysis of the spatial shape of avalanche processes. One of the main contributions of our work was to remark that the most convenient way to center the shape of avalanches (from the perspective of performing analytical calculations) is to center them around their seed. Indeed, as was shown with (\ref{Eq:SecII3:FRGDepAva14b}), the structure of the MSR action for the depinning of elastic interface allows quite naturally to isolate in any observable $O[\dot{u}_{tx}]$ the contribution of avalanches starting at a given time $t_i$ and position $x_i$. In some sense this is natural since this is the only centering which respects the causal structure of avalanche processes: the seed centering is a conditioning on the stochastic process of the velocity field inside an avalanche with respect to its initial condition. Another type of centering, e.g. the centering with respect to the maximum or to the center of mass of the avalanche, does not respect this causality since it corresponds to a conditioning on the full history of the stochastic process. It is reasonable to think that it is this absence of an adapted centering procedure that made the mean spatial shape of avalanches ignored from theoretical studies until this paper. Of course, having a centering procedure that is analytical-work friendly is almost useless if the observable cannot be measured in numerical simulations and experiments. One of the challenge of \cite{ThieryLeDoussal2016a} was therefore to devise an algorithm allowing a simple study of seed-centered shapes. We refer the reader to the paper \cite{ThieryLeDoussal2016a} for a description of this algorithm and only give the results here, everywhere presented in dimensionless units $x =\tilde{x}/m$, $t = \tau_m \tilde{t}$, $S = S_m \tilde{S}$ with $\tau_m = \eta/m^2$ and $S_m = \sigma/m^4$ (as usual $m$ is the mass of the driving spring and $\eta$ and $\sigma$ are linked to the renormalized parameters of the models, see \cite{ThieryLeDoussal2016a} for details and Sec.~\ref{subsec:avalanches} and Sec.~\ref{subsec:FRGDepAva} for the notations and definitions used in this section)

\medskip

 {\it The mean velocity field inside avalanches of fixed size in the BFM in arbitrary $d$}

We first showed that for the BFM in arbitrary $d$ with SR and LR elasticity, the scaling formula for the {\it mean value of the velocity field inside seed-centered avalanches of fixed size $S$ in the scaling regime}
\bea \label{EqRes:Shape1}
&& \langle \dot{u}_{tx} \rangle_S = S^\frac{\zeta-z}{d+\zeta} F(t/S^{\frac{z}{d+\zeta}},x/S^{\frac{1}{d+\zeta}})  \ ,
\eea
holds with a simple scaling function for SR elasticity in arbitrary $d$, $\forall S$
\bea \label{EqRes:Shape2}
%&& \langle v(x,t) \rangle_S = S^\frac{2-d}{4} F(t/S^{1/2},x/S^{1/4}) \nn \\
&& F(t,x)=2 t e^{-t^2} \frac{1}{(4 \pi t)^{d/2}} e^{-x^2/(4 t)} \ ,
\eea
and for LR elasticity (obtained in arbitrary $d$ in Fourier space and in $d=1$ in real space) $\forall S \ll S_m$

\bea \label{EqRes:Shape3} 
F(t,x) = 2 t e^{-t^2 } \int_q e^{i q y - |q| t }=_{d=1} \frac{2 t^2 e^{-t^2}}{\pi(x^2 + t^2)} \ .
\eea 

These are shown in Fig.~\ref{fig:PresShape1} but have not been measured in simulations.

\begin{figure}
\centerline{
\includegraphics[width=0.35\textwidth]{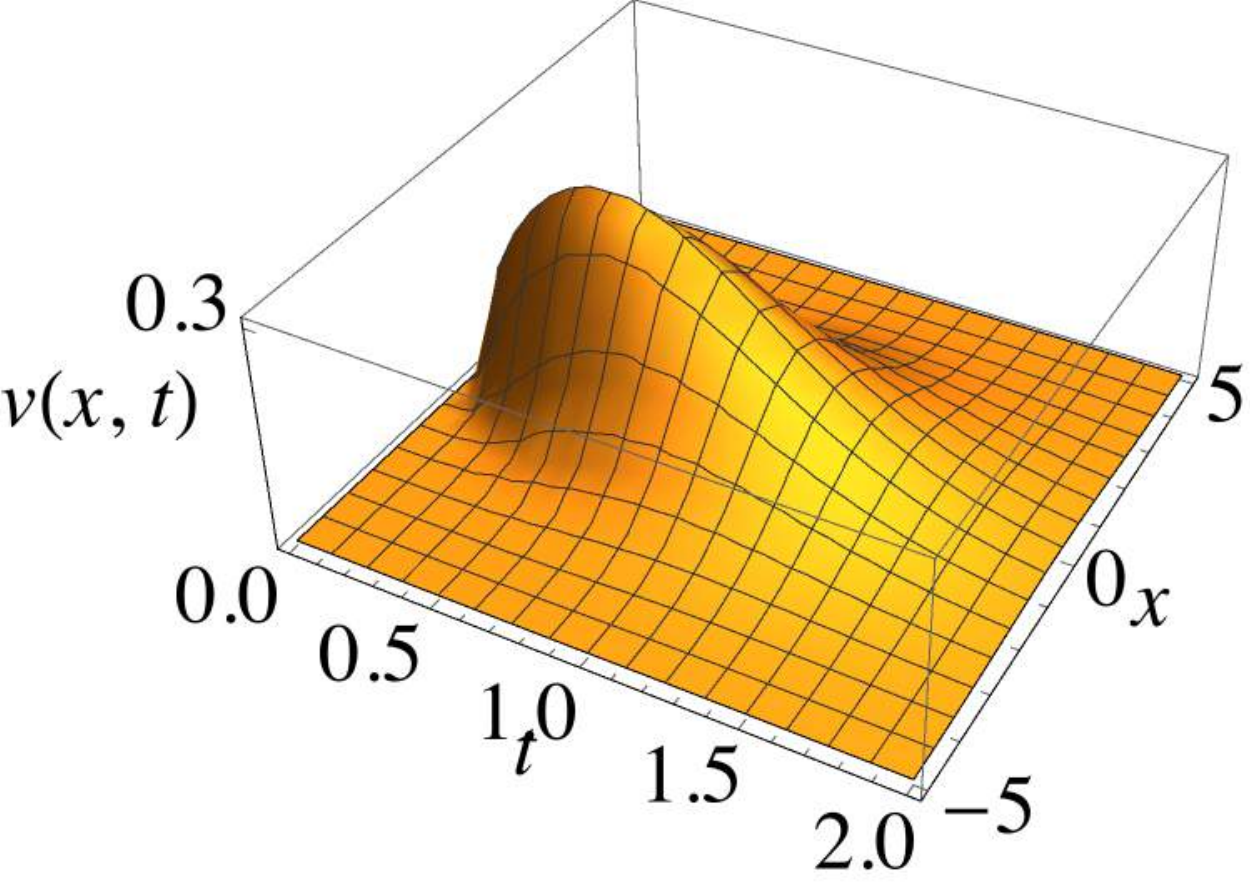} \quad \quad \includegraphics[width=0.35\textwidth]{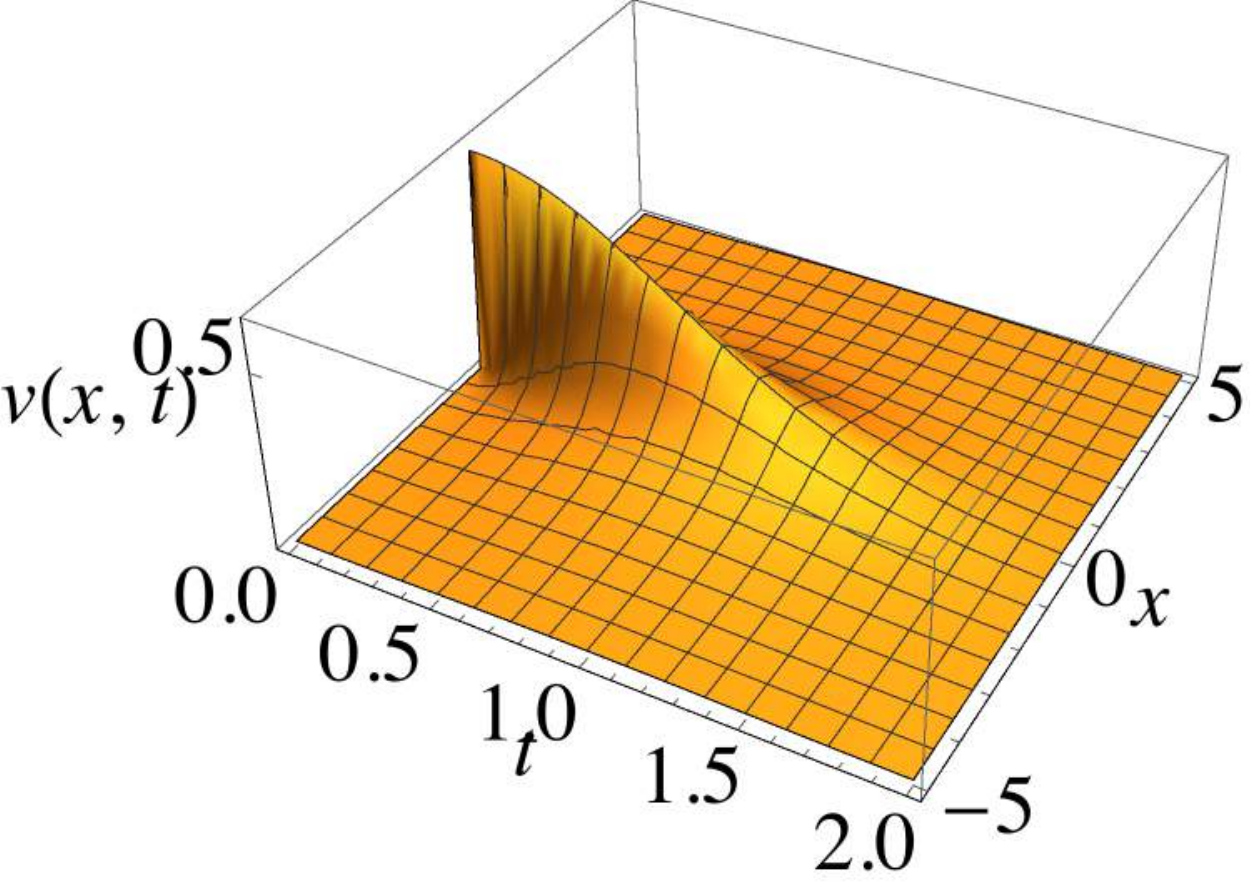} }
 \caption{Plot of the mean-field result for the space-time mean velocity profile inside an avalanche in $d=1$ for SR (left, see (\ref{EqRes:Shape2})) and LR elasticity (right, see (\ref{EqRes:Shape3})). Figures taken from \cite{ThieryLeDoussal2016a}.}
\label{fig:PresShape1}
\end{figure}
\medskip

\begin{figure}[h]
\centerline{
\includegraphics[width=0.3\textwidth]{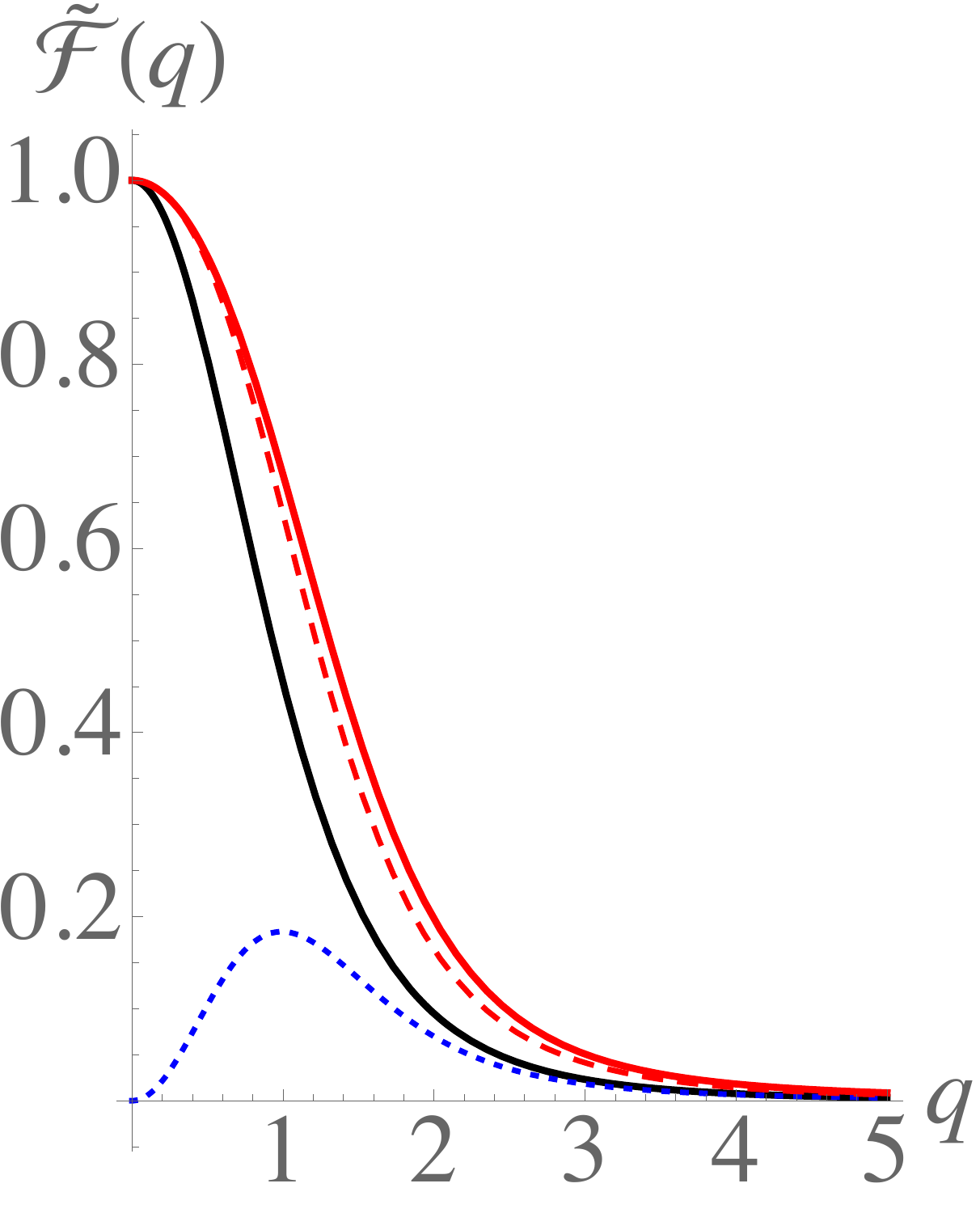} \includegraphics[width=0.3\textwidth]{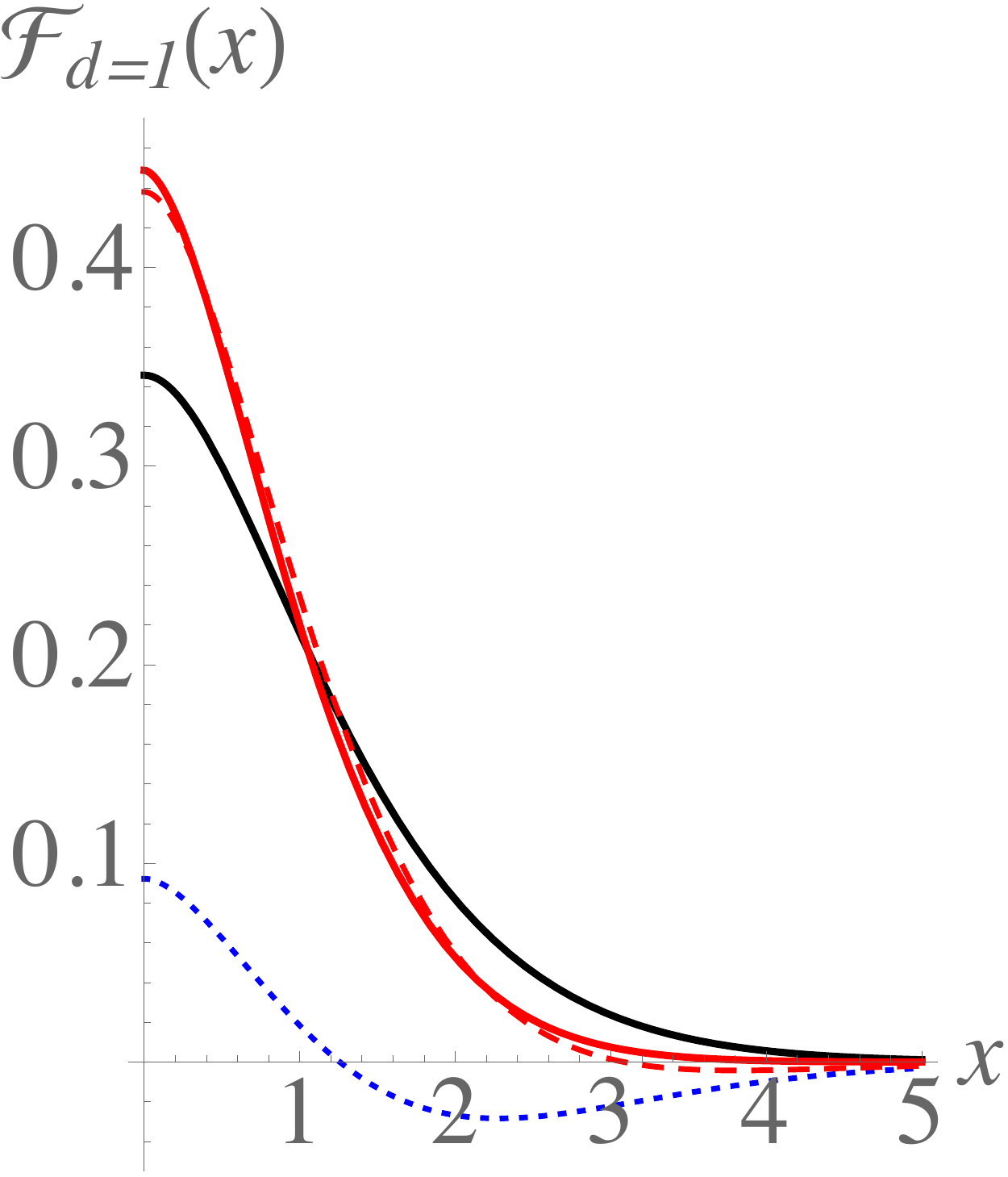} \includegraphics[width=0.3\textwidth]{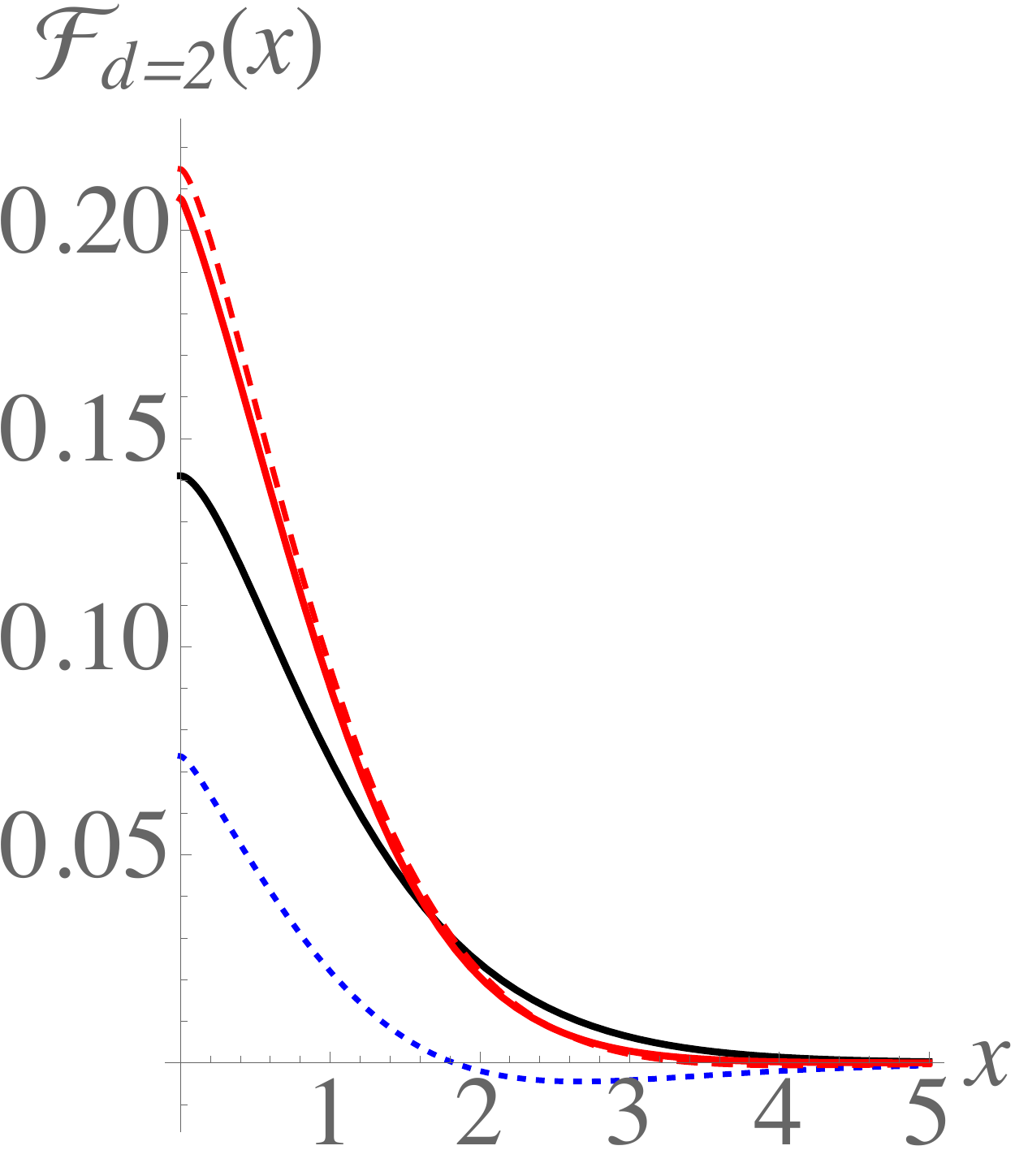}
 }
 \caption{Analytical results at MF and $O(\epsilon)$ level for the universal scaling function $\tilde{{\cal F}}_{d=1}$ in Fourier space (Left) and ${\cal F}_d$ in real space for $d=1$ (Middle) and $d=2$ (Right) for SR elasticity. Black lines: tree/mean-field results. Dotted blue lines: universal corrections, $\delta \tilde{{\cal F}}_1(q) $ (left, $O(\epsilon)$ correction in Fourier space in $d=1$), $\delta {\cal F}_1(x) $ (middle) and $\delta{\cal F}_2(x) $ (right). Red-dashed lines: $O(\epsilon)$ estimate obtained by simply adding the corrections to the MF value. 
 Red lines: improved $O(\epsilon)$ estimate, 
 which, through a re-exponentiation procedure, takes properly into account the modification of exponents (\ref{EqRes:Shape8}) and (\ref{EqRes:Shape9}) (see \cite{ThieryLeDoussal2016a}). Note that the cusp at the origin  of the avalanche shape at $O(\epsilon)$ is not obvious in this plot since the non-analyticity is rather small, but it can be emphasized using a log-log scale.  Figures taken from \cite{ThieryLeDoussal2016a}.}
\label{fig:PresShape2}
\end{figure}

{\it The mean spatial shape of avalanches of fixed size in the BFM and at one loop for SR elasticity} 

Integrating (\ref{EqRes:Shape2}) with respect to time leads to the {\it mean shape of seed centered avalanches in the BFM for SR elasticity in arbitrary $d$}. The latter can be checked to satisfy the following scaling form $\forall S$, equivalently written in Fourier or real space
\bea \label{EqRes:Shape4}
&& \langle S(x) \rangle_S=  S^{1- \frac{d}{d+\zeta}}  {\cal F}_d(\frac{x}{S^{\frac{1}{d+\zeta}}}) \quad , \quad \langle S(q) \rangle_S=  S  \tilde{{\cal F}}_d(q S^{\frac{1}{d+\zeta}}) \ssp .
\eea
${\cal F}_d(x)$ can be expressed using hypergeometric functions (see \cite{ThieryLeDoussal2016a}), and $\tilde{{\cal F}}_d(q)$ has the remarkably simple, $d-$independent form
\bea \label{EqRes:Shape5}
\tilde{{\cal F}}_d^{\rm BFM}(q) = \tilde{{\cal F}}^{\rm BFM}(q) = 1 - \frac{ \sqrt{\pi} q^2 }{ 2} e^{\frac{q^4}{4}}  \text{erfc} \left(\frac{q^2}{2}\right) \ssp .
\eea
The result for these scaling functions in the BFM in any $d$ in Fourier space and in $d=1,2$ in real space are shown in black in Fig.~\ref{fig:PresShape2}.\\
For more realistic models of interfaces in a short-range (SR) correlated disorder, the above results are the $O(\epsilon^0)$ results to the mean spatial shape of avalanches. In order to go beyond mean-field, we use the results of FRG and in \cite{ThieryLeDoussal2016a}, using the simplified action
(\ref{Eq:SecII3:FRGDepAva18}), we were able to compute the $O(\epsilon)$ corrections to the mean shape. In Fourier space we show that the scaling (\ref{EqRes:Shape4}) is still compatible with one-loop FRG in the scaling regime $S \ll S_m$ and we obtain the  $O(\epsilon)$ correction
\bea \label{EqRes:Shape6}
&& \tilde{{\cal F}}_d^{{\rm BFM}}(q) = \tilde{{\cal F}}^{{\rm MF}}(q) +\delta \tilde{{\cal F}}_d (q) + O(\epsilon^2) \ ,
\eea
where $\delta \tilde{{\cal F}}_d (q) = \epsilon \tilde {\cal F}^{(1)}(q)$. Here
$ \tilde{{\cal F}}^{(1)} (q) = \int_{{\cal C}} 
 \frac{d\mu}{2i \pi}  e^{\mu} \tilde H(\mu,q)$ is obtained as an Inverse Laplace Transform (ILT) $\mu \to 1$ of: 
\bea \label{EqRes:Shape7}
&& \tilde H(\mu,q) =  \frac{4  \sqrt{\pi} }{9}
\bigg[    \frac{2-3\gamma_E}{8} \frac{1}{q^2 + 2 \sqrt{\mu}}  -
\frac{4 \sqrt{\mu}}{(q^2  + 2 \sqrt{\mu})^2}  \\
&& \times  \bigg(\frac{q^2 + 9 \sqrt{\mu}}{q  \sqrt{q^2 + 8 \sqrt{\mu}} } 
\sinh ^{-1}\left(\frac{q}{2 \sqrt{2 \sqrt{\mu}} }  \right) -1 + \frac{3}{16} \ln(4 \mu) \bigg) \bigg] \ssp , \nn
\eea
where $\gamma_E$ is Euler's Gamma constant (see \cite{ThieryLeDoussal2016a} for the choice of the contour ${\cal C}$). 
We then define the correction to the mean shape in real
space as the $d$-dimensional Fourier transform $\delta {\cal F}_d (x) =  \int \frac{d^d q}{(2 \pi)^d} e^{-i q x} 
\delta \tilde{{\cal F}}_d (q)$. We were not able to perform the ILT of (\ref{EqRes:Shape7}) in full generality but obtained the behavior of the mean shape in Fourier space at small and large $q$ and in real space at small and large $x$. In particular we show that:\\
(i) At large $q$ in Fourier space,
 \bea \label{EqRes:Shape8}
\tilde{{\cal F}}_d (q) \simeq_{q \gg1} \tilde A_d q^{- \tilde \eta_d} \quad , \quad 
\tilde \eta_d= 4 - \frac{4 \epsilon}{9}
+ O(\epsilon^2)  \ ,
\eea
with a universal prefactor $ \tilde A_d=2 (1 -  (2 + \frac{\gamma_E}{4}) \frac{2 \epsilon}{9})$. In real space this implies, 
in the expansion of ${\cal F}_d (x)$ at small $x$,
a non-analytic term $\sim |x|^{\eta_d}$ with $\eta_d  = \tilde \eta_d - d = \frac{5 \epsilon}{9}
+ O(\epsilon^2)$. Restoring the $S$ dependence from (\ref{EqRes:Shape4})
this leads to $\langle S(q) \rangle_S \sim_{q \to +\infty} S^{1 - \frac{\tilde \eta_d}{d+\zeta}} q^{- \tilde \eta_d}$
and the non-analytic part $\langle S(x) \rangle_S^{n.a} \sim_{x \to 0} S^{1 - \frac{\tilde \eta_d}{d+\zeta}} 
|x|^{\eta_d}$. Note that in the BFM case (retrieved by taking $\epsilon = 0$) the value $\tilde \eta_d=4=d+\zeta_{BFM}$ implies that the large $q$ behavior of $\langle S(q) \rangle_S$ {\it does not depend on $S$}. This may seem natural: in the BFM the small scales
do not know about the total size of the avalanche. A generalization of this property to the SR disorder case
would suggest the guess $\tilde \eta^{{\rm guess}}_d=d+\zeta$. Our $O(\epsilon)$ result however explicitly shows that 
this property fails and $\tilde \eta_d > d+\zeta$ (at least close to $\epsilon = 0$). Hence in the SR disorder case the {\it large avalanches tend to be more smooth than small avalanches.} Note that the predicted value of $\eta_d$ 
is smaller than $2$ in all physical dimensions: {\it this non-analytic term should actually dominate the behavior of ${\cal F}_d (x)$ around $0$} (and thus lead to a cusp singularity). \\
(ii) At large $x$ in real space, we obtain that the mean shape has a stretched exponential decay as: 
\bea \label{EqRes:Shape9}
{\cal F}_d (x) \sim e^{ -C x^{\delta}} \quad , \quad \delta = \frac{4}{3}  + \frac{2}{27} \epsilon + O(\epsilon^2)  \ ,
\eea
with a universal prefactor $C =\frac{3}{4} + (\frac{7 \sqrt{3}}{36} -1  ) \frac{2}{9} \epsilon$. Remarkably, using $\zeta = \epsilon/3
+ O(\epsilon^2)$, this agrees to
$O(\epsilon)$ with the general conjecture $\delta= \frac{d+\zeta }{d+\zeta -1}$ that we justify in \cite{ThieryLeDoussal2016a}.

\begin{figure}[h]
\centerline{
\includegraphics[width=0.4\textwidth]{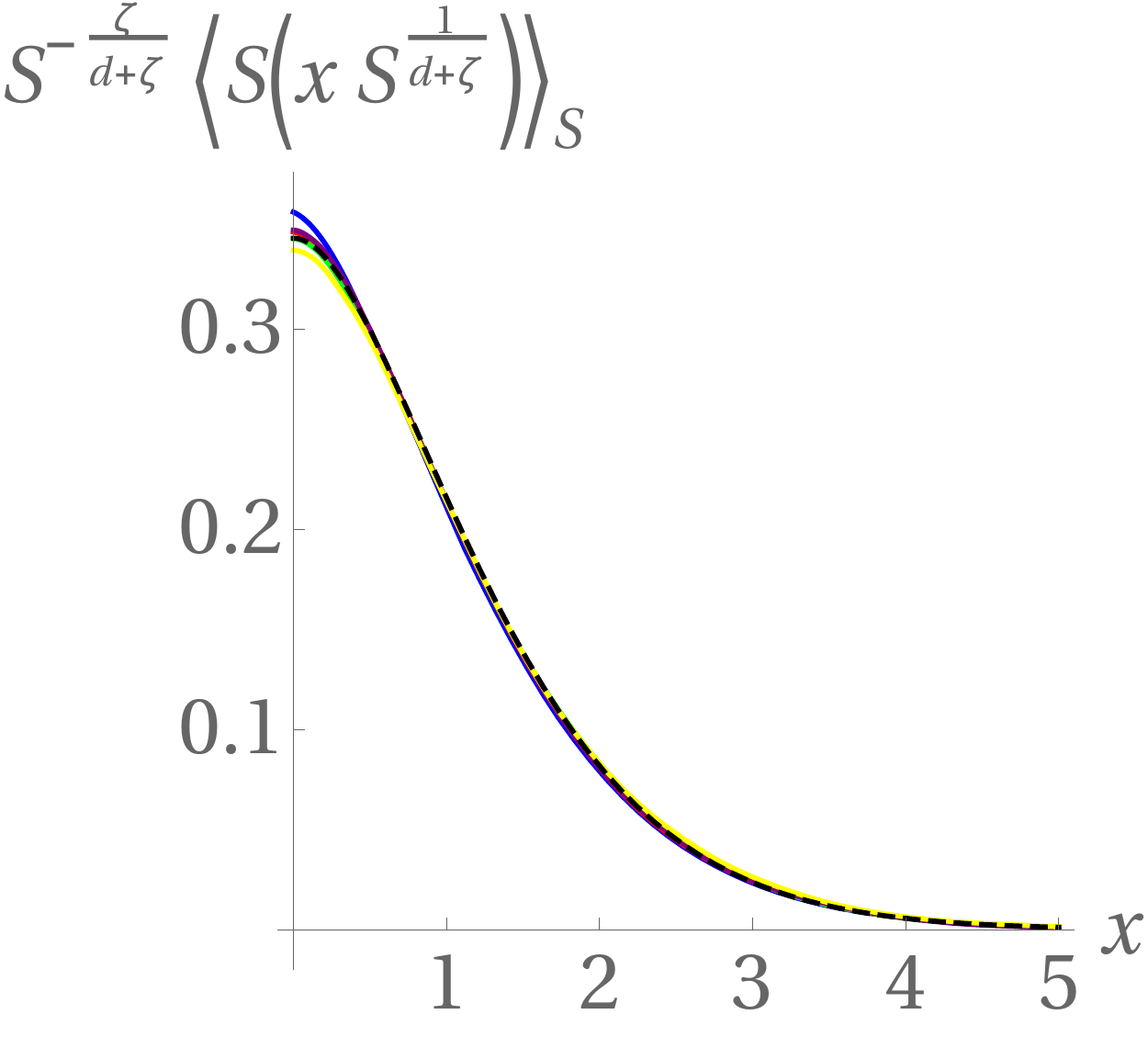} \includegraphics[width=0.5\textwidth]{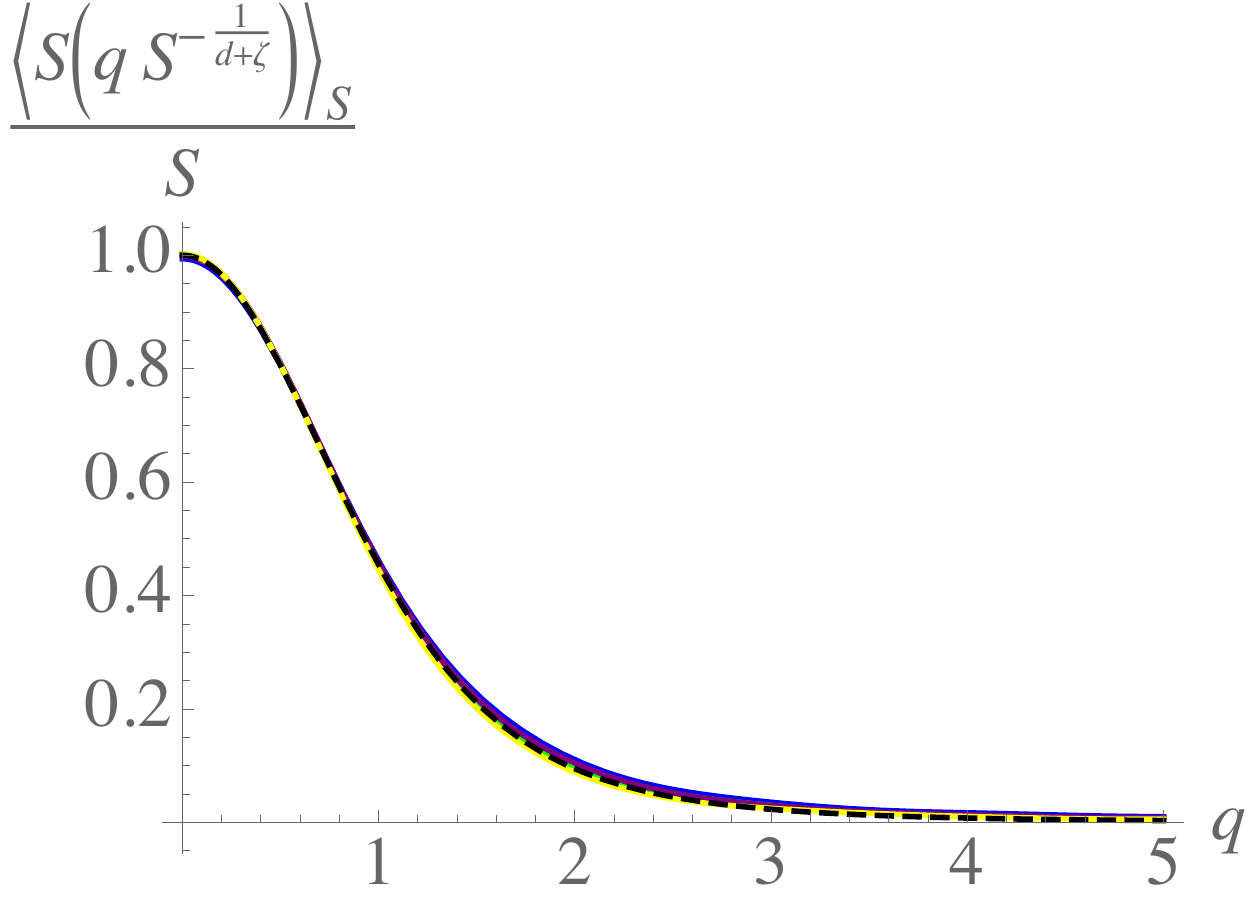}}
 \caption{Lines: rescaled mean shapes of avalanches at fixed size $S$ obtained from the simulations of the BFM model in $d=1$ in real (left) and Fourier (right) space, for $S=10$ (blue), $S=10^2$ (red), $S=10^3$ (green), $S=10^4$ (purple) and $S=10^5$ (yellow). Dashed black lines: exact theoretical results in the BFM. No fitting parameter.  Figures taken from \cite{ThieryLeDoussal2016a}.}
\label{fig:PresShape3}
\end{figure}

\begin{figure}[h]
\centerline{
\includegraphics[width=0.4\textwidth]{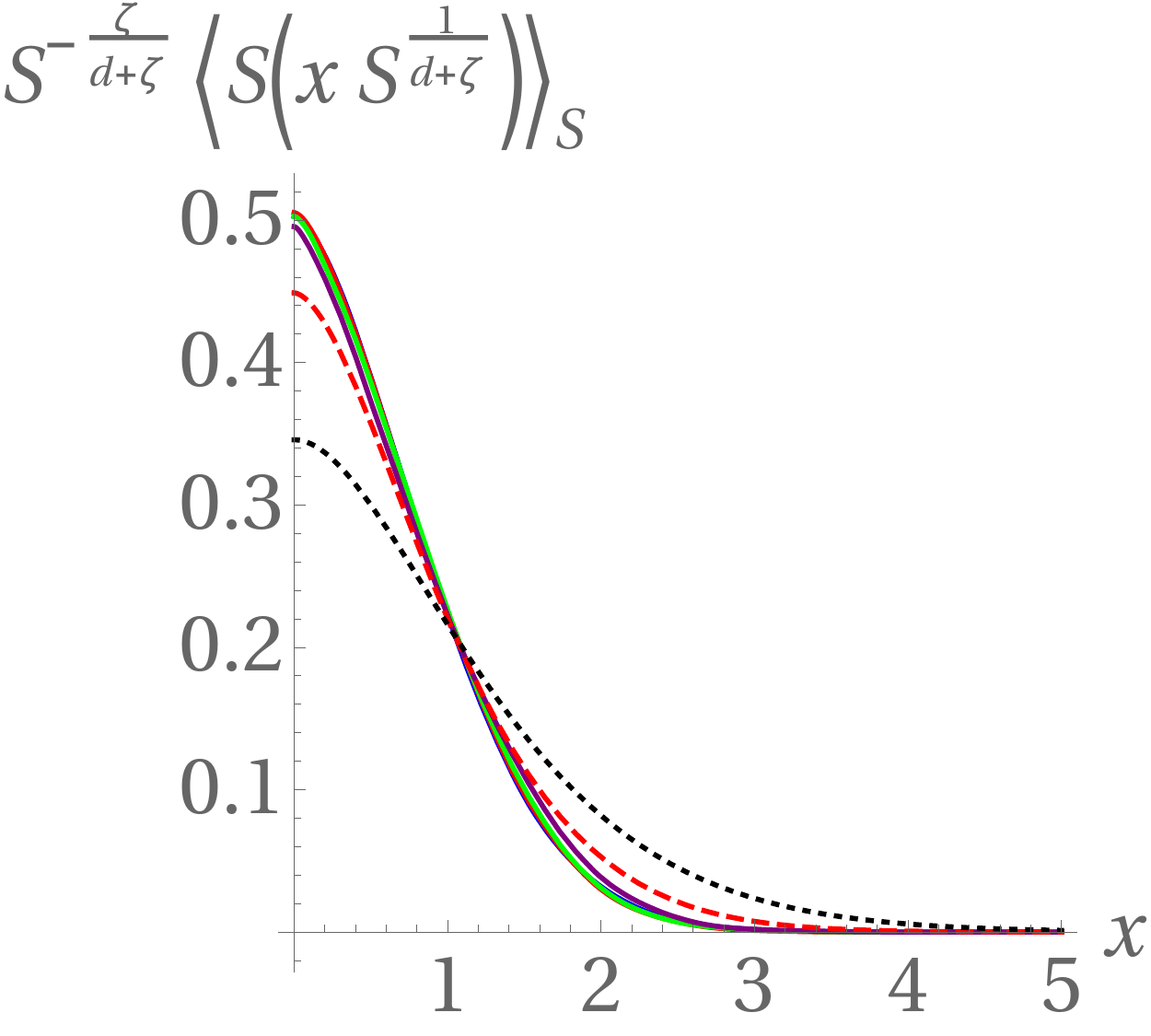} \includegraphics[width=0.5\textwidth]{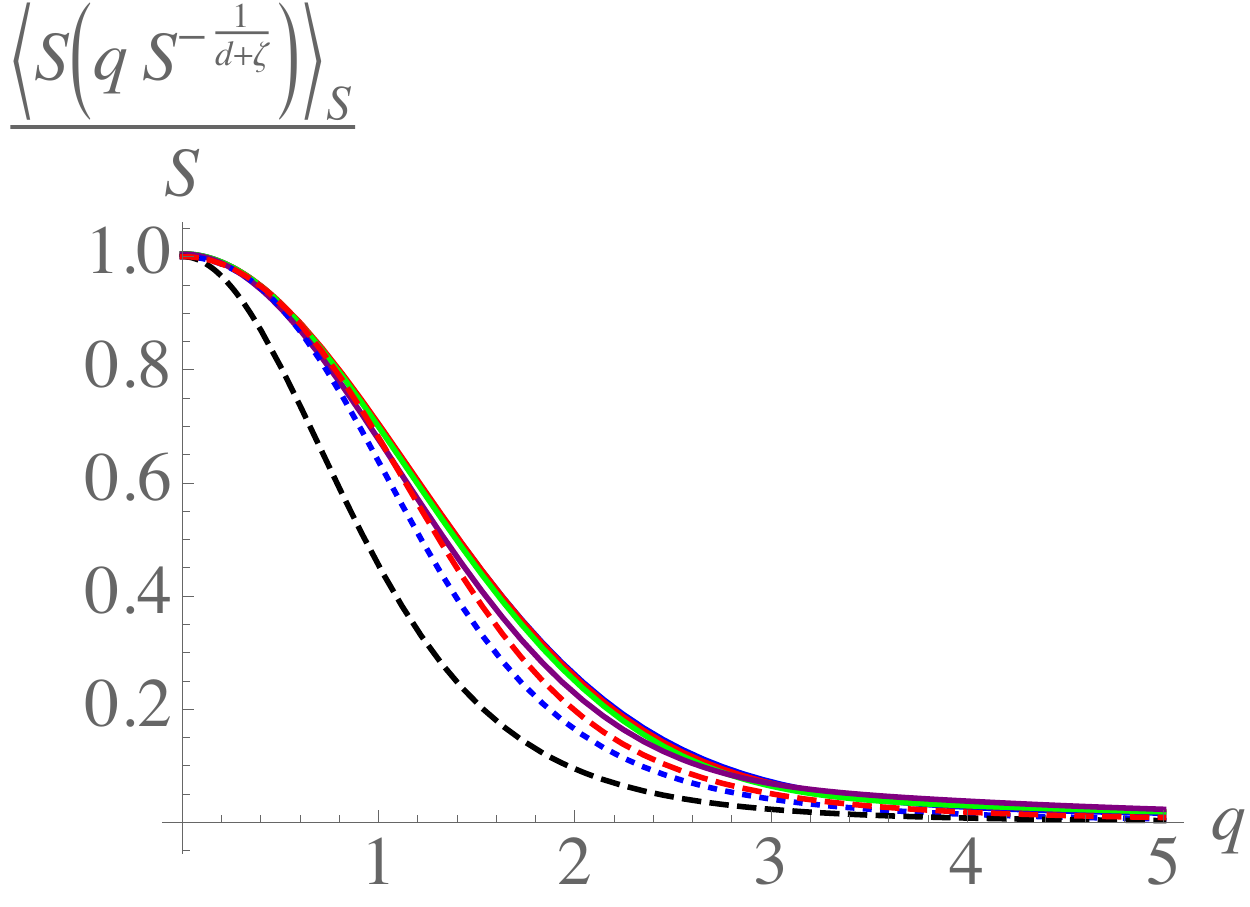}}
 \caption{Lines: rescaled mean shapes of avalanches at fixed size $S$ from the simulation of the model with SR disorder in $d=1$ in real (left) and Fourier (right) space for $S=50$ (blue), $S=10^2$ (red), $S=10^3$ (green), $S=10^4$ (purple). Dashed black lines: theoretical MF result. Red dashed line: improved $O(\epsilon)$ result taking into account the modification of exponents (\ref{EqRes:Shape8}) and (\ref{EqRes:Shape9}). Blue dashed line on the right: $O(\epsilon)$ obtained by simply adding one-loop corrections to the MF result. No fitting parameter.  Figures taken from \cite{ThieryLeDoussal2016a}.}
\label{fig:PresShape4}
\end{figure}

{\it Comparison with numerical simulations}

In \cite{ThieryLeDoussal2016a}, using an original algorithm to retrieve the seed of the avalanches we compare the above theoretical results with simulations of the BFM and of a model with SR disorder in $d=1$, both with short-range elasticity. For the BFM (as it should since our results are exact) we obtain a perfect agreement, see Fig.~\ref{fig:PresShape3}. This demonstrates that our observable is measurable in numerical simulations. For the model with SR disorder, our results compare reasonably well with the results of the numerical simulations and bring a substantial improvement compared to the mean-field results, see Fig.~\ref{fig:PresShape4}. The results look better in Fourier space: when integrated, the small discrepancy $\forall q$ in Fourier space gives a larger discrepancy around the origin in real space. Additional numerical results are shown \cite{ThieryLeDoussal2016a}. In particular it is shown that the cusp of the mean shape for model with SR disorder that is predicted by our one-loop result is indeed compatible with numerical simulations.

\medskip

Some additional results can be found in \cite{ThieryLeDoussal2016a}, in particular we introduce and compute to order $O(\epsilon)$ some {\it universal ratios}, which are quantities allowing us to efficiently compare different shape functions when one scale is unknown. This could be particularly useful for comparison with experiments.

\subsection{Presentation of the main results of \cite{ThieryLeDoussalWiese2016} } \label{subsec:PresCorrel}

\stab {\it The two-shock density}

In \cite{ThieryLeDoussalWiese2016} we investigated the presence of correlations at order $O(\epsilon)$ in the sequence of shocks of the ground state $(w_i , S^{(i)}_x))_{i \in \JZ} $ for a $d$-dimensional elastic interface in a disordered medium (we refer the reader to Sec.~\ref{subsec:shocks} for definitions). Since depinning/avalanches and statics/shocks at zero temperature for disordered elastic interfaces are, for all we know, equivalent up to order $O(\epsilon^2)$, our results are expected to apply equally well for avalanches at the depinning transition. The phenomenology is, however, as we will see, richer in the case of shocks due to the presence of one more universality class (random bond) for short-range correlated disorder. We investigate these correlations by looking at the two-shock density at a distance $W>0$, defined as,
\be
\rho_W(S_1,S_2):=  \overline{ \sum_{i\neq j } \delta(w-w_i) \delta(S_1-S^{(i)})  \delta(w+W-w_j) \delta(S_2-S^{(j)}) } \nn  \ .
\ee
Here $\int_{w_1}^{w_1'} dw \int_{w_2}^{w_2'} dw' \int_{S_1}^{S_1'} dS \int_{S_2}^{S_2'} dS' \rho_{w'-w}(S,S')$ counts the mean number of pairs of shocks such that the first  shock occurred between $w_1$ and $w_1'$, and the second between $w_2$ and $w_2'$, with sizes between $S_1$ and $S_1'$, resp. $S_2$ and $S_2'$. An absence of correlations in the sequence of shocks would imply $\rho_W(S_1 , S_2) = \rho(S_1) \rho(S_2)$ where the one-shock density $\rho(S)$ was defined in (\ref{Eq:SecII2:DefrhoS}). To investigate the presence of correlations we thus study the {\it connected two-shock size density} $\rho_W^c(S_1,S_2)$, defined as
\be  \label{EqPresCorrel1}
\rho_W^c(S_1,S_2) := \rho_W(S_1,S_2) - \rho(S_1) \rho(S_2)\ .
 \ee 
At the mean-field level in the BFM, as we know (see Sec.~\ref{subsec:PresBFM}) shocks are independent thus $\rho_W^c(S_1,S_2)=0$. At order $O(\epsilon)$ however as we show below this is not the case and $\rho_W^c(S_1,S_2)=O(\epsilon)$ is given by a universal scaling function.\\

{\it An exact formula for the first connected moment}

We first obtain an {\it exact formula} for the first connected moment
\be \label{EqPresCorrel2}
 \frac{\langle S_1 S_2 \rangle_{\rho_W^c} }{[\langle
S \rangle_\rho]^2} =    - \frac{\Delta''(W)}{L^d m^4}\ .
\ee
Here a subscript indicates the density with respect to which the average is taken. For uncorrelated shocks the right-hand side of (\ref{EqPresCorrel2}) would be $0$. Here as usual $\Delta(W)$ is the universal scaling function of the FRG, which can be measured as an observable using (\ref{Eq:SecII3:FRGStatShock7}). The above equation is a generalization to the two-shocks case of the exact formula for $S_m = \langle S^2 \rangle_{\rho}/(2 \langle S \rangle_{\rho}$ given in (\ref{Eq:SecII3:FRGStatShock11}). For $m$ close to $0$ as we know (see Sec.~\ref{subsec:FRGStatic}) $\Delta(W)$ takes a universal scaling form with
\bea \label{EqPresCorrel2b}
\Delta(u) = A_d^{\gamma} \mu^{\epsilon - 2 \zeta_d} \kappa^2 \Delta^*(\mu^{\zeta_d} u/\kappa)
\eea
where $\kappa$ is a non-universal microscopic scale. Depending on the range of correlations of the initial disorder, $\Delta^*(u)= O(\epsilon)$ is the fixed point of the FRG equation with $\Delta^*(0) = \epsilon$ of the RB or RF universality class. From \ref{EqPresCorrel2} the shocks appear positively correlated for $\Delta''(W) \leq 0$ and anti-correlated for $\Delta''(W) \geq 0$. Taking a look at the cartoon of the typical shape of the RB and RF fixed points presented in Fig.~\ref{fig:PresCorrel1}, correlations between shocks thus give an unambiguous distinction between these two universality classes: for RF shocks are always anti-correlated, while for RB they are anti-correlated at small distances and positively correlated at large distances. An exact equation similar to (\ref{EqPresCorrel2b}) can be proved for avalanches at depinning. In this case random bond bare disorder flows at large scale to random field disorder and thus avalanches at the depinning are always anti-correlated.

\begin{figure}
\centerline{\includegraphics[width=5.0cm]{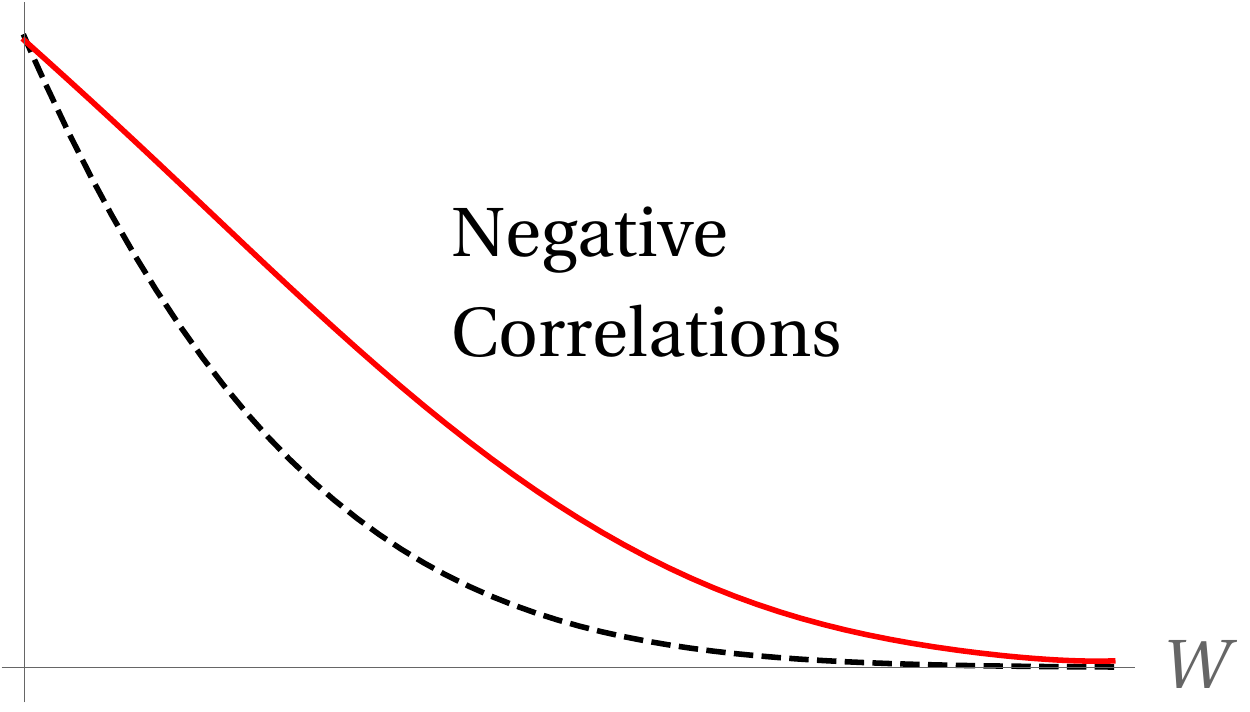} \includegraphics[width=5.0cm]{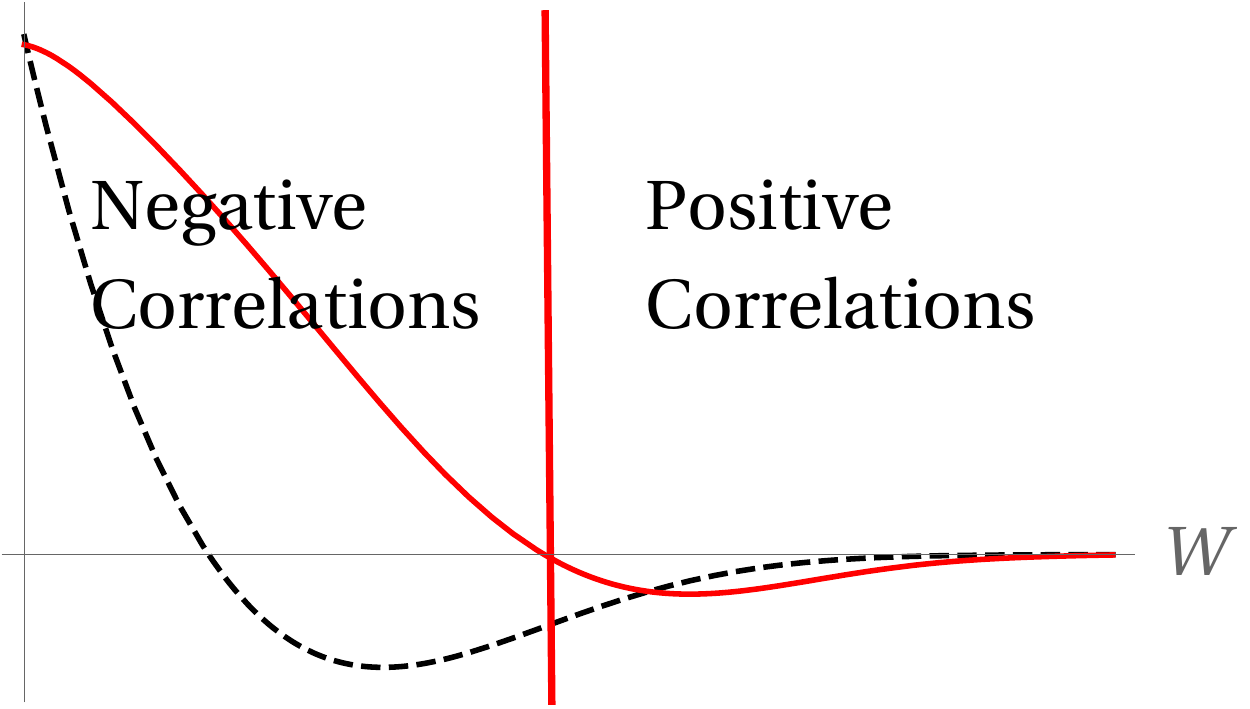}} 
\caption{Cartoons of the typical shape of the renormalized disorder correlator $\Delta(W)$ (black-dashed line) and of its second derivative $\Delta''(W)$ (red line) for the random field (left) and random bond (right) universality classes (not to scale). Our results predict that the shock sizes are always negatively correlated in the random field universality class, whereas the random bond universality class exhibits a richer structure with negatively (resp. positively) correlated shock sizes at small (resp. large) distances. Figures taken from \cite{ThieryLeDoussalWiese2016}.}
\label{fig:PresCorrel1}
\end{figure}

\medskip

{\it The two-shock density at $O(\epsilon)$}

To go beyond the exact result (\ref{EqPresCorrel2}) we use FRG in \cite{ThieryLeDoussalWiese2016} to obtain $\rho_W^c(S_1,S_2)$ at first order in $O(\epsilon)$. We obtain

\be \label{EqPresCorrel3}
 \rho_W^c(S_1,S_2) = \frac{1}{(L \mu)^d} \frac{L^{2d}}{S_m^4} {\cal F}_d\Big(\frac{W}{W_\mu}, \frac{S_1}{S_\mu} , \frac{S_2}{S_\mu}\Big) \ .
\ee  
Where $W_\mu \simeq \kappa \mu^{-\zeta}$, $S_m \simeq A_d^{\gamma} \kappa  \Delta^{* \prime}(0^+) \mu^{-(d + \zeta)}$ and the function ${\cal F}_d$ is universal and apart from its three arguments depends only on the spatial dimension and range of elasticity inside the interface (the form of the large scale cutoff, here exponential, depends also on the chosen IR cutoff scheme of the theory).
To first order in $d=d_{\rm uc} -\epsilon$, and in the limit of large $L$ and small $\mu$, it is given by 
\be \label{EqPresCorrel4}
 {\cal F}_d(w, s_1 , s_2) \simeq A_d^{\gamma} \frac{ \tilde  \Delta^{* \prime \prime}(w)}{16 \pi \sqrt{s_1 s_2} } e^{-(s_1+s_2)/4} 
 + O(\epsilon^2)\ . 
\ee 

In \cite{ThieryLeDoussalWiese2016} we use (\ref{EqPresCorrel4}) to obtain a variety of results: the normalized PDF for shocks sizes at a distance $W$, the conditional probability to observe one shock given that another one occurred, and in particular the mean-density of pairs of shocks at a distance $W$:
\bea \label{EqPresCorrel5}
\rho_2(W) = \int dS_1 dS_2 \rho_W(S_1,S_2) = \rho_0^2\left[1- \frac{\Delta''(W)}{L^d m^4} \left(\frac{\langle S \rangle_P}{2S_m} \right)^{\!2} \right]  + O(\epsilon^2) \ssp .
\eea

{\it Generalization to the local shapes} \\
The factor $\frac{1}{(L \mu)^d} $ in (\ref{EqPresCorrel3}) highlights the fact that correlations are local and avalanches are correlated only if they occur in regions of space that are elastically connected (that is at a distance $|x-x'| \ll \ell_{\mu}$). To emphasize this local structure in \cite{ThieryLeDoussalWiese2016} we investigated the correlations between the local size of the shocks measured on arbitrary subspace (see \cite{ThieryLeDoussalWiese2016} for precise definitions). We obtained a result for the generating function of all connected moments. Here we only show the results for the first connected moments for SR and LR elasticity: for SR we obtain 
\bea
  \langle \langle S_{1x_1} S_{2x_2} \rangle \rangle_{\rho^{c}_W} &=& {\cal F}^{11}_d(\frac{W}{W_\mu} ,  m|x_1-x_2| )  \\
  {\cal F}^{11}_d(w ,  x)  &=& -  2^{-\frac{d}{2}-1} \pi ^{-\frac{d}{2}}  A_d \Delta^{* \prime \prime} (w)  x^{2-\frac{d}{2}} K_{2-\frac{d}{2}}(x)  \nn\\
 && + O(\epsilon^2)    \ .
\eea
where $K_n(x)$ is a modified Bessel function of the second kind. For LR elasticity we obtain
\begin{align}
 &\langle \langle S_{1x_1} S_{2x_2} \rangle \rangle_{\rho^{c}_W} = {\cal F}^{11}_{d,{\rm LR}}(\frac{W}{W_{\mu}} ,  m^2|x_1-x_2| )  \\
 & {\cal F}^{11}_d(w ,  x)  = -  (2 \pi )^{-\frac{d}{2}}  A_d \Delta^{* \prime \prime} (w)  x^{1-\frac{d}{2}} K_{1-\frac{d}{2}}(x)  + O(\epsilon^2) \nn  \ssp .
\end{align}

\begin{figure}
\centerline{\includegraphics[width=7.0cm]{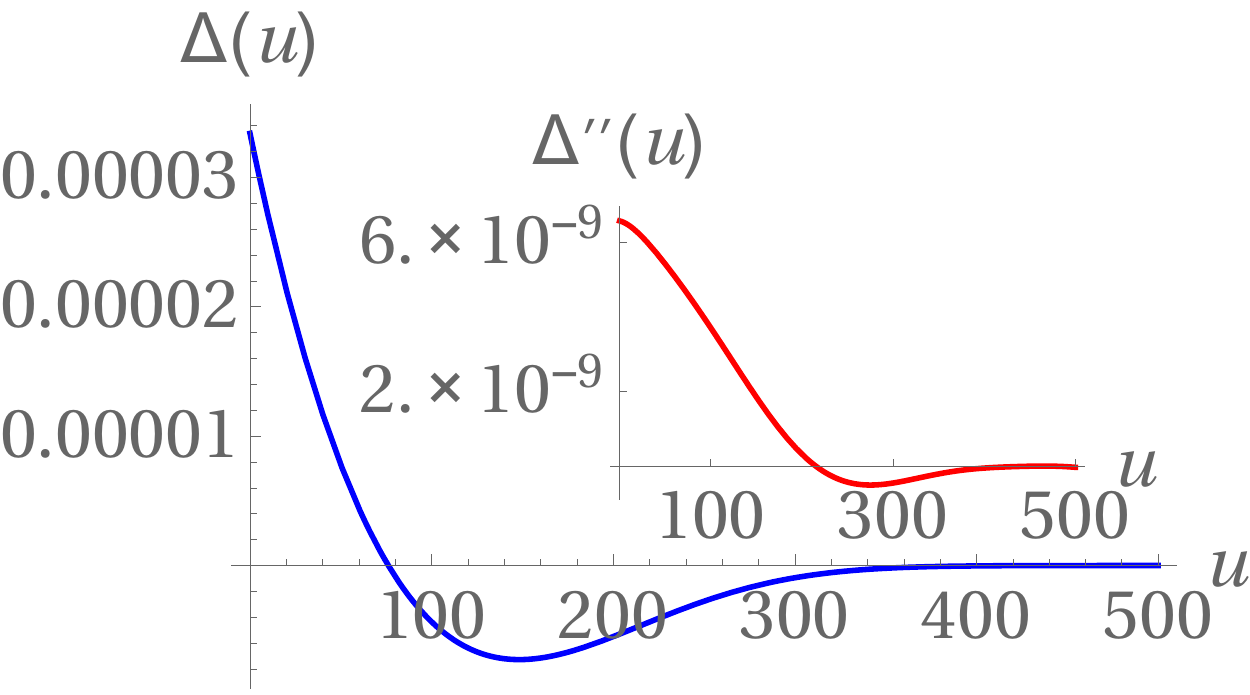} \includegraphics[width=7.0cm]{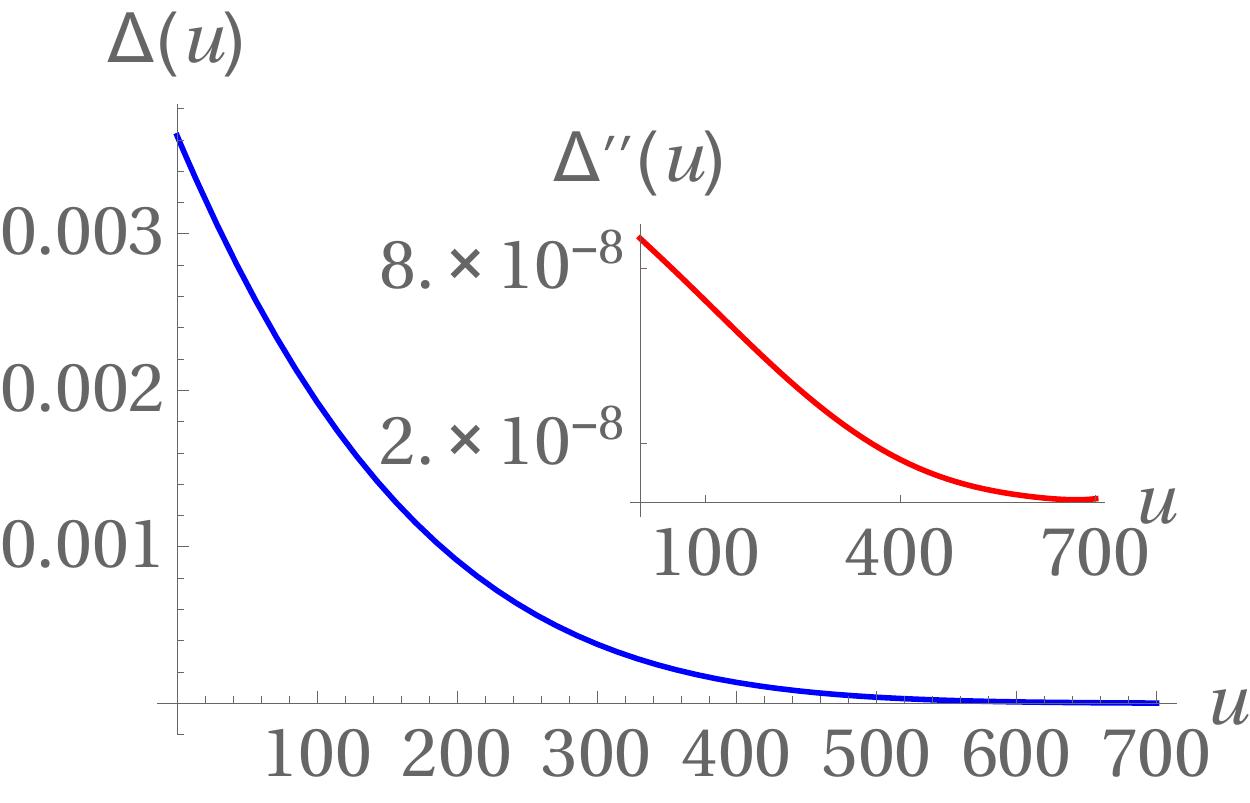} } 
\caption{Left: Renormalized disorder $\Delta(u)$ measured in the $d=0$ RB toy model. Inset: its second derivative $\Delta''(u)$, computed using a numerical fit of the measured $\Delta(u)$. Right: The same for the RF toy model. Figures taken from \cite{ThieryLeDoussalWiese2016}.}
\label{fig:PresCorrel2}
\end{figure}

\begin{figure}
\centerline{\includegraphics[width=7.0cm]{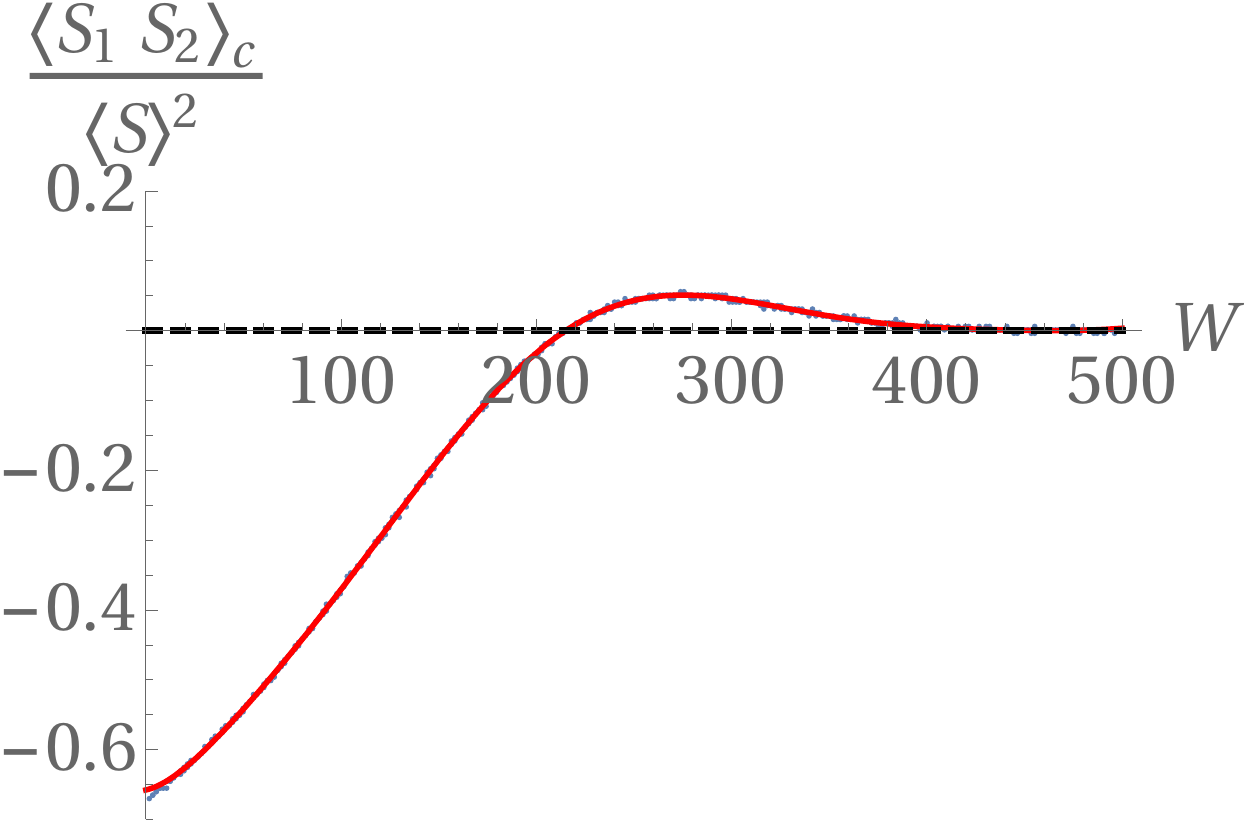} \includegraphics[width=7.0cm]{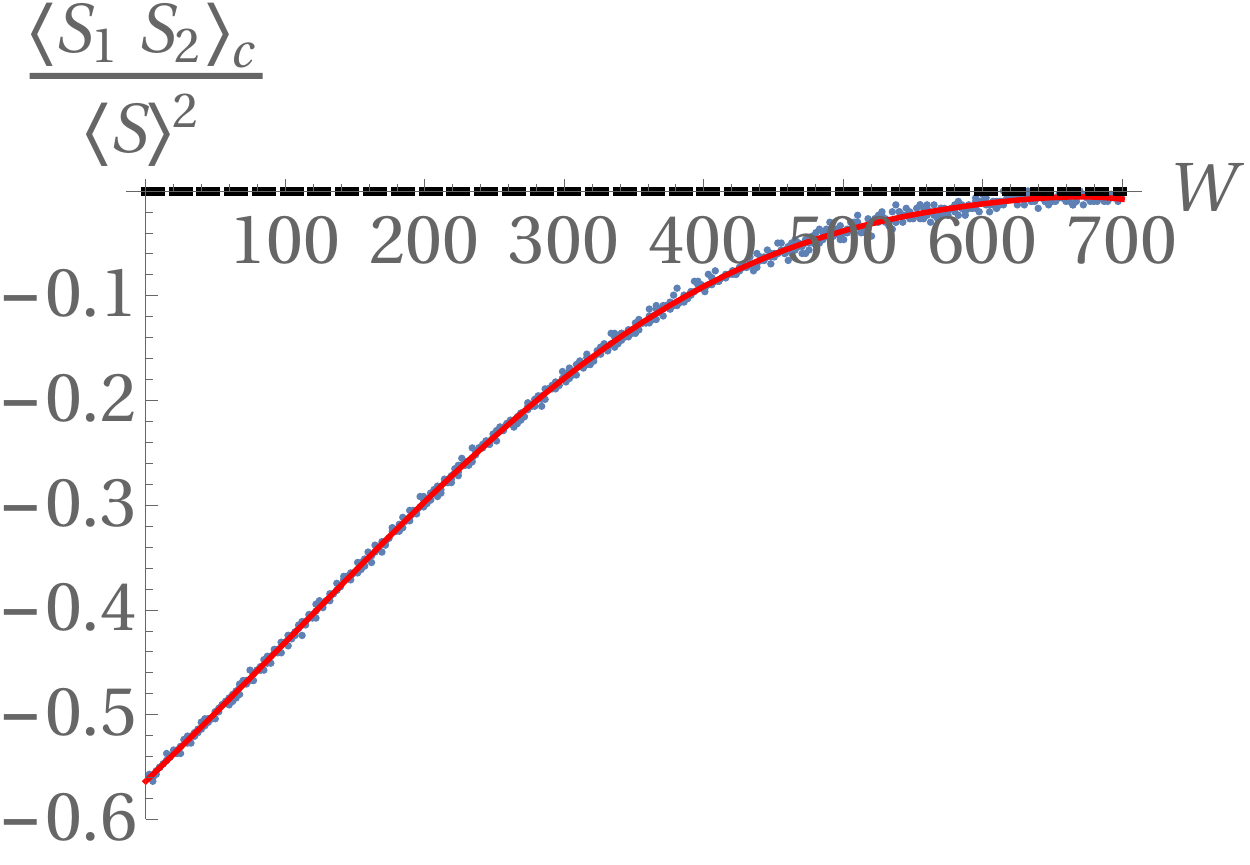}} 
\caption{Left: Comparison between the measurement of the normalized moment $\frac{ \langle S_1S_2 \rangle_{\rho_W^c} }{\langle S \rangle_{\rho}^2}$ (blue dots) and the prediction from the exact result (\ref{EqPresCorrel2}) using the measurement of $\Delta(u)$ (red curve) in the RB toy model. The agreement is perfect as expected. Right: The same for the RF toy model. Figures taken from \cite{ThieryLeDoussalWiese2016}.}
\label{fig:PresCorrel3}
\end{figure}

\begin{figure}
\centerline{ \includegraphics[width=7.0cm]{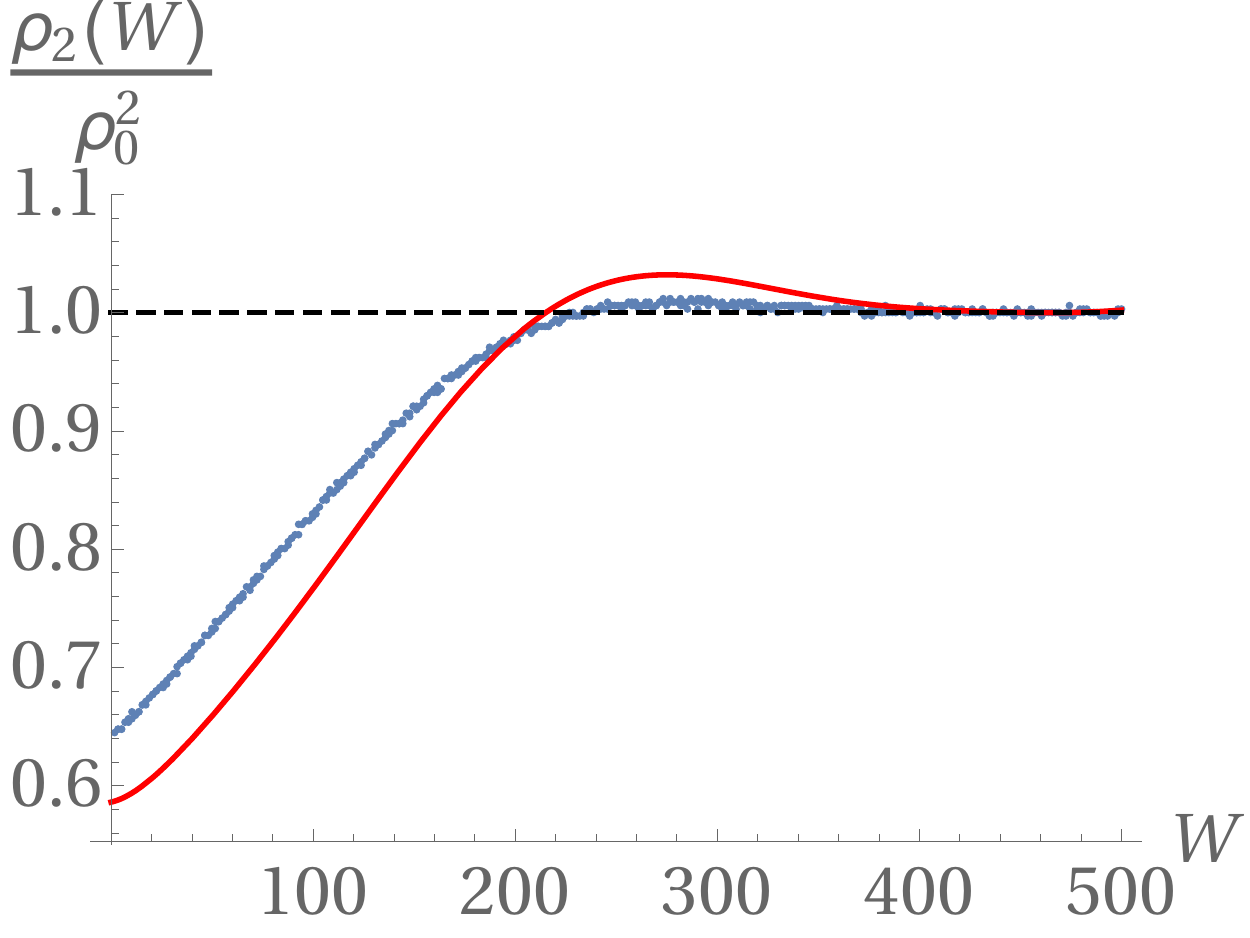} \includegraphics[width=7.0cm]{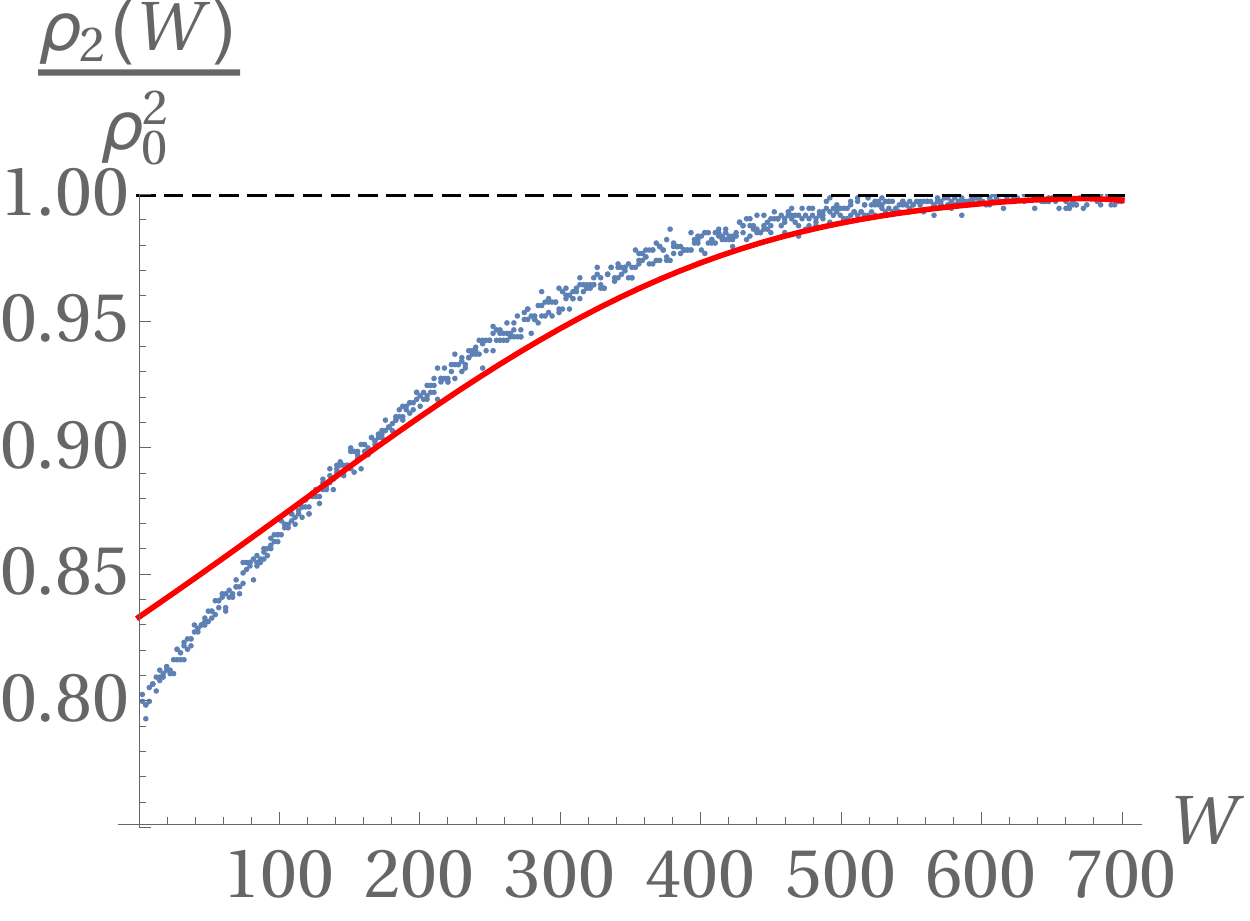} } 
\caption{Left: Comparison between the measurement of $\rho_2(W)$ (blue dots) and the prediction from the $O(\epsilon)$ result (\ref{EqPresCorrel5}) using the measurement of $\Delta(u)$ (red curve) in the RB toy model. We obtain a surprisingly good agreement. Right: the same for the RF toy model. Figures taken from \cite{ThieryLeDoussalWiese2016}.}
\label{fig:PresCorrel4}
\end{figure}

{\it Comparison with numerical simulations of toy models in $d=0$} \\
In \cite{ThieryLeDoussalWiese2016} we confronted our results with numerical simulations of toy models of a particle on $\JZ$ with either a random bond type potential, or a random field type potential. In both cases, we measure the renormalized disorder correlator $\Delta(W)$, the first and second connected moment $\langle S_1 S_2 \rangle_{\rho_W^c}$ and $\langle S_1^2 S_2 \rangle_{\rho_W^c}$, and the mean density of pairs of shocks at a distance $W$, $\rho_2(W)$. As shown in Fig.~\ref{fig:PresCorrel2} to Fig.~\ref{fig:PresCorrel4}, we obtain in both cases a perfect agreement for our exact formula (\ref{EqPresCorrel2}). We also obtain a surprisingly good agreement with our $O(\epsilon)$ formula for $\rho_2(W)$ (\ref{EqPresCorrel5}), considering that here $\epsilon=4$. The results for the second connected moments $\langle S_1^2 S_2 \rangle_{\rho_W^c}$ are poorer and not shown here. These simulations were more a proof of principle than a real test of our theory since what we are really interested in is the case of interfaces for which FRG should work better. However, we clearly see that these correlations exist, are not negligible, and as theoretically found, possess an interesting structure since they allow to distinguish between the RB and RF universality class.

\section{Conclusion} \label{conclusionAva}

The study of shocks and avalanche processes in disordered elastic interfaces is of outstanding interest. On one hand avalanche type processes are observed in nature in a variety of physical situations. From the statistical physics perspective, as complex scale free spatio-temporal processes, they represent a remarkable field of applications of universality ideas outside the standard study of continuous phase transitions in equilibrium statistical mechanics. From a more conceptual perspective, they are fundamental processes at the core of the theory of disordered elastic systems. Indeed, they are a direct consequence, and a characterization, of the presence of many metastable states in the energy landscape of the system, and they dominate both the physics of the statics and of depinning. Understanding avalanche processes has brought new light to the functional renormalization group approach to disordered elastic systems. The `curiosity' at the center of the theory, the non-analyticity of the effective action, is now directly linked to avalanches: measuring the functional renormalization group fixed point function in numerics and experiments is now possible.

\smallskip

In this thesis we have focused on the study of correlations and spatial shapes of avalanches. In both cases we have obtained results beyond mean-field that unveiled a rich structure. Many directions remain to extend these results. Concerning the spatial shape of avalanches, a natural extension would be to obtain one-loop results for the spatial shape with a long-range elastic kernel, which is of immediate experimental interest, or also one-loop results for the mean-velocity field inside avalanches. Concerning the correlations in avalanche processes, it would be interesting to extend our results to the dynamics. Although we have argued that our results obtained in the case of shocks should apply equally well to avalanches at depinning, it surely remains to be shown. More importantly, some features of the avalanche process at depinning are not present in the case of shocks, in particular the notion of seeds. Analyzing the correlation between the seeds of successive avalanches would be a very natural characterization of the non-Lévy nature of the avalanche process beyond mean-field. A complementary direction of research would be to gain a better understanding of correlations induced by mechanisms not captured by the elastic interface model, e.g. memory effects in the dynamics as in \cite{DobrinevskiLeDoussalWiese2013}. Finally, for both the shape of avalanches and correlations, it would be highly interesting to compare the results with refined simulations (particularly for the correlations where ours were performed in $d=0$), and with experiments. Comparing the temporal shape of avalanches computed in the ABBM model with the shape of Barkhausen pulses measured in soft magnets has triggered important developments that we reviewed. It would be interesting if similarly we could learn new aspects of the (eventually non-universal) physics e.g. of fracture processes by comparing the spatial shape of avalanches with our results.

\smallskip

Many other interesting open (and difficult) questions not tackled during this thesis remain. For example a better understanding of the functional renormalization group approach to the dynamics of disordered elastic interfaces at finite temperature and/or velocity would be of great interest. For avalanches this would lead to a better characterization of avalanches during the creep dynamics, as recently investigated in \cite{FerreroFoiniGiamarchiKoltonRosso2016}. In Barkhausen noise experiments, it is observed that a non-zero velocity modifies the avalanche exponents in materials with an effective LR elasticity, but not for materials with SR elasticity \cite{DurinZapperi2000}. While the velocity dependence of the exponents is known in the ABBM model, a FRG approach is surely necessary to understand such thinner effects. Similarly as for the effect of the temperature on avalanches, the effect of quantum fluctuations on avalanches (related to the notion of quantum creep) remains to be understood. Another interesting problem, of immediate experimental interest for fracture experiments, is to understand the statistics of clusters in avalanches for disordered elastic interfaces with LR elasticity (a question which was swept under the carpet during most of this chapter since we were effectively considering the total avalanche, which is eventually formed of several smaller avalanches). Finally it would be interesting to understand how to extend our results and methods to models close but not equivalent to disordered elastic interfaces, e.g. by taking into account plastic deformations or an additional conservation law as would be relevant for forced-flow imbibition experiments.

\chapter[Exactly solvable models of directed polymer]{Exact solvability, directed polymers in 1+1d random media and KPZ universality} \label{chapIII}

In this chapter we will focus on the study of the static properties of directed polymers (DPs, the $d=1$ problem) with short-range elasticity ($\gamma=2$) in a random bond type potential at a finite temperature $T$ in dimension $N=1$. Taking a look at the phase diagram of Fig.~\ref{fig:StaticPhaseDiagram} drawn in Chapter~\ref{chapI}, the large scale properties of the system are predicted to always be described by a strong disorder fixed point. Of course one can use the results of Sec.~\ref{subsec:FRGStatic} to study this FP by taking $\epsilon = 3$ (assuming the $\epsilon$ expansion has a sufficiently large radius of convergence) but that is not what we will do here. In fact in this chapter we will not focus so much on this FP. Rather, guided by the belief that it exists, we will study very specific models with {\it exact solvability properties}, i.e. for which exact analytical methods are available\footnote{To be fair with FRG, let us mention here more precisely that it is also an exact method since it provides an exact perturbative expansion of observables in an $\epsilon$ expansion. Its application in $d=1$ is, however, necessarily inexact since in practice one needs to truncate the expansion to a given order.}. The large scale analysis of exact results obtained for peculiar models will then lead to indirect information about the FP. The main issues with these methods is that they will not be robust to arbitrary small perturbations of the model. In particular we will study models of DP on the square lattice that are exactly solvable, for a given distribution of random energies, at a unique temperature. Guided by the qualitative analysis of Chapter~\ref{chapI} we however know that the temperature is irrelevant at large scales and therefore believe that the universal properties do not depend on its choice. Determining from the solution of the model which property is universal will not always be trivial, however, see in particular Sec.~\ref{subsec:PresBe}. From the point of view of universality, while the choice $\gamma=2$, $d=N=1$ and RB disorder can seem awfully restrictive compared to the more general analysis performed in  Chapter~\ref{chapI} and Chapter~\ref{chapII}, we will see that (i) by restricting to this choice we will obtain very sharp results; (ii) this universality class is actually very large; (iii) results that, to this date, have only been obtained using exact methods have been measured in modern experimental settings. For these first two points we already refer the reader to Sec.~\ref{subsec:SecI3:KPZ} where the links between the continuum DP problem and the KPZ equation was recalled (thus already bringing in models of out-of-equilibrium growth in $1+1$d in the large universality class mentioned above) together with some important results on the KPZ equation. For the last point we refer to Sec.~\ref{subsec:ExpKPZ} for good to amazing experimental verifications of properties of KPZ universality in $1+1$d. 

\smallskip

The outline of this chapter is as follows: in Sec.~\ref{Sec:ChapIII:Sec1} we will present a more complete introduction to the KPZ universality class in $1+1$d. In Sec.~\ref{Sec:ChapIII:Sec2} we will present a few exact solvability properties that played an important role over the years. In Sec.~\ref{Sec:ChapIII:Sec3} we will finally present the results obtained during the thesis.

\section{The KPZ universality class in $1+1$d}\label{Sec:ChapIII:Sec1}
In this section we review some results about the KPZ universality class in $1+1$d. We will start by presenting a few models in the KPZ universality class in Sec.~\ref{subsec:ChapIII:Sec1:Models}, with an emphasis on directed polymers models. We will then review important results obtained in some models and present the notion of {\it strong universality} and KPZ fixed point in Sec.~\ref{subsec:ChapIII:Sec1:KPZFP}. Finally in Sec.~\ref{subsec:ChapIII:Sec1:WeakU} we will discuss the notion of {\it weak universality} and the universal scaling limits of directed polymers on the square lattice. The material contained in this section is by now standard and inspired by a few excellent reviews on the subject \cite{Corwin2011Review,HalpinTakeuchi2015,Quastel2011,QuastelSpohn2015,SpohnLesHouches2016}.

\subsection{A few models in the KPZ universality class}\label{subsec:ChapIII:Sec1:Models}

In this section we present a few models believed, under some mild assumptions, to be in the KPZ universality class. We mainly focus on models of DPs: in the continuum, on the square lattice, at finite and zero temperature. But we also present some links with interacting particle systems and growing interfaces.

\subsubsection{The continuum DP and the KPZ equation Vs The Edwards-Wilkinson case}

The continuum KPZ equation is in the KPZ universality class. Behind this statement lies a rather long history that highlights the fact that many results on the KPZ universality class were obtained using exact solutions of some discrete models that resisted proof directly in the continuum setting until recently (2010). For now we just wish to make here a few more comments on the links between the continuum DP and the KPZ equation. Only in this section we keep the dimension arbitrary.\\

{\it Derivation of the Stochastic Heat Equation}\\
We recall that the partition sum of the continuum DP at temperature $T$ in a random potential $V(t,x)$ with both endpoints fixed was defined in (\ref{Eq:secIntroKPZ-1}) as the path integral (here with the change of notations $L \to t$ and $u \to x$)
\bea \label{Eq:secIntroKPZ-1bis}
Z_t(x) := \int_{u(0) = 0}^{u(t) =x} \cD[u] e^{- \frac{1}{2T} \int_0^{t} (\frac{du}{dt})^2 dt' - \frac{1}{T} \int_0^t V(t',u(t')) dt' } \ssp .
\eea
And here, keeping the notations of Chapter~\ref{chapI}, $u(t') \in \JR^N$ and the random potential is taken Gaussian with RB correlations (\ref{Eq:secIntroKPZ-1ter}), $\overline{V(t,x)V(t',x')} = \delta(t-t') R_0(x-x')$. In order to obtain the differential equation satisfied by $Z_t(x)$ (\ref{Eq:secI3:SHE}), it is already convenient to adopt a stochastic process language and note that the term in the path measure $ \cD[u] e^{- \frac{1}{2T} \int_0^{t} (\frac{du}{dt})^2 dt'} $ is just the measure on $N$-dimensional Brownian motion $u(t)$. The partition sum $Z_t(x)$ is thus written as\footnote{I thank Francis Comets for noticing a mistake in (\ref{Eq:secIntroKPZ-3}) in a preliminary version of this manuscript.}
\bea\label{Eq:secIntroKPZ-3}
Z_t(x) =  \mathbb{E}\left( e^{- \frac{1}{T} \int_{0}^{t} dt' V(t',u(t'))}  \delta^{(N)}(u(t)-x)  \right) \ ,
\eea
where here the average $ \mathbb{E}$ is over the stochastic process $u(t)$ now defined by
\bea \label{Eq:secIntroKPZ-4}
&& \partial_t u (t)= \sqrt{T} \xi(t) \nn \\
&& u(0) = 0  \ .
\eea
Here $\xi(t) =(\xi_1(t) , \cdots , \xi_N(t))$ is a vector of unit centered Gaussian white noise (GWN) with $\delta$ correlations
\bea \label{Eq:secIntroKPZ-5}
\langle \xi_i(t) \xi_j(t') \rangle_{\xi} = \delta_{ij} \delta(t-t') \ ,
\eea
and $\langle\rangle_{\xi}$ is the average over the GWN. Note that in this interpretation, the elasticity of the DP is of pure entropic origin (paths of the BM are more numerous close to the diagonal), and the fact that it is short-range comes partly from the fact that the BM satisfies a local stochastic partial differential equation (SPDE). Using the Feynman-Kac formula it is possible to show that $Z_t(x)$ solves a SPDE. Let us now give a simple derivation of this SPDE. We compute $Z_{t+dt}(x)$ as given by (\ref{Eq:secIntroKPZ-3}) and separate in the expectation value the contribution of the last time step between $t$ and $t+dt$: 
\bea \label{Eq:secIntroKPZ-6}
&& Z_{t+dt}(x)  = Z_t(x) + \frac{\partial}{\partial t} Z_t(x) dt + o(dt)  \\
&& =  \mathbb{E}\left( e^{- \frac{1}{T} \int_{0}^{t} dt' V(t',u(t'))  - \frac{dt}{T} V(t , u(t)) }    \delta^{(N)}(u(t+dt)-x)  \right) + o(dt)  \nn \\
&& = \left(1 - \frac{dt}{T} V(t,x) \right)   \mathbb{E}\left( e^{- \frac{1}{T} \int_{0}^{t} dt' V(t',u(t')) }  \delta^{(N)}(u(t+dt)-x) \right) + o(dt) \nn  .
\eea
Let us now discretize the last time step between $t$ and $t+dt$ as
\bea
u(t+dt) = u(t) - \sqrt{T dt} \xi
\eea
with $\xi$ a unit centered normal distribution. Hence we can write
\bea
&& \!\!\!\!\!\!\!\!\!\!\!\!\!\!\!\!\!\!\!\! Z_{t+dt}(x)  =\left(1 - \frac{dt}{T} V(t,x) \right)   \mathbb{E}\left( e^{- \frac{1}{T} \int_{0}^{t} dt' V(t',u(t')) }  \delta^{(N)}(u(t)-(x+\sqrt{T dt} \xi)) \right) + o(dt)  \nn \\
 && \!\!\!\!\!\!\!\!\!\!\!\!\!\!\!\!\!\!\!\! = \left(1 - \frac{dt}{T} V(t,x) \right)  \frac{1}{ (2\pi dt)^{\frac{N}{2}}} \int d^N  \xi e^{- \frac{\xi^2}{2  dt}}Z_{t}(x+\sqrt{T}  \xi )   + o(dt) \nn \\
&& \!\!\!\!\!\!\!\!\!\!\!\!\!\!\!\!\!\!\!\! = \left(1 - \frac{dt}{T} V(t,x) \right) \frac{1}{ (2\pi)^{\frac{N}{2}}}   \int d^N  \xi e^{- \frac{\xi^2}{2 }}  \left(  Z_t(x) +\sum_{i=1}^{N} \left( \sqrt{T dt} \xi_i \frac{\partial}{\partial x_i} Z_t(x)  + \frac{T dt }{2} \xi_i^2 \frac{\partial^2}{\partial x_i^2} Z_t(x)  \right) \right)  + o(dt) \nn \\
&& \!\!\!\!\!\!\!\!\!\!\!\!\!\!\!\!\!\!\!\! = \left(1 - \frac{dt}{T} V(t,x) \right) \left( Z_t(x) + \frac{T dt}{2}( \nabla_x)^2 Z_t(x) \right) + o (dt)  \nn 
\eea
Hence, comparing the terms of order $dt$ in the first and last line of the above calculation, we obtain,
\bea \label{Eq:secI3:SHEbis}
\frac{\partial}{\partial t} Z_t(x)  = \left( \frac{T}{2} ( \nabla_x)^2 - \frac{1}{T} V(t,x) \right) Z_t(x) \ ,
\eea
with the initial condition $Z_{t=0}(x) = \delta^{(N)}(x)$.

\medskip

{\it A subtlety}\\
There is a non-trivial subtlety in the equivalence between (\ref{Eq:secIntroKPZ-1bis}), (\ref{Eq:secIntroKPZ-3}) and (\ref{Eq:secI3:SHEbis}) on which we now comment. On one hand, looking at (\ref{Eq:secI3:SHEbis}), for a centered potential, it seems that the mean value of $Z_t(x)$ is just the transition probability for a random walk on $\JR^N$ (since we are using Ito's convention). On the other hand, looking at (\ref{Eq:secIntroKPZ-1bis}) or (\ref{Eq:secIntroKPZ-3}) it seems that this mean value should contain a term involving $R_0(0)$. The equivalence between (\ref{Eq:secIntroKPZ-3}) and (\ref{Eq:secI3:SHEbis}) is ensured if one takes as a definition of the exponential the so-called time/Wick ordered exponential as:
\be
e^{-\frac{1}{T} \int_{0}^tV(t,u(t))  } := \sum_{n=0}^{\infty}  (-1/T)^n \int_{0 \leq t_1  < t_2 < \cdots < t_n \leq t}  V(t_1,u(t_1)) \cdots V(t_1,u(t_n) )\ssp .
\ee
The difference with the ordinary exponential is obviously immaterial for non-random smooth potentials $V(t,x)$. For the case of a random potential with $\delta$ interaction in the $t$ direction it however makes a big difference: here the ordering $t_i < t_{i+1}$ ensures the equivalence between (\ref{Eq:secIntroKPZ-3}) and (\ref{Eq:secI3:SHEbis}), which was actually implicit in the derivation of (\ref{Eq:secIntroKPZ-3})\footnote{To see this, consider for example the exponential of the integral of a GWN $Y(t)=e^{\int_{0}^{t} \xi(t')dt'}$. Using a similar derivation as above one obtains $\partial_t Y = \xi(t) Y(t)$ and thus Ito's convention imposes $\partial_t \overline{Y(t)}=0$. This is true only if $Y(t)$ is interpreted as $Y(t) = \sum_n \int_{0<t_1 < \cdots <t_n <t} \xi(t_1) \cdots \xi(t_n)$. Using the regular interpretation of the exponential one obtains $\overline{Y(t)} = e^{t/2}$.}. Finally, denoting now by $\check{Z}_t(x)$ the object defined by the path integral (\ref{Eq:secIntroKPZ-1bis}), the equivalence with $Z_t(x)$ defined by (\ref{Eq:secIntroKPZ-3}) (with the time-ordered exponential) or (\ref{Eq:secI3:SHEbis}) (interpreted in the Ito sense) is
\bea  \label{Eq:secIntroKPZ-1bisreplaced}
&& Z_t(x)  = \check{Z}_t(x)  e^{ - \frac{1}{2T^2}   \int_0^t dt'  R_0(0)} \nn \\
&& \check{Z}_t(x) = \int_{u(0)=0}^{u(t)=x} {\cal D}[u] e^{ - \frac{1}{2T} \int_{0}^{t} dt' (\frac{du}{dt})^2  - \frac{1}{T} \int_0^t dt' V(t',u(t'))} \ssp .
\eea
 This subtlety will be particularly important to make sense of the path integral formula (\ref{Eq:secIntroKPZ-1bis}) in the important case where $V(t,u)$ is taken as a centered Gaussian white noise and $R_0(u-u') \sim \delta^{(N)}(u-u')$. From now on we will adopt this convention.

\medskip

{\it From the MSHE to the KPZ equation}\\
In Sec.~\ref{subsec:SecI3:KPZ} we already saw that taking the logarithm of the MSHE (\ref{Eq:secI3:SHEbis}), assuming that $V(t,x)$ is smooth, one obtains that $h(t,x) = T \log(Z_t(x))$ satisfies the KPZ equation (\ref{Eq:secI3:KPZ}). This derivation, however, assumed a smooth random potential $V(t,x)$. In the same spirit as above, let us take into account the fact that the random potential is rough in the time direction and use Ito's lemma (see e.g. \cite{gardiner2004handbook}) to compute the time derivative of $h(t,x) = T \log(Z_t(x))$. We obtain
\bea \label{Eq:ChapIII:Intro:1}
\partial_t h(t,x)  && =  \frac{T}{Z_t(x)} \left(  \left( \frac{T}{2} ( \nabla_x)^2 - \frac{1}{T} V(t,x) \right) Z_t(x) \right) - \frac{1}{2Z_t(x)^2} R_0(0) (Z_t(x))^2 \nn \\
&& =  \frac{1}{2} ( \nabla_x  h )^2 +\frac{T}{2}  \nabla_x^2  h   - V(t,x) - \frac{R_0(0)}{2} \ssp .
\eea
Hence we see that the Ito's lemma precisely makes the short-distance correlations of the disorder play a role again. Making a change of variables
\bea \label{Eq:ChapIII:Intro:1ter}
&& \check h(t,x) = h(t,x) + \frac{R_0(0)t}{2} \ssp , 
\eea
$\check h(t,x) $ solves
\bea \label{Eq:ChapIII:Intro:1bis}
&& \partial_t \check h(t,x) =  \frac{1}{2} ( \nabla_x  \check h )^2 +\frac{1}{2}  \nabla_x^2  \check h   - V(t,x)  \ssp .
\eea
And note that $\check h(t,x)=  T \log \check Z_t(x)$ where $\check Z_t(x)$ was defined through the path-integral formula (\ref{Eq:secIntroKPZ-1bisreplaced}). The KPZ equation usually refers to the equation (\ref{Eq:ChapIII:Intro:1bis}) satisfied by $\check h$. In the following we will drop the different checkmarks, knowing that $h(t,x) = \log Z_t(x)$ is a (sometimes dangerous) shortcut. From now on we will also restrict ourselves to the case $N=1$, i.e. the directed polymer in a two dimensional random environment, or the one-dimensional KPZ equation, hereafter referred to as the $1+1$d case. Before we continue we remind the reader that we already commented in Sec.~\ref{subsec:SecI3:KPZ} on the interpretation of the KPZ equation as a model of out-of-equilibrium growth.

\medskip

{\it The continuum DP and KPZ equation}\\
 The continuum directed polymer and KPZ equation generally refer to the case where $V(t,x) = -\xi(t,x)$ with $\xi$ a Gaussian white noise (GWN) with correlations
\bea  \label{Eq:ChapIII:Intro:3}
\overline{\xi(t,x) \xi(t',x')} = 2 \sigma \delta(t-t') \delta(x-x') \ssp .
\eea
In this case, let us emphasize that the MSHE (\ref{Eq:secI3:SHEbis}) and the KPZ equation (\ref{Eq:ChapIII:Intro:1bis}) have a very different status. Focusing now on the case $N=1$, while the MSHE, interpreted in the Ito sense, remains well defined, the KPZ equation a priori is not: we will see that $\log(Z_t(x))$ looks locally like a BM and thus taking the square of its derivative is ill advised. The $R_0(0)$ ($=+\infty$ in this case) wandering around in between (\ref{Eq:secI3:SHEbis}) and (\ref{Eq:ChapIII:Intro:1bis}) is a sign of this issue. This calls for a regularization technique and making sense of the `solution of the KPZ equation' is a hard problem. In fact well before a precise sense was given to solving the KPZ equation \cite{Hairer2013} people believed that the right way to interpret the KPZ equation is through the Cole-Hopf transform, and thus $\log(Z_t(x))$ with $Z_t(x)$ the solution of the MSHE has long been thought of as the solution of the KPZ equation. Accordingly in this chapter when we mention the solution of the KPZ equation, we actually mean the Cole-Hopf solution. \\

Ignoring from now on these issues and taking $V(t,x) = -\xi(t,x)$ a GWN with correlations (\ref{Eq:ChapIII:Intro:3}), let us first note that the change of variables,
\bea \label{Eq:ChapIII:Intro:4}
&& t = a \tilde{t} \quad , \quad  x = b \tilde x  \quad  , \quad  a = \frac{T^{\frac{5}{3}}}{(2\sigma)^{\frac{2}{3}}}  \quad  , \quad b =(2\sigma/T)^{\frac{2}{3}} \ssp , \nn \\
&& \tilde{Z}_{\tilde{t}}(\tilde x) = Z_{ t = a \tilde{t}}( x = b \tilde x) \quad , \quad \tilde{h}(\tilde t , \tilde x) = h(t= a \tilde t , x = b \tilde x) \ssp ,
\eea
make the equations (\ref{Eq:secI3:SHEbis}) and (\ref{Eq:ChapIII:Intro:1bis}) equivalent to (dropping the tildes)

\bea \label{Eq:ChapIII:Intro:5}
&& \frac{\partial}{\partial t} Z_t(x)  =  \frac{1}{2} ( \nabla_x)^2 Z_t(x) + \xi(t,x) Z_t(x)  \ssp , \nn \\ 
&& \partial_t h(t,x) = \frac{1}{2} (\nabla_x h(t,x))^2 + \frac{1}{2} (\nabla_x)^2 h(t,x) + \xi(t,x) \ssp .
\eea
where now $\xi(t,x)$ is a centered GWN with $\overline{\xi(t,x) \xi(t',x')} = \delta(t-t') \delta(x-x') $. Note that this simple calculation shows that there cannot be any phase transition in the large scale properties of the DP at a finite critical value of the temperature or of the noise strength: the system is always in the same phase. This was expected from the static phase diagram Fig.~\ref{fig:StaticPhaseDiagram} and we know that this phase corresponds to a strong disorder, zero temperature phase for the DP.

\medskip

{\it The Edwards-Wilkinson equation}\\
 Only at e.g. vanishing non-linearity can one obtain a different large scale behavior. The resulting equation in this case is known as the Edwards-Wilkinson equation. It reads 
\bea
 \partial_t h(t,x) = \frac{1}{2} (\nabla_x)^2 h(t,x) + \xi(t,x) \ssp .
\eea
Note that in this case the growth favors neither direction. The critical exponents are easily extracted in this case by a simple scaling argument, leading to $z=2$, $\alpha= 1/2$ and $\beta = 1/4$ (see Sec.~\ref{subsec:SecI3:KPZ} for definitions). The full solution is easily obtained by going to Fourier space $x \to q$: each Fourier component performs an independent Ornstein-Uhlenbeck process with diffusivity $\sim q^2$ and
\bea
h(t,q) = h(0,q)e^{-\frac{1}{2} t q^2} + \int_{0}^{t} e^{-\frac{1}{2} q^2(t-t')} \xi(t,q) \ssp .
\eea
The problem is thus essentially solved and fluctuations at the Edwards-Wilkinson fixed point are Gaussian. The solution of the KPZ equation at non-zero linearity and the description of the associated FP will be much more difficult. Before we discuss some known properties of this FP, let us now present a few models in the KPZ universality class in $1+1$d (KPZUC).

\subsubsection{Models of DP on the square lattice at finite temperature}

The definition (\ref{Eq:secIntroKPZ-3}) of the partition sum of the continuum DP makes it transparent how to discretize the DP on any graph: given an underlying `free' measure on directed paths $\pi$ on the graph, one associates to each path a random energy $E(\pi)$ that is the sum of all energies encountered by the path along the way. In this thesis we will be interested in models of DP on the square lattice which is a natural discretization of the continuum DP problem in $d=1+1$. We thus consider the square lattice $\JZ^2$, with Euclidean coordinates $(x_1,x_2)$. Directed paths on $\JZ^2$ are up-right paths: they jump either to the right $(x_1,x_2) \to (x_1+1,x_2)$, or upward $(x_1,x_2) \to (x_1,x_2+1)$ (see Fig.~\ref{fig:IntroToDPOnZ2}). Given a temperature $T$ and an ensemble of random energies $\{ \cE_{x_1,x_2} , (x_1,x_2) \in \JZ^2 \}$ that are drawn from a given PDF, the {\it point-to-point partition sum} of the DP with starting point $(0,0)$ and endpoint $(x_1,x_2) \in \JN^2$ is
\bea \label{Eq:ChapIII:Sec1:DiscreteDP1}
Z_{x_1,x_2} := \sum_{\pi :(0,0) \to (x_1,x_2)} e^{-\frac{1}{T} \sum_{(x_1',x_2') \in \pi} \cE_{x_1',x_2'} } \ssp .
\eea
Here $\sum_{\pi :(0,0) \to (x_1,x_2)}$ denotes the sum over all up-right paths on $\JN^2$ from $(0,0)$ to $(x_1,x_2) \in \JN^2$. Introducing the random Boltzmann weights $W_{x_1,x_2} := e^{-\frac{1}{T} \cE_{x_1,x_2}} $, (\ref{Eq:ChapIII:Sec1:DiscreteDP1}) is equivalently rewritten as
\bea \label{Eq:ChapIII:Sec1:DiscreteDP2}
Z_{x_1,x_2} := \sum_{\pi :(0,0) \to (x_1,x_2)} \prod_{(x_1',x_2') \in \pi}  W_{x_1',x_2'} \ssp .
\eea
Alternatively we will use, in analogy with the continuum DP case, the coordinate $t$ defined by
\bea \label{Eq:ChapIII:Sec1:DiscreteDP3}
t = x_1+ x_2 \ssp .
\eea
The latter is thus the length of the DPs. When using $t$, the space coordinate will be taken either as
\bea \label{Eq:ChapIII:Sec1:DiscreteDP4}
x = x_1, \quad {\rm or } \quad \hat x = \frac{x_1-x_2}{2} \ssp ,
\eea
as schematized in Fig.~\ref{fig:IntroToDPOnZ2}. The factor $1/2$ in the definition of $\hat x$ is to ensure that neighboring lattice sites at the same time coordinate $t$ are distant from $1$ in units of $\hat x$. The coordinate $t$ is thus strictly increasing along a DP path, $x$ is weakly increasing, while $\hat x$ decreases or increases and is the coordinate that is the closest in spirit to the $x$ coordinate of the continuum DP. Using these coordinates, directed paths $\pi$ from $(0,0)$ to $(x_1,x_2)$ identify with functions $x(t) =\hat x(t)+ \frac{t}{2}  $ such that $x(0) = 0$, $x(x_1+x_2) = x_1$ and $x(t+1)-x(t) \in \{0,1 \}$ (and similarly for $\hat x(t)$). Using these coordinates we will note the partition sum equivalently as
\bea \label{Eq:ChapIII:Sec1:DiscreteDP5}
Z_t(x) = Z_{x_1=x ,  x_2 = t-x_1} \quad , \quad Z_t(\hat x) = Z_{x_1 = (t+2 \hat x)/2 , x_2 = (t - 2 \hat x)/2} \ssp .
\eea
We will adopt a similar notation for any function on the lattice.

\begin{figure}
\centerline{\includegraphics[width=9.0cm]{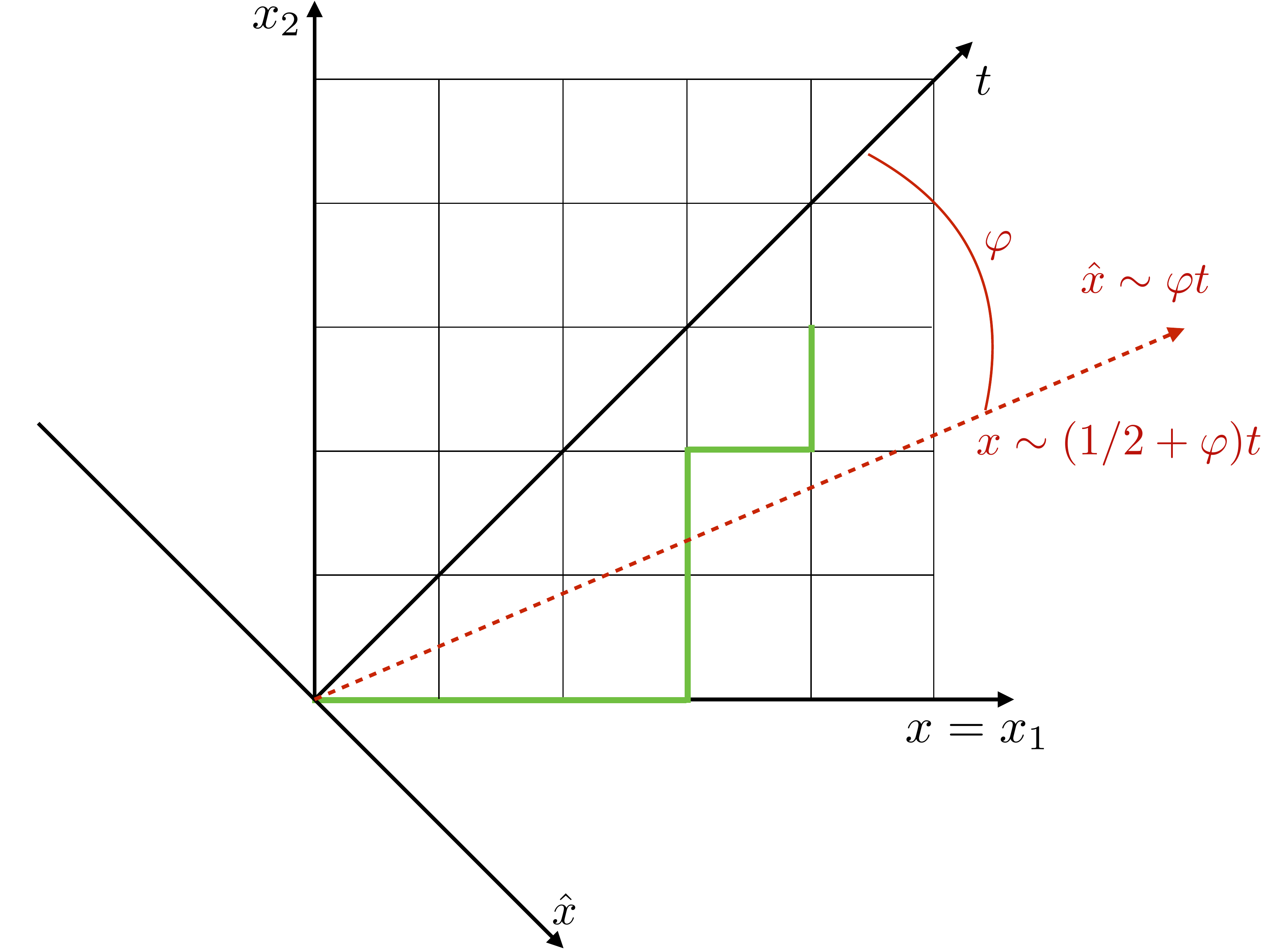} } 
\caption{Different coordinate systems for directed polymers on the square lattice. Green: an admissible directed path, i.e. up-right path, from $(0,0)$ to $(x_1,x_2) = (4,3)  $ i.e. $ (t,x) =(7,4) $ i.e. $ (t,\hat x)=(7,1/2)$. Red, convention used for the asymptotic analysis of polymers of large length $t \gg 1$ in a given direction $\varphi$.}
\label{fig:IntroToDPOnZ2}
\end{figure}

In each random environment and for $t \geq 0$, the partition sum of the DP can also be defined recursively as
\bea \label{Eq:ChapIII:Sec1:DiscreteDP6}
&& Z_{t+1}(x) = W_{t+1}(x)\left(Z_{t}(x) + Z_{t}(x-1) \right) \quad {\rm with} \quad Z_{t=0}(x) = \delta_{x,0}  \\
\Longleftrightarrow &&  Z_{t+1}(\hat x) = W_{t+1}(\hat x)\left(Z_{t}(\hat x +1/2) + Z_{t}(\hat x-1/2) \right) \quad {\rm with} \quad Z_{t=0}(\hat x) = \delta_{\hat x,0} \nn 
\eea
Note that (\ref{Eq:ChapIII:Sec1:DiscreteDP6}) appears as a discrete analogue of the MSHE (\ref{Eq:ChapIII:Intro:5}) (this will be made more precise in Sec.~\ref{subsec:ChapIII:Sec1:WeakU}). The KPZ universality hypothesis then basically states that, for sufficiently nice disorder (such that all the moments of the random energy are finite, $\overline{(\cE_{x_1,x_2})n}<\infty$, and the disorder has short-range correlations), the large scale properties of $\log Z_t(\hat x)$ are similar to those of $h(t,x)$ in the KPZ equation (this will be made more precise in Sec.~\ref{subsec:ChapIII:Sec1:WeakU}). In this thesis we will often look at the large scale properties in an arbitrary direction $\varphi$, referring to $t \gg 1$ with the ballistic scaling
\bea
x \sim (1/2 + \varphi )t \quad \Longleftrightarrow \quad \hat x \sim \varphi t \ssp ,
\eea
i.e. the `angle' $\varphi$ is measured with respect to the diagonal of the square lattice see Fig.~\ref{fig:IntroToDPOnZ2}.

\smallskip

We should stress here that for most models in the KPZUC, there is often one observable whose large scale properties is similar to the height in the KPZ equation, but that does not mean that the properties of any observable of any model in the KPZUC are related to some observable in the KPZ equation. More generally interesting observables in one language may not necessarily be relevant in the other and vice-versa. In particular in the DP framework, the `KPZ-height' like variable is the free-energy of the DP. The latter is certainly interesting, but does not contain all DP properties. Before being a model in the KPZUC, the DP is the statistical mechanics of directed paths in a random environment, and $Z_{x_1,x_2}$ is the normalization factor that allows to define the {\it quenched measure on paths as}, for all paths from $(0,0)$ to $(x_1,x_2) \in \JN^2$,
\bea \label{Eq:ChapIII:Sec1:DiscreteDP7}
Q_{x_1,x_2}(\pi) := \frac{e^{- \frac{1}{T} \sum_{(x_1',x_2')\in \pi} \cE_{x_1',x_2'}}}{Z_{x_1,x_2}}  \ssp .
\eea
The latter is a deformed version of the underlying free measure on directed paths that favors the energy, it is a probability measure on paths from $(0,0)$ to $(x_1,x_2)$ and it is itself a random (disorder dependent) object. The {\it annealed measured } is the average over disorder of the quenched measure:
\bea \label{Eq:ChapIII:Sec1:DiscreteDP8}
P_{x_1,x_2}(\pi) := \overline{Q_{x_1,x_2}(\pi)} \ssp .
\eea
With these definitions, for a given observable on paths $O(\pi)$, one is interested in the quenched and annealed averages as 
\bea \label{Eq:ChapIII:Sec1:DiscreteDP9}
\langle O(\pi) \rangle_Q := \sum_\pi O(\pi) Q_{x_1,x_2}(\pi)   \quad , \quad  \overline{ \langle O(\pi) \rangle_Q}:= \sum_\pi O(\pi) P_{x_1,x_2}(\pi)  \ssp .
\eea
In the DP framework, understanding the properties of the quenched and annealed measure are the most challenging questions. Some of these properties are indeed contained in KPZ universality: taking $(x_1,x_2)$ along the diagonal $ (x_1,x_2)=(\sT/2,\sT/2)$ with $\sT \to \infty $, parametrizing paths by functions $\hat x(t)$, one expects that (i) with probability $1$ at large $\sT$ the support of the quenched measure $Q$ is on paths scaling like $\hat x(t) \sim \sT^{\zeta} \tilde{\hat x}(t/\sT)$ with $\zeta$ the roughness exponent of the DP related to the dynamic exponent of KPZ as $\zeta= 1/z = 2/3$ (see Sec.~\ref{subsec:SecI3:KPZ}); (ii) a similar (less strong) statement for the annealed measure $P$. More subtle properties like the shape of the rescaled path $\tilde{\hat x}(t)$ or its localization properties are not trivially related to observables in the growing interface language. In the DP framework, localization refers to the fact that, even in the limit of infinite polymer $\sT \to \infty$, with probability $1$ (i.e. for almost any drawing of the random environment), there exists a point at time $t = \sT/2$ which is visited by the polymer with a non-zero probability. This should be compared to a standard random walk where the point visited with maximum probability is on the diagonal and the probability is of order $1/\sqrt{\sT}$ in $d=1+1$. A stronger statement of localization of the full path in the DP case is the fact that there are paths for which the quenched measure $Q_{T/2 , T/2}(\pi)$ remains non-zero in the limit $T\to \infty$. This is consistent with the idea of Chapter~\ref{chapI} and Chapter~\ref{chapII} that temperature is (although dangerously \cite{LeDoussal2008}) irrelevant at large scales.
\smallskip

Let us finally mention that such models of DPs on the square lattice and in higher dimensions have received a considerable amount of attention from the mathematical community, independently of any exact solvability properties and using purely probabilistic approaches, starting with the work of Imbrie and Spencer \cite{ImbrieSpencer1988} and Bolthausen \cite{Bolthausen1989}. Rigorous results in particular confirm the static phase diagram of Fig.~\ref{fig:StaticPhaseDiagram}. We refer the reader to \cite{comets2004probabilistic} for a review, in particular for the mathematical definition of the strong disorder regime in terms of martingales or localization properties of the path \cite{comets2003directed}.

\subsubsection{Models of DPs on the square lattice at zero temperature}

As temperature is irrelevant at the strong disorder FP of the DP, it is natural that models of DPs at zero temperature are also in the KPZ universality class. A model of DP at zero temperature can be obtained as the limit of the model previously considered as 
\bea \label{Eq:ChapIII:Sec1:DiscreteDP10}
\sE_{x_1,x_2} := \lim_{T \to \infty} -T \log Z_{x_1,x_2}  = {\rm min}_{\pi : (0,0) \to (x_1,x_2)  }  \sum_{(x_1',x_2') \in \pi} \cE_{x_1',x_2'} \ssp .
\eea
$\sE_{x_1,x_2}$ is thus the energy of the optimal path from $(0,0)$ to $(x_1,x_2)$.  KPZ universality says here that the large scale properties of $\sE_{x_1,x_2}$ should be the same as those of the height in the growing interface language. At a given time the profile $\sE_t(x)$ is interpreted as an interface. For $t\geq 0$, it evolves according to
\bea \label{Eq:ChapIII:Sec1:DiscreteDP11}
\sE_{t+1}(\hat x) = \cE_t(\hat x)+{\rm min}\left(\sE_t(\hat x-1/2),\sE_t(\hat x +1/2) \right) \ssp ,
\eea
with the initial condition $\sE_{t=0}(0) = 0$ and $\sE_{t=0}(\hat x) = - \infty$ for $\hat x \neq 0$. Let us denote here for future use
\bea
h^{(1)}(t,\hat x) = -\sE_{t}(\hat x) \ssp ,
\eea
the `growing interface' defined in this way. Two particular subclasses of zero temperature DP models have been much studied, each having links with interesting other models. 
\smallskip

The first is the case where the random energies are bounded from below, say by $0$. In this case the DP model is usually referred to as a {\it directed first passage percolation problem} (FPP). The random energies $\cE_{x_1,x_2}$ are interpreted as waiting times $\st_{x_1,x_2}:=\cE_{x_1,x_2}$ and the optimal energy $\sE_{x_1,x_2}$ as a first passage time $\sT_{x_1,x_2}$. This type of model was originally introduced in \cite{hammersley1965first} to describe the invasion of a fluid into a porous medium. These models can provide examples of the fact that KPZ universality should always be applied with caution: if the $\st_{x_1,x_2}$ can be zero with a finite probability $p$, there can be a region of space where the first passage time converges to $0$ in the large time limit with probability $1$. This precisely occurs if $p$ is large enough and that a percolation threshold is reached. Of course in this case the fluctuations of $T_{x_1,x_2}$ do not scale with the KPZ exponent $t^{1/3}$. We will see an example of such a model in Sec.~\ref{subsec:PresStat}. 

\smallskip

The second subclass is the case where the random energies are bounded from above, say by $0$. In this case the DP model is usually referred to as a {\it directed last passage percolation problem} (LPP). The random energies $\cE_{x_1,x_2}$ are interpreted as the opposite of waiting times $\st_{x_1,x_2}:=-\cE_{x_1,x_2}$ and the optimal energy $\sE_{x_1,x_2}$ as the opposite of the last passage time $\sT_{x_1,x_2}:=-\sE_{x_1,x_2} = {\rm max}_{\pi : (0,0) \to (x_1,x_2)  }  \sum_{(x_1',x_2') \in \pi} \st_{x_1',x_2'} $. See \cite{martin2006last} for a review. In this case it is usual to define a growing interface different from the one mentioned previously by looking at the boundary of the set $B(t) := \{ (x_1,x_2) \in \JN^2 , \sT_{x_1,x_2} \leq t \}$ see Fig.~\ref{fig:LPPSimu}. This growing interface fall in the more general class of so-called corner growth models as we will show in the next section. Note that the recursion equation for the last passage time reads
\bea
T_{x_1+1,x_2+1}  = \st_{x_1+1,x_2+1} + {\rm max}(T_{x_1+1,x_2},T_{x_1,x_2+1} ) \ssp .
\eea
And the growing interface thus necessarily has the shape shown in Fig.~\ref{fig:LPPSimu}: the boundary is a {\it down-right} path $(x_1(i) , x_2(i))_{i=0, \cdots, N}$ on $\mathbb{N}^2$: its starting point on the vertical edge is $(x_1(0), x_2(0) ) = (x_1^{\rm max} , 0)$, with $x_1^{\rm max} = {\rm max}_{x_1 \in \JN | T_{x_1,0} \leq t}  ~ x_1 $. Switching to the $t,\hat x$ coordinate, the growing interface is thus defined now as
\bea
h^{(2)}(t,\hat x) := {\rm max}\{t' | T_{\sf t'}(\hat x) \leq t \} \ssp .
\eea
 The fluctuations of this growing interface are believed to be in the KPZUC for sufficiently nice distributions of weights \cite{martin2006last}. We will see in the next section that growth models defined in this way fall in a class of models known as corner growth models, and it is actually for a model of LPP with geometric or exponentially distributed weights that the emergence of Tracy-Widom type fluctuations in the DP context was first shown in \cite{johansson2000}. Note that the two growing interfaces $h^{(1)}$ and $h^{(2)}$ are $\forall \hat x$ the inverse of one another: $\forall (t,\hat x)$, $h^{(1)}(h^{(2)}(t , \hat x) , \hat x ) = t$, and
 \bea
&& Prob(h^{(1)}(t , \hat x) \leq H ) = Prob( T_{t}(\hat x) \leq H )  \nn \\
&& Prob(h^{(2)}(t , \hat x) \leq H ) = Prob( T_{H}(\hat x) \geq t ) \ssp .
\eea

Their large scale fluctuations are thus linked. Assuming that, for $\hat x$ fixed, $h^{(1)}(t,\hat x) \sim c_1 t + \lambda_1 t^{1/3} X $ with $X$ a $O(1)$ RV and $c_1 \geq 0$, then $ \lim_{t \to \infty} Prob(h^{(1)}(t , \hat x) \leq c_1 t + \lambda_1 t^{1/3} z ) = F(z)  $ with $F(z)$ the CDF of $X$. We thus must have
\bea
F(z) && = \lim_{t \to \infty} Prob( T_t(\hat x) \leq  c_1 t + \lambda_1 t^{1/3} z) \nn \\
&& = 1 - \lim_{t' \to \infty} Prob( T_{\frac{t'}{c_1} - \frac{\lambda_1}{c_1^{1/3}} (t')^{1/3} z}(\hat x) \geq t' ) \nn \\
&& = 1 - \lim_{t' \to \infty}  Prob(h^{(2)}(t' , \hat x) \leq  \frac{t'}{c_1} - \frac{\lambda_1}{c_1^{1/3}} (t')^{1/3}z   ) 
\eea
Hence $h^{(2)}(t,\hat x) \sim c_2 t - \lambda_2 t^{1/3} X' $ with $c_2 = 1/c_1$, $\lambda_2 =   \frac{\lambda_1}{c_1^{1/3}}$ and the CDF of $X'$ is $Prob(X'\leq z) = 1 - F(z)$.

\begin{figure}
\centerline{\includegraphics[width=11.0cm]{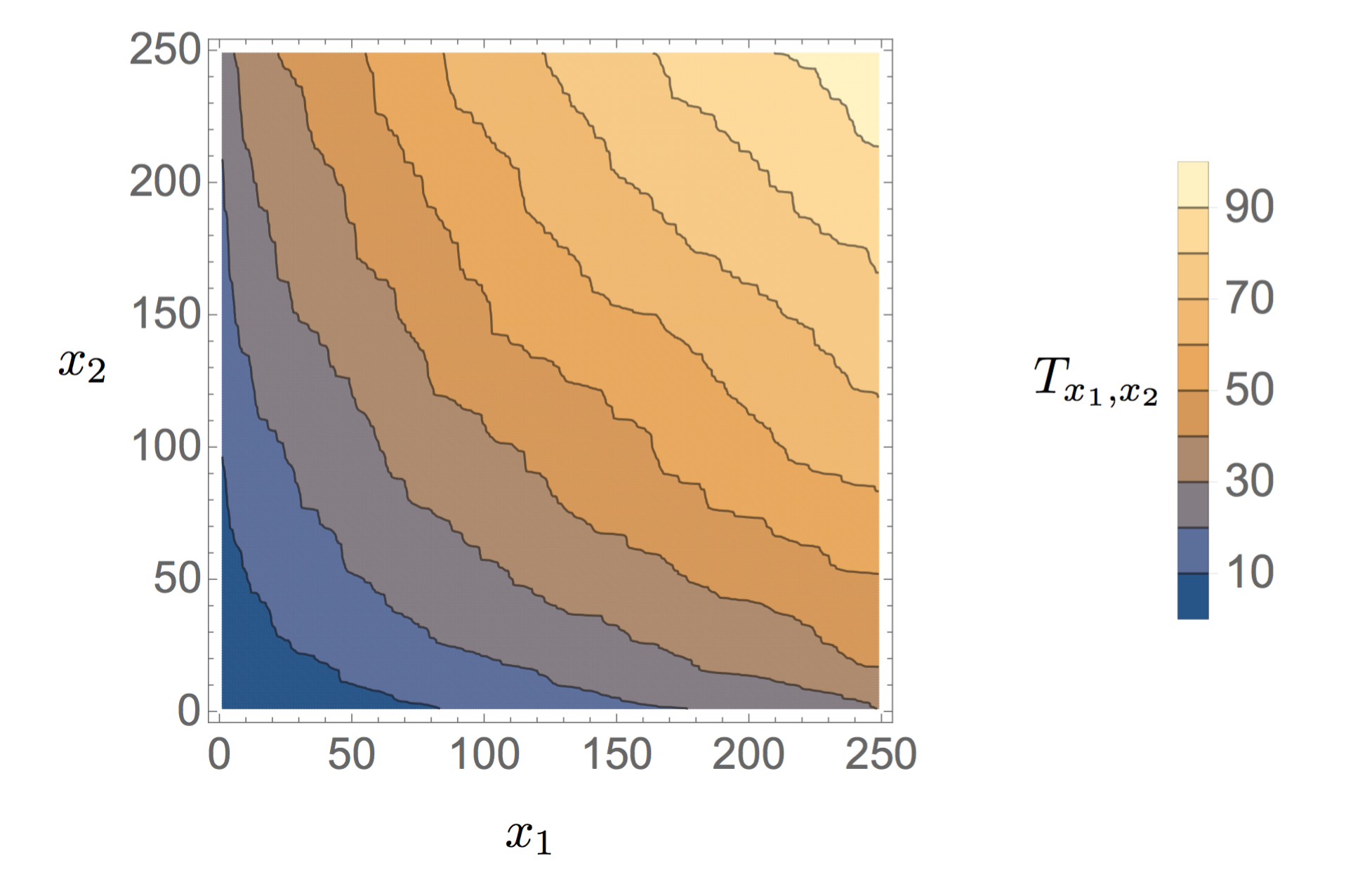} } 
\caption{Starting from a model of last passage percolation, a growing interface is defined as the level lines of $T_{x_1,x_2}$. The interface evolves according to a corner growth model type dynamics. Figure obtained using a simulation of LPP on a square lattice of $256\times 256$ sites and using exponentially distributed waiting times with rate $10$.}
\label{fig:LPPSimu}
\end{figure}

\subsubsection{Interacting particle systems}

\stab {\it The simple exclusion process and the corner growth model}

`Some' interacting particle systems on $\JZ$ are in the KPZ universality class. To make the discussion clearer we consider the simple exclusion process (SEP) with exponential waiting times (see Fig.~\ref{fig:ASEP}). Particles are labeled by $i \in \JN$ or $\JZ$ depending on the setting and their position is denoted $x_i(t)$. Each particle carries two exponential clocks and attempt jumps to the right with rate $p$ and jumps to the left with rate $q$. Jumps are suppressed if the target site is already occupied. Let us introduce the `spin' variable $\hat \eta(t,y)$, which is taken as $+1$ if $y$ is occupied by a particle at time $t$ and $-1$ if it is empty. Denoting $N(t)$ the number of particles that jumped from $0$ to $1$ up to time $t$, a growing interface is defined as (see e.g. \cite{Corwin2011Review})
\bea \label{Eq:ChapIII:Sec1:IPP1}
&& h(0,t) = N(t)  \nn \\ 
&& h(x>0,t) := N(t)-\sum_{y=1}^x \hat \eta(t,y)    \nn \\
&&  h(x<0 ,t):=N(t) + \sum_{y= x}^{-1} \hat \eta(t,y) \ssp .
\eea
Which basically consists in saying that $-\hat \eta(t,y)$ is, at each time, the discrete derivative of the height field: $-\hat \eta(t,y) = h(t,x)- h(t,x-1)$. From this point of view the spin variables in the exclusion process are related to a discretization of the noisy Burgers equation and $\eta(t,y)$ is the velocity of the fluids particles (see Sec.~\ref{subsec:ChapIII:SecII:StatMeas} for the continuum Burgers equation). Note that the velocity field is conserved locally: $\sum_{y=a}^{b} \hat \eta(t,y) $ only evolves when events at the boundaries $a,b$ occur. From the exclusion rule and the fact that particle jumps are local, the interface evolves only at points where its slope changes: a local valley is transformed into a local hill with rate $p$ and a local hill is transformed into a local valley with rate $q$. It is clear that the interface grows upward if $p>q$ and downward if $p<q$. For this reason the case $p=q$ (symmetric simple exclusion process) is very special and the associated field does not display fluctuations in the KPZUC, but rather in the Edwards-Wilkinson universality class. All other cases $p\neq q$ (asymmetric simple exclusion process, ASEP) are expected to be in the KPZUC. Let us now draw the connection with last passage percolation previously defined by considering the case of the totally-asymmetric exclusion process (TASEP), i.e. the SEP with $q=0$ and $p \neq 0$ (see e.g. \cite{KriecherbauerKrug2008}). \\

\begin{figure}
\centerline{\includegraphics[width=11.0cm]{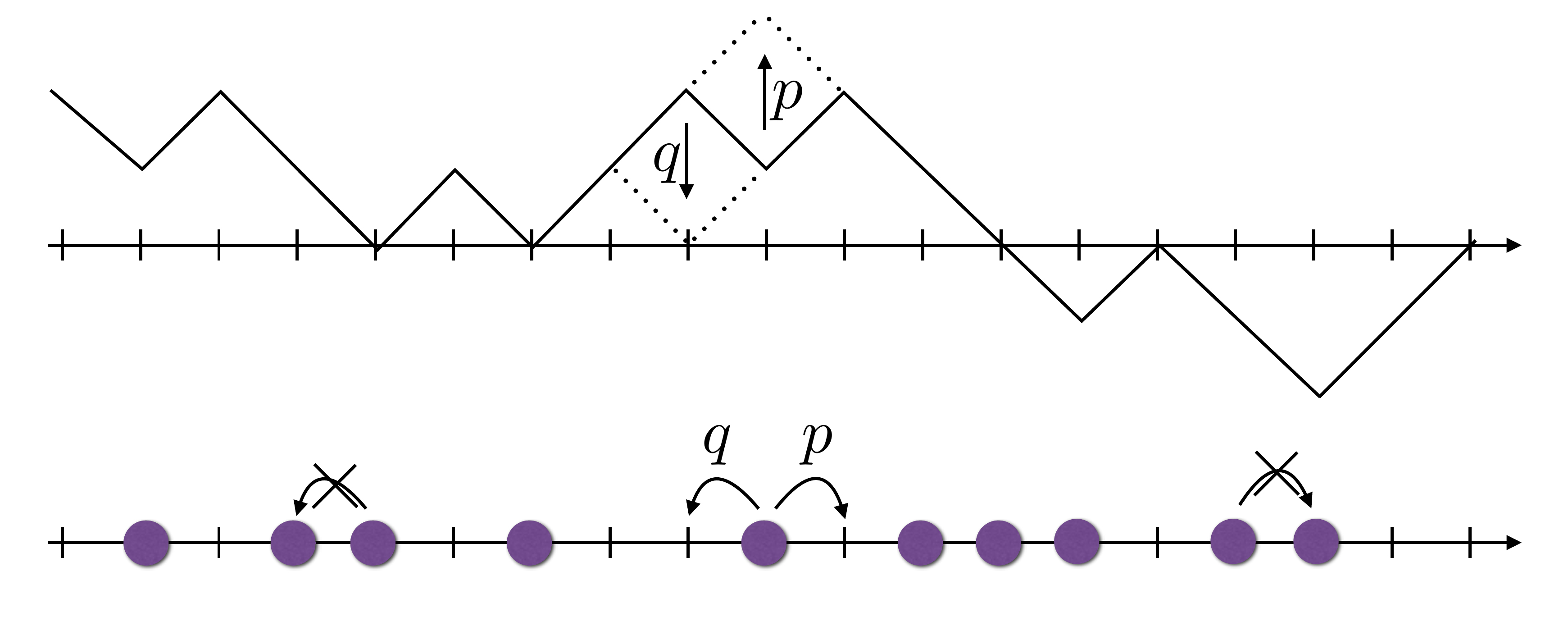} } 
\caption{Illustration of the mapping between the ASEP and the corner growth model, that is also related in the totally asymmetric case $q=0$ to LPP.}
\label{fig:ASEP}
\end{figure}

{\it From the TASEP to LPP}

To do so we will consider the so-called {\it step initial condition}: particles are labelled by $i\in \JN$ and at $t=0$ we take
\bea \label{Eq:ChapIII:Sec1:IPP2}
x_i(t=0) = -i \ssp . 
\eea
And let us denote by $T(i,j)$ the time where the particle $i$ makes her $(j+1)^{th}$ jump. The height at the origin $N(t)$ defined in (\ref{Eq:ChapIII:Sec1:IPP1}) is thus equal to 
\bea \label{Eq:ChapIII:Sec1:IPP3}
N(t) = i_{\rm max}(t) +1 \quad   i_{\rm max}(t) := {\rm max}_{i \in N}\{i \ , \  T(i,i) \leq t \} \ssp .
\eea
On the other hand we have that, for $(i,j) \in \JN^2$
\bea \label{Eq:ChapIII:Sec1:IPP4}
&& T(0,j+1) = T(0,j) + \st_{0,j+1} \ssp ,  \nn \\
&& T(i+1,0) = T(i,0)+ \st_{i+1,0} \ssp , \nn \\ 
&& T(i+1,j+1) = \st_{i+1,j+1} + {\rm max}( T(i+1,j) , T(i,j+1) ) \ssp .
\eea
Here the $\st_{i,j}$ are independent, exponentially distributed RVs with parameter $q$. Indeed the first line corresponds to saying that the time it takes for the $0^{th}$ particle to make her $(j+2)^{th}$ jump is just the time it takes her to make her $(j+1)^{th}$ jump $+$ an exponentially distributed RV. The second line corresponds to saying that the time it takes for the $(i+1)^{th}$ particle to make her first jump is just the time it takes 
to the $i^{th}$ particle to make her first jump $+$ an exponentially distributed RV. Finally the last line corresponds to saying that, before she makes her $(j+2)^{ th}$ jump, the $(i+1)^{th}$ particle has to wait that the $i^{th}$ particle makes her $(j+2)^{ th}$ jump (so that the arrival site is empty) and that she first has to do her $(j+1)^{ th}$ jump and finally wait for an exponentially distributed amount of time. Note that all this is true with the updating rules that were given because an exponential clock is memoryless. This will be more generally true for arbitrary waiting time distributions if the clock is started each time it is possible to make a jump. We have thus shown here that the waiting times of the TASEP (\ref{Eq:ChapIII:Sec1:IPP4}) corresponds to the last passage time in a directed last passage percolation problem. The height in the corner growth model (\ref{Eq:ChapIII:Sec1:IPP1}) is exactly the height $h^{(2)}$ defined in the LPP context. The TASEP with exponentially distributed waiting times with step initial condition thus provides an example where it is possible to switch between the interacting particle language, the language of directed polymer at zero temperature and the growing interface language.

\subsection{The KPZ fixed point - strong universality} \label{subsec:ChapIII:Sec1:KPZFP}

There is various degrees of precision which can be achieved when describing the so-called KPZ fixed point and we refer the reader to \cite{CorwinQuastelRemenik2011} for the most complete description. Once the identification of a `growing interface' $h(t,x)$ in a model has been made, the KPZUC hypothesis is that the appropriately rescaled large time fluctuations of the interface are completely universal. More precisely, starting from a given initial condition, one first has to subtract the deterministic (angle-dependent) growth rate of the interface
\bea \label{Eq:ChapIII:Sec1:StrongU1}
v_{\infty}(\varphi):= \lim_{t \to \infty} \frac{h(t, \varphi t)}{t } \ssp,
\eea
which is generally expected to be well-defined and non random (i.e. the above convergence holds with probability $1$). One then defines 
\bea \label{Eq:ChapIII:Sec1:StrongU2}
\underline{h}(t,x) := h(t,x) - t  v_{\infty}(x/t) \ssp .
\eea
And the rescaled process that involves the roughness exponent $\alpha = 1/2$  and dynamic exponent $z = 3/2$.

\bea \label{Eq:ChapIII:Sec1:StrongU3}
\sh(\st , \sx ) := \lim_{b \to \infty} b^{-\alpha} \underline{h}(t= b^z \st ,x  = b \sx) \ssp .
\eea 

Then, \\
(i) The universality of critical exponents amounts to saying that $\sh(\st , \sx )$ is a well defined, non trivial stochastic process. \\
(ii) The stronger universality property is that, up to three non-universal constants associated with the measure of time, space, and height $C_{\st}$, $C_{\sx}$ and $C_{\sh}$, the random process $\sh(\st , \sx )$ is model independent. i.e. $\tilde{\sh}(\st , \sx) :=  \sh(\st/C_{\st} , \sx/C_{\sx} )/C_{\sh} $ is fully universal and depends only on the initial condition. Furthermore there are only a few attractive subclasses of initial conditions (in the sense that the large-time statistics for an arbitrary initial condition will fall into one of these subclasses). The classification of these subclasses is not complete and does not rely on rigorous results but the three main classes that are commonly considered are the so called droplet initial condition, the flat initial condition and the stationary initial condition (see below). For each of these classes the limiting process at a fixed time $\tilde{\sh}(\st , \sx)$ is known:
\begin{enumerate}
\item{The droplet case: in the interface language, it corresponds to an initial condition such that the interface stays curved for all time (i.e. the mean profile $v_{\infty}(\varphi)$ is not flat). For the KPZ equation itself, the appropriate initial condition is often taken as $h(t=0,x) := -\lim_{ w \to \infty} w |x|$. In the DP case the latter is clearer and corresponds to directed polymers with fixed starting point $Z_{t=0}(x) = \delta(x)$, and $Z_t(x)$ is the {\it point to point} partition sum. For the corner growth model it corresponds to ASEP with a step initial condition. In this case, for a given time $\st$ and point $\sx$, the one point distribution of $\sh(\st , \sx )$ is the GUE Tracy-Widom (TW) distribution (introduced in \cite{TracyWidom1993}). Moreover at a fixed time $\st$, as a function of $\sx$, $\tilde{\sh}(\st , \sx)$ is a process known as the Airy2 process $\cA_2(\sx)$ (introduced in \cite{PraehoferSpohn2001}). It is a process that is stationary in $\sx$, that satisfies $\overline{\cA_2(\sx)\cA_2(0)}^c \to_{\sx \to \infty} =0 $, whose one point distribution is the GUE-TW distribution and which locally looks like a Brownian motion, $\overline{(\cA_2(\sx)-\cA_2(0))^2}  \sim_{\sx \to 0 } |x| $.}
\item{The flat case: in the interface language, it corresponds to a flat initial condition $h(t=0,x) = cst$. In this case $v_{\infty}(\varphi) = cst'$. In the DP framework this corresponds to a directed polymer with the starting point free to move on a line $Z_{t=0}(x) =1$, and $Z_t(x)$ is the {\it point to line} partition sum. For the corner growth model it corresponds to ASEP with particles only on even sites. In this case, for a given time $\st$ and point $\sx$, the one point distribution of $\sh(\st , \sx )$ is the GOE Tracy-Widom (TW) distribution (introduced in \cite{TracyWidom1996}). Moreover at a fixed time $\st$, as a function of $\sx$, $\tilde{\sh}(\st , \sx)$ is a process known as the Airy 1 process $\cA_1(\sx)$ (introduced in \cite{Sasamoto2005}). It has properties similar to the Airy 2 process: stationarity in $\sx$, $\overline{\cA_1(\sx)\cA_1(0)}^c \to_{\sx \to \infty} =0 $, and it locally looks like a Brownian motion.}
\item{The stationary initial condition is a random initial condition that ensures a stationarity property in the model dynamics. Examples and the nature of this stationarity property will be discussed in Sec.~\ref{subsec:ChapIII:SecII:StatMeas}. In this case at large time as a function of $\sx$, $\tilde{\sh}(\st , \sx)$ is given by $\tilde{\sh}(\st , \sx )=  X_{BR} + B(x)$ where $X_{BR}$ is a RV distributed according to the the Baik-Rains distribution (introduced in \cite{BaikRains2000}) and $B(x)$ is a two-sided Brownian motion. $X_{BR}$ and $B(x)$ have non-trivial correlations.}
\end{enumerate}
Many other things are known about these limiting spatial processes, including for example other initial `crossover initial conditions' (half flat, Brownian-flat, half Brownian), see \cite{QuastelRemenik2014}. In contrast to this accurate description of spatial correlations at large time, much less is known about two-time correlations. In an abstract setting these are related to the notion of `Airy sheet' \cite{CorwinQuastelRemenik2011}, but explicit formulae are rare (see however \cite{Dotsenko2010,Dotsenko2016,Johansson2015,FerrariSpohn2016}). Finally we should mention here the related question of growth from a given initial condition in a restricted geometry, for example directed polymers in a half-space, which, when both starting points are taken `close to the wall', exhibit large scale fluctuations of free energy scaling as $t^{1/3}$ and governed by the GSE Tracy-Widom distribution \cite{TracyWidom1996}, see \cite{GueudreLeDoussal2012} for DP in the continuum.
\smallskip

The belief in these remarkable properties of the KPZUC comes from exact solutions of peculiar models. At the level of one point distribution for the droplet case the emergence of the TW-GUE distribution was first shown in the LPP model with geometric weights in \cite{johansson2000} (a similar result was shown just before in the related context of the longest increasing subsequence of a random permutation in \cite{BaikDeiftJohansson1998}). The corresponding result for the KPZ equation itself is more recent \cite{AmirCorwinQuastel2010,SasamotoSpohn2010,CalabreseLeDoussalRosso2010,Dotsenko2010} (thus the fact that the KPZ equation is in its own universality class is a recent result). Extension to multi-points were first shown for the PNG model \cite{PraehoferSpohn2001,Johansson2003}, and later for the KPZ equation in \cite{ProlhacSpohn2011}. For the flat case, one-point formulae were again first shown in discrete models \cite{BaikRains2001} then for the KPZ equation \cite{CalabreseLeDoussal2011,LeDoussalCalabrese2012,GuedreLeDoussalRossoHenryCalabrese2012,OrtmannQuastelRemenik2014}, while multi-point correlations were only studied in a discrete setting \cite{Sasamoto2005}. A similar story goes for the stationary case \cite{FerrariSpohn2006,BaikFerrariPeche2011,ImamuraSasamoto2012,ImamuraSasamoto2013,BorodinCorwinFerrariVeto2015}. We should here that the works cited here are not all at the same level of rigor and many other works could be cited. We will review later the appropriate references for the scope of this chapter, namely discrete models of directed polymers, and refer to \cite{Corwin2011Review} for other references.

\subsection{Universality of the KPZ equation: notion of weak universality and universal scaling limits of DP on the square lattice} \label{subsec:ChapIII:Sec1:WeakU}

The limiting spatial process (\ref{Eq:ChapIII:Sec1:StrongU3}) does not obey the KPZ equation: the KPZ equation is not the KPZ fixed point. Under a general rescaling, if $h(t,x)$ solves the KPZ equation (\ref{Eq:ChapIII:Intro:5}), $\tilde h (\tilde t, \tilde x ) := b^{-\alpha} h( t=b \tilde t , x=b^z \tilde x )$ solves (dropping the tildes)
\be \label{Eq:ChapIII:Sec1:WeakU1}
\partial_t h(t,x) = \frac{b^{\alpha+z-2}}{2} c_1 (\nabla_x h(t,x))^2 + \frac{b^{z-2}}{2} (\nabla_x)^2  c_2 h(t,x) + b^{(z-2\alpha-1)/2} c_3 \xi(t,x) \ssp ,
\ee
where we have reintroduced explicit constants in front of each term on the right hand side. There is no way to fix $z$ and $\alpha$ to get a scale invariant equation. The values $\alpha = 1/2$ and $z=3/2$ of the KPZUC, a property of the KPZ FP, are not trivially obtained through scaling.\\

 In all the models in the KPZUC, however, the KPZ equation/the continuum DP plays a peculiar role that is linked with the notion of weak universality \cite{HairerQuastel2015}. The latter refers to two limits:\\

\smallskip

 (i) The weak asymmetry limit. Symmetric growth models ($c_1=0$) are in the Edwards-Wilkinson universality class and are characterized by exponents $\alpha=1/2$ and $z=2$. If an asymmetry is present $c_1 \neq 0$, the large scale properties are those of the KPZ FP. Note that rescaling (\ref{Eq:ChapIII:Sec1:WeakU1}) with $\alpha = 1/2$, $z=2$ and $c_1 \sim b^{-1/2}$ leaves the KPZ equation itself invariant. More generally it is conjectured that (under the usual assumptions such as locality, etc...)  the KPZ equation itself is the universal scaling limit of weakly asymmetric growth models in $1+1$d when rescaling space $x \sim b$, $t \sim b^2$ and the asymmetry as $b^{-1/2}$. For this reason the KPZ equation is sometimes referred to as implementing the universal crossover between the EW FP and the KPZ FP. An important example of this weak-universality property is provided by the work of Bertini and Giacomin \cite{BertiniGiacomin1997} that states that, upon scaling space as $x \sim b$, $t \sim b^2$ and the asymmetry $p-q$ as $b^{-1/2}$ in the corner growth model previously defined, the corner growth model height profile converges to the Cole-Hopf solution of the KPZ equation.\\

 \smallskip

 (ii) On the other hand the weak noise limit is linked with the DP in a disordered medium. At zero disorder, $c_3 = 0$, DPs are equivalent to random walks, are diffusive $z=2$ and have no disorder fluctuations $\alpha = 0$. For $c_3 \neq 0$, the large scale properties of DPs are described by the KPZ FP. Noting that the rescaling (\ref{Eq:ChapIII:Sec1:WeakU1}) with $\alpha = 0$, $z=2$ and $c_3 \sim b^{-1/2}$ leaves the KPZ equation invariant, it is conjectured that the KPZ/MSHE is the universal scaling limit of weakly disordered DPs. This scaling has also been called the intermediate disorder regime in the literature \cite{AlbertsKhaninQuastel2012}. Let us illustrate it heuristically on the case of DPs on the square lattice, using the notations of Sec.~\ref{subsec:ChapIII:Sec1:Models}. We start with the discrete version of the MSHE in the variables $t, \hat x$ given in (\ref{Eq:ChapIII:Sec1:DiscreteDP6}) which we recall here
 \bea \label{Eq:ChapIII:Sec1:WeakU2}
 Z_{t+1}(\hat x) = e^{-\frac{1}{T} \cE_{t+1}(\hat x)}\left(Z_{t}(\hat x +1/2) + Z_{t}(\hat x-1/2) \right) \ssp .
 \eea
 We thus scale
 \bea \label{Eq:ChapIII:Sec1:WeakU3}
t = b^2 D \tilde t \quad , \quad  \hat x = b \tilde  x \quad , \quad \cE_{t+1}( \hat x)  = 1/\sqrt{b}   V(t ,\hat x )  \quad , 
\eea
with $V$ a centered $O(1)$ RV, take $b$ very large, $b \gg 1$, and consider the limiting partition sum
\bea \label{Eq:ChapIII:Sec1:WeakU4}
\tilde{Z}_{\tilde t}(\tilde x) := \lim_{b \to  \infty} A^{b^2 D \tilde t} Z_{t = b^2 D \tilde t }(\hat x=  b \tilde  x) \ssp ,
\eea
with $A$ left undetermined for now. We thus obtain from \ref{Eq:ChapIII:Sec1:WeakU2}
\be \label{Eq:ChapIII:Sec1:WeakU5}
\frac{1}{A} \left( \tilde{Z}_{\tilde t}(\tilde x) + \frac{1}{b^2 D} \partial_{\tilde t} \tilde{Z}_{\tilde t}(\tilde x) \right) = \left( 1- \frac{1}{\sqrt{b} T} V(t ,\hat x)  \right) \left( 2 \tilde{Z}_{\tilde t}(\tilde x) + \frac{1}{4 b^2} (\partial_{\tilde x})^2 \tilde{Z}_{\tilde t}(\tilde x)  \right) \ssp .
\ee
This suggests to take $A=1/2$ and we obtain
\bea \label{Eq:ChapIII:Sec1:WeakU6}
 \partial_{\tilde t} \tilde{Z}_{\tilde t}(\tilde x) =  \frac{D}{8} (\partial_{\tilde x})^2 \tilde{Z}_{\tilde t}(\tilde x) + \frac{D}{T} b^{3/2} V( b^2 D \tilde{t} , b \tilde{x} )   \tilde{Z}_{\tilde t}(\tilde x)\ssp . 
\eea
In the limit $b\to 0$, $D  b^{3/2} V( b^2 D \tilde{t} , b \tilde{x} )$ converges to a GWN. To see this explicitly, consider for example a random potential which is uncorrelated from site to site with cumulant $\overline{(V(t,x))^n}^c = c_n$ and the cumulant generating function
\bea \label{Eq:ChapIII:Sec1:WeakU7}
G[\lambda]&& := \overline{e^{\int_{\tilde x = 0}^{1} \int_{\tilde t=0}^{1} D b^{3/2} V( b^2 D \tilde{t} , b \tilde{x} ) \lambda( \tilde t , \tilde x ) }} \nn \\
&&  = \overline{e^{  D b^{3/2}   \frac{1}{b^3 D} \sum_{\hat x=0}^{b} \sum_{t =0}^{b^2 D} V( b^2 D \tilde{t} , b \tilde{x} ) \lambda(t/(D b^2) , x/b ) } } \nn \\
&& =   e^{ \frac{c_2}{2 b^3}  \sum_{\hat x=0}^{b} \sum_{t =0}^{b^2 D}  \lambda^2(t/(D b^2) , x/b )  + \frac{c_3}{3!b^{9/2} }  \sum_{\hat x=0}^{b} \sum_{t =0}^{b^2 D}  \lambda^3(t/(D b^2) , x/b ) + \dots } \nn \\
&& = e^{ \frac{c_2 D}{2} \int_{\tilde x = 0}^{1} \int_{\tilde t=0}^{1} \lambda^2( \tilde t , \tilde x )   + \frac{c_3}{3!b^{3/2} } \int_{\tilde t=0}^{1} \lambda^3( \tilde t , \tilde x ) + \dots } \nn \\
&& =_{b \to \infty}  e^{ \frac{c_2 D}{2} \int_{\tilde x = 0}^{1} \int_{\tilde t=0}^{1} \lambda^2( \tilde t , \tilde x )} \ssp .
\eea 
Hence in the limit $b \to \infty$ the cumulant generating function $G$ is equal to 
\bea \label{Eq:ChapIII:Sec1:WeakU8}
G[\lambda] = \overline{  e^{ \int_{\tilde x = 0}^{1} \int_{\tilde t=0}^{1} \sqrt{c_2 D} \xi(\tilde t , \tilde x)  }} \ssp ,
\eea
where $\xi(\tilde t , \tilde x)  $ is a unit GWN with correlations $ \overline{\xi(\tilde t , \tilde x)   \xi(\tilde t' , \tilde x') }  = \delta(\tilde t'-\tilde t) \delta(\tilde x'-\tilde x)$. Hence the rescaled partition sum (\ref{Eq:ChapIII:Sec1:WeakU4}) with $A=1/2$ converges, as $b \to \infty$, to the rescaled solution of the MSHE
\bea \label{Eq:ChapIII:Sec1:WeakU9}
 \partial_{\tilde t} \tilde{Z}_{\tilde t}(\tilde x) =  \frac{D}{8} (\partial_{\tilde x})^2 \tilde{Z}_{\tilde t}(\tilde x) +  \sqrt{2 c} \xi(\tilde t , \tilde x)   \tilde{Z}_{\tilde t}(\tilde x)\ssp ,
\eea
with $c = D \overline{V^2}^c/(2 T^2) $. This establishes that the MSHE is the universal scaling limit of DPs on the square lattice with weak disorder under diffusive scaling. We refer the reader to \cite{AlbertsKhaninQuastel2012} for a rigorous approach. Note that the weak disorder regime is equivalent to the regime of high temperature $T \sim \sqrt{b}$ at fixed disorder. In this phrasing, the continuum DP is the universal high temperature limit of DPs, see in particular \cite{CalabreseLeDoussalRosso2010,BustingorryLeDoussalRosso2010} for a discussion of this property and an extension to disorder with non-zero correlation length. \\

Before we close this chapter let us mention here that there exists another universal scaling limit of DPs on the square lattice that has played an important role in recent developments. Taking $\sT$ and $N$ finite, it is obtained by taking $x_1 =  b \sT$, $x_2 = N$ and $\cE_{x_1,x_2} \sim 1/ \sqrt{b} \tilde{\cE}_{x_1,x_2} $ and taking $b \to \infty$ with $N$ and the distribution of $\tilde{\cE}_{x_1,x_2}$ fixed so that the $\tilde{\cE}_{x_1,x_2}$ are independent $O(1)$ centered random variables with at least a well defined second moment. This is the limit of long directed polymers on the square lattice conditioned to stay close to the horizontal axes. The polymer makes an infinite number ($\sim b$) of horizontal jumps, and a finite number ($N$) of vertical jumps.  In this case the possible polymer paths can be indexed by $(s_1 , s_2 , \cdots , s_N ) \in [0,\sT]^N$ where $x_1 =  b s_i$ is the position of the jump from $x_2 = i$ to $x_2 = i+1$. Between two jumps the random energy of the DP is a sum of all the random energies encountered on the horizontal line at $x_2 = i-1$. This sum is equivalent in law to a centered Gaussian random variable that we can write down as the difference of two Brownian motions in $s$:
\bea \label{Eq:ChapIII:Sec1:SD1}
\sum_{x_1 = N s_i }^{x_1 = N s_{i+1}} \cE_{x_1 , i-1} \sim B_i(s_{i+1}) - B_i(s_i)
\eea
where $B_i(s)$ is a BM with $B_i(0) = 0$ and $B_i'(s) = \sqrt{\overline{\tilde{\cE}^2}} \xi_{i}(s)$ where $\xi_i(s)$ are a collection of independent GWN with $\overline{\xi_{i}(s) \xi_{i'}(s') } = \delta_{i,i'}\delta(s-s')$ . Thus in this limit the partition sum is rewritten as
\bea \label{Eq:ChapIII:Sec1:SD2}
Z_T^{s.d.}(N) && := \lim_{b \to \infty} Z_{x_1 =  b \sT,x_2=N} \nn \\
&& = \int_{ 0 \leq s_1 , \cdots , \leq s_N \leq \sT} e^{-\frac{1}{T} \sum_{i=0}^{N} B_i(s_{i+1}) -B_i(s_{i}) } \ssp ,
\eea
with by definition $s_0 = 0$ and $s_{N+1} = \sT$ . This is the definition of the O'Connell-Yor semi-discrete directed polymer which was introduced in \cite{OConnellYor2001} and for which KPZ universality (more precisely GUE TW fluctuations) in the limit $N \to \infty$ was shown in $\cite{OConnell2009,BorodinCorwinMacDo2014}$. This was later used in \cite{AuffingerBaikCorwin2012} to prove a universality result of TW-GUE fluctuations for point to point partition sum of directed polymer in `thin rectangles' \cite{AuffingerBaikCorwin2012}.

\section[Exact solvability properties]{A partial selection of analytical miracles in models in the KPZ universality class} \label{Sec:ChapIII:Sec2}

In the previous section we discussed the notion of weak and strong KPZ universality and introduced a few models in the KPZUC. In this chapter we now present a few {\it exact solvability properties} that permitted over the years to build the belief in the remarkable properties of the KPZUC and focus on DPs. We will discuss: (i) in Sec.~\ref{subsec:ChapIII:SecII:Sym} the symmetries of the KPZ equation; (ii) in Sec.~\ref{subsec:ChapIII:SecII:StatMeas} the stationary measure of the the KPZ equation and some models of DPs on $\JZ^2$; (iii) in Sec.~\ref{subsec:ChapIII:SecII:BA} the Bethe-ansatz solvability of the continuum DP; (iv) in Sec.~\ref{subsec:ChapIII:SecII:OtherExact} some other exact solvability properties: the RSK and gRSK correspondences and Macdonald processes (briefly discussed).

\subsection{Symmetries of the continuum KPZ equation} \label{subsec:ChapIII:SecII:Sym}

\stab {\it Hydrodynamic point of view: Galilean symmetry}\\
The KPZ equation (\ref{Eq:ChapIII:Intro:5}) enjoys Galilean invariance: for a given realization of the GWN $\xi(t,x)$, if $h(t,x)$ is a solution of the KPZ equation, then
\bea \label{Eq:ChapIII:Sec2:Sym1}
h_{v}(t,x) = h(t,x-tv) - v x + \frac{v^2 t}{2} \ssp ,
\eea
is a solution of the KPZ equation in the noise $\xi_v(t,x) = \xi(t,x-tv)$. Defining $u(t,x) = \partial_x h(t,x)$, $u(t,x)$ solves the equation
\bea \label{Eq:ChapIII:Sec2:Sym2}
\partial_t u(t,x) = u(t,x) \partial_x u(t,x) + \frac{1}{2}\partial_x^2 u(t,x) + \partial_x \xi(t,x) \ssp .
\eea
Which is the Burgers's equation for a randomly forced fluid, much studied in the literature in the context of turbulence: here $u(t,x)$ is interpreted as the velocity field of a one-dimensional fluid (see \cite{BecKhanin2007} for a review). In this framework the above symmetry reads $u_v(t,x) = u(t,x-tv) - v$ and really is the Galilean symmetry associated with a change of Galilean referential.\\

{\it Directed polymer point of view: STS}\\
From the DP point of view this symmetry is easily seen using the path integral formula for the point to point partition sum (\ref{Eq:secIntroKPZ-1}) of the DP in a potential $V(t,x)$ (here rewritten using dimensionless units, i.e. $T=1$)
\bea \label{Eq:ChapIII:Sec2:Sym3bis}
Z_t(x - t v) && := \int_{u(0)=0}^{u(t) = x-tv} \cD[u] e^{ - \frac{1}{2} \int_{0}^{t} dt' (\partial_{t'} u(t'))^2  - \frac{1}{2} \int_{0}^t dt' V(t',u(t'))} \nn \\
&&  = \int_{u(0)=0}^{u(t) = x} \cD[u] e^{ - \frac{1}{2} \int_{0}^{t} dt' (\partial_{t'} u(t'))^2  +v\int_{0}^{t} dt' \partial_{t'} u(t') - \frac{v^2}{2} \int_{0}^t dt' - \frac{1}{2} \int_{0}^t dt' V(t',u(t') - t' v)} \nn \\
&& =  e^{v x - \frac{v^2}{2} t} \int_{u(0)=0}^{u(t) = x} \cD[u] e^{ - \frac{1}{2} \int_{0}^{t} dt' (\partial_{t'} u(t'))^2 - \frac{1}{2} \int_{0}^t dt' V(t',u(t') - t' v)}
\eea
and by taking the logarithm, one obtains the symmetry (\ref{Eq:ChapIII:Sec2:Sym1}). For a random potential which satisfies the symmetry in law $\forall v$, $V(t,x) \sim V(t,x-tv)$, we then have a statistical symmetry and in law we have $h_v(t,x) \sim h(t,x)$. In particular this holds for a GWN $V(t,x) = \xi(t,x)$. In Chapter~\ref{chapI} and \ref{chapII} this symmetry was called the statistical tilt symmetry. In the DES context we saw that it implies a non-renormalization of the elastic coefficient. In particular, the effective action of the theory contains the term $- \frac{1}{2 T}\sum_{a=1}^{n}  \int_x (\nabla u_x^a)^2$. This implied the symmetry between the roughness exponent of the DP $\zeta_s$ and the exponent $\theta$ of the fluctuations of the free energy as.
\bea \label{Eq:ChapIII:Sec2:Sym3}
\theta = 1-2 + 2 \zeta_s = -1 + 2 \zeta_s \ssp .
\eea
Interpreted in the KPZ context, the roughness exponent $\zeta_s$ of the DP is $1/z$ with $z$ the dynamic exponent, and $\theta$ is the growth exponent $\beta$, related to the roughness exponent of the interface $\alpha$ as $\beta = \alpha/z$. Hence we have
\bea \label{Eq:ChapIII:Sec2:Sym4}
\frac{\alpha}{z} = - 1 + \frac{2}{z} \quad \Longleftrightarrow \quad \alpha + z = 2 \ssp .
\eea
The KPZUC is thus characterized by a single critical exponent. Note that this symmetry holds in any dimension $N$.

\subsection{An analytical exact solvability property: the stationary measure} \label{subsec:ChapIII:SecII:StatMeas}

\subsubsection{Stationary measure of the $1d$ Burgers equation: the continuum KPZ case}

In any model in the KPZUC, the appropriate height field is never stationary itself but grows with time. The height gradients, however, can be stationary and the stationary measure is known since \cite{ForsterNelsonStephen1977,KPZ}. For the KPZ equation in particular, one studies the Burgers's equation for $u(t,x) = \partial_x h(t,x)$ already written in (\ref{Eq:ChapIII:Sec2:Sym2}). Let us first study the case where the non-linearity is zero (Edwards-Wilkinson case), the equation for $u(t,x)$ is now
\bea  \label{Eq:ChapIII:Sec2:Stat1}
\partial_t u(t,x) = \frac{1}{2} \partial_x^2 u(t,x) +   \partial_x \xi(t,x) \ssp .
\eea
where $\xi(t,x)$ is a unit centered GWN. Let us now show that a family of stationary processes is obtained by taking $u(t,x)$ as $u(t,x) = \mu + \eta_t(x)$ where $\mu$ is the average slope of the KPZ interface (this labels the family) and $\eta_t(x)$ is distributed as a unit centered GWN for each $t$. Let us start from an initial condition $u(t=0,x) = \mu + \eta(x)$ with $\eta(x)$ a GWN independent of $\xi(t,x)$ and show that $\forall t$, $u(t,x) = \mu + \eta_t(x)$ with $\eta_t(x)$ a unit centered GWN. (\ref{Eq:ChapIII:Sec2:Stat1}) is easily solved in Fourier space as
\bea \label{Eq:ChapIII:Sec2:Stat2}
u(t,q) = e^{-1/2 q^2 t}( \mu  + \eta(q)) + \int_{0}^{t} e^{-\frac{1}{2}q^2(t-t') } i q \xi(t,q) dt' \ssp .
\eea
This is sufficient to show that $u(t,x)$ is Gaussian distributed and one easily concludes the calculation by computing the first moment $\overline{u(t,q)}= \mu$ and the first cumulant $\overline{u(t,q) u(t,q')}^c = \hat \delta(q+q')$. Of course a priori this concerns only the EW universality class but remarkably the same stationary measure works for the KPZ equation. Checking this property directly in the continuum setting is, however, non-trivial due to the non-linear term and one has to be more precise here about what is meant by the continuum KPZ equation. Writing a functional Fokker-Planck equation for the PDF of $u(t,x)$, ${\cal P}_t[u]$, one obtains
\be \label{Eq:ChapIII:Sec2:Stat3}
\frac{\partial}{\partial_t} {\cal P}_t[f] = - \int_x \frac{\delta }{\delta f(x)} \left( (f \partial_x f + \frac{1}{2}\partial_x^2 f ) {\cal P}_t[f] \right) + \frac{1}{2} \int_{x ,y } \frac{\delta^2}{\delta f(x) \delta f(y) } ( \partial_x \partial_y \delta(x-y)  {\cal P}_t[f]  ) 
\ee
And since
\bea \label{Eq:ChapIII:Sec2:Stat4}
{\cal P}_{stat}[f] \sim e^{- \frac{1}{2} \int dx f(x)^2  + \mu f(x)}
\eea
is already stationary for the Edwards-Wilkinson case, it is stationary for the KPZ case iff
\bea \label{Eq:ChapIII:Sec2:Stat5}
- \int_x \frac{\delta }{\delta f(x)} \left( f \partial_x f {\cal P}_{stat}[f] \right) = 0 \ssp .
\eea
Acting with the functional derivative on the $f \partial_x f$ is however rather ambiguous and one has to define a discretization procedure. It was argued in \cite{KrugSpohnGodreche1991} that the proper way to discretize the non linear term in the KPZ equation is, taking $h(x,t) \to h_i(t)$,
\bea \label{Eq:ChapIII:Sec2:Stat6}
(\partial_x h(x,t))^2 \to \frac{1}{3}((\nabla_i h )^2 +(\nabla_i h ) (\nabla_{i-1} h ) + (\nabla_{i-1} h )^2 ) \ssp .
\eea
With $\nabla_i h(t) := h_{i+1}(t) -h_{i}(t) $. For the corresponding discretized Burgers equation, this implies the discretization, writing $u_i(t) := \nabla_i h(t)$
\bea \label{Eq:ChapIII:Sec2:Stat7}
u \partial_x u \to \frac{1}{3} ( (u_{i+1} - u_{i-1})(u_{i+1} + u_{i-1}) + u_i (u_{i+1} - u_{i-1} )  ) \ssp .
\eea
With this discretization, the functional derivative is thus interpreted as
\bea \label{Eq:ChapIII:Sec2:Stat8}
 \int_x \frac{\delta }{\delta f(x)} \left( f \partial_x f \right) && \sim  \sum_i \frac{d}{df_{i}} \frac{1}{3} ( (f_{i+1} - f_{i-1})(f_{i+1} + f_{i-1}) + f_i (f_{i+1} - f_{i-1} )   \nn \\
 && \sim \sum_i \frac{1}{3} (f_{i+1} - f_{i-1} ) \nn \\
 && \sim \int_x \frac{2}{3}  \partial_x f_x  \ssp . 
\eea
And hence
\be \label{Eq:ChapIII:Sec2:Stat8bis}
- \int_x \frac{\delta }{\delta f(x)} \left( f \partial_x f {\cal P}_{stat}[f] \right) = - \int_x  \frac{2}{3}  \partial_x f_x  {\cal P}_{stat}[f]  -  \int_x (f \partial_x f)(-f + \mu){\cal P}_{stat}[f] \ssp .
\ee
Note that all three terms in this expression are total derivatives: hence ${\cal P}_{stat}[f]$ is indeed a stationary measure of the continuum KPZ equation (provided appropriate boundary conditions are assumed). Note that this property only holds because the discretization (\ref{Eq:ChapIII:Sec2:Stat6}) was chosen and it is thus legitimate to ask what the KPZ equation actually means? This reflects the difficulties related to tackling the KPZ equation directly in the continuum. The right way to interpret it is really through the Cole-Hopf transform. Note that at the level of the Burgers velocity field, the discretization (\ref{Eq:ChapIII:Sec2:Stat7}) precisely ensures that the velocity field is locally conserved: the right hand side of (\ref{Eq:ChapIII:Sec2:Stat7}) can be written as a difference. We already noticed this property of the velocity field (i.e. height gradients) when we discussed the case of the corner growth model in Sec.~\ref{subsec:ChapIII:Sec1:Models}. Since the derivation of the stationary measure is at this stage quite unsatisfactory we will recover the result below in an unambiguous way starting from a model of DP on the square lattice. Let us first draw the consequences of the existence of this stationary measure. In the interface language, let us start from an initial condition
\bea \label{Eq:ChapIII:Sec2:Stat9}
h(t=0,x) = \mu x + B(x) \ssp ,
\eea
where $B(x)$ is a two-sided Brownian motion. Although the interface grows, the slope field is stationary and
\bea \label{Eq:ChapIII:Sec2:Stat10}
h(t,x) = h(t,x=0) + \mu x + B_t(x) \ssp ,
\eea
where for each $t$, $B_t(x)$ is a two-sided BM. This invariance of the Brownian motion implies that the roughness exponent of the interface is $\alpha = 1/2$, and together with (\ref{Eq:ChapIII:Sec2:Sym4}) this entirely determines the critical exponents of the KPZUC. Note that the interface dynamics itself is not stationary and, as mentioned in Sec.~\ref{subsec:ChapIII:Sec1:KPZFP} (i) the fluctuations of $h(t,x=0)$ grows at large time as $t^{1/3}$ and are distributed with the Baik-Rains distribution; (ii) $h(t,x=0)$ and $B_t(x)$ are non-trivially correlated. In the DP language the free-energy of the polymer thus performs a Brownian motion.
\medskip

It is expected that, starting from any initial condition, the slope field of the interface at large time will be stationary on local scales. Denoting as before $v_{\infty}(\varphi)$  the large time asymptotic velocity of the interface in the direction $\varphi$, it is expected that, in law, we have
\bea \label{Eq:ChapIII:Sec2:Stat11}
h(T + t , \varphi T +x) - h(T  , \varphi T ) \sim_{T \to \infty} X_t + \mu x + B_t(x)
\eea
where $\mu = \partial_{\varphi} v_{\infty}(\varphi)$ (to ensure the equality of the mean value on both sides of (\ref{Eq:ChapIII:Sec2:Stat11})), $B_t(x)$ is a two-sided BM, $X_t$ is a RV whose fluctuations grow at large time as $t^{1/3}$ and are distributed according to the Baik-Rains distribution, and $x,t = O(1)$. Of course if one scales $x$ as $T^{2/3}$ one will see non-trivial correlations in the spatial direction and the process in $x$ is not anymore a BM but, as discussed before, the Airy process. Conversely this explains why the Airy process and the interface at large time locally looks like a BM.

\subsubsection{Stationary measure of the Log-Gamma polymer}

We now recover the stationary measure of the continuum KPZ equation in the DP framework. To this aim we first consider a model of DP on the square lattice introduced by Sepp\"al\"ainen in the landmark paper \cite{Seppalainen2009}, the Log-Gamma polymer. It is defined as in Sec. \ref{subsec:ChapIII:Sec1:Models} through the recursion equation (\ref{Eq:ChapIII:Sec1:DiscreteDP6})
\bea \label{Eq:ChapIII:Sec2:Stat12}
Z_{t+1}(\hat x) = W_{t+1}(\hat x)\left(Z_{t}(\hat x +1/2) + Z_{t}(\hat x-1/2) \right)  \ssp  ,
\eea
where the random Boltzmann weights are all independent and distributed as the inverse of a Gamma RV with parameter $\gamma>0$:
\bea \label{Eq:ChapIII:Sec2:Stat13}
W_{t}(x) \sim Gamma(\gamma)^{-1} \ssp .
\eea
And we recall that a RV $X$ is distributed as  a Gamma RV with parameter $\gamma$ iff its PDF is
\bea \label{Eq:ChapIII:Sec2:Stat14}
X \sim Gamma(\gamma) \Longleftrightarrow p(X) = \frac{1}{\Gamma(\gamma)} X^{1+ \gamma} e^{-X} \ssp . 
\eea
In \cite{Seppalainen2009} it was shown that, now interpreting (\ref{Eq:ChapIII:Sec2:Stat12}) $\forall t$ as the definition of a Markov process, if at $t=0$ the successive ratios of partition sums are chosen to be distributed as quotients of independent Gamma RV with
\bea \label{Eq:ChapIII:Sec2:Stat15}
\frac{Z_{t=0}(\hat x +1)}{ Z_{t=0}(\hat x)} \sim \frac{Gamma(\gamma-\lambda)}{ Gamma(\lambda)} \ssp ,
\eea
where $0< \lambda < \gamma$, then they remain so for all time. This model provided the first example of a discrete model of DP on the square lattice at finite temperature where the stationary measure is known exactly - a sign of the existence of {\it exact solvability properties}. The model was later shown to be exactly solvable using the gRSK correspondence (see Sec.~\ref{subsec:ChapIII:SecII:OtherExact}) in \cite{CorwinOConnellSeppalainenZygouras2014} and Bethe ansatz in \cite{ThieryLeDoussal2014}. As in the continuum KPZ case, the partition sum $Z_t(\hat x)$ is not stationary but $\log Z_t(\hat x)$ performs a random walk $\forall t$. The random walk is in general biased, except if $\lambda= \gamma/2$. As already shown in \cite{ThieryLeDoussal2014} and as we now recall, this model has a weak noise limit in the $\gamma \to \infty$ limit where the partition sum of the Log-Gamma polymer converges to the solution of the MSHE. Normalizing by definition the temperature $T$ to $1$, the random energies of the model are
\bea \label{Eq:ChapIII:Sec2:Stat16}
\cE_{t}(\hat x) = -Log(W_{t}(x)) \sim \log(Gamma(\gamma)) \ssp.
\eea
And the first two moments are given in terms of the diGamma function $\psi = \Gamma'/\Gamma$ as
\be \label{Eq:ChapIII:Sec2:Stat17}
\overline{\cE} = \psi(\gamma) \sim_{\gamma \gg 1} \log(\gamma) + O(1/\gamma) \quad , \quad \overline{\cE^2}^c \sim \psi'(\gamma) \sim_{\gamma \gg 1} \frac{1}{\gamma} + O(1/\gamma^2) \ssp .
\ee
This shows that, taking $\gamma = b \gamma'$, we can use the weak universality of the MSHE as discussed in Sec.~\ref{subsec:ChapIII:Sec1:WeakU} \footnote{This is a slight adaptation of the weak universality since we are taking a weak noise limit while changing the shape of the distribution at the same time. The ideas are, however, identical.}. More precisely, adapting (\ref{Eq:ChapIII:Sec1:WeakU9}) to take into account the mean value of the energies, we now define
\bea  \label{Eq:ChapIII:Sec2:Stat18}
\tilde{Z}_{\tilde t}(\tilde x) := \lim_{b \to  \infty}  e^{b^2 D \tilde t  \log(\gamma/2) } Z_{t = b^2 D \tilde t }(\hat x=  b \tilde  x) \ssp ,
\eea
and $\tilde{Z}_{\tilde t}(\tilde x)$ satisfies the stochastic equation
\bea \label{Eq:ChapIII:Sec2:Stat19}
 \partial_{\tilde t} \tilde{Z}_{\tilde t}(\tilde x) =  \frac{D}{8} (\partial_{\tilde x})^2 \tilde{Z}_{\tilde t}(\tilde x) +  \sqrt{2 c} \xi(\tilde t , \tilde x) \ssp . 
\eea
with $c = D /(2 \gamma') $. To compare with our previous results on the stationary measure we thus take $D=4$ and $c=1/2$, i.e. $\gamma' = 4$. In order for the stationary initial condition (\ref{Eq:ChapIII:Sec2:Stat15}) of the discrete model to have a well-defined limit as $b \to \infty$ we take $\lambda = \gamma/2 -\mu = b \gamma'/2   -\mu$ with $\mu = O(1)$. In this limit $\log \tilde{Z}_{t}(\tilde x)$ performs a drifting Brownian motion with
\bea \label{Eq:ChapIII:Sec2:Stat20}
\overline{\log \tilde{Z}_{\tilde t}(\tilde x) - \log \tilde{Z}_{\tilde t}(0)} && \sim b \tilde{x} (\psi(\gamma - \lambda) - \psi(\lambda))  \nn \\
&&  \sim  \mu \tilde x
\eea
and
\bea \label{Eq:ChapIII:Sec2:Stat21}
\overline{(\log \tilde{Z}_{\tilde t}(\tilde x) - \log \tilde{Z}_{\tilde t}(0))^2}^c   && \sim  b \tilde{x}(\psi'(\gamma - \lambda) + \psi'(\lambda))  \nn \\
&& \sim  \tilde{x} \ssp .
\eea
And thus we obtain in an unambiguous way that the family of stationary measures of the MSHE corresponds to $\log \tilde{Z}_{\tilde t}$ performing a drifting BM (one easily checks that higher cumulants $\overline{(\log \tilde{Z}_{\tilde t}(\tilde x) - \log \tilde{Z}_{\tilde t}(0))^n}^c$ with $n \geq 3$ are $O(1/b^{n-2})$).

\subsubsection{Stationary measure of exponential and geometric Last-Passage-Percolation}

An elegant way to obtain results on last-passage-percolation with exponential waiting times is to use the zero-temperature limit of the Log-Gamma polymer. Although the Log-Gamma polymer does not contain a parameter corresponding to the temperature (which is one by definition) we saw in the last section that sending $\gamma \to \infty$ corresponds to a weak-noise limit, similar to a large temperature limit. Conversely, sending $\gamma \to 0^+$ yields a zero-temperature limit. More precisely, setting $\gamma = \epsilon \gamma'$ with $(\epsilon, \gamma') \in \JR_+^2 $, one easily shows that rescaled random energies in the Log-Gamma model (\ref{Eq:ChapIII:Sec2:Stat16}) converge in law to (minus) exponential random variables:
\bea \label{Eq:ChapIII:Sec2:Stat22}
 \tilde{\cE}_t(\hat x) := \frac{\cE_t(\hat x)}{\epsilon}  = - \frac{Log(W_t(\hat x))}{\epsilon} \sim_{\epsilon \to 0^+}  - Exp(\gamma') \ssp .
\eea
We recall that the PDF of an exponential random variable is
\bea \label{Eq:ChapIII:Sec2:Stat23}
x \sim Exp(\gamma') \Longleftrightarrow  p(x) = \frac{1}{\gamma'} e^{- \gamma' x} \ssp .
\eea
Hence, introducing 
\bea \label{Eq:ChapIII:Sec2:Stat24}
\sE_{t}(\hat x) = \lim_{\epsilon  \to 0^+} - \frac{\log Z_t(\hat x)}{\epsilon} \ssp ,
\eea
the linear recursion relation of the Log-Gamma polymer becomes, in the limit $\epsilon \to 0$, equivalent to 
\bea \label{Eq:ChapIII:Sec2:Stat25}
\sE_{t+ 1} (\hat x) = \tilde{\cE}_t(\hat x) + {\rm min}(\sE_{t} (\hat x -1/2) ,  \sE_{t} (\hat x + 1/2) ) \ssp .
\eea
Switching to the waiting-time language as in Sec.~\ref{subsec:ChapIII:Sec1:Models}, one sees that this recursion equation corresponds to a problem of last passage percolation with exponential waiting times. We can now take the limit of the stationary initial condition of the Log-Gamma polymer (\ref{Eq:ChapIII:Sec2:Stat15}). Setting $\lambda = \epsilon \lambda'$ we obtain that, if at $t=0$ the energy is taken as a two-sided random walk with increments distributed as differences of independent random variables with
\bea \label{Eq:ChapIII:Sec2:Stat26}
\sE_{t=0}(\hat x +1) - \sE_{t=0}(\hat x)   = Exp(\lambda') - Exp(\gamma'- \lambda') \ssp ,
\eea
then they remain so at all time. This exact solvability property of last passage percolation with exponential waiting times was first proved by Burke in \cite{burke1956} in the language of stochastic queuing systems. As for the Log-Gamma case, this exact solvability property is only the tip of the iceberg and last passage percolation with exponential weights enjoy other remarkable properties. These notably include: (i) the possibility to use the RSK (see Sec.~\ref{subsec:ChapIII:SecII:OtherExact}) correspondence \cite{johansson2000}; (ii) Bethe ansatz solvability of the associated particle system, the TASEP with exponential clocks (see \cite{schutz1997} and references therein). Let us finally mention here that these different exact solvability properties can be generalized to the geometric case where the random energies are discrete and distributed as $\tilde{\cE}_t(\hat x)  \sim -Geo(q) $ where $0<q<1$ and we use the convention
\bea \label{Eq:ChapIII:Sec2:Stat27}
X = Geo(q) \Longleftrightarrow Proba(X = k) = (1-q) q^k \ssp .
\eea  
In particular in this case the stationary initial condition is obtained by setting
\bea \label{Eq:ChapIII:Sec2:Stat28}
\sE_{t=0}(\hat x +1) - \sE_{t=0}(\hat x)   = Geo(q_b) - Geo(q/q_b) \ssp 
\eea
with $q<q_b<1$. This is a generalization of the exponential case since the limit $q = 1 - \gamma' \epsilon$ of the geometric distribution is the exponential distribution.

\subsection{An algebraic exact solvability property: Bethe ansatz integrability of the continuum DP} \label{subsec:ChapIII:SecII:BA}

In this section we discuss the Bethe ansatz integrability of the continuum DP and recall the main steps that led to the results for the fluctuations of the point-to-point free-energy of the DP obtained in \cite{CalabreseLeDoussalRosso2010} (as well as in \cite{Dotsenko2010}).

\subsubsection{From the stochastic-heat-equation to the attractive Lieb-Liniger model }

We now discuss the replica Bethe ansatz solution of the continuum DP which was first initiated by Kardar in \cite{Kardar1987}. In this section, to conform with recent works on the subject, we will study the MSHE with the following normalization,
\bea   \label{Eq:ChapIII:Sec2:SheToLL:1}
 \partial_{t} Z_t(x) =   (\partial_{x})^2 Z_t(x)  +  \sqrt{2 \bar c}  \xi(t,x) Z_t(x)  \ssp ,
\eea
with as usual $\xi(t,x)$ a unit Gaussian white noise with correlation $\overline{\xi(t,x)\xi(t',x')} = \delta(t-t') \delta(x-x')$. It can be obtained from the universal scaling limit of DP on the square lattice discussed in (\ref{Eq:ChapIII:Sec1:WeakU6}) with $D=8$. For concreteness we also consider the initial condition
\bea \label{Eq:ChapIII:Sec2:SheToLL:2}
Z_{t=0}(x) = \delta(x=0) \ssp ,
\eea
and $Z_t(x)$ is thus a point to point partition sum. Let us introduce the `wave-function'
\bea \label{Eq:ChapIII:Sec2:SheToLL:3}
\psi_t(x_1,\cdots,x_n) = \overline{Z_t(x_1)  \cdots Z_t(x_n)} \ssp ,
\eea
from which the $n^{th}$ integer moment of the partition sum is obtained by taking coinciding points. Introducing $n$ auxiliary independent Brownian motions
\bea \label{Eq:ChapIII:Sec2:SheToLL:4}
b_i'(t) =  \sqrt{2} \eta_i(t) \quad , \quad b_i(0)=0 \ssp,
\eea
with $\eta_i(t) $ independents unit centered GWN, $\psi_t(x_1,\cdots,x_n)$ is obtained as a conditional expectation value through the Feynman-Kac representation of the solution of the MSHE (see (\ref{Eq:secIntroKPZ-3})) as
\bea \label{Eq:ChapIII:Sec2:SheToLL:5}
\psi_t(x_1,\cdots,x_n)  =  \overline{\mathbb{E}\left( e^{\sum_{i=1}^{n} \int_{0}^{t} dt' \sqrt{2 \bar c} \xi(t,b_i(t'))}  \prod_{i=1}^n \delta(b_i(t) - x_i )  \right)}  \ssp .
\eea
Exchanging the two averages we obtain\footnote{Note that here there is no contribution from the term $i=j$ since we are using a time-ordered exponential, see the discussion in Sec.~\ref{subsec:ChapIII:Sec1:Models}.}
\bea \label{Eq:ChapIII:Sec2:SheToLL:6}
\psi_t(x_1,\cdots,x_n)  = \mathbb{E}\left( e^{ \sum_{1 \leq i < j \leq n} \int_{0}^{t} dt' 2 \bar c \delta(b_i(t) - b_i(t')) }  \prod_{i=1}^n \delta(b_i(t) - x_i ) \right) \ssp .
\eea
Using again the Feynmac-Kac type formula (or an elementary derivation as in Sec.~\ref{subsec:SecI3:KPZ}) one obtains that $\psi_t$ satisfies the PDE
\bea \label{Eq:ChapIII:Sec2:SheToLL:7}
&& \partial_t \psi_t = - H_n \psi_t \quad , \quad  H_n := -\sum_{j=1}^{n} \frac{\partial^2}{\partial x_j^2} - 2 \bar c \sum_{1 \leq i < j \leq n} \delta(x_i-x_j) \ssp , \nn \\
&&  \psi_{t=0}(x_1, \cdots, x_n) = \prod_{i=1}^{n} \delta(x_i) \ssp .
\eea
Remarkably, this equation corresponds to the Schr\"odinger equation written in imaginary time for bosons (the wave function (\ref{Eq:ChapIII:Sec2:SheToLL:5}) is manifestly symmetric) in the attractive ($\bar c \geq 0$ is the strength of the white noise) Lieb-Liniger (LL) model \cite{LiebLiniger1963}. Note that comparing (\ref{Eq:ChapIII:Sec2:SheToLL:3}) with (\ref{Eq:ChapIII:Sec2:SheToLL:6}), we have replaced the average over disorder by an average over interacting Brownian paths. This is an example of stochastic duality relations which relates the observables of two a priori unrelated stochastic processes and the LL model is dual to the MSHE. The particles of the LL model correspond to replica of the partition sum and we will use indifferently both denominations.

\subsubsection{Bethe ansatz solution of the LL model}
The strategy adopted to solve (\ref{Eq:ChapIII:Sec2:SheToLL:7}) is to first compute all the symmetric eigenfunctions of the Lieb-Liniger Hamiltonian $H_n$:
\bea \label{Eq:ChapIII:Sec2:BALL:1}
H_n \psi_\mu = \Lambda_\mu \psi_\mu \ssp,
\eea
where $\mu$ labels the different eigenvectors. As already noticed in \cite{LiebLiniger1963}, the symmetric eigenfunctions of $H_n$ are obtained using the Bethe ansatz (see \cite{franchini2011notes} for a review) as we now recall. Since we are looking for symmetric eigenfunctions of $H_n$ it is sufficient to specify their values in the so-called {\it Weyl chamber} $x_1 \leq x_2 \leq \cdots \leq x_n$ as
\bea \label{Eq:ChapIII:Sec2:BALL:2}
\psi_\mu(x_1 \leq \cdots \leq x_n) = \tilde{\psi}_{\mu}(x_1 , \cdots , x_n) \ssp,
\eea
where $\tilde{\psi}_{\mu}$ is a priori not symmetric and the other sectors are obtained by using the symmetry of $\psi_\mu$. In the interior of the Weyl chamber $x_1 <x_2 < \cdots < x_n$, the interaction has no influence and the spectral equation (\ref{Eq:ChapIII:Sec2:BALL:1}) just amounts to the {\it free spectral equation} 
\bea \label{Eq:ChapIII:Sec2:BALL:3}
H_n^{{\rm free}} \tilde{\psi}_{\mu} = \Lambda_\mu \tilde{\psi_{\mu}} \quad , \quad H_n^{{\rm free}} = -\sum_i \frac{\partial^2}{\partial x_{i}^2} \ssp .
\eea
And since the values of $\tilde{\psi}_{\mu}(x_1,\cdots,x_n)$ outside the Weyl chamber have no influence on $\psi_{\mu}$, we might as well take (\ref{Eq:ChapIII:Sec2:BALL:2}) to hold $\forall (x_1 , \cdots , x_n) \in \JR^n$. We thus look for $\tilde{\psi}_{\mu}$ as a superposition of plane waves:
\bea \label{Eq:ChapIII:Sec2:BALL:4}
\tilde{\psi}_{\mu} := \sum_{\sigma \in S_n} A_{\sigma} \prod_{i=1}^{n} z_{\sigma(i)}^{x_i} \ssp ,
\eea
where $S_n$ is the group of permutations of $\{1, \cdots , n \}$, the $n!$ complex numbers $A_{\sigma}$ are called {\it amplitudes} and the $n$ complex numbers $z_i$ are called {\it rapidities}. Alternatively we will consider the {\it quasi-momenta} $\lambda_i$ defined by
\bea \label{Eq:ChapIII:Sec2:BALL:5}
z_j = e^{i \lambda_j} \ssp .
\eea
The eigenvalue of the spectral problem (\ref{Eq:ChapIII:Sec2:BALL:1}) is completely specified by the free equation (\ref{Eq:ChapIII:Sec2:BALL:3}) as
\bea \label{Eq:ChapIII:Sec2:BALL:6}
\Lambda_\mu = \sum_{i=1}^{n} \lambda_i^2 \ssp .
\eea
Note that each of the $n!$ terms in $\tilde{\psi}_\mu$ independently solve (\ref{Eq:ChapIII:Sec2:BALL:3}) with the same eigenvalue. Here we are considering for now the most general superposition of plane waves to try to maintain the maximum liberty with the aim of solving the full spectral problem (\ref{Eq:ChapIII:Sec2:BALL:1}). For now the spectral problem is already solved in the interior of the Weyl chamber. Let us now consider the condition of solvability at the `two particles boundary' of the Weyl chamber $x_1< x_2 < \cdots < x_i = x_{i+1} <x_{i+2} <\cdots  < x_n$. From (\ref{Eq:ChapIII:Sec2:BALL:1}) and the form of the LL Hamiltonian (\ref{Eq:ChapIII:Sec2:SheToLL:7}) it is clear that we are looking for solutions $\psi_{\mu}$ that are continuous everywhere but not ${\cal C}^1$. More precisely the derivatives of $\psi_{\mu}$ must exhibit jumps at coinciding points to compensate the $\delta$ interaction. Assuming that these discontinuities only concern the derivatives on coinciding points we must have
\bea \label{Eq:ChapIII:Sec2:BALL:8}
&& \frac{\partial^2 \psi_\mu}{\partial x_i^2} = \left( \frac{\partial \psi_\mu}{\partial x_i}|_{x_i = x_{i+1}^+} - \frac{\partial \psi_\mu}{\partial x_i}|_{x_i = x_{i+1}^-} \right) \delta(x_{i+1} - x_i) + \cdots \nn \\
&& \frac{\partial^2 \psi_\mu}{\partial x_{i+1}^2} = \left( \frac{\partial \psi_\mu}{\partial x_{i+1}}|_{x_{i+1} = x_{i}^+} - \frac{\partial \psi_\mu}{\partial x_{i+1}}|_{x_{i+1} = x_{i}^-} \right) \delta(x_{i+1} - x_i) + \cdots 
\eea
where the dots denote more regular terms. Using the symmetry of the wave-function one easily obtains that these singular parts are actually equals and expressed in terms of $\tilde{\psi}_{\mu}$ as, considering only the most singular terms
\bea \label{Eq:ChapIII:Sec2:BALL:9}
\frac{\partial^2 \psi_\mu}{\partial x_i^2}   \simeq    \frac{\partial^2 \psi_\mu}{\partial x_{i+1}^2}  \simeq  \left(\frac{\partial}{\partial_{x_{i+1}}} -\frac{\partial}{\partial_{x_{i}}} \right)\tilde{\psi_{\mu}}|_{x_{i+1} = x_i}  \delta(x_{i+1}-x_i) + \cdots 
\eea

Let us now integrate, for all $x_j$ fixed, distinct and ordered except $x_i$ and $x_{i+1}$, the spectral equation (\ref{Eq:ChapIII:Sec2:BALL:1}) from $x_i =x_{i+1} - \epsilon$ to $x_i =x_{i+1} + \epsilon$. We obtain,
\bea \label{Eq:ChapIII:Sec2:BALL:10}
\int_{x_i = x_{i+1} - \epsilon}^{x_i = x_{i+1} + \epsilon} (- \frac{\partial^2}{\partial_{x_i}^2} -\frac{\partial^2}{\partial_{x_{i+1}}^2}) \psi_{\mu} - 2 \bar c \psi_{\mu} + O(\epsilon) = O(\epsilon) \ssp .
\eea \label{Eq:ChapIII:Sec2:BALL:11}
implying, using (\ref{Eq:ChapIII:Sec2:BALL:8}) and (\ref{Eq:ChapIII:Sec2:BALL:9})
\bea \label{Eq:ChapIII:Sec2:BALL:12}
-\left(\frac{\partial}{\partial_{x_{i+1}}} -\frac{\partial}{\partial_{x_{i}}} \right)\tilde{\psi_{\mu}}|_{x_{i+1} = x_i} = \bar c \tilde{\psi}_{\mu} \ssp .
\eea
And this implies for the amplitudes $A_{\sigma}$, that any amplitudes differing from one another by a transposition of $i \leftrightarrow j$, noted $\tau_{ij}$, must satisfy
\bea \label{Eq:ChapIII:Sec2:BALL:13}
\frac{A_{\sigma \circ \tau_{ij}}}{A_{\sigma }} = \frac{\lambda_{\sigma(j)} - \lambda_{\sigma(i)} - i \bar c}{\lambda_{\sigma(j)} - \lambda_{\sigma(i)} + i \bar c} \ssp .
\eea
There are thus $\frac{1}{2} C^{2}_n n!$ equations for $n!$ variables, but they are mutually consistent \cite{LiebLiniger1963}. The solution is defined up to a constant (i.e. the choice of $A_{Id}$) and we will choose here, in agreement with e.g. \cite{LeDoussalCalabrese2012},
\bea \label{Eq:ChapIII:Sec2:BALL:14}
A_{\sigma} = \prod_{1 \leq \alpha < \beta \leq n} \left( 1 + \frac{i \bar c }{\lambda_{\sigma(\beta)}-\lambda_{\sigma(\alpha)}  } \right) \ssp .
\eea
We have now obtained a complete solution of the spectral problem (\ref{Eq:ChapIII:Sec2:BALL:1}). Here we have only taken care of the two-body interaction, but the Lieb-Liniger Hamiltonian (\ref{Eq:ChapIII:Sec2:SheToLL:7}) has indeed only two-bodies interactions. Note that we have not yet specified the values of the rapidities $z_i$. We now must do so in such a way that we obtain a complete basis of symmetric functions.

\subsubsection{The string solution}
 
 Here we will adopt the same strategy as e.g. in \cite{LeDoussalCalabrese2012} and adopt periodic boundary conditions on a line of length $L$. More precisely we look for solutions such that $\forall (x_1 , \cdots , x_n) \in \JR^n$ we have
 \bea \label{Eq:ChapIII:Sec2:BALLString:1}
\psi_{\mu}(x_1 , \cdots x_{i-1}, x_i + L , \cdots , x_n) = \psi_{\mu}(x_1 , \cdots , x_n) \ssp .
 \eea
 This implies the Bethe equations for the quasi-momenta:
\bea \label{Eq:ChapIII:Sec2:BALLString:2}
e^{i \lambda_{\alpha } L} = \prod_{\beta \neq \alpha} \frac{\lambda_{\alpha} - \lambda_{\beta} -i \bar c}{\lambda_{\alpha} - \lambda_{\beta} + i \bar c} \ssp .
\eea

 These boundary conditions are appropriate to study the problem of the continuum DP on a cylinder as in \cite{BrunetDerrida2000a,BrunetDerrida2000b}. For our purpose they are only a trick to obtain a well defined complete basis\footnote{Proving the completeness of the Bethe solutions is in general a very non-trivial problem, see \cite{ProlhacSpohn2011b} and references therein for the case studied here.} of eigenfunctions and we will study the limit $L \to \infty$. The solution in this case was first described in \cite{McGuire1964} and here we recall the main features. In the large $L$ limit a set of solutions $\lambda_{\alpha}$ can be decomposed in $n_{s}$ packets, also called {\it strings}, of $m_{i}$ particles (the string {\it multiplicities}) with $i=1, \cdots , n_s$ and $n = \sum_{i=1}^{n_s} m_j$. Inside the $j^{th}$ string, the quasi-momenta are labeled by $a= 1 , \cdots , m_j$ and take the form
 \bea \label{Eq:ChapIII:Sec2:BALLString:3}
\lambda^{j,a} := k_j + \frac{i \bar c}{2} (m_j +1 - 2 a) + i \delta^{j a} \ssp , 
 \eea
where $k_j$ is quantized as for free-particles, $k_j = 2 \pi I_j/L$ with $I_j \in \JZ$ and $\delta^{ja}$ are corrections to this leading behavior that decay exponentially with $L$. Note that the quasi-momenta inside a string are symmetric with respect to the real axis. To understand this structure let us consider the two-particle case and take the log of the Bethe equations (\ref{Eq:ChapIII:Sec2:BALLString:2}): we obtain
\bea \label{Eq:ChapIII:Sec2:BALLString:4}
&& \lambda_1 = \frac{2  \pi I_1}{L} + \frac{1}{iL} \left( \log(\lambda_1 - \lambda_2 - i \bar c)  - \log(\lambda_1 - \lambda_2 + i \bar c) \right) \nn \\
&& \lambda_2 = \frac{2  \pi I_2}{L} - \frac{1}{iL} \left( \log(\lambda_1 - \lambda_2 - i \bar c)  - \log(\lambda_1 - \lambda_2 + i \bar c) \right)  \ssp ,
\eea
with $(I_1, I_2) \in \JZ^2$ and note that $\lambda_1 + \lambda_2$ is always real. The naive solution of this equation in an expansion in $1/L$ would thus be $\lambda_j = \frac{2 \pi I_j}{L} + O(1/L^2)$, i.e. an asymptotic solution equivalent to free-particles. This however neglects the fact that the singularity of the logarithm at $0$ can kill the $1/L$ decay before it. More precisely if $\lambda_1 = \lambda_2 + i \bar c + O(e^{- \delta L}) $ with $\delta >0$, then $I_1 = I_2$ and we obtain
\bea \label{Eq:ChapIII:Sec2:BALLString:5}
\lambda_1 = \frac{2 \pi I_1}{L} + \frac{1}{iL} (- \delta L)  \quad , \quad \lambda_2 = \frac{2 \pi I_1}{L} - \frac{1}{i L } (- \delta L)  \ssp .
\eea
By consistency we must have $\lambda_1 - \lambda_2 \simeq i \bar c = 2 i \delta $. Hence this solution is only consistent if $\bar c >0$: strings only exist in the attractive phase of the Lieb-Liniger model. Finally since $\lambda_1 + \lambda_2$ must be real this implies the string form (\ref{Eq:ChapIII:Sec2:BALLString:3}) for $m=2$. This reasoning can be generalized to any $m$ and leads to (\ref{Eq:ChapIII:Sec2:BALLString:3}). Note that these considerations do not constitute in any case a proof that the string states form a complete basis in the large $L$ limit, but they indeed do, see \cite{BorodinCorwinPetrovSasamoto2015}. In the large $L$ limit the dynamics of the system is thus relatively simple and the particles form string states that move essentially independently. In particular, the wave function associated with $n$ particles forming a single $n$-string is
\bea \label{Eq:ChapIII:Sec2:BALLString:6}
\psi_{\mu}(x_1 , \cdots , x_n):= n! e^{i k \sum_i x_i - \frac{\bar c}{2} \sum_{1 \leq i < j \leq n} |x_i-x_j|  } \ssp .
\eea
String states thus correspond to bound states. The contribution to the eigenvalue $\Lambda_\mu$ of a string of $m_j$ particles is the energy of the string
\bea \label{Eq:ChapIII:Sec2:BALLString:7}
E_j := \sum_{a=1}^{m_j} (\lambda^{j,a})^2 = m_j k_j^2 - \frac{(\bar c)^2}{12} m_j (m_j^2 - 1) \ssp .
\eea
The ground state of the system is thus obtained by forming a single $n$-string. Finally, noting that the Bethe wavefunction (\ref{Eq:ChapIII:Sec2:BALL:4}) with (\ref{Eq:ChapIII:Sec2:BALL:14}) is symmetric by exchange $z_i \leftrightarrow z_j$ the sum over eigenstates can be computed as
\bea \label{Eq:ChapIII:Sec2:BALLString:8}
\sum_{\mu} = \sum_{n_s=1}^{n}  \frac{1}{n_s!} \sum_{(m_1 , \cdots , m_{n_s})_n} \prod_{j=1}^{n_s} \int_{\JR}  \frac{ m_j L dk_j}{2 \pi} \ssp ,
\eea
where the $n_s!$ avoids a double counting of the states, the sum $\sum_{(m_1 , \cdots , m_{n_s})_n}$ is over all $n_s-$uplets such that $\sum_{j=1}^{n_s} m_j = n$ and the last terms comes from the fact that each string-state should be considered as a free-particle with total moment $K_j := \sum_{a=1}^{m_j} \lambda^{j,a} = m_j k_j$ (see \cite{CalabreseCaux2007}).

\subsubsection{Norm of the states}
As the eigenfunctions of a symmetric operator, the Bethe eigenfunctions are orthogonal. They are however not normalized and for many practical applications it is important to know their norm. A remarkable formula due to Gaudin \cite{gaudin1983} is that the norm of the eigenstates of the system with periodic boundary conditions can be computed, at finite $L$, as
\bea \label{Eq:ChapIII:Sec2:BALLNorm:1}
||\psi_{\mu}||^2 && = \int_{[0,L]^n} dx_1  \cdots dx_n \psi_{\mu}^*(x_1,\cdots , c_n)\psi_{\mu}(x_1,\cdots , c_n) \nn \\
&& = n!  \prod_{1 \leq \alpha < \beta \leq n }  \frac{   (\lambda_{\alpha}^{LL} - \lambda_{\beta}^{LL}  ) ^2 + (\bar{c}^{LL} )^2  }   {    (\lambda_{\alpha}^{LL}  - \lambda_{\beta}^{LL}  ) ^2} \det{ G^{LL}  }
\eea 
where $G^{LL}$ is the Gaudin matrix whose entries are:
 \bea \label{Eq:ChapIII:Sec2:BALLNorm:2}
&& G_{\alpha \beta}^{LL} = \delta_{\alpha  \beta} \left( L + \sum_{\gamma=1}^n K (\lambda_{\alpha}^{LL} - \lambda_{\gamma}^{LL}) \right) - K ( \lambda_{\alpha}^{LL} - \lambda_{\beta}^{LL}) \\
&& K(x) =  \frac{- 2 \bar c^{LL}}{x^2 + (\bar c^{LL})^2} 
\eea
Note that the entries of the Gaudin matrix in the LL case 
are the derivatives of the logarithm of the Bethe equations. Performing the asymptotic analysis of this formula at large $L$ for string states is a non-trivial problem (due to the divergences or zeros both in the prefactor and in the Gaudin kernel) which was only accomplished recently in \cite{CalabreseCaux2007}. The result is
\be \label{Eq:ChapIII:Sec2:BALLNorm:3}
||\psi_{\mu}||^2 = \frac{n! L^{n_s}}{(\bar c)^{n-n_s}}  \prod_{j=1}^n m_j^2   \prod_{1 \leq i < j \leq n_s }  \frac{4 (k_i - k_j)^2 + (m_i + m_j)^2 \bar c^2 }{4 (k_i - k_j)^2 + (m_i -m_j)^2 \bar c^2 }  + O(L^{n_s -1})\ssp .
\ee
Note in particular that this norm scales as $L^{n_s}$, a scaling that would be obtained for $n_s$ free particles.

\subsubsection{Point-to-point free-energy fluctuations of the continuum DP}
Starting from
\bea \label{Eq:ChapIII:Sec2:BALLFinal:1}
\overline{Z_t(x)^n} && = \sum_{\mu } \frac{\langle \psi_\mu , \psi_{t=0} \rangle }{|| \psi_{\mu} ||^2} \Lambda_{\mu}^t \psi_{\mu}(x,\cdots,x)  \ssp , 
\eea
and combining the ingredients presented with the previous section together with the fact that $\psi_{\mu}(x,\cdots , x)  = n! e^{ i \sum_{j=1}^{n_s} m_j k_j x}$ one obtains the formula that first appeared in \cite{CalabreseLeDoussalRosso2010,Dotsenko2010}:
\bea \label{Eq:ChapIII:Sec2:BALLFinal:2}
\overline{Z_t(x)^n} = && \sum_{n_s =1}^{n} \frac{n! \bar c^n}{n_s! (2 \pi \bar c)^{n_s}} \sum_{(m_1 , \cdots , m_{n_s})_n} \prod_{j=1}^{n_s} \frac{dk_j}{m_j} \prod_{1 \leq i < j \leq n_s }  \frac{4 (k_i - k_j)^2 + (m_i - m_j)^2 \bar c^2 }{4 (k_i - k_j)^2 + (m_i +m_j)^2 \bar c^2 } \nn \\
&& \times  \prod_{j=1}^{n_s}  e^{ i \sum_{j=1}^{n_s} m_j k_j x -t \left( m_j k_j^2 - \frac{(\bar c)^2}{12} m_j (m_j^2 - 1) \right) } \ssp .
\eea
This formula is exact, but does not determine the probability distribution of $Z_t(x)$: the moments grow as $e^{t n^3 \frac{(\bar c)^2}{12}}$ (contribution from the ground state), i.e. too fast to determine the distribution. Hence determining the Laplace transform
\bea \label{Eq:ChapIII:Sec2:BALLFinal:3}
g_{t,x}(u) := \overline{e^{- u Z_t(x)} } 
\eea
from (\ref{Eq:ChapIII:Sec2:BALLFinal:1}) is impossible from a mathematical point of view. Exchanging the average over disorder and the series expansion of the exponential in (\ref{Eq:ChapIII:Sec2:BALLFinal:3}) leads to a diverging series. This is a caveat of the replica method for the continuum DP and one has to devise a recipe to resum the diverging series. The correct way used for example in \cite{CalabreseLeDoussalRosso2010,Dotsenko2010} is to use the Airy trick  $\int_{\mathbb{R}} dy Ai(y) e^{ys} = e^{\frac{s^3}{3}}$ (valid for $Re(s)>0$) to `linearize' the energies of the string. Let us now give the result for the Laplace transform at $x=0$ (which is not restrictive since STS holds, see Sec.~\ref{subsec:ChapIII:SecII:Sym}). Introducing the rescaled Laplace transform
\bea \label{Eq:ChapIII:Sec2:BALLFinal:4}
\tilde{g}(s) = \overline{  \exp\left( -e^{\lambda s}  \frac{Z_t(0)}{\bar c e^{- \frac{\bar c^2 t}{12}}} \right) } = \overline{  \exp\left( -e^{\lambda (s-f_t)}\right) } 
\eea
where we have introduced the parameter $\lambda$ and the rescaled free-energy
\bea \label{Eq:ChapIII:Sec2:BALLFinal:5}
\lambda = (\frac{\bar c^2 t}{4})^{\frac{1}{3}} \quad , \quad f_t := \frac{ - \log Z_t(0) + \log(\bar c) + \frac{\bar c^2 t}{12}}{\lambda}  \ssp .
\eea
Then $\tilde{g}(s) $ can be rewritten as a Fredholm determinant (see \cite{Bornemann2008} and references therein for background on Fredholm determinants):
\bea \label{Eq:ChapIII:Sec2:BALLFinal:6}
\tilde{g}(s) = {\rm Det}\left( I + K \right)
\eea
with the kernel
\bea \label{Eq:ChapIII:Sec2:BALLFinal:7}
K(v,v') = - \int_{\JR} \frac{dk}{2\pi} dy Ai(y + k^2 - s + v+v') \frac{e^{\lambda y - i k(v-v')}}{1 + e^{\lambda y}} \ssp . 
\eea
$K$ is an operator $K : L^2(\JR^+) \to L^2(\JR^+)$, i.e. the $v,v'$ variables above live on $\JR_+$. The essential steps to go from (\ref{Eq:ChapIII:Sec2:BALLFinal:4}) to (\ref{Eq:ChapIII:Sec2:BALLFinal:4}) are (i) (wrongfully) inverting the average with respect to disorder and the series expansion of the exponential in (\ref{Eq:ChapIII:Sec2:BALLFinal:4}); (ii) use the exact formula (\ref{Eq:ChapIII:Sec2:BALLFinal:1}); (iii) exchange in the resulting (diverging) expression the sum over $n$ (originating from step (i)) and the sum over $n_s$ in (\ref{Eq:ChapIII:Sec2:BALLFinal:1}); (iv) use the Airy trick; (v) notice the determinantal structure using the formula
\be \label{Eq:ChapIII:Sec2:BALLFinal:8}
{\rm det} \left[ \frac{1 }{i(k_i -k_j) + (m_i +m_j)/2}  \right]_{n_s \times n_s } = \prod_{i=1}^{n_s} \frac{1}{m_i} \prod_{1 \leq i < j \leq n_s} \frac{4(k_i-k_j)^2 + (m_i - m_j)^2}{4(k_i-k_j)^2 + (m_i +m_j)^2} \ssp .
\ee
The emergence of a determinant thus in the end comes from the remarkable formula for the norm of string states in the large $L$ limit. We refer the reader to \cite{CalabreseLeDoussalRosso2010} for more details on the derivation of (\ref{Eq:ChapIII:Sec2:BALLFinal:6})-(\ref{Eq:ChapIII:Sec2:BALLFinal:7}). Finally, starting from (\ref{Eq:ChapIII:Sec2:BALLFinal:6}) it is possible to perform the large time (i.e. large $\lambda$) limit of the Fredholm determinant. Assuming that the rescaled free-energy $f_t$ defined in (\ref{Eq:ChapIII:Sec2:BALLFinal:5}) is a $O(1)$ RV (and thus $\overline{\frac{\bar c^2}{12}}$ is the extensive part of the free-energy of the DP while $\lambda$ is the scale of the fluctuations of the free-energy), one obtains \cite{CalabreseLeDoussalRosso2010} (using $\lim_{\lambda \to \infty} \exp( -e^{\lambda(s- f)}) = \theta(f-s) $ )
\bea
\lim_{t \to \infty} Prob\left( \frac{ - \log Z_t(0) + \frac{\bar c^2 t}{12}}{\lambda}  > s \right) = Prob(f_{\infty} > s) = F_2\left(- \frac{s}{2^{2/3}}\right) \ssp, 
\eea
where $F_2(s)$ is the CDF of the Tracy-Widom GUE distribution which also admits an expression as a Fredholm determinant \cite{TracyWidom1993}.

\subsubsection{A few results obtained using Bethe ansatz}

The replica Bethe ansatz approach to DP has led to a variety of exact results. Known since the work of Kardar \cite{Kardar1987}, it was first applied for technical reasons, to the study of DP properties that can be deduced more or less from the sole knowledge of the ground state energy, i.e. the limit of DPs of large length $t \gg 1 $ on a finite cylinder $L$ (the limit $L \to \infty$ being eventually taken afterwards). This was used to already determine the critical exponents \cite{Kardar1987,BouchaudOrland1990} or the large deviation function for the fluctuations of the free-energy of the DP on the cylinder \cite{BrunetDerrida2000a,BrunetDerrida2000b}. Obtaining the universal distribution of fluctuations for the growth of an interface in an infinite space, however, requires to consider the limit $t\to \infty$ with at least $ L \gg t^{2/3}$. The study of this limit from BA requires a summation over all excited states. This was only achieved recently, partly thanks to the work of Calabrese and Caux \cite{CalabreseCaux2007} who managed to compute the norm of string states (\ref{Eq:ChapIII:Sec2:BALLNorm:3}). 

\medskip

Even with this knowledge it is still far from trivial to obtain exact results. Additionally since the method is not rigorous from a mathematical point of view due to the too rapid growth of moments, it requires a large number of tricks. Once a solid recipe to tackle this issue has been devised (a recipe that sometimes appears retrospectively to be the shadow of a rigorous derivation, as e.g. by considering a $q-$deformed model, see \cite{BorodinCorwinPetrovSasamoto2015}), the replica Bethe ansatz approach has led a variety of new (presumably exact) results. Here we name a few: (i) TW-GUE distribution of fluctuations for the point-to-point free energy \cite{CalabreseLeDoussalRosso2010,Dotsenko2010}; (ii) TW-GOE distribution of fluctuations for the point-to-line free energy \cite{CalabreseLeDoussal2011,LeDoussalCalabrese2012,GuedreLeDoussalRossoHenryCalabrese2012}; (iii) multi-point correlations for the point-to-point free-energy and the Airy process \cite{ProlhacSpohn2011,Dotsenko2points}; (iv) one point (Baik-Rains) and multi-point distributions of fluctuations for the point-to-Brownian (i.e. the DP with stationary initial condition) free-energy \cite{ImamuraSasamoto2012,ImamuraSasamoto2013}; (v) TW-GSE distribution of fluctuations for the point-to-point free energy of a directed polymer in a half-space \cite{GueudreLeDoussal2012}; (vi) fluctuations of free-energy in the crossover from droplet to stationary initial condition \cite{ImamuraSasamoto2013}; (vii) fluctuations of free-energy in the crossover from droplet to flat initial condition \cite{LeDoussal2014}; (viii) distribution of the endpoint of the polymer \cite{Dotsenko2013}; (ix) extension to two-times \cite{Dotsenko2016}. Some of these results have been shown rigorously since then (see \cite{Corwin2011Review}), giving credit to the replica method.

\subsection{A few words on other exact solvability properties} \label{subsec:ChapIII:SecII:OtherExact}

In the next chapter we will review the recent progresses that have been made on applying the replica Bethe ansatz to models of DPs on $\JZ^2$, which are one of the major focus of this thesis. Additionally we will see that these BA exactly solvable models of DPs have another exact-solvability property that has also been discussed previously, namely their stationary measure can be written down exactly. Before we do so we now mention other exact solvability properties that have played an important role in the study of DPs, particularly for discrete models.

\subsubsection{A combinatorial exact solvability property: RSK and gRSK correspondence}

As we mentioned, the first proof of TW-GUE fluctuations for a point-to-point directed polymer was in \cite{johansson2000} for a model of DP at zero temperature with (minus) exponential or geometric distribution of random energies, i.e. equivalently a model of LPP with exponential or geometric distribution of waiting times (see Sec.~\ref{subsec:ChapIII:Sec1:Models}). The exact solvability used there appears to be of purely combinatorial origin and proceeds in three steps. It consists, for the geometric case, in: (i) mapping the problem of finding the last passage time $T(x_1,x_2)$ to the problem of finding the length of the longest increasing subsequence in a sequence of pairs of integers $(x_1',x_2')$ where the number of times $(x_1,x_2)$ appears in the sequence is given by the value of the random waiting time $\st_{x_1,x_2}$; (ii) using the Robinson-Schensted-Kuth (RSK) algorithm to map this problem onto the problem of finding the length of the first row in a pair of `semi-standard Young tableaux'; (iii) obtain a representation of the PDF of the latter using Schur polynomials. The formula obtained in the last step shares striking similarities with formulae obtained in random matrix theory. The asymptotic analysis is carried out using the theory of symmetric polynomials, and the role played by the Hermite polynomials for the GUE is played by the Meixner polynomials. We refer to \cite{KriecherbauerKrug2008} for a pedagogical review of this approach to LPP with geometric waiting times. This approach was later adapted to study a model of first passage percolation with geometric waiting times on horizontal edges of $\JZ^2$ only \cite{DraiefMairesseOConnell2005}. The RSK correspondence was later `tropicalized' in \cite{OConnel2009} to obtain the proof of GUE-TW distribution of free-energy fluctuations for the semi-discrete directed polymer. This new combinatorial mapping has since then been referred to as the geometric RSK (gRSK) correspondence. LPP with geometric and exponential waiting times was a precursor of the first discovered exactly solvable model of DP on the square lattice at finite temperature, the Log-Gamma polymer. Introduced in \cite{Seppalainen2009} for the possibility of writing down exactly its stationary measure, it was later shown that the gRSK correspondence could be used to tackle this finite temperature case as well \cite{CorwinOConnellSeppalainenZygouras2014}, the results were then later used in \cite{BorodinCorwinRemenik2013} to obtain the first proof of the emergence of TW-GUE fluctuations in a finite temperature model of DP on $\JZ^2$. Similarly, the LPP model discussed previously was a precursor of the `Strict-Weak' polymer, the second exactly solvable model of finite temperature DP on $\JZ^2$. Introduced in \cite{CorwinSeappalainenShen2015,OConnellOrtmann2015}, three exactly solvable properties were shown simultaneously for this model: exact stationary measure, Bethe ansatz solvability and gRSK correspondence. TW-GUE fluctuations were shown too. The links between RSK and gRSK correspondence was recently clarified in \cite{MateevPetrov2015} where the authors obtained a general correspondence that interpolates between both.

\subsubsection{Macdonald processes}

The theory of Macdonald processes, developed in \cite{BorodinCorwinMacDo2014}, has provided important results for various models in the KPZUC. Macdonald processes are a two-parameter family of stochastic processes which, in several limits, converge to models now known to belong to the KPZUC in particular the q-TASEP, the semi-discrete DP and the continuum DP. The exact solvability property of Macdonald processes notably relies on results from the theory of symmetric functions (Macdonald functions) and Macdonald processes in general constitute a class of processes different from BA solvable processes (but their intersection is not empty).

\section[Summary of the thesis]{Summary of (and more context around) the results obtained during the thesis} \label{Sec:ChapIII:Sec3}

\subsection{Introduction}

In the last sections we have thus discussed some aspects of the KPZUC and presented some exact solvability properties that, in the DP context, contributed to the statement of the KPZ universality hypothesis. Before we continue, let us emphasize here that there is still no simple explanation for the values of the exponents in the KPZUC, and even less for the emergence of universal distributions related to extreme value statistics of RMT\footnote{Note that the Baik-Rains distribution has no equivalent in RMT.}. Despite this, this universality class seems remarkably robust. {\it The emergence of the KPZ exponents and of the Tracy-Widom GUE distribution in so many different models has the flavor of an analogue of the central limit theorem in the case of strongly correlated RVs.} Understanding simple mechanisms explaining this universality is one of the major objectives of research in this field. This task, however, still appears beyond reach, and most of the research in the KPZUC still relies on the study of exactly solvable models. It is in this spirit that in this thesis we studied directed polymers on the square lattice and tried to understand how KPZUC appears in these models (see Sec.~\ref{subsec:PresBe}).

\smallskip

As we discussed in the last section, the RSK and gRSK correspondences have played a major role in the study of DPs on $\JZ^2$. While the RSK correspondence provided the first proof of TW-GUE type fluctuations of free-energy for LPP with geometric waiting times \cite{johansson2000}, the gRSK correspondence led to a similar result, for the first time, in a model of a DP on $\JZ^2$ at finite temperature, the Log-Gamma polymer \cite{Seppalainen2009,CorwinOConnellSeppalainenZygouras2014,BorodinCorwinRemenik2013}.

\smallskip

On the other hand, for the continuum DP, as reviewed in the last section, the Bethe ansatz approach, although non-rigorous, is a powerful and versatile technique that was recently applied to obtain a variety of {\it exact} results. This state of affairs provided the motivation to obtain a Bethe ansatz approach to the Log-Gamma polymer. Similarly to the continuum case, the moment problem is mapped into a discrete-time and discrete-space dynamics of replica on $\JZ$. It was found by Éric Brunet that the transfer matrix of the problem (equivalent to the LL Hamiltonian), could be diagonalized using the Bethe ansatz. As we showed in the last section, this BA solvability is, however, only the beginning of the route to the `proof' of GUE-TW-type fluctuations for the DP free-energy. The paper \cite{ThieryLeDoussal2014} showed how the route used for the continuum DP could be successfully adapted to this discrete case. The results obtained in \cite{ThieryLeDoussal2014} will be presented in Sec.~\ref{subsec:PresLG}.

\smallskip

The dynamics of the replica on $\JZ$ for the Log-Gamma polymer is very similar to the dynamics of interacting particle systems referred to as zero-range-processes (ZRP). In the seminal paper \cite{Povolotsky2013}, Povolotsky obtained a classification of BA solvable models of ZRP on $\JZ$ with parallel updates\footnote{A classification later extended in \cite{CorwinPetrov2015} to the case of non-parallel updates.}. The purpose of the paper \cite{ThieryLeDoussal2015} was to understand whether or not this classification could be adapted to obtain a classification of finite temperature BA solvable models of DP on $\JZ^2$. As we will discuss in Sec.~\ref{subsec:PresIB}, with some specific hypothesis, this classification is very close to being complete. In particular it encompasses all known models of exactly solvable models of DP on $\JZ^2$: the already discussed Log-Gamma and Strict-Weak polymer, but also (i) the Beta polymer introduced shortly before by Barraquand and Corwin in \cite{BarraquandCorwin2015}; (ii) the Inverse-Beta polymer, a new integrable model of DP on the square lattice that remarkably interpolates between the Inverse-Beta and Log-Gamma polymer, and for which we showed TW-GUE fluctuations for the point-to-point free energy. The results obtained in \cite{ThieryLeDoussal2015} will be presented in Sec.~\ref{subsec:PresIB}.

\smallskip

Among the classification of BA exactly solvable models of DP obtained in \cite{ThieryLeDoussal2015}, the Beta polymer has the remarkable peculiarity that it can also be interpreted as a model of a random walk on $\JZ$ in a time-dependent random environment (TD-RWRE). This is the first example of an exactly solvable model of TD-RWRE and it brings to this field the possibility of using exact techniques and the scope of KPZUC. In \cite{BarraquandCorwin2015} the authors obtained exact results for the point to half-line partition sum of the DP. In the correspondence with TD-RWRE these correspond to results for the cumulative distribution function (CDF) of the TD-RWRE transition probability. They notably showed a convergence at large time of the fluctuations of the point to half-line free-energy of the DP, in the large deviations regime of the TD-RWRE, to the GUE-TW distribution. This suggests that KPZ universality does apply in the large deviations regime, but the behavior of the TD-RWRE in the diffusive regime was not considered in \cite{BarraquandCorwin2015}. In \cite{ThieryLeDoussal2016b}, using the techniques developed in \cite{ThieryLeDoussal2014,ThieryLeDoussal2015}, we obtained complementary exact results for the point to point partition sum of the DP, equivalent in the correspondence with TD-RWRE to the PDF of the TD-RWRE transition probability. We performed the asymptotic analysis of these formulae both in the large deviations regime and in the diffusive regime. While in the large deviations regime we obtain TW-GUE type fluctuations for the point to point DP free-energy, in the diffusive regime we obtain that the fluctuations of the partition sum are Gamma distributed. This permits a discussion of the crossover between both regimes. The results obtained in \cite{ThieryLeDoussal2016b} will be presented in Sec.~\ref{subsec:PresBe}.

\smallskip

Finally in \cite{Thiery2016}, on one hand we pursued the analysis of the Inverse-Beta polymer and obtained its stationary measure exactly. Using the stationary measure we additionally recovered rigorously some results obtained in \cite{ThieryLeDoussal2015}. On the other hand we used this knowledge to go back to zero temperature models of DPs on the square lattice and we introduced a new exactly solvable model, the Bernoulli-Geometric polymer. The motivation to look for this model came from the fact that (i) the Inverse-Beta polymer appears as a general model encompassing the two gRSK solvable finite temperature models of DP on $\JZ^2$, the Log-Gamma and the Strict-Weak polymer; (ii) the Log-Gamma and the Strict-Weak polymer are both linked with RSK solvable zero-temperature models of DP on $\JZ^2$ with discrete energies, namely models of first and last passage percolation with geometric waiting times. It was thus natural to conjecture that a zero-temperature model of DP on $\JZ^2$, linked with the Inverse-Beta polymer and with discrete random energies should exist. The Bernoulli-Geometric model introduced in this paper appears as this missing model and we obtained its stationary measure exactly and deduced from it several results. The results obtained in \cite{Thiery2016} will be presented in Sec.~\ref{subsec:PresStat}.

\medskip

We now give a more detailed overview of the main results obtained in \cite{ThieryLeDoussal2014,ThieryLeDoussal2015,ThieryLeDoussal2016b,Thiery2016}. We only focus on the main results and begin with \cite{ThieryLeDoussal2014} on which we will be quite exhaustive since many of the methods developed in \cite{ThieryLeDoussal2014} are used in \cite{ThieryLeDoussal2015,ThieryLeDoussal2016b}.

\subsection{Presentation of the main results of  \cite{ThieryLeDoussal2014}} \label{subsec:PresLG}

\stab {\it Introduction, our strategy and an issue}\\
In \cite{ThieryLeDoussal2014} we attempted to solve the Log-Gamma polymer model, already presented in Sec.~\ref{subsec:ChapIII:SecII:StatMeas}, using the coordinate Bethe ansatz. The model is defined through the recursion equation for the partition sum, for $t \geq 0$,
\be \label{Eq:PresLG:1}
Z_{t}(\hat x) = W_{t+1}(\hat x) \left(Z_{t}(\hat x-1/2)+ Z_{t}(\hat x+1/2) \right) \quad , \quad Z_{0}(\hat x) =  \delta_{\hat x,0} W_0(0) \ssp .
\ee
Where the random Boltzmann weights $W_{t}(\hat x)$ are independent and distributed according to the inverse of Gamma random variables with parameter $\gamma >0$ (see (\ref{Eq:ChapIII:Sec2:Stat14})). Note that here we have added a Boltzmann weight on the first site of the DP. The moments of $W_{t}(\hat x)$ are well defined for $n \leq \gamma$ and are given by
\bea \label{Eq:PresLG:2}
\overline{(W_{t}(\hat x))^n} = \frac{\Gamma(\gamma-n)}{\Gamma(\gamma)} = \frac{(-1))^n}{(1-\gamma)_n} \ssp, 
\eea
where we introduced the Pochammer symbol $(a)_n = \prod_{k=0}^{n-1} (a+k)$. For $n> \gamma$, $\overline{(W_{t}(\hat x))^n} = + \infty$. In this paper, our goal is to compute the PDF of $Z_t(\hat x)$ from the knowledge of its integer moments $\overline{(Z_t(\hat x))^n}$ that we will compute using BA. This problem is obviously ill-defined since only a finite number (the first $\gamma$) of moments of $Z_t(\hat x)$ exist. The problem here is thus in some sense even worse than in the continuum DP where the moments were `only' growing too fast.\\

{\it A way out}\\
Note, however, that using the analytical continuation of the Gamma function, the right-hand side of (\ref{Eq:PresLG:2}) is well-defined $\forall n$, although these are not the moments of the Inverse-Gamma distribution (the moments only exist for $n$ in the complex plane with $Re(n)< \gamma$). The question we ask here is: `can we still somehow use these analytically continued moments to obtain (non-rigorously) the Laplace transform of $W_t(\hat x)$', which is defined by
\bea \label{Eq:PresLG:3}
g(u) := \overline{e^{- W_t(\hat x)} } = \overline{ \sum_{n=0}^{\infty} \frac{(-u)^n}{n!}( W_t(\hat x))^n } \ssp .
\eea
If we find a `solution' to this ill-defined problem, then our goal will be to adapt this solution to obtain the Laplace transform of $Z_t(\hat x)$ from similarly analytically continued moments. First note that in the expression (\ref{Eq:PresLG:3}), it is not possible to exchange the series expansion of the exponential and the average over disorder since only a finite number of moments of $W_t(\hat x)$ exist. We can, however, rewrite this series expansion using a Mellin-Barnes representation:
\bea \label{Eq:PresLG:4}
g(u) =  \overline{ \frac{-1}{2 i \pi} \int_{s \in {\cal C}} \frac{\pi ds}{\sin(\pi s)} u^s \frac{(W_t(\hat x))^s}{\Gamma(1+s)}  } \ssp ,
\eea
where ${\cal C}$ is a vertical contour oriented from down to top with ${\cal C} = -a + i \JR$ and $0<a<1$. The identity between (\ref{Eq:PresLG:3}) and (\ref{Eq:PresLG:4}) follows from the (legitimate) application of the residue theorem when closing the contour on the right and taking the poles of the sine function in the denominator. Using (\ref{Eq:PresLG:2}), $g(u)$ can thus be obtained as
\bea \label{Eq:PresLG:5}
g(u) = \frac{-1}{2 i \pi} \int_{s \in {\cal C}} \frac{\pi ds}{\sin(\pi s)} u^s \frac{\Gamma(\gamma-s)}{\Gamma(\gamma) \Gamma(1+s)}  \ssp .
\eea
This integral converges and the inversion of the integral and of the average over disorder is legitimate. We can now `un-do' the Mellin-Barnes transform. Closing the contour ${\cal C}$ on the right and taking into account the poles of the sine function {\it and} of the $\Gamma$ function at the numerator, we obtain (using Euler's reflection formula)
\bea \label{Eq:PresLG:6}
&& g(u) = g_{\rm mom}(u)+ g_{\rm non-analytic}(u)  \ssp , \\
&& g_{\rm mom}(u) := \sum_{n=0}^{\infty} \frac{(-u)^n}{n!} \frac{\Gamma(\gamma-n)}{\Gamma(\gamma)}  \ssp ,  \nn \\
&& g_{\rm non-analytic}(u) := \sum_{n=0}^{\infty} \frac{(-1)^n}{n!} u^{\gamma + n} \frac{\Gamma(-\gamma-n)}{\Gamma(\gamma)}  \ssp .\nn 
\eea
The Laplace Transform thus admits a non-analytic series expansion. Note that the analytic part of the series expansion $g_{\rm mom}(u)$ can also be obtained by naively (and wrongfully) inverting the series expansion of the exponential and the average over disorder in (\ref{Eq:PresLG:3}) and using (\ref{Eq:PresLG:2}) $\forall n$. The question now is {\it how one can guess $g(u)$ from the sole knowledge of $g_{\rm mom}(u)$?}. The answer is: wrongfully (again) apply a Mellin-Barnes transform using the proper analytical continuation of the coefficients of the series expansion of $g_{\rm mom}(u)$ to $n \in \mathbb{C}$:
\be \label{Eq:PresLG:7}
g_{\rm mom}(u) = \sum_{n=0}^{\infty} \frac{(-u)^n}{\Gamma(1+n)} \frac{\Gamma(\gamma-n)}{\Gamma(\gamma)} \to  \frac{-1}{2 i \pi} \int_{s \in {\cal C}} \frac{\pi ds}{\sin(\pi s)} u^s \frac{\Gamma(\gamma-s)}{\Gamma(\gamma) \Gamma(1+s)} = g(u) \ssp .
\ee
Note that there is a single analytical continuation that provides the right answer and it has to be guessed. This `trick' gives us hope to solve the problem for the Log-Gamma polymer as follows: (i) compute the `moments'  $\overline{(Z_t(\hat x))^n}$ $\forall n$ using BA, as if the moments of $W_t(\hat x)$ were given by the right hand side of (\ref{Eq:PresLG:2}) $\forall n$; (ii) compute the `moment generating function' $g_{t,x}^{{\rm mom}}(u) = \sum_{n=0}^{\infty} \overline{Z_t(\hat x)^n}$; (iii) perform an ``illegal'' Mellin-Barnes transform on $g_{t,x}^{{\rm mom}}(u)$ using an analytical continuation and obtain a function $g_{t,x}(u)$; (iv) check that $g_{t,x}(u)$ is indeed the Laplace transform of $Z_t(\hat x)$ for low values of $t$ and $\hat x$ and hope that it holds $\forall (t, \hat x)$.\\

{\it Bethe-Brunet ansatz} \\
Introducing the mean value of the Boltzmann weights $w_0 = \overline{W_{t}(\hat x)} = \frac{1}{\gamma-1} $
we consider the `wave-function' (denoting from now on $\hat x$ by $x$ as in \cite{ThieryLeDoussal2014})
\bea \label{Eq:PresLG:8}
\psi_t( x_1 , \dots ,  x_n) = \frac{1}{(2 )^{nt} (w_0)^{n(t+1)}} \overline{Z_t(x_1) \cdots Z_t( x_n)} \ssp .
\eea
The normalization of the wave-function ensures an easy comparison with the continuum DP case (see (\ref{Eq:ChapIII:Sec2:Stat18})) that is also discussed independently in the paper. Defining
\bea \label{Eq:PresLG:9}
h_n = \frac{\overline{(W_{t}( x))^n}}{w_0^n} \ssp ,
\eea
the transfer matrix is obtained as
\bea \label{Eq:PresLG:10}
&& \psi_{t+1}(x_i) = (T_n \psi_t)( x_i) = \frac{1}{2^n} a_{x_1 , \cdots ,x_n}  \sum_{(\delta_1,\cdots,\delta_n) \in \{-\frac{1}{2},\frac{1}{2}\}^n} \psi_t(x_1 - \delta_1,\cdots,x_n - \delta_n) \nn \\
&& a_{x_1 , \cdots,x_n}   = \prod_{x} h_{ \sum_{\alpha=1}^n \delta_{x,x_\alpha}} \ssp.
\eea
As shown by Éric Brunet, the symmetric eigenfunction of $T_n$
\bea \label{Eq:PresLG:11}
T_n \psi_\mu = \theta_\mu \psi_{\mu}  \ssp ,
\eea
takes a Bethe-Ansatz form: for $x_1 \leq \cdots \leq x_n$, $\psi_{\mu}(x_1 , \cdots , x_n) = \tilde{\psi}_{\mu}(x_1, \cdots , x_n)$ with
\bea \label{Eq:PresLG:12}
&& \tilde{\psi}_\mu(x_1,\cdots,x_n) = \sum_{\sigma \in S_n} A_\sigma \prod_{\alpha=1}^n z_{\sigma(\alpha)}^{x_\alpha}  \hspace{0.3 cm},\hspace{0.3 cm}  A_\sigma = \prod_{1 \leq \alpha < \beta \leq n } \left[1+ \frac{\bar{c}}{2(t_{\sigma(\alpha)} - t_{\sigma(\beta)})} \right] \ssp , \nn \\
&& z_\alpha = e^{ i \lambda_\alpha}   \hspace{0.3 cm},\hspace{0.3 cm} t_\alpha = i \tan (\frac{\lambda_\alpha}{2}) =\frac{z_\alpha-1}{z_\alpha+1}  \ssp ,
\eea
and 
\bea \label{Eq:PresLG:13}
\theta_\mu = \prod_{i=1}^{n} z_\alpha^{\frac{1}{2}} \frac{1+ z_\alpha^{-1}}{2}  \ssp .
\eea
The form (\ref{Eq:PresLG:12}) should be compared with (\ref{Eq:ChapIII:Sec2:BALL:14}). In an appropriate scaling limit (the $\gamma \to \infty$ weak noise limit, see Sec.~\ref{subsec:ChapIII:SecII:StatMeas}) (\ref{Eq:PresLG:12}) converges to (\ref{Eq:ChapIII:Sec2:BALL:14}). Imposing periodic boundary conditions on a line of length $L$ (immaterial in the computation of a  moment as long as $t \geq L$) one obtains the following Bethe equations
\be \label{Eq:PresLG:14}
e^{i \lambda_{\alpha}L} = \prod_{1 \leq \beta \leq n, \beta \neq \alpha} \frac{2 t_\alpha- 2 t_\beta+\bar{c}}{2 t_\alpha-2 t_\beta-\bar{c}} =  \prod_{1 \leq \beta \leq n, \beta \neq \alpha} 
 \frac{2 \tan (\frac{\lambda_\alpha}{2})-2 \tan (\frac{\lambda_\beta}{2})- i \bar{c} }{2 \tan ( \frac{\lambda_\alpha}{2})-2 \tan (\frac{\lambda_\beta}{2}) + i \bar{c}} 
\ee
which should be compared with (\ref{Eq:ChapIII:Sec2:BALLString:2}).\\

{\it Symmetric transfer matrix, weighted scalar product and norm formula} \\
The above transfer matrix is not symmetric and the eigenfunctions (\ref{Eq:PresLG:12}) are not orthogonal with respect to the canonical scalar product on $(\JZ/(L \JZ))^n$. In \cite{ThieryLeDoussal2014} we argue that the eigenfunctions (\ref{Eq:PresLG:12}) are orthogonal with respect to the following weighted scalar product:
\begin{equation} \label{Eq:PresLG:15}
\langle \phi , \psi \rangle = \sum_{ (x_1 , \cdots , x_n ) \in  \{0,\cdots, L-1\}^n } \frac{1}{a_{x_1,\cdots,x_n}} \phi^*(x_1,\cdots,x_n)  \psi(x_1,\cdots,x_n)
\end{equation}
We then conjecture a generalization of the Gaudin formula for the norm of the eigenfunctions (see \cite{ThieryLeDoussal2014} for some checks)
\begin{equation} \label{Eq:PresLG:16}
||\psi_{\mu}||^2 := \langle \psi_{\mu} , \psi_{\mu} \rangle  = n!  \prod_{1 \leq \alpha < \beta \leq n }  \frac{   (2 t_{\alpha} -2  t_{\beta} ) ^2 - \bar{c}^2  }   {   (2 t_{\alpha} - 2 t_{\beta} ) ^2  } \det{ G }
\end{equation}
with
\bea \label{Eq:PresLG:17}
&& G_{\alpha \beta} = \delta_{\alpha  \beta} \left( L + (1 - t_{\alpha}^2) \sum_{\gamma=1}^n 
\tilde K (t_{\alpha} - t_{\gamma}) \right) - (1- t_{\beta}^2)  \tilde K ( t_{\alpha} - t_{\beta})  \nn \\
&& \tilde K (t)  = \frac{- 2 \bar{c} }{ - 4 t^2 + \bar{c}^2}  \ssp .
\eea

This formula should be compared with (\ref{Eq:ChapIII:Sec2:BALLNorm:1}). It is rather remarkable since the Gaudin formula does not a priori seem to know that we have defined the weighted scalar product (\ref{Eq:PresLG:15}).\\

\stab {\it The large $L$ limit and the string solution} \\
We assume $\bar c >0$ (i.e. $\gamma>1$, which is true if the first moment of the partition sum is well defined so that it makes sense to use the Bethe ansatz at least for small $n$). We then argue in \cite{ThieryLeDoussal2014} that the Bethe equations are solved in the large $L$ limit similarly as in the continuum case (see Sec.~\ref{subsec:ChapIII:SecII:BA}). Namely, a general eigenstate is given by partitioning $n$ into $n_s$ strings, each string containing $m_j$ particles where the index $j=1,\cdots,n_s$ labels the string and 
\begin{equation} \label{Eq:PresLG:19}
t_\alpha = t_{j,a} = i \frac{ k_j}{2} + \frac{\bar{c}}{4} ( m_j + 1 - 2a ) +  \frac{\delta_{j,a}}{2}  \ssp ,
\end{equation}
where we introduce an index $a=1,\cdots, m_j$ that labels the rapidity inside a string, and 
$\delta_{j,a}$ are deviations that fall off exponentially with $L$. This formula should be compared with (\ref{Eq:ChapIII:Sec2:BALLString:3}). To compute the norm of the string states, we adapt the derivation of the Calabrese-Caux formula \cite{CalabreseCaux2007} to our formula (\ref{Eq:PresLG:16}) and obtain
\be  \label{Eq:PresLG:20}
||\psi_{\mu}||^2 = n! L^{n_s}   \prod_{1\leq i < j  \leq n_s} \frac{4 (k_i-k_j)^2 + \bar{c}^2 (m_i + m_j)^2}{4(k_i-k_j)^2 + \bar{c}^2 (m_i - m_j)^2} \prod_{j=1}^{n_s} [ \frac{m_j}{ \bar{c}^{m_j-1} } (\sum_{a=1}^{m_j}  \frac{1}{1-t_{j,a}^2}) \prod_{b=1}^{m_j} (1-t_{j,b}^2) ] \ssp ,
\ee
which should be compared with (\ref{Eq:ChapIII:Sec2:BALLNorm:3}). Note that as in the LL case the norm is almost a determinant. Here however, the additional factor $\sum_{a=1}^{m_j}  \frac{1}{1-t_{j,a}^2}$ which comes from the careful large $L$ analysis of (\ref{Eq:PresLG:16}) spoils the algebraic structure\footnote{It could obviously be integrated into a determinant, but not in a symmetric form.}. `Luckily' it will be canceled out in the final calculation by a factor coming from the phase space: we argue that the sum over eigenstates can be computed as
\begin{equation} \label{Eq:PresLG:21}
\sum_{\mu} =  \sum_{n_s=1}^{n}  \frac{1}{n_s!} \sum_{(m_1 , \cdots , m_{n_s})_n} \prod_{j=1}^{n_s} \int_{\JR}  \frac{L dk_j}{2 \pi} \sum_{a=1}^{m_j} \frac{1}{ 1 - t_{j,a}^2} \ssp ,
\end{equation} 
which should be compared with (\ref{Eq:ChapIII:Sec2:BALLString:8}). Finally noting that the contribution to the eigenvalue associated with the unit translation in time and on the lattice $\JZ$ are $ \theta_{\mu} = \prod_{j=1}^{n_s} \theta_{m_j,k_j}$ with
\begin{eqnarray} \label{Eq:PresLG:22}
 \theta_{m_j,k_j}  = \left(\frac{2}{ \bar{c}} \right)^{ m_j}  \left(\frac{  \Gamma(-\frac{m_j}{2} + \frac{ \gamma}{2} - i\frac{k_j}{\bar{c} } ) \Gamma(-\frac{m_j}{2} + \frac{ \gamma}{2} + i\frac{k_j}{\bar{c} })}{  \Gamma(\frac{m_j}{2} + \frac{ \gamma}{2} -i\frac{k_j}{\bar{c} } ) \Gamma(\frac{m_j}{2} + \frac{ \gamma}{2} + i\frac{k_j}{\bar{c} }) } \right)^{\frac{1}{2}} \ssp ,
\end{eqnarray}
and $\prod_{\alpha} z_\alpha = \prod_{j=1}^{n_s} \prod_{a=1}^{m_j} \frac{1+t_{j,a}}{1-t_{j,a}}$ with
\begin{equation} \label{Eq:PresLG:23}
\prod_{a=1}^{m_j}  \frac{1+t_{j,a}}{1-t_{j,a}} =\frac{ \Gamma( -\frac{m_j}{2} + \frac{\gamma}{2} - i\frac{k_j}{\bar{c} }) \Gamma( \frac{m_j}{2} + \frac{\gamma}{2} +i\frac{k_j}{\bar{c} })}{\Gamma( \frac{m_j}{2} + \frac{\gamma}{2} - i\frac{k_j}{\bar{c} }) \Gamma( -\frac{m_j}{2} + \frac{\gamma}{2} + i\frac{k_j}{\bar{c} })}  \ssp .
\end{equation} 

{\it An exact formula for the moments of the Log-Gamma polymer} \\
Combining the precedent results, we obtain, for $\gamma < n$:
\bea  \label{Eq:PresLG:24}
&&\!\!\!\!\!\!\!\!\!\!\!\!\!\!\!\!\!\!  \overline{Z_t(x)^n} =  n!  \sum_{n_s=1}^n  \frac{1}{n_s!} \sum_{(m_1,..m_{n_s})_n} 
\prod_{j=1}^{n_s}  \int_{-
 \infty}^{+\infty} \frac{dk_j}{2 \pi} 
\prod_{1\leq i < j  \leq n_s} \frac{4(k_i-k_j)^2 +  (m_i - m_j)^2}{4(k_i-k_j)^2 + (m_i + m_j)^2} \nn  \\
&&
\prod_{j=1}^{n_s} \frac{1 }{m_j } 
\left( \frac{  \Gamma(-\frac{m_j}{2} + \frac{ \gamma}{2} - i k_j ) }{  \Gamma(\frac{m_j}{2} + \frac{ \gamma}{2} - i k_j )  } \right)^{\frac{t}{2} +1 +x} \left( \frac{   \Gamma(-\frac{m_j}{2} + \frac{ \gamma}{2} +i k_j )}{ \Gamma(\frac{m_j}{2} + \frac{ \gamma}{2} + i k_j ) } \right)^{\frac{t}{2} +1-x} \ssp ,
\eea 
which should be compared with (\ref{Eq:ChapIII:Sec2:BALLFinal:2}). Note that the right hand side of (\ref{Eq:PresLG:24}) makes sense $\forall n$. We can use this fact to perform the rest of the program announced earlier.\\

{\it Fredholm determinant formulae for the Laplace transform of the partition sum in the Log-Gamma polymer} \\
Let us first perform the step (ii): we compute
\bea \label{Eq:PresLG:25}
g_{t,x}^{{\rm mom}}(u) = \sum_{n \in \JN} \frac{(-u)^n}{n!}` \overline{Z_t(x)^n}  ` \ssp ,
\eea
where $`\overline{Z_t(x)^n}`$ denotes $\forall n$ the right hand side of (\ref{Eq:PresLG:24}). Adapting the route followed in the continuum DP case (see Sec.~\ref{subsec:ChapIII:SecII:BA} ) we obtain a Fredholm determinant formula for $g_{t,x}^{{\rm mom}}(u) $ as 
\begin{equation} \label{Eq:PresLG:26}
 g_{t,x}^{{\rm mom}}(u) = {\rm Det} \left( I  + K_{t,x}^{{\rm mom}} \right) \ssp ,
\end{equation}
with the kernel:
\begin{eqnarray} \label{Eq:PresLG:27}
 K_{t,x}^{{\rm mom}}(v_1,v_2) = &&   \sum_{m=1}^{\infty} \int_{-
 \infty}^{+\infty}   \frac{dk}{ \pi}  (-u)^m   e^{ -  2 i k(v_1-v_2) -  m (v_1+v_2) }   \\
  && \left( \frac{  \Gamma(-\frac{m}{2} + \frac{ \gamma}{2} - i k) }{  \Gamma(\frac{m}{2} + \frac{ \gamma}{2} - i k )  } \right)^{\frac{t}{2} +1+x}  \left( \frac{   \Gamma(-\frac{m}{2} + \frac{ \gamma}{2} +i k )}{ \Gamma(\frac{m}{2} + \frac{ \gamma}{2} + i k ) } \right)^{\frac{t}{2} +1-x}  \nonumber
\end{eqnarray}
and $ K_{t,x}^{{\rm mom}} : L^2 ( \mathbb{R}_+) \to L^2 ( \mathbb{R}_+) $. Performing now the step (iii), we conjecture a formula for the Laplace transform of the Log-Gamma polymer by changing the sum over $m$ in the above kernel to an integral on the complex plane as for a Mellin-Barnes transform. We obtain
\begin{equation} \label{Eq:PresLG:28}
g_{t,x}(u) = \overline{ \exp{ - u Z_t(x) } }= {\rm Det} \left( I + K_{t,x} \right)
\end{equation}
with
\begin{eqnarray} \label{Eq:PresLG:29}
 K_{t,x}(v_1,v_2) = && \int_{-\infty}^{+\infty}   \frac{dk}{ \pi}  \frac{-1}{2i} \int_C \frac{ds}{ \sin( \pi s ) }   u^s  e^{ -  2 i k(v_1-v_2) -  s (v_1+v_2) } \\
 &&  \left( \frac{  \Gamma(-\frac{s}{2} + \frac{ \gamma}{2} - i k ) }{  \Gamma(\frac{s}{2} + \frac{ \gamma}{2} - i k )  } \right)^{\frac{t}{2} +1+x} \left( \frac{   \Gamma(-\frac{s}{2} + \frac{ \gamma}{2} +i k )}{ \Gamma(\frac{s}{2} + \frac{ \gamma}{2} + i k ) } \right)^{\frac{t}{2} +1-x} \nonumber \ssp ,
\end{eqnarray}
where ${\cal C} = a + i \JR$ with $0<a<{\rm min}(1,\gamma)$ (note that the sum over $m$ in (\ref{Eq:PresLG:27}) starts at $m=1$) and $ K_{t,x}: L^2 ( \mathbb{R}_+) \to L^2 ( \mathbb{R}_+)$.\\

{\it KPZUC in the Log-Gamma polymer} \\
Performing the asymptotic analysis of (\ref{Eq:PresLG:29}) we finally obtain
\begin{equation}
\lim_{t \to \infty} Prob\left( \frac{ \log Z_t( \varphi t) + tc_{\varphi}}{\lambda_{\varphi} } <2^{\frac{2}{3}} z \right) = F_2(z)
\end{equation}
where $F_2(z)$ is the standard GUE Tracy-Widom cumulative distribution function, and the (angle-dependent)
constants are determined by the system of equations:
\begin{eqnarray}
&&0=(\frac{1}{2} + \varphi) \psi'(\frac{\gamma}{2} - k_\varphi)-(\frac{1}{2} - \varphi) \psi' (\frac{\gamma}{2} +k_\varphi )\\
&&c_{\varphi}= (\frac{1}{2} + \varphi) \psi(\frac{\gamma}{2} - k_\varphi)+(\frac{1}{2} - \varphi) \psi (\frac{\gamma}{2} +k_\varphi )\\
&&\lambda_{\varphi}=\left( -\frac{t}{8} \left( (\frac{1}{2} + \varphi) \psi''(\frac{\gamma}{2} - k_\varphi)+(\frac{1}{2} - \varphi) \psi''(\frac{\gamma}{2} +k_\varphi ) \right) \right)^{\frac{1}{3}} \ssp .
\end{eqnarray}
We recall that $\psi= \Gamma'/\Gamma$ is the diGamma function. Here $k_{\varphi}$, which is implicitly defined by the first equation, encodes the position of the saddle-point at $(s,k)=(0,k_{\varphi})$ in the kernel (\ref{Eq:PresLG:29}). This formula reproduces the results obtained from the gRSK correspondence in \cite{BorodinCorwinRemenik2013} for $\varphi= \varphi^* = 0$, that is $c_{\varphi^*} = \psi(\gamma/2)$ and $\lambda_{\varphi^*}= (\frac{-t}{8} \psi''(\gamma/2))^{1/3}$, and generalize it to arbitrary angles. It is successfully confronted to numerical simulations in \cite{ThieryLeDoussal2014}. \\

Other results contained in \cite{ThieryLeDoussal2014} are (i) additional formulae for the PDF of $Z_t(x)$ at any $t,x$ as differences of two Fredholm determinant formulae; (ii) additional Fredholm determinant formulae for $g_{t,x}(u)$ closer to those usually encountered in the mathematical literature; (iii) the comparison at each step of our Bethe ansatz approach with the BA approach to the continuum DP using the weak-universality of the continuum DP and thus the convergence of our results to the continuum case; (iv) the study of the limit to the semi-discrete DP; (v) many checks of the above formulae.

\subsection{Presentation of the main results of \cite{ThieryLeDoussal2015}} \label{subsec:PresIB}

 \stab {\it Classification of BA solvable models of a DP on $\JZ^2$: step 1} \\
The BA solvability of the Log-Gamma polymer is not an accident and comes from the algebraic structure of the model, more precisely as we will see below, it comes from the structure of its moments (\ref{Eq:PresLG:2}) that satisfy the simple recursion relation $\overline{(W_t(\hat x))^{n+1}}= \overline{(W_t(\hat x))^{n}}/(\gamma-n-1)$. It is a natural question to understand whether or not the Inverse-Gamma distribution is the only distribution that permits BA solvability. For this purpose in \cite{ThieryLeDoussal2015} we investigated the conditions of BA solvability of a DP at finite temperature on $\JZ^2$\footnote{By BA solvability here we mean BA solvability of the moments problem. There could be other types of BA solvability. For example the partition sum of the Log-Gamma polymer can be obtained as the limit of an observable of a BA solvable interacting particle system on $\JZ$, the q-Push TASEP \cite{MateevPetrov2015}. This BA solvability does not seem trivially related to the one studied in  \cite{ThieryLeDoussal2014}.}. For this purpose we consider a `general' model of DP on $\JZ^2$ where (see Fig.~\ref{fig:PresIB1}) (i) the Boltzmann weights (BW) live on the edges; (ii) the couple of BWs leading to the vertex $(t,x)$ is noted $(u_{t,x} , v_{t,x})$ where $u_{t,x}$ is the BW on the vertical edge and $v_{t,x}$ is the BW on the horizontal edge; (iii) BWs leading to different vertices are not correlated; (iv) the BWs are homogeneously distributed as $(u_{t,x} , v_{t,x}) \sim (u,v)$ with $(u,v) \in \JR_+^2$ two (a priori correlated) positive RVs. Note that this class of models contains models with on-site Boltzmann weights since $u_{t,x}$ and $v_{t,x}$ can be correlated. For example the Log-Gamma case is reproduced taking $u_{t,x} =v_{t,x}\sim Gamma(\gamma)^{-1}$. The extent of the `generality' of this class of models is precisely the one that allows us to perform the classification (see below). Other models outside this class could be considered and are not covered by our results.

\begin{figure}
\centerline{\includegraphics[width=7.0cm]{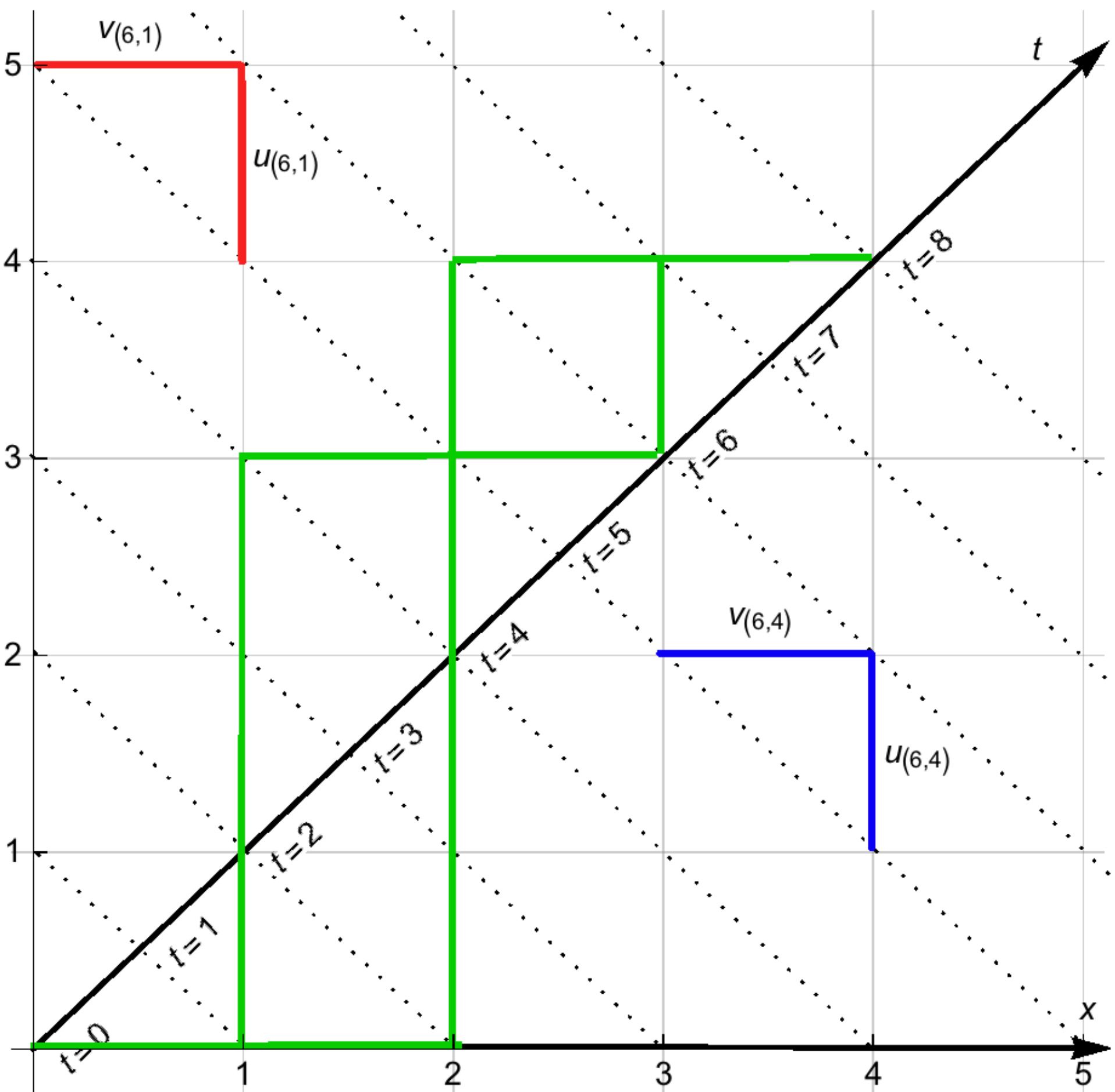}} 
\caption{General scheme for the models of directed polymer in the classification of \cite{ThieryLeDoussal2015}. Blue (resp. Red) : couple of correlated Boltzmann weight on edges arriving at $(t=6,x=1)$ (resp. $(t=6,x=4)$). Green: two admissible (i.e. up/right) paths for polymers with starting point $(0,0)$ and endpoint $(8,4)$. Figure taken from \cite{ThieryLeDoussal2015}.}
\label{fig:PresIB1}
\end{figure}

 In this framework the point-to-point partition sum satisfies the recursion relation
\bea \label{Eq:PresIB:1}
&& Z_{t = 0}(x) = \delta_{x , 0} \nn \\
&& Z_{t+1}(x) = u_{t+1 , x} Z_t(x) +  v_{t+1 , x} Z_t(x-1)  \ .
\eea
And this can be translated to a recursive (i.e. transfer matrix) equation for $\psi_t$:
\bea  \label{Eq:PresIB:2}
&& \psi_{t=0}(x_1 \dots , x_n ) = \delta_{x_1  ,0} \dots \delta_{x_n  ,0} \nn \\
&& \psi_{t+1}(x_1 \dots , x_n ) = \sum_{ \{ \delta_1 , \cdots , \delta_n \}  \in \{ 0 ,1 \}^n } a_{x_1 , \cdots , x_n}^{\delta_1 , \cdots , \delta_n} \psi_{t}(x_1 - \delta_1, \cdots , x_n - \delta_n)  = (T_n \psi_t) (x_1 \dots , x_n ) \nn \\
&& a_{x_1 , \cdots , x_n}^{\delta_1 , \cdots , \delta_n} = \prod_{y \in \mathbb{Z}}\overline{ (u)^{\sum_{i=1}^n \delta_{x_i,y} \delta_{\delta_i ,0}} (v)^{\sum_{i=1}^n \delta_{x_i,y} \delta_{\delta_i ,1}} } \ .
\eea
The latter generalizes the recursion equation of the Log-Gamma case (\ref{Eq:PresLG:10}) and we are interested in models for which all the symmetric eigenfunctions of the transfer matrix
\bea \label{Eq:PresIB:3}
T_n \psi_\mu = \Lambda_{\mu} \psi_\mu
\eea
can be obtained in the BA form
\bea  \label{Eq:PresIB:4}
&& \psi_{\mu}(x_1 , \dots , x_n) = \tilde{\psi}_{\mu}(x_1 , \dots ,x_n) \text{     if    } x_1 \leq \dots \leq x_n  \ssp , \nn \\
&& \tilde \psi_{\mu}(x_1, \dots , x_n) =\sum_{\sigma \in S_n} A_{\sigma} \prod_{i=1}^n z_{\sigma(i)}^{x_i}  \ssp .
\eea 
The class of models we consider is precisely the one that makes (\ref{Eq:PresIB:2}) similar to the recursion equation for the PDF for the position of $n$ particles moving on $\JZ$ with a ZRP (with parallel updates) type dynamics. In this interpretation the transfer matrix must be stochastic (i.e. conserve the probability) and the recursion equation (\ref{Eq:PresIB:2}) is often called the Master equation. In a recent work \cite{Povolotsky2013}, Povolotsky managed to classify all the ZRP with parallel updates for which the transfer matrix $T_n$ can be diagonalized by the Bethe ansatz. This construction precisely encodes the fact that `the $n$ particle problem must be reminiscent of the $2$ particle problem'\footnote{A feature that was not present in the continuum DP model where there are only two-bodies interactions.} in the form of a deformed Binomial formula for non-commutative variables. For our purpose this classification must be slightly adapted since we are not a priori only interested in stochastic transfer matrix $T_n$. We will not repeat here the arguments that lead to the classification (see \cite{ThieryLeDoussal2015}), but only give the result: the spectral problem (\ref{Eq:PresIB:3}) is solved by the BA (\ref{Eq:PresIB:4}) iff the moments take the form
\bea \label{Eq:PresIB:5b}
\overline{u^{n_1} v^{n_2} } = (\epsilon_1)^{n_1}(\epsilon_2)^{n_2}  \frac{(\frac{\nu}{\mu};q)_{n_1} (\mu;q)_{n_2}}{(\nu;q)_{n_1+n_2}}  \frac{(q;q)_{n_1+n_2} }{(q;q)_{n_1} (q;q)_{{n_2}}}  \frac{1}{ C^{n_1}_{n_1+n_2} }  \ssp ,
\eea
with $(\epsilon_1, \epsilon_2 , q , \mu , \nu) \in \JR^5$ and $(a;q)_n = \prod_{k=0}^{n-1} (1 - a q^k)$. In this case the symmetric eigenfunctions of $T_n$ are obtained as (\ref{Eq:PresIB:4}) with the condition (which can be solved)
\bea  \label{Eq:PresIB:6b}
\frac{A_{\sigma \circ \tau_{ij}}}{A_{\sigma }} = -\frac{{\sf c}+ {\sf b} z_{\sigma(j)}+ {\sf a} z_{\sigma(i)} z_{\sigma(j)} - z_{\sigma(i)}}{{\sf c} + {\sf b} z_{\sigma(i)}+ {\sf a} z_{\sigma(i)} z_{\sigma(j)} - z_{\sigma(j)}} \ .
\eea
with 
\bea \label{Eq:PresIB:7b}
{\sf a} =  \frac{\overline{u^2} - (\overline{u})^2}{ (\overline{u})(\overline{v})}  \quad {\sf b} =   \frac{2 \overline{ u v} -(\overline{u})( \overline{v} ) }{(\overline{u} )( \overline{v} )} \quad {\sf c} = \frac{\overline{v^2} - (\overline{v})^2}{ ( \overline{u} )(\overline{v})}  \ssp .
\eea

{\it Classification of BA solvable models of DP on $\JZ^2$: step 2} \\
It remains to understand whether or not (\ref{Eq:PresIB:5b}) are indeed the moments of positive random variables for some choice of the parameters. In (\cite{ThieryLeDoussal2015}) we consider, for $(\epsilon_1, \epsilon_2 , q , \mu , \nu) \in \JR^5$  fixed and $x \in \JR$
\bea \label{Eq:PresIB:8b}
P(x) = \overline{(u + x v)^2}^c \ssp .
\eea
If (\ref{Eq:PresIB:5b}) are moments of real variables, then $P(x)$ must be positive $\forall x \in \JR$ (and maybe $0$ at some $x_c$ if $u$ and $v$ are correlated as $u+x_c v = 0$). $P(x)$ being a degree $2$ polynomial, its root can easily be studied and we arrive to the conclusion that the only possibility for (\ref{Eq:PresIB:5b}) to be the moments of real variables is to consider the degenerate limit
\bea \label{Eq:PresIB:9b}
\nu = q^{\alpha + \beta} \quad ,  \quad \mu = q^{\beta} \quad,  \quad q \to 1 \ssp .
\eea
Taking this limit using $(q^a;q)_n \simeq_{q \to 1} (1-q)^n (a)_n$ (where $(a)_n=\prod_{k=0}^{n-1}(a+k)$) we thus restricted our search for integrable models of DP to those with moments as
\bea \label{Eq:PresIB:5}
\overline{u^{n_1} v^{n_2} } = (\epsilon_1)^{n_1}(\epsilon_2)^{n_2}  \frac{(\alpha)_{n_1} (\beta)_{n_2}}{ (\alpha + \beta)_{n_1+n_2}} \ssp .
\eea
A combinatorial identity shows that in this case $\forall n$, $\overline{(u/\epsilon_1 + v/\epsilon_2)^n} = 1$ and the BWs are thus strongly correlated as $u/\epsilon_1 + v/\epsilon_2 = 1$. At this point we cannot obtain more general results and only exhibit models for which the moments are given by (\ref{Eq:PresIB:5}). We consider both (i) models for which the moments of the BWs are indeed given by (\ref{Eq:PresIB:5}) $\forall (n_1,n_2)$; (ii) models for which only a finite number of moments exist, namely for $n_1+ n_2 \leq n_{\rm max}$ with some $n_{\rm max} \in \JN^*$. In the second case (inspired by the Log-Gamma case) the BA a priori only allows us to compute the first $n_{\rm max}$ moments of the partition sum. First, this classification indeed contains the Log-Gamma case as a degenerate limit: taking $(\epsilon_1 , \epsilon_2) = (+1 , -1)$, $\alpha + \beta = 1 - \gamma$ with $\gamma \geq 0$, $\beta \to \infty$ and rescaling the Boltzmann weights as (implying a corresponding rescaling of the partition sum) $u =  \beta u^{LG}$, $v = \beta  v^{LG} $ we obtain
\be \label{Eq:PresIB:6}
((u^{LG})^{n_1} ,(v^{LG})^{n_2} ) = \lim_{\beta \to \infty} \frac{1}{\beta^{n_1+n_2}} (-1)^{n_2}  \frac{(1- \gamma - \beta)_{n_1} (\beta)_{n_2}}{ (1 - \gamma)_{n_1+n_2}} = \frac{(-1)^{n_1+n_2}}{(1- \gamma)_{n_1+n_2}} \ssp ,
\ee
which indeed reproduces the moments of the Log-Gamma polymer (\ref{Eq:PresLG:2}). In between our work on the Log-Gamma polymer \cite{ThieryLeDoussal2014} and this work \cite{ThieryLeDoussal2015}, two new exactly solvable models of DP on $\JZ^2$ were obtained: the Strict-Weak (SW) polymer and the Beta polymer. Both were shown to be solvable by BA in \cite{CorwinSeappalainenShen2015} and \cite{BarraquandCorwin2015}. In the SW case Boltzmann weights are given by $v^{SW}= 1$ and $u^{SW} \sim Gamma(\alpha)$. It is obtained from our framework by taking the limit $(\epsilon_1 , \epsilon_2)=(1,1)$ and $\beta \to \infty$ with $\alpha >0$ fixed and $u^{SW} = \beta u$, $v^{SW} = v$. Indeed in this limit the moments are:
\bea \label{Eq:PresIB:7}
((u^{SW})^{n_1} ,(v^{SW})^{n_2} ) = \lim_{\beta \to \infty}  \beta^{n_1} \frac{(\alpha)_{n_1} (\beta)_{n_2}}{ (\alpha + \beta)_{n_1+n_2}} =  (\alpha)_{n_1} \ssp .
\eea
This corresponds to the distribution described above. In the Beta polymer the Boltzmann weights are distributed as $u^{B}+v^{B} =1$ and $u \in [0,1]$ is a Beta random variable with parameters $(\alpha , \beta)$. Its PDF is
\bea \label{Eq:PresIB:8}
u \sim Beta(\alpha , \beta) \Longleftrightarrow p(u) = \frac{\Gamma(\alpha + \beta)}{\Gamma(\alpha) \Gamma(\beta)} u^{\alpha-1} (1-u)^{\beta -1} \ssp .
\eea
and the moments are as above with $(\epsilon_1 , \epsilon_2) = (1,1)$. We will come back to this model in the next section. Finally in \cite{ThieryLeDoussal2015} we introduce a new exactly solvable model of DPs on $\JZ^2$, the Inverse-Beta polymer. In this model $(\epsilon_1 , \epsilon_2) =(1,-1)$, $v = u-1$ and $u$ is distributed as the inverse of a Beta random variable: $u \sim Beta(\gamma , \beta)^{-1} > 1$. The moments are as above with $\alpha = 1-\beta-\gamma$. In the limit $\beta \to \infty$, this model converges to the Log-Gamma polymer, while in the limit $\gamma \to \infty$ it converges to the Strict-Weak model (this limit is different from the one mentioned above, which corresponds to the degeneration from the Beta to the Strict-Weak polymer). In \cite{ThieryLeDoussal2015} we attempted a more systematic study of possible models of DPs, but it remains inconclusive. In any case, for now, all exactly solvable models of DPs on $\JZ^2$ are thus BA solvable and can be regrouped using our notations as in Fig.~\ref{fig:PresIB2}. Note that in Fig.~\ref{fig:PresIB2} we also include the symmetrized version (with respect to the diagonal of $\JZ^2$) of the Inverse-Beta and Strict-Weak polymer which are both anisotropic models that favor one edge. The rest of this section is devoted to the Inverse-Beta polymer, which thus generalizes both the Log-Gamma and Strict-Weak polymer.\\

\begin{figure}
\centerline{\includegraphics[width=8.0cm]{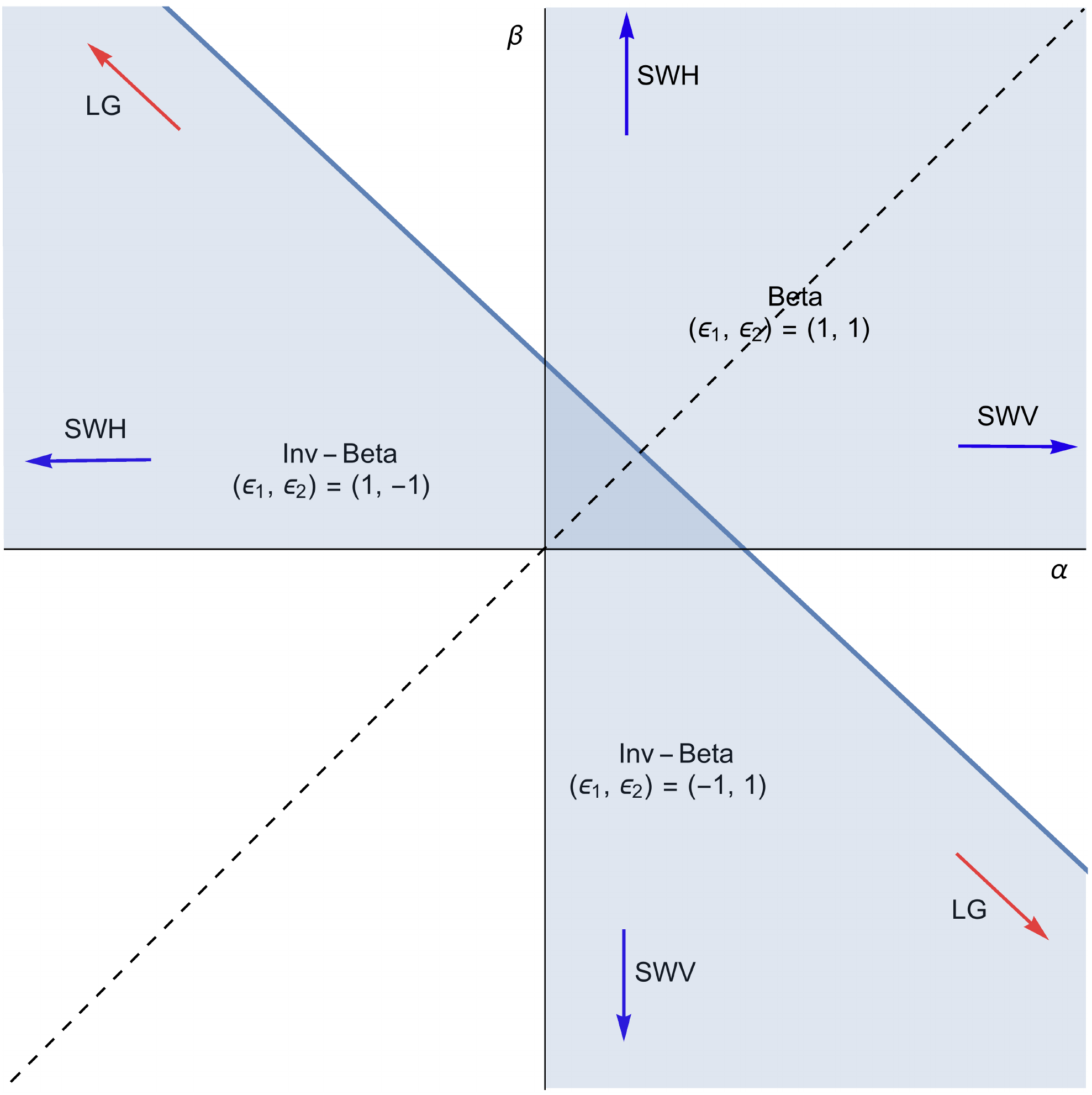}} 
\caption{Classification of exactly solvable finite temperature models of DPs on $\JZ^2$ following the notations of \cite{ThieryLeDoussal2015}. The dashed line represents the axis of symmetry $\alpha \leftrightarrow \beta$, or equivalently the symmetry between vertical and horizontal edges. The blue line indicates the line $\alpha + \beta =1$ or equivalently $\gamma = 1- (\alpha + \beta )= 0$. Limiting polymer models are indicated by red arrows for the log-Gamma (LG) and blue arrows for the Strict-Weak (with weights either on horizontal edges (SWH) or vertical edges (SWV) ).  We also emphasize the values of $(\epsilon_1 , \epsilon_2)$ which corresponds to the polymer considered. Notice that the region $\alpha>0$, $\beta>0$ and $\gamma<1$ is a region of coexistence of the Inverse-Beta and the Beta polymer, only distinguished by the value of $(\epsilon_1 , \epsilon_2)$. Figure taken from \cite{ThieryLeDoussal2015}.}
\label{fig:PresIB2}
\end{figure}

{\it BA solution of the Inverse-Beta polymer} \\
In the Inverse-Beta polymer, the moments of the Boltzmann weights are well defined for $(s_1 , s_2) \in \mathbb{C}^2$ with $Re(s_1+s_2) \leq \gamma$ and $Re(s_2) > - \beta$ and are given by
\bea  \label{Eq:PresIB:9}
\overline{u^{s_1} v^{s_2}} =  \frac{\Gamma(\gamma+\beta) }{\Gamma(\gamma) \Gamma(\beta)} \frac{\Gamma(\gamma-s_1-s_2)  \Gamma(\beta + s_2) }{\Gamma(\gamma+ \beta -s_1)} \ssp .
\eea
The replica Bethe ansatz approach to this model suffers from exactly the same problem as the one for the Log-Gamma case: only a finite number of integer moments of the partition sum exist. It is however possible to use the same strategy as for the Log-Gamma case. The two approaches are actually remarkably similar: defining again $\bar c = 4/(\gamma- 1)$, we showed in \cite{ThieryLeDoussal2015} that {\it the eigenfunctions of the transfer matrix $T_n$ in the Log-Gamma and Inverse-Beta case are equal.} A simple incarnation of this remarkable property is that the quotient of two amplitudes of the Bethe wave-function in (\ref{Eq:PresIB:6b}), which controls completely the structure of the wave-function and the Bethe equations, does not depend on $\beta$. In particular they are equal to those in the Log-Gamma limit $\beta \to \infty$. We can thus use the same results as in the Log-Gamma case: string solution at large $L$, Gaudin formula... The only things that differs is the eigenvalue associated with the unit translation in time: the Inverse-Beta polymer is an anisotropic model with $v = u-1$, and the vertical direction is thus favored in this model. In \cite{ThieryLeDoussal2015} we showed that this change leads to the following formula for the integer moments of the DP partition sum:
\bea \label{Eq:PresIB:10}
&&  \!\!\!\!\!\!\!\!\! \overline{Z_t(x)^n} =  n! \frac{\Gamma(\gamma)}{\Gamma(\gamma-n)} \sum_{n_s=1}^n  \frac{1}{n_s!} \sum_{(m_1,..m_{n_s})_n} 
\prod_{j=1}^{n_s}  \int_{-
 \infty}^{+\infty} \frac{dk_j}{2 \pi}
\prod_{1\leq i < j  \leq n_s} \frac{4(k_i-k_j)^2 + (m_i - m_j)^2}{4(k_i-k_j)^2 + (m_i + m_j)^2}  \nn \\
&&
\!\!\!\!\!\!\! \prod_{j=1}^{n_s} \frac{1}{m_j} 
 \left( \frac{  \Gamma(-\frac{m_j}{2} + \frac{ \gamma}{2} - i k_j ) }{  \Gamma(\frac{m_j}{2} + \frac{ \gamma}{2} - i k_j )  } \right)^{1 +x} \left( \frac{   \Gamma(-\frac{m_j}{2} + \frac{ \gamma}{2} +i k_j )}{ \Gamma(\frac{m_j}{2} + \frac{ \gamma}{2} + i k_j ) } \right)^{ 1-x + t} \left( \frac{ \Gamma ( \beta +i k_j+\frac{\gamma }{2}+\frac{m_j}{2})}{\Gamma( \beta +i k_j+\frac{\gamma }{2}-\frac{m_j}{2}   )}  \right)^t  \ssp ,  \nn \\
\eea
which is valid for $n \leq \gamma$ and very similar to the corresponding formula for the Log-Gamma case (\ref{Eq:PresLG:24}). The first factor $\frac{\Gamma(\gamma)}{\Gamma(\gamma-n)}$, which comes out of the structure of the BA, forbids to express $\overline{e^{-u Z_t(x)}}$ as a Fredholm determinant. As in the Log-Gamma case we thus consider a partition sum with a BW added at the origin:
\bea \label{Eq:PresIB:10bis}
\tilde{Z}_t(x) = w_{00} Z_t(x) \ssp ,
\eea
where $w_{00} \sim Gamma(\gamma)^{-1}$ is independent of $Z_t(x)$. The moments of $\tilde{Z}_t(x)$ are given by (\ref{Eq:PresIB:10}), but without the factor $\frac{\Gamma(\gamma)}{\Gamma(\gamma-n)}$. Following the same route as in the Log-Gamma case we express the moment generating function of $\tilde{Z}_t(x)$ as a Fredholm determinant. Using a Mellin-Barnes transform inside the associated kernel, we conjecture a formula for the Laplace transform $g_{tx}(u) =  {\rm Det} \left( I + K_{tx} \right)$ with
\begin{eqnarray} \label{Eq:PresIB:11}
 K_{t,x}(v_1,v_2) = && \int_{-\infty}^{+\infty}   \frac{dk}{ \pi}  \frac{-1}{2i} \int_C \frac{ds}{ \sin( \pi s ) }   u^s  e^{ -  2 i k(v_1-v_2) -  s (v_1+v_2) } \\
 &&  \left( \frac{  \Gamma(-\frac{s}{2} + \frac{ \gamma}{2} - i k ) }{  \Gamma(\frac{s}{2} + \frac{ \gamma}{2} - i k )  } \right)^{1 +x} \left( \frac{   \Gamma(-\frac{s}{2} + \frac{ \gamma}{2} +i k )}{ \Gamma(\frac{s}{2} + \frac{ \gamma}{2} + i k ) } \right)^{ 1-x + t} \left( \frac{ \Gamma ( \beta +i k+\frac{\gamma }{2}+\frac{s}{2})}{\Gamma( \beta +i k+\frac{\gamma }{2}-\frac{s}{2})} \right)^t \nonumber ,
\end{eqnarray}
where $C = a + i \mathbb{R}$ with $0<a<{\rm min}(1,\gamma)$ and $K_{t,x} : L^2 ( \mathbb{R}_+) \to L^2 ( \mathbb{R}_+) $. Performing the asymptotic analysis of this formula we obtain TW-GUE fluctuations in the Inverse-Beta polymer:
\begin{equation}\label{Eq:PresIB:12}
\lim_{t \to \infty} Prob\left( \frac{ \log Z_t((1/2+ \varphi) t) + tc_{\varphi}}{\lambda_{\varphi} } <2^{\frac{2}{3}} z \right) = F_2(z) \ssp ,
\end{equation}
where the ($\varphi$-dependent)
constants are determined by the system of equations:
\begin{eqnarray} \label{Eq:PresIB:13}
&&0=(\frac{1}{2} + \varphi) \psi'(\frac{\gamma}{2} - k_\varphi)-(\frac{1}{2} - \varphi) \psi' (\frac{\gamma}{2} +k_\varphi ) + \psi'(\beta + \frac{\gamma}{2} + k_\varphi )  \ssp , \\
&&c_{\varphi}= (\frac{1}{2} + \varphi) \psi(\frac{\gamma}{2} - k_\varphi)+(\frac{1}{2} - \varphi) \psi (\frac{\gamma}{2} +k_\varphi ) - \psi( \beta + \frac{\gamma}{2} +  k_\varphi )  \ssp , \nn \\
&&\lambda_{\varphi}=\left( -\frac{t}{8} \left( (\frac{1}{2} + \varphi) \psi''(\frac{\gamma}{2} - k_\varphi)+(\frac{1}{2} - \varphi) \psi''(\frac{\gamma}{2} +k_\varphi )  - \psi''( \beta + \frac{\gamma}{2} + k_\varphi )   \right) \right)^{\frac{1}{3}}  \ssp . \nn \ .
\end{eqnarray}
 These cannot be solved in full generality, except for the optimal angle $\varphi^*$, defined by $\partial_{\varphi} c_{\varphi} = 0$ for which we find
\bea \label{Eq:PresIB:14}
&& \varphi* = -\frac{1}{2} \frac{ \psi'(\beta+\gamma/2)}{\psi'(\gamma/2)}  <0  \nn \\
&& c^* = c_{\varphi^*}  = \psi(\gamma/2) - \psi(\beta+ \gamma/2) \nn \\ 
&& \lambda_{\varphi*}  = \left(  \frac{t}{8} ( \psi''(\beta+\gamma/2) - \psi''(\gamma/2) ) \right)^{1/3}  \ssp .
\eea
The optimal angle is the angle of maximum probability chosen by the endpoint of the polymer with one end free to move on the line. Note that its value is non-trivial (it is different from the value expected in an averaged environment). This thus generalizes the explicit results obtained in the Log-Gamma case where $\varphi^*=0$ (note that here $\varphi^* \to_{\beta \to \infty} 0$ and $\varphi^* \to_{\gamma \to \infty} -1/2$, corresponding to the Log-Gamma and Strict-Weak limits). \\

{\it Zero-temperature limit of the Inverse-Beta polymer: Bernoulli-Exponential polymer}\\
As we saw in Sec.~\ref{subsec:ChapIII:SecII:OtherExact}, the $\gamma \to 0$ limit of the Log-Gamma polymer leads to LPP with exponential waiting times. Similarly, the Inverse-Beta polymer admits a zero temperature limit obtained by setting $\gamma = \epsilon \gamma'$, $\beta  = \epsilon \beta'$. In this limit one shows \cite{ThieryLeDoussal2015} that the rescaled random energies of the model $({\cal E}^u , {\cal E}^v) =(- \epsilon \log(u) , -\epsilon \log(v))$ converge in law to
\bea  \label{Eq:PresIB:16}
({\cal E}^u , {\cal E}^v) \sim_{\epsilon \to 0}  \left( - \zeta E_{\gamma'} , (1-\zeta) E_{\beta'} - \zeta E_{\gamma'} \right) = ({\cal E}'_u , {\cal E}'_v) ,
\eea
where $\zeta$ is a Bernoulli random variable of parameter $p=\beta'/(\gamma' + \beta')$, $E_{\gamma'}$ and $E_{\beta'}$ are exponential random variables of parameter $\gamma'>0$ and $\beta'>0$, independent of $\zeta$ (the PDF of exponential RVs was given in (\ref{Eq:ChapIII:Sec2:Stat23}) and the Bernoulli RV is by definition $1$ with probability $p$). The optimal energy in the model satisfies the recursion
\bea \label{Eq:PresIB:17}
\sE_{t+1}(x) = {\rm min}\left(\sE_{t}(x) + \cE^u_{t+1}(x) ,\sE_{t}(x-1) + \cE^v_{t+1}(x) \right)  \ssp .
\eea
And the initial condition is $\sE_{t}(x=0)=0$ and $\sE_{t}(x)=- \infty$ for $x \neq 0$. In the limit $\beta' \to \infty$ the model corresponds to LPP with exponential waiting times, while in the limit $\gamma' \to \infty$ it converges to the zero temperature limit of the Strict-Weak polymer, which is a model of FPP with exponentially distributed waiting times on horizontal edges only. This model thus remarkably interpolates between a model of first and last passage percolation on $\JZ^2$. Based on our exact results for the Inverse-Beta polymer, we obtain various exact results for this zero-temperature model. In particular we show that $Prob(   \sE_{t}(x) > r) =  {\rm Det} \left( I + K^{T=0}_{tx} \right)$ with
\begin{eqnarray} \label{Eq:PresIB:18}
 K^{T=0}_{t,x}(v_1,v_2) = && -\int_{-\infty}^{+\infty}   \frac{dk}{ \pi}   \int_{\tilde C} \frac{ds}{ 2i \pi s  }   e^{ s r-  2 i k(v_1-v_2) -  s (v_1+v_2) } \\
 &&  \left( \frac{  \frac{s}{2} + \frac{ \gamma'}{2} - i k  }{  -\frac{s}{2} + \frac{ \gamma' }{2} - i k  } \right)^{1 +x} \left( \frac{   \frac{s}{2} + \frac{ \gamma'}{2} +i k }{-\frac{s}{2} + \frac{ \gamma'}{2} + i k  } \right)^{ 1-x + t} \left( \frac{ \beta' +i k+\frac{\gamma' }{2}-\frac{s}{2}}{ \beta' +i k+\frac{\gamma'}{2}+\frac{s}{2}} \right)^t \ , \nonumber 
\end{eqnarray}
where $\tilde C = a + i \mathbb{R}$ with $0<a<\gamma'$  and $K^{T=0}_{t,x} : L^2(\mathbb{R}_+) \to L^2(\mathbb{R}_+)$. The asymptotic analysis then leads to Tracy-Widom fluctuations for the optimal energy in this zero temperature model:
\bea  \label{Eq:PresIB:19}
\lim_{t \to \infty}  Prob\left( \frac{\sE_{t}(x=(1/2+ \varphi)t)) - t \tilde c_{\varphi} }{ \tilde \lambda_{\varphi}} > -2^{\frac{2}{3}} \tilde z   \right) = F_2 (\tilde z)
\eea
with 
\bea \label{Eq:PresIB:20}
&& 0 = \frac{(\frac{1}{2} + \varphi )}{(\frac{\gamma'}{2} - \tilde k_{\varphi})^2} - \frac{(\frac{1}{2} - \varphi )}{(\frac{\gamma'}{2} + \tilde k_{\varphi})^2}+ \frac{1}{(\beta' + \frac{\gamma'}{2} + \tilde k_{\varphi})^2} \ssp , \\
&& \tilde c_{\varphi} = - \frac{(\frac{1}{2} + \varphi )}{\frac{\gamma'}{2} - \tilde k_{\varphi}} - \frac{(\frac{1}{2} - \varphi )}{\frac{\gamma'}{2} + \tilde k_{\varphi}}+ \frac{1}{\beta' + \frac{\gamma'}{2} + \tilde k_{\varphi}} \ssp , \\
&& \tilde \lambda_{\varphi} = \left( \frac{t}{8} \left(  \frac{(1 + 2 \varphi )}{(\frac{\gamma'}{2} - \tilde k_{\varphi})^3} 
 + \frac{(1 - 2 \varphi )}{(\frac{\gamma'}{2} + \tilde k_{\varphi})^3}- \frac{2}{(\beta' + \frac{\gamma'}{2} + \tilde k_{\varphi})^{ 3} }  \right) 
\right)^{\frac{1}{3}}  \ssp.
\eea
In this case the system can be solved exactly since the equation for $\tilde{k}_{\varphi}$ is a quartic equation.

\medskip

Let us conclude this section by mentioning that other results and many details are given in \cite{ThieryLeDoussal2015}. In particular we conjecture interesting $n$-fold integral formulae for the Laplace transform of the partition sum/optimal energy PDF of the Inverse-Beta/Bernoulli-Exponential model that are reminiscent of those obtained using the gRSK/RSK correspondence for the Log-Gamma/LPP model.

\subsection{Presentation of the main results of  \cite{ThieryLeDoussal2016b}} \label{subsec:PresBe}

\stab {\it The Beta Polymer and its TD-RWRE interpretation}\\
The Beta polymer is a BA solvable model of DP on $\JZ^2$ introduced in \cite{BarraquandCorwin2015} and included in the classification of \cite{ThieryLeDoussal2015}. In \cite{BarraquandCorwin2015} the authors studied the half-line to point partition sum and our initial motivation was to use the results of \cite{ThieryLeDoussal2015} to study the point to point partition sum. The Beta polymer is, however, a very peculiar model and in doing so we have notably unveiled a novel fluctuation behavior (see below). The Beta polymer has two parameters $(\alpha,\beta) \in \JR_+^2$ and its random BWs are correlated as $u+v= 1$ with $u \sim Beta(\alpha,\beta)$ as in (\ref{Eq:PresIB:8}). Thanks to these correlations and as already noticed in \cite{BarraquandCorwin2015}, given a random environment specified by a drawing of the $(u_{tx} , v_{tx} = 1- u_{tx})$, the partition sum of the point to point Beta polymer can also be interpreted as a transition probability for a random walk on $\JZ$ in a time-dependent random environment (TD-RWRE).  Introducing the time coordinate $\st$ and the hopping probabilities
\bea \label{Eq:PresB:1}
\st = -t \quad , \quad  {\sf p}_{\st ,x} = u_{t,x} \in [0,1] \ssp ,
\eea
the TD-RWRE is defined as follows: denoting $X_{\st}$ the position of the particle at time $\st$, the particle performs a RW 
on $\mathbb{Z}$ with the following transition probabilities
\bea \label{Eq:PresB:2bis}
&& X_{\st} \to X_{\st+1}=X_{\st}  \text{ with probability } {\sf p}_{\st ,X_{\st}} =  u_{t= -\st ,X_{\st}}    \   ,   \nn \\
&&  X_{\st} \to X_{\st+1}=X_{\st}-1  \text{ with probability } 1 - {\sf p}_{\st ,X_{\st}} =  v_{t=-\st ,X_{\st}}    \   .    
\eea
In the RWRE language, the point to point partition sum of the Beta polymer $Z_t(x)$ is the probability, given that a particle starts at position $x$ at time $\st = -t \leq 0$, that it arrives at position $0$ at time $\st = t = 0$:
\bea \label{Eq:PresB:2}
Z_t(x) = \sP( X_0 = 0 | X_{\st = -t} = x)  \ssp .
\eea
In this interpretation, the recursion equation for the polymer partition sum (\ref{Eq:PresIB:1}) is a Backward equation for the probability $\sP( X_0 = 0 | X_{\st } = x)$:
\bea  \label{Eq:PresB:3}
&&  \sP(X_0 = 0 | X_{\st -1} = x) = p_{\st-1 x} \sP( X_0 = 0| X_{\st } = x)  
+ (1 - p_{\st-1 x})\sP( X_0 = 0| X_{\st } = x-1)\nn \\
&& \sP( 0,0 | 0, x) = \delta_{x,0}    .
\eea 
Note finally that the starting point of the polymer corresponds to the endpoint of the RW and vice-versa. \\

{\it Definition of the optimal direction} \\
The optimal direction can be defined by considering the {\it annealed} PDF, defined as
\bea \label{Pann}
\sP_{{\rm ann}}(X_0 = 0 | X_{-t} = x) := \overline{ \sP(X_0 = 0 | X_{-t} = x)} = \overline{Z_t(x) } \ssp ,
\eea
which is the transition PDF for a RW defined as above with $p_{\st ,x}$ replaced by its average: $p_{\st,x} \to \overline{p_{\st,x}} = \overline{u} = \alpha/(\alpha+\beta)$. By translational invariance of the averaged environment we have $\sP_{{\rm ann}}(X_0 = 0 | X_{-t} = x) = \sP_{{\rm ann}}(X_t = -x | X_{0} = 0) = \overline{Z_t(x) } = \frac{\alpha^x \beta^{t-x}}{(\alpha + \beta)^t} $. In general, the annealed PDF decreases at large time in a given direction as, scaling $x =(1/ 2 + \varphi t)$,
\be \label{Eq:PresB:4bis}
\overline{Z_t(x=(1/ 2 + \varphi t)) } = \sqrt{\frac{2}{\pi t (1 -4 \varphi^2)}} \left( \frac{2}{\sqrt{1- 4 \varphi^2}}  \left(\frac{1-2 \varphi }{2 \varphi +1}\right)^{ \varphi }  \frac{ r^{(1/2 +\varphi)}}{1+r} \right)^t  \left( 1 + O(1/\sqrt{t}) \right) \ .
\ee
where we have introduced the asymmetry parameter
\bea
r = \beta/\alpha \in \mathbb{R}_+       \  .
\eea
It is easily seen that $\overline{Z_t(x=(1/ 2 + \varphi t)) }$ decreases exponentially in every direction, except at its maximum $\varphi=\varphi_{opt}(r)$ which defines the {\it optimal angle}
\bea 
\varphi_{opt}(r) = \frac{r-1}{2 (r+1)} \in ]-1/2 , 1/2 [ \ ,
\eea
The optimal angle thus appears as the most probable space-time direction taken by a RW in an averaged environment. Below we will see that the fluctuations of $Z_t(x)$ will depend on the chosen direction and we will mainly consider the {\it large deviations regime}, corresponding to the scaling $x=(1/ 2 + \varphi t)$ with $\varphi \neq \varphi_{opt}(r)$, and the {\it diffusive regime around the optimal direction}, corresponding to the scaling  $x=(1/ 2 + \varphi_{opt}(r) t) + \kappa \sqrt{t}$ where (\ref{Eq:PresB:4bis}) takes a Gaussian form. \\

{\it Bethe ansatz solution of the Beta polymer} \\
In \cite{ThieryLeDoussal2016a} we show that, defining $c  = \frac{4}{\alpha+\beta} > 0$ and
\be \label{Eq:PresB:4}
z_j = e^{ i \lambda_j}   \hspace{0.3 cm},\hspace{0.3 cm} \tilde t_j = - i \cot (\frac{\lambda_j}{2}) =\frac{z_j+1}{z_j-1}  \quad , \quad z_j = -  \frac{1 + \tilde t_j}{1- \tilde t_j} \ ,
\ee
{\it the eigenfunctions and the Bethe equations of the Beta polymer are identical to those of the Log-Gamma and Inverse-Beta polymer (\ref{Eq:PresLG:12})-(\ref{Eq:PresLG:14}), up to the change $t_i \to \tilde{t}_i$ and $\bar c \to c$}. However, in \cite{ThieryLeDoussal2016a}, we show that this change has important consequences: the string solutions are not stable and $c>0$ can be interpreted as a repulsive interaction parameter. In the large $L$ limit the replica thus behave as free particles and do not form bound states. The repulsive nature of the model is interpreted as a consequence of the TD-RWRE nature of the model. In this case we therefore obtain a formula for the moments of the partition sum which is simpler (compared to (\ref{Eq:PresLG:24}) and (\ref{Eq:PresIB:10})) and does not contain a summation over string states:
\bea  \label{Eq:PresB:5}
&&  \overline{Z_t(x)^n}  =  (-1)^n  \frac{\Gamma(\alpha+\beta+n)}{\Gamma(\alpha + \beta  )}    \\
&& 
\prod_{j=1}^{n}  \int_{- \infty}^{+\infty} \frac{dk_j}{2 \pi}
\prod_{1\leq i < j  \leq n} \frac{(k_i-k_j)^2}{(k_i-k_j)^2 + 1} \prod_{j=1}^{n} 
\frac{(i k_j + \frac{\beta - \alpha}{2} )^t}{(i k_j + \frac{\alpha+\beta}{2})^{1 +x} (i k_j - \frac{\alpha+\beta}{2})^{1-x+ t}} \nn \ssp . 
\eea 
Making the link with the nested contour integral approach to BA used in \cite{BarraquandCorwin2015} we also obtain a formula for the multi-point moments: for $0 \leq x_1 \leq \cdots \leq x_n $:
\bea \label{Eq:PresB:6}
&& \overline{Z_t(x_1) \cdots Z_t(x_n)}  =  (-1)^n \frac{\Gamma(\alpha+\beta+n)}{\Gamma(\alpha + \beta  )}  \\
 && \prod_{j=1}^{n}  \int_{\mathbb{R}} \frac{dk_j}{2 \pi  }
\prod_{1\leq i < j  \leq n} \frac{k_i-k_j}{k_i-k_j+i } \prod_{j=1}^{n} 
\frac{(i k_j + \frac{\beta - \alpha}{2} )^t}{(i k_j + \frac{\alpha+\beta}{2})^{1 +x_j} (i k_j - \frac{\alpha+\beta}{2})^{ 1-x_j + t} }  \nn \ssp .
\eea

{\it Cauchy-type Fredholm determinant formula} \\
A further simplification that is specific to the Beta polymer is that, since the Boltzmann weights are bounded, the moments of the partition sum $\overline{(Z_t(x))^n}$ do not grow too fast in this case and indeed determine unambiguously the distribution of $Z_t(x)$. As in the Inverse-Beta case, the first factor $\frac{\Gamma(\alpha+\beta+n)}{\Gamma(\alpha + \beta  )}$ in front of (\ref{Eq:PresB:5}) forbids to express the LT of $Z_t(x)$ as a Fredholm determinant. For this reason we consider the generating function $g_{t,x}(u) = \sum_{n=0}^{\infty} \frac{(-u)^n}{n!} Z_n$ with $Z_n =\frac{\Gamma(\alpha+\beta)}{\Gamma(\alpha + \beta +n )} \overline{(Z_t(x))^n} $. We obtain several equivalent Fredholm determinant formulae for $g_{t,x}(u)$, and notably
\begin{equation}  \label{Eq:PresB:7}
 g_{t,x}(u) = {\rm Det} \left( I  +  u \hat{K}_{t,x} \right)
\end{equation}
with the kernel $ \hat K_{t,x} : L^2 ( \mathbb{R}) \to L^2 ( \mathbb{R}) $:
\be \label{Eq:PresB:8}
\hat K_{t,x}(q_1,q_2) =  - \frac{2}{\pi} \frac{(1+i q_1 (\alpha-\beta))^{-x+t}}{(1+i q_1 (\alpha+\beta) )^{1-x+t}}  \frac{(1+i q_2 (\alpha-\beta))^{x}}{(1-i q_2 (\alpha+\beta))^{1+x}} \frac{1}{2 + i (q_2^{-1}-q_1^{-1})}  \ssp .
\ee

The procedure to go from $g_{t,x}(u)$ to the PDF of $Z_t(x)$ is discussed in \cite{ThieryLeDoussal2016b}. Note that compared to the Fredholm-determinant formula presented up to now (\ref{Eq:ChapIII:Sec2:BALLFinal:7})-(\ref{Eq:PresLG:29})-(\ref{Eq:PresIB:11}), (\ref{Eq:PresB:8}) has the distinguishing feature that the Laplace Transform variable appears simply linearly in front of the kernel. This type of Fredholm determinant formula is known in the literature as Cauchy-type formulae and first appeared in the KPZ-related literature in the work of Tracy and Widom on the ASEP \cite{TracyWidom2008}. \\

\begin{figure}
\centerline{\includegraphics[width=9.0cm]{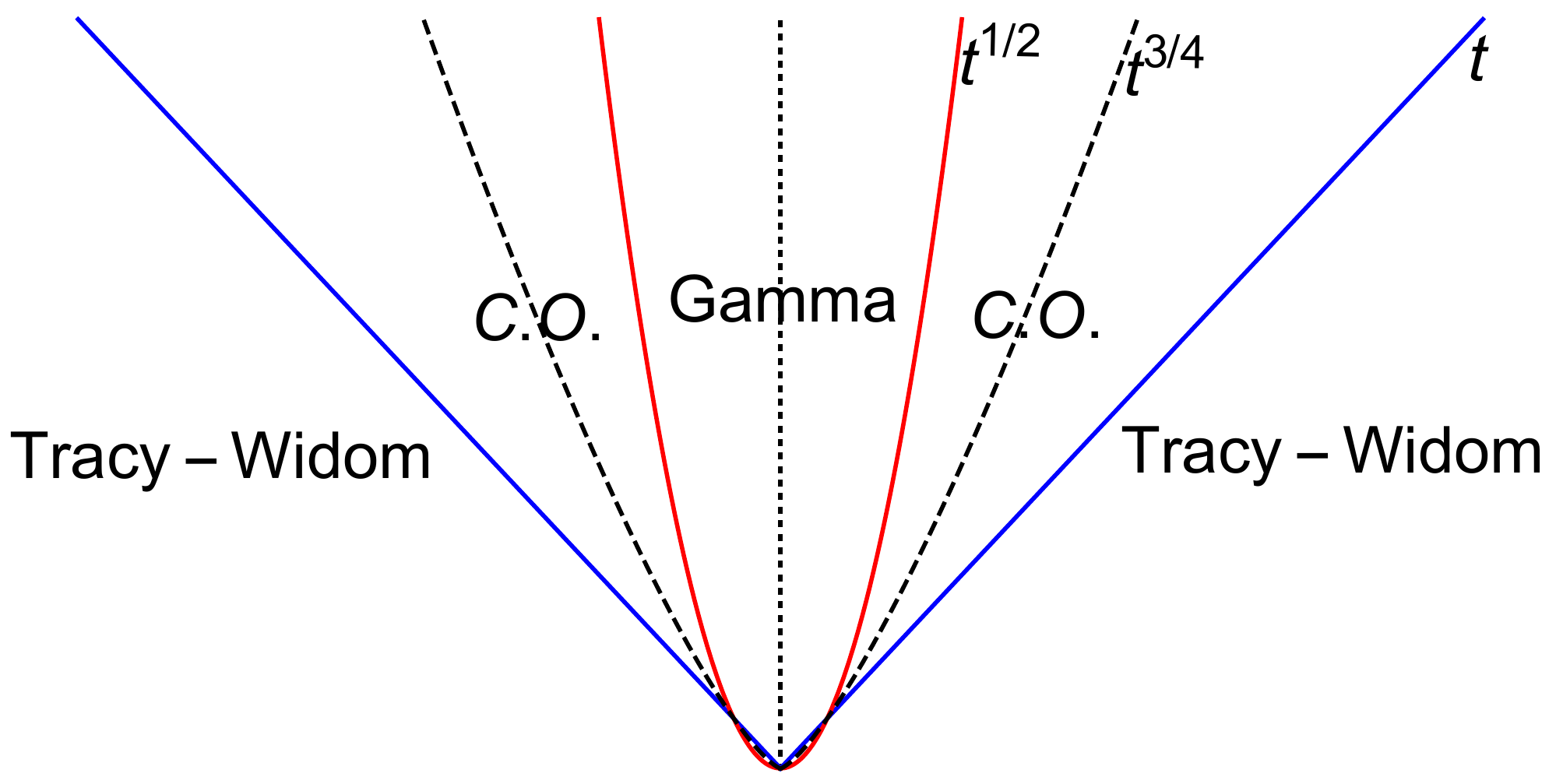}} 
\caption{The different regimes of sample to sample fluctuations of the PDF in the Beta TDRWRE problem around the optimal direction (indicated by a dotted line) for different scaling of the deviation with respect to the optimal direction $\check x = x- (1/2+\varphi_{opt}) t$. In the diffusive regime $\check x \sim \sqrt{t}$ the fluctuations of the PDF are Gamma distributed. In the large deviations regime $\check x \sim t$, fluctuations of the logarithm of the PDF are distributed according to the GUE Tracy-Widom distribution with exponents in agreement with the usual KPZ universality expected in point to point directed polymers problem. These two regimes are connected by a cross-over regime (C.O.) at a scale $\check x \sim t^{3/4}$. Figure taken from \cite{ThieryLeDoussal2016b}.}
\label{fig:PresB1}
\end{figure}

{\it Asymptotic analysis in the optimal direction} \\
In \cite{ThieryLeDoussal2016b} we show that
\bea  \label{Eq:PresB:9}
{\cal Z}_t(\kappa) =  \alpha \sqrt{2 \pi rt} e^{\frac{(r+1)^2}{2r} \kappa^2} Z_t\left(x = (\frac{1}{2} +  \varphi_{opt}(r)) t + \kappa \sqrt{t}\right) \ssp ,
\eea
converges at fixed $t$, in the large time limit to a {\it process, constant in $\kappa$, with marginal distribution a Gamma distribution with parameter $\alpha+\beta$}
\bea \label{Eq:PresB:10}
{\cal Z}_{\infty}(\kappa) \sim Gamma(\alpha+ \beta)  \ .
\eea
From the point of view of DPs on $\JZ^2$, this result can be thought of as a breaking of KPZUC in the optimal direction, due to the presence of an additional conservation law, namely the conservation of the probability, encoded in the correlations of the random BWs as $u+v=1$. From the point of view of TD-RWRE this result shows that, in a given environment, $\sP(X_0 = 0 | X_{\st -1} = x = (1/2 + \varphi_{opt}(r))t + \kappa \sqrt{t})$ converges to a Gaussian distribution, which is modulated by a $\kappa$ (= starting-point) independent Gamma distributed RV. The origin of the Gamma distribution can be traced back to the first factor in (\ref{Eq:PresB:5}), which thus plays a very important role here. \\

{\it Asymptotic analysis in the large deviations regime} \\
To perform the asymptotic analysis in the large deviations regime we find that our formula (\ref{Eq:PresB:7})-(\ref{Eq:PresB:8}) is not adapted. This was already remarked on the Cauchy-type Fredholm determinant formula obtained by Tracy and Widom in \cite{TracyWidom2008} and performing the asymptotic analysis required to obtain another Fredholm determinant representation \cite{TracyWidom2009}. The Beta polymer is thus an example of a model where performing the asymptotic analysis using Cauchy-type formulae is well adapted to the study of the diffusive regime of the TD-RWRE. For the large deviations regime, we thus first obtain a {\it formal} Fredholm determinant formula for $g_{t,x}(u)$: $g_{t,x}(u) =  {\rm Det} \left( I + \check K_{t,x} \right)$ with
\begin{eqnarray} \label{Eq:PresB:11}
 \check K_{t,x}(v_1,v_2) = && \int_{L}   \frac{dk}{ \pi}  \frac{-1}{2i} \int_C \frac{ds}{ \sin( \pi s ) }   u^s  e^{ -  2 i k(v_1-v_2) -  s (v_1+v_2) } \\
 &&  \left( \frac{  \Gamma(-\frac{s}{2} + \alpha+\beta + i k ) }{  \Gamma(\frac{s}{2}  + \alpha+ \beta +i k )  } \right)^{1 +x} \left( \frac{   \Gamma(-\frac{s}{2} +i k )}{ \Gamma(\frac{s}{2}  + i k ) } \right)^{ 1-x + t} \left( \frac{ \Gamma ( \beta +i k+\frac{s}{2})}{\Gamma( \beta +i k-\frac{s}{2})} \right)^t \nonumber  \nn \ssp ,
\end{eqnarray}
which now appears rather similar to those for the Log-Gamma and Inverse-Beta polymer (\ref{Eq:PresLG:29})-(\ref{Eq:PresIB:11}) (see \cite{ThieryLeDoussal2016b} for the precise sense in which this formula is formal). Performing the asymptotic analysis of (\ref{Eq:PresB:11}) we obtain, for $\varphi_{opt}(r) < \varphi < 1/2$ (the other case being obtained by symmetry)
\begin{equation} \label{Eq:PresB:12}
\lim_{t \to \infty} Prob\left( \frac{ \log Z_t((1/2+ \varphi) t) + tc_{\varphi}}{\lambda_{\varphi} } <2^{\frac{2}{3}} z \right) = F_2(z) \ ,
\end{equation}
with
\bea \label{SDPeqnBETA}
&& \varphi = \frac{\psi'(\beta + k_{\varphi})- \frac{1}{2} \left(\psi'(k_{\varphi}) + \psi'(\alpha+ \beta+ k_{\varphi}) \right)}{\psi'(\alpha+ \beta+ k_{\varphi}) -\psi'(k_{\varphi}) }   \ssp , \\
&&  c_{\varphi} = - G_{\varphi}'( k_\varphi ) = \left(\varphi +\frac{1}{2}\right) \psi(k_{\varphi}+\alpha +\beta )-\psi(k_{\varphi}+\beta )+\left(\frac{1}{2}-\varphi \right) \psi(k_{\varphi}) \ssp ,  \nn \\
 &&  \frac{8 \lambda_{\varphi}^3}{t}=  G_{\varphi}'''( k_\varphi ) =-\left(\varphi +\frac{1}{2}\right) \psi''(k_{\varphi}+\alpha +\beta )+\psi''(k_{\varphi}+\beta )-\left(\frac{1}{2}-\varphi \right) \psi''(k_{\varphi})  \ssp . \nn 
\eea
From the point of view of DPs, this result shows that KPZUC is restored away from the optimal direction of the TD-RWRE. The value of $c_{\varphi}$ was already determined in \cite{BarraquandCorwin2015} using a general theorem of \cite{RASY2013}, the value of $\lambda_{\varphi}$ was, however, still unknown. Our result (\ref{SDPeqnBETA}) actually appears equivalent to the one of \cite{BarraquandCorwin2015} in the large deviations regime if one replaces the point to point partition sum in (\ref{Eq:PresB:12}) by the half-line to point partition sum that is studied in \cite{BarraquandCorwin2015}.

\medskip

Let us conclude this section by mentioning other results obtained in \cite{ThieryLeDoussal2016b} (i) alternative Fredholm determinant formulae for $g_{t,x}(u)$, in particular a formula that gives a rigorous meaning to the formal formula (\ref{Eq:PresB:11}); (ii) relations of our approach with the nested contour integral approach to the Bethe ansatz; (iii) formula for the PDF of $Z_t(x)$ at any time; (iv) a discussion of the crossover between the diffusive region and the large deviations regime, identified with deviations from the optimal direction of order $t^{3/4}$ (see Fig.~\ref{fig:PresB1}); (v) an extensive numerical study of the validity of our results using simulations of the Beta polymer.

\subsection{Presentation of the main results of \cite{Thiery2016}} \label{subsec:PresStat}

In Sec.~\ref{subsec:ChapIII:SecII:StatMeas} we recalled that the Log-Gamma polymer was first introduced for the possibility of writing down exactly its stationary measure, an example of an exact solvability property that led to many developments. The initial motivation of \cite{Thiery2016} was to investigate whether or not the stationary measure of the Inverse-Beta polymer could also be obtained. This is interesting from several points of view, in particular (i) the links between the different types of exact solvability properties mentioned up to now are not yet understood; (ii) the stationary measure encodes the space-time correlations of the DP free-energy at large $t$ on distances $\ll t^{2/3}$, an information which is notoriously hard to obtain from the BA; (iii) the results obtained using BA for the Inverse-Beta polymer required the use of several mathematically non-rigorous tricks and some of them can be obtain rigorously from the stationary measure, thus partially confirming the approach of \cite{ThieryLeDoussal2015}; (iv) the knowledge of the stationary measure of the Log-Gamma is at the basis of the derivation of a variety of results, e.g. the fluctuation exponents of the DP \cite{Seppalainen2009}, the large deviation function of the partition sum \cite{GeorgiouSeppalainen2011}, localization properties of the DP \cite{CometsNguyen2015}. These results can thus probably be generalized to a richer model using the results of \cite{Thiery2016}. On the other hand the possibility to write down exactly the stationary measure indicates the presence of exact solvability properties. Considering the fact that LPP is both exactly solvable for exponential and geometric distributions of waiting times, it was natural to conjecture that the zero temperature limit of the Inverse-Beta polymer, the Bernoulli-Exponential polymer, could be generalized to an exactly solvable model with discrete energies. In \cite{Thiery2016} the Bernoulli-Geometric polymer is introduced and corresponds to this model: we obtain its stationary measure and deduce from it several results.\\

\begin{figure}
\centerline{\includegraphics[width=9.0cm]{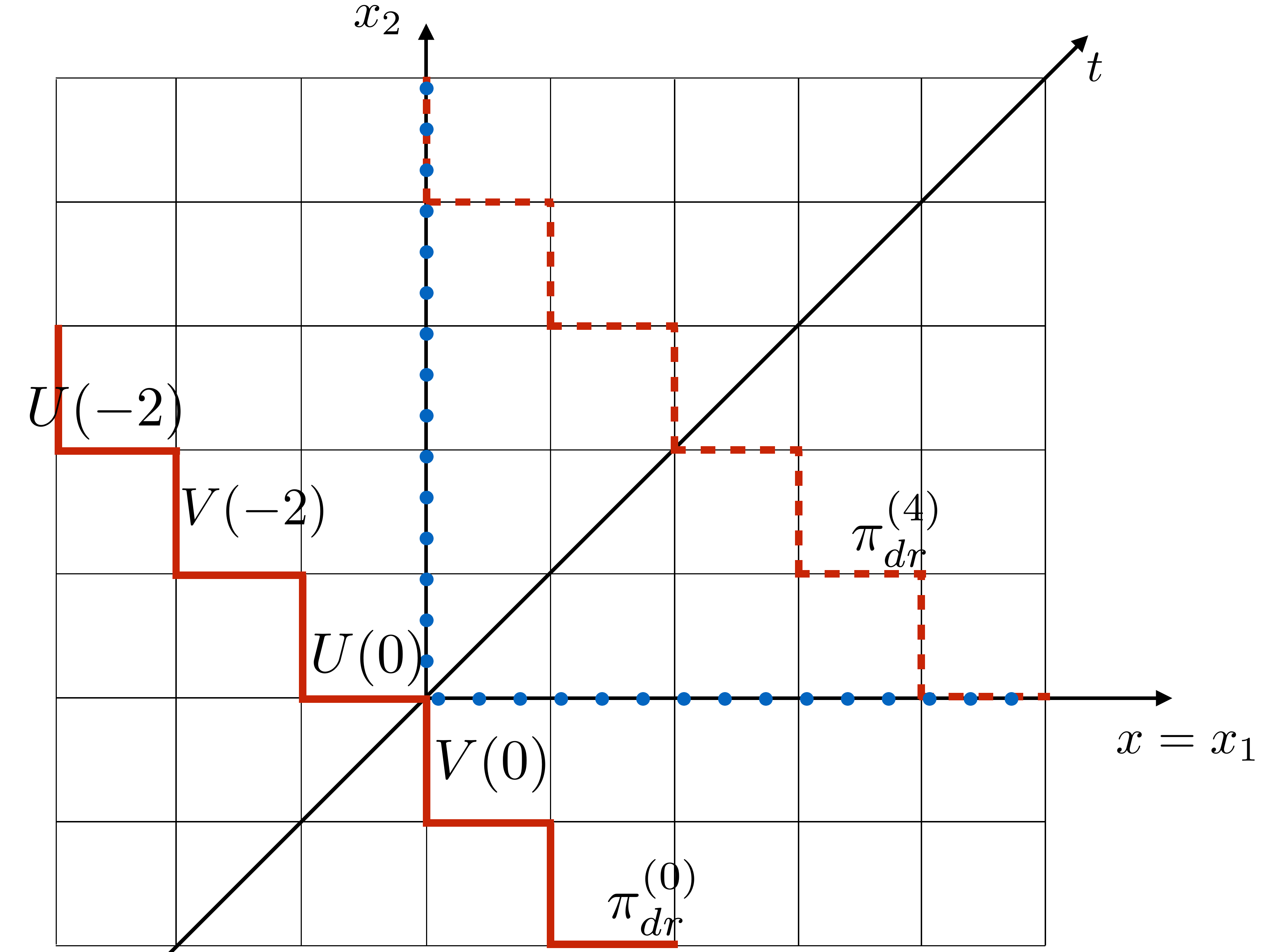}} 
\caption{Different down-right paths on $\JZ^2$: the dotted-blue down-right path (boundary of $\JN^2$) and dashed-red down-right path $\pi_{dr}^{(4)}$ can both be obtained by a sequence of down-left to top-right transformation ($\phi$ arrow above) from $\pi_{dr}^{0}$.}
\label{fig:PresStat1}
\end{figure}

{\it Stationary measure of the Inverse-Beta polymer}\\
Keeping the notations of \cite{ThieryLeDoussal2015} and Sec.~\ref{subsec:PresIB}\footnote{These notations have been changed in \cite{Thiery2016} so that the $v$ (resp. $u$) BWs live on the vertical (resp. horizontal) edges of $\JZ^2$.}, we thus consider the Inverse-Beta polymer defined by the recursion equation
\bea \label{Eq:PresStat:1}
\check Z_{t+1}(x) = u_{t+1 , x} \check Z_t(x) +  v_{t+1 , x}  \check Z_t(x-1)  \ ,
\eea
where $v_{t+1,x} =u_{t+1 , x} -1$ and $u \sim Beta(\gamma, \beta)^{-1}$ and the BWs are defined on the full square lattice $\JZ^2$. We will specify later the stationary initial condition and have added a checkmark on $\check Z_{t}(x)$ to emphasize the distinction with the point-to-point partition sum of the Inverse-Beta polymer $Z_{t}(x)$ defined in Sec.~\ref{subsec:PresIB}. As we saw in Sec.~\ref{subsec:ChapIII:SecII:StatMeas}, the partition sum itself is never stationary and one has to consider ratios of partition sums. Defining $\forall t \geq 0$, $\forall x \in \JZ$, the ratios of partition sum on horizontal and vertical edges as
\bea  \label{Eq:PresStat:3}
&& \check U_{t} (x) := \frac{\check Z_{t} ( x) }{\check Z_{t-1} ( x-1)} \quad , \quad  \check V_{t} (x) :=\frac{ \check Z_{t} ( x)}{  \check Z_{t-1} ( x)} \ ,
\eea
these satisfy the recursion relation 
\bea  \label{Eq:PresStat:3b}
\check U_{t+1}(x)  =  \phi^{(1)}(\check U_{t}(x) ,\check V_{t}(x-1)   , W_{t+1}(x))   \quad , \quad\check V_{t+1}(x)  =  \phi^{(2)}(\check U_{t}(x) ,\check V_{t}(x-1)   , W_{t+1}(x))     \    .   \nn \\
\eea
where $W_{t+1}(x) = v_{t+1,x}$ and $\phi^{(i)}$ are the components of the {\it stationarity-reversibility map} defined as $\phi: (U,V,W) \to (U',V',W')$ with
\bea \label{Eq:PresStat:4b}
U' = W  + (W+1) \frac{U}{V}  \quad , \quad   V' = W \frac{V}{U} + W +1  \quad , \quad  W' = \frac{U(V-1) }{U+V} \ssp .
\eea
Elementary (but non-trivial) properties of $\phi$ then show the following. Taking (\ref{Eq:PresStat:3b}) as the definition, $\forall t$, of a stochastic process for the $\check U_{t}(x)$ and $\check V_{t}(x)$ variables, we consider an initial condition such that at $t=0$ they are independent and distributed as $\check U_{t=0}(x) \sim U $ and $\check V_{t=0}(x) \sim V $ with
\bea \label{Eq:PresStat:4c}
 U \sim (Beta(\gamma- \lambda , \beta + \lambda))^{-1} -1 \quad , \quad  V \sim (Beta(\lambda , \beta ))^{-1} \ssp . 
\eea
Here $0 < \lambda < \gamma$ is a parameter that labels a family of stationary measures. In \cite{Thiery2016} we show that for all down-right paths $\pi_{dr}$ on $\JZ^2$ that can be obtained from the down-right path 
\be \label{Eq:PresStat:4}
\pi_{dr}^{(0)} = \{ (x_1 ,x_2) = (m , -m) \to (m , -m-1)  \to (m+1 , -m-1)         , m \in \mathbb{Z} \}
\ee
by a sequence of down-left to top-right transformation (which amounts to changing a down-left corner of a down-right path to a top-right corner, see Fig.~\ref{fig:PresStat1} for a self-explanatory definition of these notions), the variables $\check U_{t}(x)$ and $\check V_{t}(x)$ living on the down-right path are independent and distributed as in (\ref{Eq:PresStat:4c}). Furthermore, this stationary measure is reversible in the following sense. Considering the stationary process (\ref{Eq:PresStat:3b}) during a finite time window $T$, the time-reversed process
\be \label{Eq:PresStat:7}
\check U_{t_R}^R(x_R ) = \check U_{t = T -t_R} (x = -x_R +1) \quad , \quad \check V_{t_R}^R(x_R ) =\check V_{t = T -t_R} (x = -x_R)  \ssp ,
\ee 
satisfies the identity in law
\be \label{Eq:PresStat:8}
\left(\check U_t (x) , \check V_t(x) \right)_{t = 0, \dots , T ; x \in \mathbb{Z}} \sim \left(\check U^R_{t_R} (x_R) , \check V^R_{t_R}(x_R) \right)_{t_R = 0, \dots , T ; x_R \in \mathbb{Z}}  \ssp .
\ee

For the original process of the partition sum (\ref{Eq:PresStat:1}) this implies that, if one starts from an initial condition such that successive partition sum quotients are random and distributed as
\bea \label{Eq:PresStat:2}
\frac{\check Z_{t=0}(x +1)}{\check Z_{t=0}(x)} \sim \frac{(Beta(\gamma- \lambda , \beta + \lambda))^{-1} -1}{(Beta(\lambda , \beta))^{-1}} \ssp ,
\eea
where the different Beta RVs appearing in this initial condition are all independent, then these quotients remain distributed as so for all time (and are independent at $t$ fixed). Finally, adding for convenience the initial condition $\check Z_{t=0}(0) = 1$ we show in \cite{Thiery2016} that the partition sum in the stationary state in the upper-right quadrant of $\JZ^2$,  $( \check Z_t(x) )_{x \in \JN}$, are equivalent in law to the {\it point to point} partition sum $\hat Z_t(x)$ of a model defined on $\JN^2$ with peculiar boundary conditions: $( \check Z_t(x) )_{x \in \JN} \sim ( \hat Z_t(x) )_{x \in \JN}$. We refer the reader to \cite{Thiery2016} for the precise definition of this model.\\

{\it The Bernoulli-Geometric polymer} \\
In \cite{Thiery2016} we define the Bernoulli-Geometric polymer as a geometric discretization of the Bernoulli-Exponential polymer defined in Sec.~\ref{subsec:PresIB} and introduced in \cite{ThieryLeDoussal2015}. It is a zero-temperature model of DP on $\JZ^2$ with random energies on-edges and where the optimal energy satisfies the relation
\bea  \label{Eq:PresStat:9}
\sE_{t+1}(x) = {\rm min}\left(\sE_{t}(x) + \cE^u_{t+1}(x) ,\sE_{t}(x-1) + \cE^v_{t+1}(x) \right)  \ssp ,
\eea
with the initial condition $\sE_{t=0}(0)=0$ and $\sE_{t=0}(x)=-\infty$ for $x \neq 0$. The random energies are distributed as\footnote{Here we keep the notations adopted for the Bernoulli-Exponential polymer in Sec.~\ref{subsec:PresIB}, which differs from those adopted in \cite{Thiery2016}.}
\bea  \label{Eq:PresStat:10}
&& \cE^u \sim -\zeta_{uv} G_{q} \in \JZ_-   \ssp , \nn \\
&& \cE^v \sim  (1- \zeta_{uv}) (1+ G_{q'})  -\zeta_{u v} G_{q} \in \JZ  \ssp ,
\eea
where $(q,q') \in ]0,1[^2$ are the two parameters of the model, and $G_q$ generally denotes a Geometric RV $G_q \sim Geo(q)$ (see (\ref{Eq:ChapIII:Sec2:Stat27})). $\zeta_{u v}$ is a Bernoulli RV with parameter
\bea \label{Eq:PresStat:11}
p_{u v} = \frac{1-q'}{1-q q'} \in ] 0, 1[  \ssp .
\eea
This value ensures an exact solvability property. This model generalizes the Bernoulli-Exponential polymer which is now retrieved in a limit $q = 1- \gamma' \epsilon$, $q' = 1 - \beta' \epsilon$ and $\epsilon \to 0^+$. The case $q' =0$ corresponds to LPP with geometric waiting times as studied in \cite{johansson2000}, while the case $q =0$ is FPP with geometric waiting times on the horizontal edges only as in \cite{DraiefMairesseOConnell2005}. \\

{\it Stationary Bernoulli-Geometric polymer} \\
We now discuss the stationary measure of the Bernoulli-Geometric polymer. The stationary optimal energy $\check E_t(x)$ satisfies the recursion equation (\ref{Eq:PresStat:9}) but with a different initial condition. Similarly as for the Inverse-Beta polymer, to describe the stationary measure, we consider the horizontal and vertical energy differences variables defined as 
\bea \label{Eq:PresStat:14}
\check {\sU}_t(x) = \check \sE_{t}(x) - \check \sE_{t-1}(x-1)  \quad , \quad \check {\sV}_t(x) = \check \sE_{t}(x) - \check \sE_{t-1}(x)  \ssp .
\eea
These satisfy the recursion relation
\bea  \label{Eq:PresStat:15}
&& \check \sU_{t+1}(x)  =  \phi_{T=0}^{(1)}(\check \sU_{t}(x) ,\check \sV_{t}(x-1) , \cE^v_{t+1}(x) ,\cE^u_{t+1}(x) )   \nn \\
&& \check \sV_{t+1}(x)  =  \phi_{T=0}^{(2)}(\check \sU_{t}(x) ,\check \sV_{t}(x-1) , \cE^v_{t+1}(x) ,\cE^u_{t+1}(x) )    \ssp ,
\eea
where $\phi_{T=0}$ is the {\it $T=0$ stationarity map} defined as: $ \phi_{T=0}:(\sU ,\sV  , \su, \sv) \to (\sU' , \sV')$ with
\bea \label{T0StationarityMap}
&& \sU' = {\rm min}\left( \su , \sv + \sU - \sV \right)   \quad , \quad \sV' = {\rm min}\left( \su+\sV-\sU , \sv \right)  \ssp . 
\eea
An elementary (but non-trivial) property of $\phi_{T=0}$ then shows the following. Taking (\ref{Eq:PresStat:15}) as the definition, $\forall t$, of a stochastic process for the $\check \sU_{t}(x)$ and $\check \sV_{t}(x)$ variables, we consider an initial condition such that at $t=0$ they are independent and distributed as $\check \sU_{t=0}(x) \sim \sU $ and $\check \sV_{t=0}(x) \sim \sV $ with
\bea \label{Eq:PresStat:16}
&& \sU \sim (1- \zeta_{\sU}) (1+ G_{q_b q'})  -\zeta_{\sU} G_{q/q_b}  \quad , \quad  \sV \sim - \zeta_{\sV} G_{q_b}  \ssp .
\eea
where $q<q_b<1$ is a parameter that labels a family of stationary measure and $\zeta_{\sU}$ and $\zeta_{\sV}$ are Bernoulli RVs with parameters 
\bea \label{Eq:PresStat:13}
p_{\sU} =  \frac{1 - q_b q'}{1-qq'} \quad, \quad p_{\sV} =  \frac{1 - q'}{1-q_b q'} \ssp .
\eea
Similarly as before, in \cite{Thiery2016} we then show that for all down-right paths $\pi_{dr}$ on $\JZ^2$ that can be obtained from $\pi_{dr}^{(0)}$ by a sequence of down-left to top-right transformation, the variables $\check \sU_{t}(x)$ and $\check \sV_{t}(x)$ living on the down-right path are independent and distributed as in (\ref{Eq:PresStat:16}). Finally we show that the stationary measure is reversible through the equality in law, similarly as for (\ref{Eq:PresStat:8})
\bea 
\left(\check \sU_t (x) , \check  \sV_t(x) \right)_{t = 0, \dots , T ; x \in \mathbb{Z}} \sim \left(\check  \sU^R_{t_R} (x_R) , \check  \sV^R_{t_R}(x_R) \right)_{t_R = 0, \dots , T ; x \in \JZ } 
\eea
where the time-reversed process on a finite time window $T$ is now
\bea
\check \sU^R_{t_R}(x_R) = \check \sU_{t=T-t_R}(x=-x_R+1)  \quad , \quad \check \sV^R_{t_R}(x_R) = \check \sV_{t=T-t_R}(x=-x_R) \ssp .
\eea
This implies for the optimal energy $\check \sE_{t}(x)$ that, taking for initial condition $\check \sE_{t=0}(0) =0$ and independent energy increments distributed as
\bea \label{Eq:PresStat:12}
\check \sE_{t=0}(x+1) - \check \sE_{t=0}(x) \sim (1- \zeta_{\sU}) (1+ G_{q_b q'})  -\zeta_{\sU} G_{q/q_b} + \zeta_{\sV} G_{q_b} \ssp ,
\eea
then they remain distributed as so for all time. As before, the optimal energy in the model with stationary initial condition in the upper-right quadrant is shown to be identical in law to a point to point optimal energy $ \hat \sE_t(x) $ in a model with special boundaries: $(\check \sE_{t}(x))_{x \in \JN} \sim (\hat \sE_{t}(x))_{x \in \JN} $. Note that in the case $q'=0$ this reproduces the known result for LPP with geometric waiting times (\ref{Eq:ChapIII:Sec2:Stat28}). \\

{\it Optimal energy per unit length in the Bernoulli-Geometric polymer}

Defining the mean energy of the horizontal and vertical energy differences in the stationary state of the Bernoulli-Geometric polymer
\bea \label{Eq:PresStat:18}
&&   \sff_\sU^{q, q'} (q_b) :=\overline{\sU} =  \frac{q_b^2 q'-q}{\left(q_b-q\right) \left(1-q_b q'\right)}   \  , \nn \\
&& \sff_\sV^{q, q'} (q_b):=\overline{\sV} =  - \frac{1-q'}{1-q_b q'}  \frac{q_b}{1-q_b}  \ssp .
\eea 
We show that the mean optimal energy per unit length in the direction $\varphi \in ]-1/2,1/2[$ in the stationary Bernoulli-Geometric polymer is linear in $\varphi$ with
\bea \label{Eq:PresStat:19}
\check \sff^{p.u.l.}(\varphi , q_b) && := \lim_{t \to \infty} \frac{1}{t} \overline{ \check \sE_{ t} (x = (1/2 +\varphi) t )} \nn \\
&&  = (1/2 + \varphi)\sff_\sU^{q, q'} (q_b) + (1/2 - \varphi)\sff_\sV^{q, q'} (q_b)  \ssp .
\eea
And using the model with boundaries we obtain a formula for the mean optimal energy per unit length in the direction $\varphi$ in the point to point Bernoulli-Geometric polymer:
\bea \label{Eq:PresStat:20}
\sff^{p.u.l.}(\varphi) && := \lim_{t \to \infty} \frac{1}{t} \overline{ \sE_{ t} (x = (1/2 +\varphi) t )} \nn \\
&&  = \check \sff^{p.u.l.}(\varphi , q_b^*(\varphi))  \ssp .
\eea
where $q_b^*(\varphi)$ is the solution of the equation 
\bea \label{Eq:PresStat:21}
\partial_{q_b} \check \sff^{p.u.l.}(\varphi , q_b) |_{q_b = q_b^*(\varphi) } = 0 \ssp .
\eea
This is a quartic equation for $q_b^*(\varphi)$ that can be solved exactly, leading to an explicit expression for $\sff^{p.u.l.}(\varphi)$ which is plotted in Fig.~\ref{fig:PresStat2} for various parameters $q,q'$, and in particular close to the LPP and FPP limits $q' \to 0$ and $q \to 0$. Note the non-analytic behavior in the FPP limit where $\sff^{p.u.l.}(\varphi) = 0$ for $-1/2 \leq \varphi \leq \varphi_{q'}= 1/2 - q'$ and $\sff^{p.u.l.}(\varphi) > 0$ for $\varphi > 1/2 - q'$. This is interpreted as a percolation threshold where for $\varphi < 1/2 -q'$, the optimal path manages to passes with probability $1$ only on edges with $0$ energy (a feature that can only be observed in the Bernoulli-Geometric polymer and not in the Bernoulli-Exponential polymer). Fluctuations in this region of space should differ from naive KPZUC expectations. We obtain results similar to (\ref{Eq:PresStat:19})-(\ref{Eq:PresStat:20}) for the Inverse-Beta polymer that bring a rigorous confirmation of the value of $c_{\varphi}$ obtained from the Bethe Ansatz in \cite{ThieryLeDoussal2015}, see (\ref{Eq:PresIB:13}).

\begin{figure}
\centerline{\includegraphics[width=7cm]{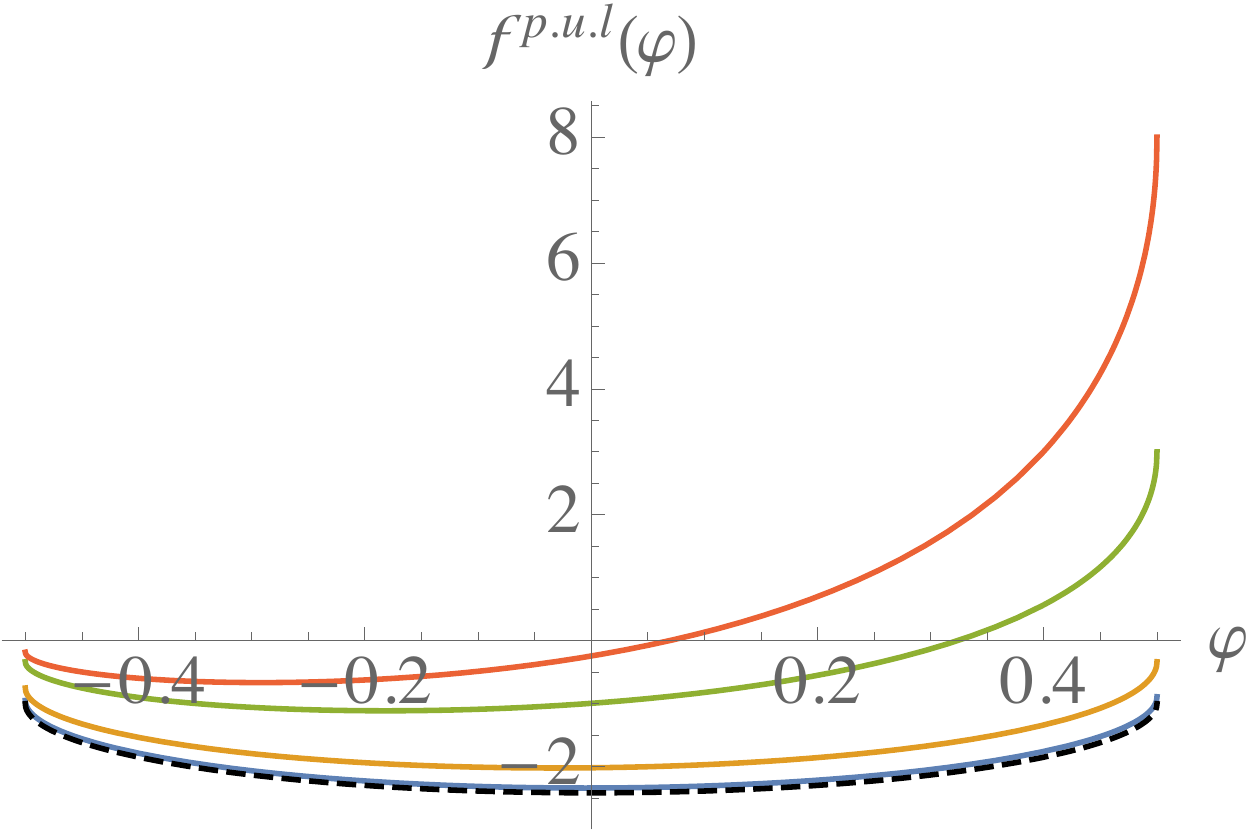} \includegraphics[width=7cm]{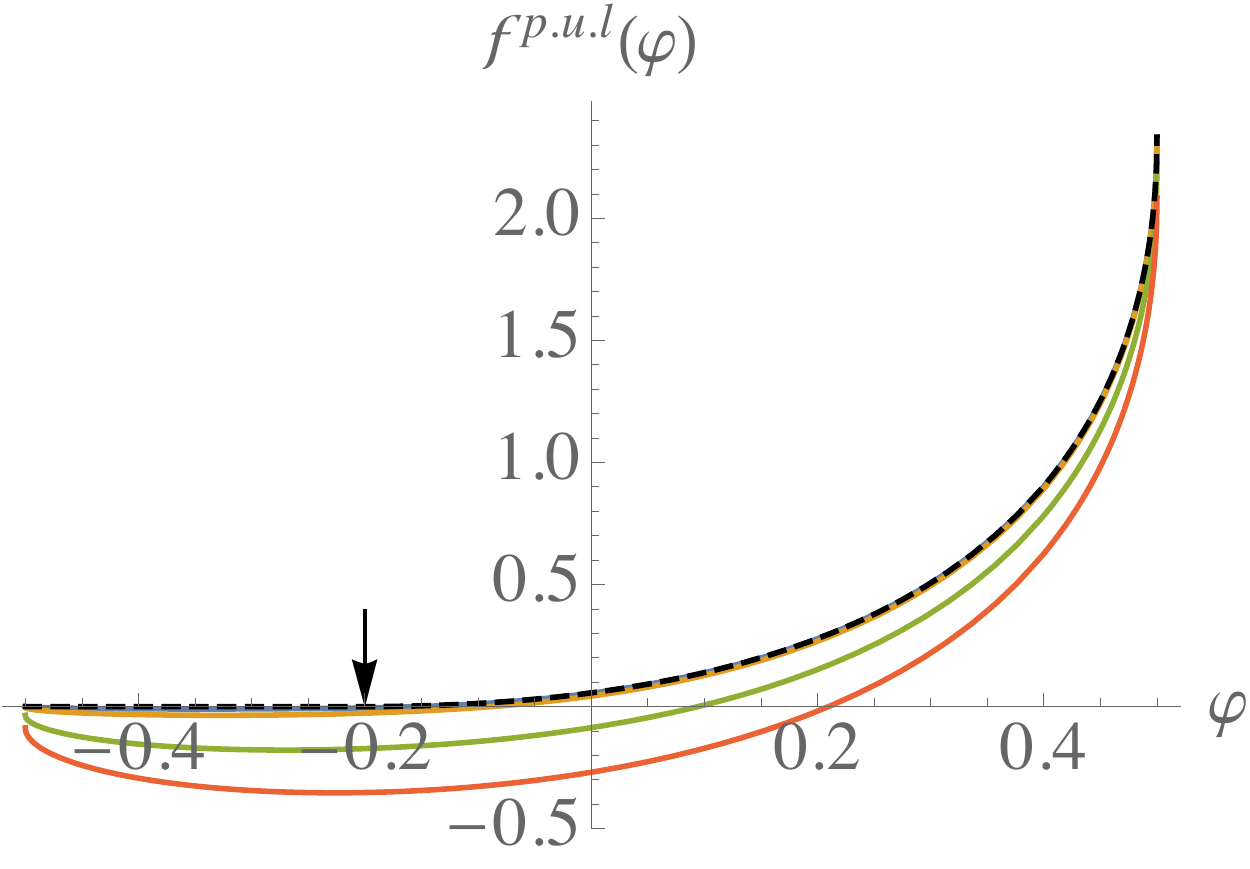} } 
\caption{Left: Optimal energy per-unit-length $\sff^{p.u.l.}(\varphi)$ (\ref{Eq:PresStat:20}) in the Bernoulli-Geometric polymer for $q=0.5$ and $q'=0.1, 0.4, 0.8, 0.9$ (plain lines, blue, orange, green and red) and in the last passage percolation limit $q' \to 0$ (black dashed line). Right: Optimal energy per-unit-length $\sff^{p.u.l.}(\varphi)$ in the Bernoulli-Geometric polymer for $q'=0.7$ and $q'=0.001, 0.01, 0.1, 0.2$ (plain lines, blue, orange, green and red) and in the first passage percolation limit $q \to 0$ (black dashed line). The arrow indicates the percolation threshold of the $q\to 0$ limit $\varphi_{q' = 0.7} = -0.2$. Figures taken from \cite{Thiery2016}.}
\label{fig:PresStat2}
\end{figure}

\smallskip 

{\it Convergence to the stationary measure} \\
Finally we discuss more qualitatively the convergence of the point-to-point partition sum $Z_t(x)$ in the Inverse-Beta polymer / point-to-point optimal energy $\sE_t(x)$ in the Bernoulli-Geometric polymer and we conjecture that the following limit holds in law (for $x,t = O(1)$)
\bea \label{Eq:PresStat:22}
\lim_{T \to \infty} \sE_{T+t}(\varphi T + x)-\sE_{T}(\varphi T) \sim \check \sE_{t}(x) \ssp ,
\eea
where as before $\check \sE_{t}(x)$ is the optimal energy in the stationary Bernoulli-Geometric polymer (i.e. with initial condition \ref{Eq:PresStat:12}), with the stationary parameter $q_b$ chosen as $q_b=q^*(\varphi)$, the solution of the quartic equation (\ref{Eq:PresStat:21}). A similar conjecture is proposed for the Inverse-Beta polymer.

\medskip

Finally, in \cite{Thiery2016}, we successfully check our main results (\ref{Eq:PresStat:20}) and (\ref{Eq:PresStat:22}) for the Bernoulli-Geometric polymer using simulations of the model. Many details on the above results are given, in particular the definition and the properties of the Inverse-Beta and Bernoulli-Geometric polymer with boundaries briefly mentioned here, that are actually at the center of \cite{Thiery2016}.

\section{Conclusion}\label{conclusionDP}

In this chapter we have reviewed some recent progress in the understanding of the KPZ universality class in $1+1$d based on the existence of models with exact solvability properties. In particular we reported the results obtained in this thesis on exactly solvable models of directed polymers on the square lattice. We have showed how the Bethe ansatz approach developed for the continuum case could be adapted in the discrete setting. The Bethe ansatz approach was also used to classify exactly solvable models of DP at finite temperature containing all known models and a new one, the Inverse-Beta polymer. This `world' of exactly solvable models of directed polymers contain models with very different properties. In the Inverse-Beta and Log-Gamma polymer we could show, using the Bethe ansatz, that the model have fluctuations of free-energy scaling with $t^{1/3}$ and distributed according to the Tracy-Widom GUE distribution with explicit non-universal constants. In the Beta polymer, also interpreted as a model of random walk in a time-dependent one-dimensional random environment, we obtained similar results for the fluctuations in the large deviations regime, and completely different ones in the diffusive regime. This model teaches us a lot about TD-RWRE and DPs: on one hand special short-range correlations of the disorder lead to an additional conservation law and break KPZ universality in the diffusive region, on the other-hand in the other directions KPZ universality is recovered in the TD-RWRE framework. Finally in a complementary work we studied the stationary measure of models of DPs on the square lattice and obtained the one of the Inverse-Beta polymer. With this knowledge, we returned to zero temperature models and introduced the Bernoulli-Geometric polymer. We showed that the latter has an exact solvability property, namely we obtained its stationary measure exactly, and deduced from it several exact results. A tentative cartoon of the relations between the models of directed polymers considered in this manuscript is presented in Fig.~\ref{fig:ConclKPZ}.

\begin{figure}
\centerline{\includegraphics[width=9cm]{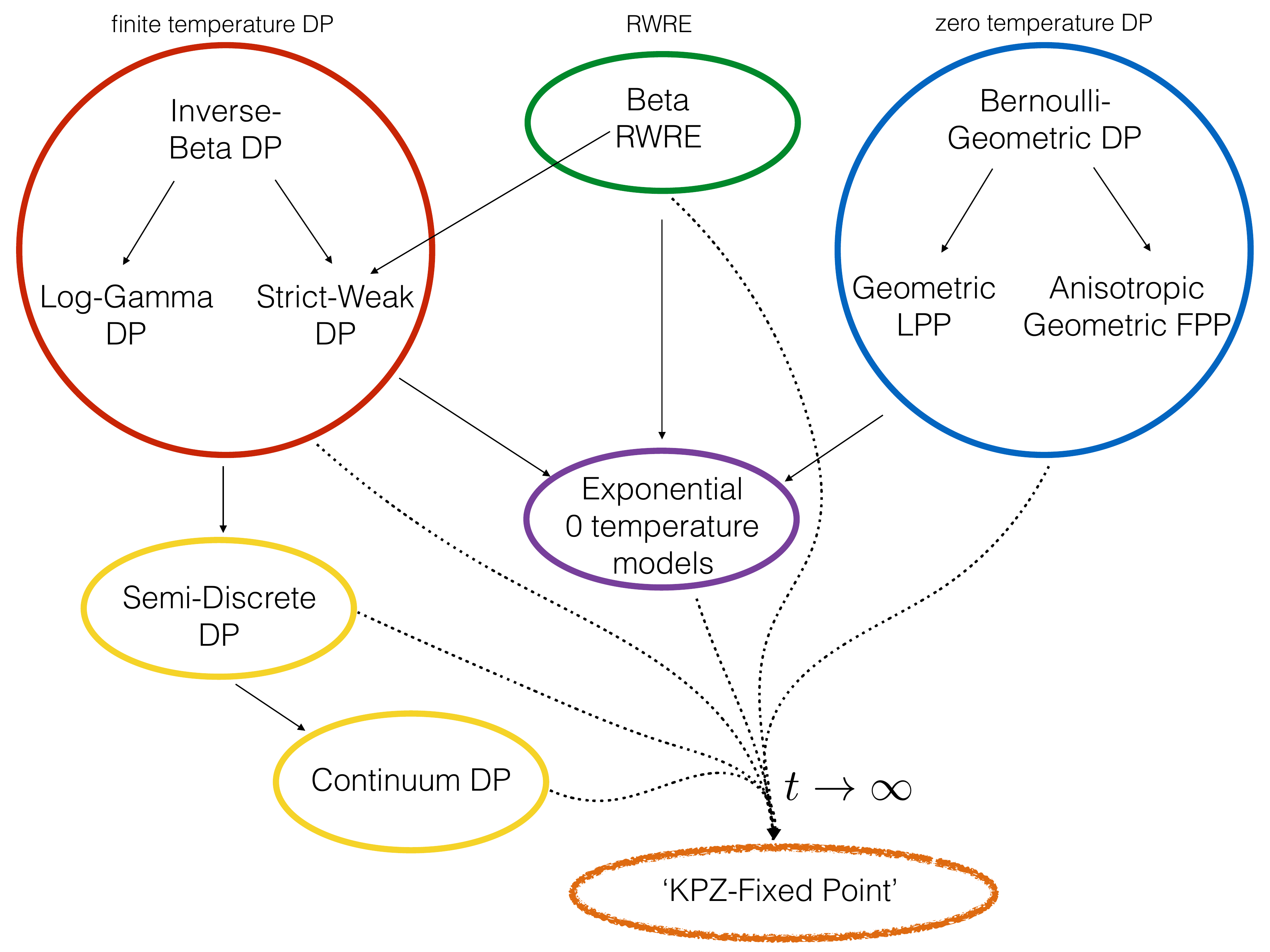} } 
\caption{The different models of DPs encountered during this thesis can be regrouped in several families. Solid arrows represent various scaling limits of the models, while the dotted arrows represent the conjectured large time convergence of the fluctuations of all these models to the KPZ FP. These dotted arrows have to be taken with caution and sometimes miss important properties: the fluctuations of the free-energy in the Beta RWRE in the diffusive regime, or of the first passage time in the anisotropic Geometric FPP model considered above in the percolating region, do not converge to those of the KPZ FP.}
\label{fig:ConclKPZ}
\end{figure}

\medskip

At this stage many directions of research remain. Understanding the remarkable universality unveiled by models in the KPZUC, and more particularly directed polymers, is still a work in progress for which the models we have studied and the techniques we have developed provide valuable tools. Obvious extensions are the study of DPs on the square lattice with different boundary conditions and extension of our results to multi-point statistics. It would also be interesting to gain a better understanding of the localization properties of DPs on the square lattice and in the continuum using a Bethe ansatz approach. The Beta polymer also brought exact solvability techniques to the TD-RWRE field and much work in this direction also remains, in particular testing the universality of our results for more general models of TD-RWRE. Another question is to gain a better understanding of models of DPs with complex weights, which are related to problems of Anderson localization. In this question we already made progress since at least some part of the classification of \cite{ThieryLeDoussal2015} also applies to this case, and if an exactly solvable model of DP on $\JZ^2$ with complex weights exists, then under some mild assumptions some signs of its existence should already be visible in our work (no such signs were found). A more conceptual issue is the understanding of the links between the different exact solvability properties discussed in the manuscript. Finally a long-standing issue is to build techniques allowing to understand the properties of the KPZ FP without relying on the use of exactly solvable models. Indeed, although exactly solvable models allow a remarkable description of the properties of the KPZ FP, it would be highly desirable to get a simple explanation for the emergence of boundary condition dependent extreme value RMT type statistics, or even of the critical exponents. This is particularly important in the aim of understanding the higher-dimensional case where (at least for now) no exactly solvable model exists.

\chapter*{Conclusion}
\phantomsection
\addcontentsline{toc}{chapter}{Conclusion}

In this thesis we have made progress in the understanding of the properties of elastic interfaces in disordered media in their strong disorder regime. In Chapter~\ref{chapII} we have been interested in characterizing the universal properties of avalanches and shocks for disordered elastic interfaces with arbitrary elastic kernels in arbitrary dimensions, working directly at zero temperature. In Chapter~\ref{chapIII} we focused on the study of the statics of a directed polymer in a $1+1d$ random media at finite temperature and were interested in properties related to the KPZ universality class. We refer the reader to Sec.~\ref{conclusionAva} and Sec.~\ref{conclusionDP} for a short summary of our results and conclusions on both subjects and now conclude the thesis with a few more general considerations on the thesis. While in Chapter~\ref{chapII} we used an analytical approach based on the functional renormalization group, leading to results perturbative in $\epsilon = d_{\rm uc}  -d$, in Chapter~\ref{chapIII} we focused on (mostly Bethe ansatz) exactly solvable models for the $d=1$ case. In both cases the aim was to gain information on universal properties of the underlying renormalization group fixed point, but the approach was completely different. While the functional renormalization group approach mostly ignore the microscopic properties of the model, but rather aims at directly describing the fixed point, the approach based on exactly solvable models is all about finding models whose microscopic properties ensure an exact solvability property. Information of great precision about the KPZ fixed point were later obtained through the large scale analysis of exact results.

\medskip

Both these methods have their pros and cons. They were used to characterize different observables mostly for technical reasons. While it is an exact method of outstanding interest, studying the statistics of shocks for the directed polymer in $1+1$d with the Bethe ansatz appears very difficult from the technical point of view. Furthermore, exact solvability methods are up to now restricted to the case of $1+1$ dimension. In this respect the functional renormalization group approach appears much more versatile. However, in the end, it only leads to results that are perturbative in a dimensional expansion below the upper-critical dimension of the interface. While it would certainly be interesting to obtain the (equivalent of) the Tracy-Widom distribution in an expansion in $\epsilon = 4 -d$, it is clear that the result would be far from the remarkable properties observed in the $d=1$ case. Those are not only theoretical gems since they are also nowadays measured in experiments.

\medskip

Overall both these methods permit an advanced understanding of different aspects of the physics of disordered elastics interfaces and of related subjects, making these systems remarkable example of disordered systems for which already existing analytical techniques permit important theoretical progress. While at the quantitative level the only physics that is accessible through their study is the physics of the universality class of disordered elastic interfaces -which already contains a variety of systems-, at the qualitative level the range of application of ideas emerging from their study might be much broader. In particular, investigating the presence of avalanches in the zero temperature physics of various disordered systems is an interesting goal as it is a neat characterization of the presence of many metastable states in the energy landscape, a property which has a strong influence on many aspects of the physics of disordered systems. This has been done e.g. in amorphous solids at the yielding transition \cite{LinLernerRossoWyart2014}, in random field systems \cite{TarjusBaczykTissier2013} or in spin glasses \cite{LeDoussalMullerWiese2010}, but much work in this direction certainly remains. Even more generally, searching for non-analyticities might be a fruitful angle of approach to the study of various disordered systems, as it has clearly been the case for disordered elastic interfaces.

\chapter*{Remerciements}
\phantomsection
\addcontentsline{toc}{chapter}{Remerciements}

Ces trois années de thèse écoulées furent riches en rencontres et je voudrais ici remercier les personnes impliquées de près\footnote{Désolé pour ceux que j'oublie!} dans mon expérience de thésard qui touche à sa fin.

\medskip

Je voudrais tout d'abord remercier Markus M\"uller et Alexander Povolotsky pour avoir accepté d'être les rapporteurs de ma thèse et pour leurs nombreuses remarques qui ont permis d'améliorer ce manuscrit. Merci aussi à Giulio Biroli, Francis Comets et Alberto Rosso pour avoir accepté de compléter le jury de la soutenance.

\medskip

Un grand merci à Pierre Le Doussal et Kay Wiese pour avoir accepté de diriger mon stage de M2 puis ma thèse. J'ai énormément appris à leurs cotés, beaucoup de belle physique, les techniques parfois arides permettant d'extraire la dite physique des équations, à être persévérant devant des problèmes complexes et à avoir le flair pour les attaquer du bon angle. J'étais bien loin d'être un chercheur au début de cette thèse et j'espère en être plus proche aujourd'hui grâce à eux. Le temps passant au cours de cette thèse j'ai été amené à travailler essentiellement avec Pierre que j'aimerais remercier plus particulièrement, pour l'attention soutenue qu'il a portée à mon travail; pour tout le temps passé à des discussions dans son bureau, qui m'ont souvent été immensément bénéfiques grâce à sa capacité à jauger mon niveau et à s'y adapter pour mieux l'élever; pour sa velléité à m'envoyer dans des endroits ensoleillés; pour la justesse qu'il a montrée dans sa manière de me guider à travers une belle aventure scientifique à la recherche de Tracy Widom et autres merveilles... J'espère que nous continuerons à l'avenir nos recherches communes.

\medskip

Je tiens aussi à remercier Viviane Sebille et Sandrine Patacchini pour tout le travail fourni au cours de ces trois ans, ainsi que Marc-Thierry Jaekel pour m'avoir aidé à de maintes reprises à apprivoiser les équipements informatiques du LPT. Merci aussi à mes nombreux collégues de bureau Antoine Bourget, Matthieu Delorme, Alexander Dobrinevski, Bruno Le Floch, Thomas Gueudré, Thibaud Maimbourg, Antoine Tilloy, Romain Vasseur et Éric Vernier, pour la convivialité constante au cours de ces années de pérégrinations dans un laboratoire au décor parfois post-apocalyptique. Je voudrais remercier plus particulièrement mes illustres prédécesseurs Alexander et Thomas pour m'avoir guidé avec beaucoup de gentillesse au début de ma thèse et pendant le stage de M2 dans les méandres des avalanches, FRG, polymère dirigé et ansatz de Bethe. Leur impact positif sur mon travail et ma compréhension de ces sujets à ces stades précoces de ma thèse a été énorme. Outre les apports scientifiques je les remercie également évidemment pour leur amitié qui m'a été précieuse durant ces années. Je remercie également mes collégues thésards et post-docs de l'ens par delà mon bureau Jacopo De Nardis et Andrea De Luca, ainsi que Jean Barbier, Ralph Bourdoukan, Alice Coucke, Christophe Gardella et Alaa Saade, pour tous les moments partagés autour de gros dwichs, fat flans, d'eau municipale et de ramens.

\medskip

Je voudrais également remercier toutes les personnes avec qui j'ai partagé des moments sympas pendant ces nombreuses \sout{vacances} écoles d'été et conférences, je pense notamment à Nicolas Posé et Mathias Van Regemortel pour ces durs moments de dolce vita à Trieste; Grégory Schehr et l'équipe des Houches: Nicola Allegra, Charlie Duclut, Tony Prat et Malo Tarpin; ainsi qu'Élizabeth Agoristas et Vivien Lecomte, notamment pour le marathon gastronomique qui nous a permis de mieux accepter le rythme harassant de la vie en Californie.

\medskip

Je remercie également mes professeurs de sciences du lycée qui ont su éveiller en moi un intérêt pour les sciences pourtant au début très profondément enfoui, en particulier Mme Capron qui a été le principal artisan de cet éveil, mais aussi Mr Morand et les autres, ainsi que Mme Biolet et Mme Guest dont les cours en prépa m'ont permis d'en arriver là.

\medskip

Finalement je tiens à profiter de ce premier espace formel de remerciements qui m'est offert pour redire mon affection aux personnes qui me sont chères. Merci ainsi à tous mes amis parisiens et tourangeaux pour toutes ces années. Merci à ma famille. Merci Suzanne.

%%%%%%%%%%%%%% les pages finales %%%%%%%%%%%%%%%%
\backmatter
%--------------------------------------------------
% \cleardoublepage
% \phantomsection
% \addcontentsline{toc}{chapter}{\listfigurename}
% \listoffigures

%%% Bibliographie 
\cleardoublepage
\phantomsection
\addcontentsline{toc}{chapter}{Bibliographie}
%%% utilisation de BiBTeX
\bibliography{Citations-Thimothee-24-06}

\begin{thebibliography}{100}
\newcommand{\enquote}[1]{``#1''}

\bibitem{ThieryLeDoussalWiese2015}
T.~Thiery, P.~{Le~Doussal}, et K.~J. Wiese, \enquote{Spatial shape of
  avalanches in the brownian force model,} Journal of Statistical Mechanics:
  Theory and Experiment \textbf{2015}, P08019 (2015).
\newblock URL~: \url{http://stacks.iop.org/1742-5468/2015/i=8/a=P08019}.

\bibitem{ThieryLeDoussal2016a}
T.~{Thiery} et P.~{Le Doussal}, \enquote{{Universality in the mean spatial
  shape of avalanches},} ArXiv e-prints  (2016).

\bibitem{ThieryLeDoussalWiese2016}
T.~{Thiery}, P.~{Le Doussal}, et K.~{J{\"o}rg Wiese}, \enquote{{Universal
  correlations between shocks in the ground state of elastic interfaces in
  disordered media},} ArXiv e-prints  (2016).

\bibitem{ThieryLeDoussal2014}
T.~{Thiery} et P.~{Le Doussal}, \enquote{{Log-gamma directed polymer with fixed
  endpoints via the replica Bethe Ansatz},} Journal of Statistical Mechanics:
  Theory and Experiment \textbf{10}, 10018 (2014).

\bibitem{ThieryLeDoussal2015}
T.~{Thiery} et P.~{Le Doussal}, \enquote{{On integrable directed polymer models
  on the square lattice},} Journal of Physics A Mathematical General
  \textbf{48}, 465001 (2015).

\bibitem{ThieryLeDoussal2016b}
T.~{Thiery} et P.~{Le Doussal}, \enquote{{Exact solution for a random walk in a
  time-dependent 1D random environment: the point-to-point Beta polymer},}
  ArXiv e-prints  (2016).

\bibitem{Thiery2016}
T.~{Thiery}, \enquote{{Stationary measures for two dual families of finite and
  zero temperature models of directed polymers on the square lattice},} ArXiv
  e-prints  (2016).

\bibitem{Balents1996}
L.~Balents, \enquote{Glassy phases and dynamics of randomly pinned elastic
  media,}   (1996).
\newblock URL~:
  \url{http://online.kitp.ucsb.edu/online/lnotes/balents/bignotes.html}.

\bibitem{ChauveThesis}
P.~Chauve, \enquote{Dynamique des syst\`emes \'elastiques d\'sordonn\'es,} PhD
  thesis, Universit\'e Paris XI Orsay  (2000).

\bibitem{GiamarchiKoltonRosso2006}
T.~{Giamarchi}, A.~B. {Kolton}, et A.~{Rosso}, \enquote{{Dynamics of Disordered
  Elastic Systems},} dans \enquote{Jamming, Yielding, and Irreversible
  Deformation in Condensed Matter,} , vol. 688 de \emph{Lecture Notes in
  Physics, Berlin Springer Verlag}, M.~C. {Miguel} et M.~{Rubi}, éds. (2006),
  vol. 688 de \emph{Lecture Notes in Physics, Berlin Springer Verlag}, p.
  91--108.

\bibitem{DSFisher1985}
D.~Fisher, \enquote{Sliding charge-density waves as a dynamical critical
  phenomena,} Phys. Rev. \textbf{B 31}, 1396--1427 (1985).

\bibitem{FedorenkoLeDoussalWiese2006b}
A.~Fedorenko, P.~{Le~Doussal}, et K.~Wiese, \enquote{Statics and dynamics of
  elastic manifolds in media with long-range correlated disorder,} Phys. Rev. E
  \textbf{74}, 061109 (2006).

\bibitem{BiroliBouchaudPotters2007}
G.~Biroli, J.-P. Bouchaud, et M.~Potters, \enquote{Extreme value problems in
  random matrix theory and other disordered systems,} Journal of Statistical
  Mechanics: Theory and Experiment \textbf{2007}, P07019 (2007).
\newblock URL~: \url{http://stacks.iop.org/1742-5468/2007/i=07/a=P07019}.

\bibitem{GueudreLeDoussalBouchaudRosso2015}
T.~{Gueudre}, P.~{Le Doussal}, J.-P. {Bouchaud}, et A.~{Rosso},
  \enquote{{Ground-state statistics of directed polymers with heavy-tailed
  disorder},} Phys. Rev. E. \textbf{91}, 062110 (2015).

\bibitem{Larkin1970}
A.~Larkin, Sov. Phys. JETP \textbf{31}, 784 (1970).

\bibitem{EfetovLarkin1977}
K.~Efetov et A.~Larkin, Sov. Phys. JETP \textbf{45}, 1236 (1977).

\bibitem{ParisiSourlas1979}
G.~Parisi et N.~Sourlas, \enquote{Random magnetic fields, supersymmetry, and
  negative dimensions,} Phys. Rev. Lett. \textbf{43}, 744--5 (1979).

\bibitem{ChauveLeDoussal2001}
P.~Chauve et P.~{Le~Doussal}, \enquote{Exact multilocal renormalization group
  and applications to disordered problems,} Phys. Rev. E \textbf{64},
  051102/1--27 (2001).

\bibitem{HohenbergHalperin1977}
P.~Hohenberg et B.~Halperin, \enquote{Theory of dynamical critical phenomena,}
  Rev. Mod. Phys. \textbf{49}, 435 (1977).

\bibitem{Middleton1992}
A.~Middleton, \enquote{Asymptotic uniqueness of the sliding state for
  charge-density waves,} Phys. Rev. Lett. \textbf{68}, 670--673 (1992).

\bibitem{LeschhornTang1994}
H.~Leschhorn et L.-H. Tang, \enquote{Avalanches and correlations in driven
  interface depinning,} Phys. Rev. E \textbf{49}, 1238--1245 (1994).
\newblock URL~: \url{http://link.aps.org/doi/10.1103/PhysRevE.49.1238}.

\bibitem{Nattermann1987}
T.~Nattermann, \enquote{Interface roughening in systems with quenched random
  impurities,} EPL (Europhysics Letters) \textbf{4}, 1241 (1987).
\newblock URL~: \url{http://stacks.iop.org/0295-5075/4/i=11/a=005}.

\bibitem{IoffeVinokur1987}
L.~B. Ioffe et V.~M. Vinokur, \enquote{Dynamics of interfaces and dislocations
  in disordered media,} Journal of Physics C: Solid State Physics \textbf{20},
  6149 (1987).
\newblock URL~: \url{http://stacks.iop.org/0022-3719/20/i=36/a=016}.

\bibitem{NattermannShapirVilfan1990}
T.~Nattermann, Y.~Shapir, et I.~Vilfan, \enquote{Interface pinning and dynamics
  in random systems,} Phys. Rev. B \textbf{42}, 8577--8586 (1990).
\newblock URL~: \url{http://link.aps.org/doi/10.1103/PhysRevB.42.8577}.

\bibitem{FeigelmandGeshkenbeinLarkinVinokur1989}
M.~V. Feigel'man, V.~B. Geshkenbein, A.~I. Larkin, et V.~M. Vinokur,
  \enquote{Theory of collective flux creep,} Phys. Rev. Lett. \textbf{63},
  2303--2306 (1989).
\newblock URL~: \url{http://link.aps.org/doi/10.1103/PhysRevLett.63.2303}.

\bibitem{ChauveGiamarchiLeDoussal1998}
P.~Chauve, T.~Giamarchi, et P.~{Le~Doussal}, \enquote{Creep via dynamical
  functional renormalization group,} Europhys. Lett. \textbf{44}, 110--15
  (1998).

\bibitem{ChauveGiamarchiLeDoussal2000}
P.~Chauve, T.~Giamarchi, et P.~{Le~Doussal}, \enquote{Creep and depinning in
  disordered media,} Phys. Rev. B \textbf{62}, 6241--67 (2000).

\bibitem{KoltonRossoGiamarchi2005}
A.~Kolton, A.~Rosso, et T.~Giamarchi, \enquote{Creep motion of an elastic
  string in a random potential,} Phys. Rev. Lett. \textbf{94}, 047002 (2005).

\bibitem{LemerleFerreChappertMathetGiamarchiLeDoussal1998}
S.~Lemerle, J.~{Ferr\'e}, C.~Chappert, V.~Mathet, T.~Giamarchi, et P.~{Le
  Doussal}, \enquote{Domain wall creep in an {Ising} ultrathin magnetic film,}
  Phys. Rev. Lett. \textbf{80}, 849 (1998).

\bibitem{AgoristasLecomte2016}
E.~{Agoritsas}, R.~{Garc{\'{\i}}a-Garc{\'{\i}}a}, V.~{Lecomte},
  L.~{Truskinovsky}, et D.~{Vandembroucq}, \enquote{{Driven interfaces: from
  flow to creep through model reduction},} ArXiv e-prints  (2016).

\bibitem{Jagla2016}
E.~A. {Jagla}, \enquote{{Scaling theory of thermal rounding phenomena in
  depinning and related transitions},} ArXiv e-prints  (2016).

\bibitem{FerreroFoiniGiamarchiKoltonRosso2016}
E.~E. {Ferrero}, L.~{Foini}, T.~{Giamarchi}, A.~B. {Kolton}, et A.~{Rosso},
  \enquote{{Spatio-temporal patterns in ultra-slow domain wall creep
  dynamics},} ArXiv e-prints  (2016).

\bibitem{KPZ}
M.~Kardar, G.~Parisi, et Y.-C. Zhang, \enquote{Dynamic scaling of growing
  interfaces,} Phys. Rev. Lett. \textbf{56}, 889--892 (1986).

\bibitem{Corwin2011Review}
I.~{Corwin}, \enquote{The kardar-parisi-zhang equation and universality class,}
  Random Matrices: Theory and Applications \textbf{01}, 1130001 (2012).
\newblock URL~:
  \url{http://www.worldscientific.com/doi/abs/10.1142/S2010326311300014}.

\bibitem{HalpinTakeuchi2015}
T.~{Halpin-Healy} et K.~A. {Takeuchi}, \enquote{{A KPZ Cocktail-Shaken, not
  Stirred...}} Journal of Statistical Physics \textbf{160}, 794--814 (2015).

\bibitem{QuastelSpohn2015}
J.~{Quastel} et H.~{Spohn}, \enquote{{The One-Dimensional KPZ Equation and Its
  Universality Class},} Journal of Statistical Physics \textbf{160}, 965--984
  (2015).

\bibitem{TracyWidom1993}
C.~A. {Tracy} et H.~{Widom}, \enquote{{Level-spacing distributions and the Airy
  kernel},} Physics Letters B \textbf{305}, 115--118 (1993).

\bibitem{TracyWidom1996}
C.~A. {Tracy} et H.~{Widom}, \enquote{{On orthogonal and symplectic matrix
  ensembles},} Communications in Mathematical Physics \textbf{177}, 727--754
  (1996).

\bibitem{BaikRains2000}
J.~{Baik} et E.~{Rains}, \enquote{{Limiting distributions for a polynuclear
  growth model with external sources},} ArXiv Mathematics e-prints  (2000).

\bibitem{HuseHenley1985}
D.~A. Huse et C.~L. Henley, \enquote{Pinning and roughening of domain walls in
  {Ising} systems due to random impurities,} Phys. Rev. Lett. \textbf{54},
  2708--2711 (1985).

\bibitem{JiRobbins1991}
H.~Ji et M.~O. Robbins, \enquote{Transition from compact to self-similar growth
  in disordered systems: Fluid invasion and magnetic-domain growth,} Phys. Rev.
  A \textbf{44}, 2538--2542 (1991).
\newblock URL~: \url{http://link.aps.org/doi/10.1103/PhysRevA.44.2538}.

\bibitem{Colaiori2008}
F.~Colaiori, \enquote{Exactly solvable model of avalanches dynamics for
  barkhausen crackling noise,} Advances in Physics \textbf{57}, 287 (2008).
\newblock URL~: \url{doi:10.1080/00018730802420614}.

\bibitem{DobrinevskiPhD}
A.~Dobrinevski, \enquote{Field theory of disordered systems -- avalanches of an
  elastic interface in a random medium,} arXiv:\null \textbf{1312.7156} (2013).

\bibitem{Barkhausen1919}
H.~Barkhausen, \enquote{{No Title},} Phys. Z. \textbf{20}, 401--403 (1919).

\bibitem{DurinZapperi2000}
G.~Durin et S.~Zapperi, \enquote{Scaling exponents for {Barkhausen} avalanches
  in polycrystalline and amorphous ferromagnets,} Phys. Rev. Lett. \textbf{84},
  4705--4708 (2000).

\bibitem{DurinZapperi2006b}
G.~Durin et S.~Zapperi, \enquote{{The Barkhausen effect},} dans \enquote{The
  Science of Hysteresis,} , G.~Bertotti et I.~Mayergoyz, éds. (Amsterdam,
  2006), p.~51.
\newblock URL~: \url{http://arxiv.org/abs/cond-mat/0404512}.

\bibitem{CizeauZapperiDurinStanley1997}
P.~Cizeau, S.~Zapperi, G.~Durin, et H.~Stanley, \enquote{{Dynamics of a
  Ferromagnetic Domain Wall and the Barkhausen Effect},} Phys. Rev. Lett.
  \textbf{79}, 4669--4672 (1997).
\newblock URL~: \url{http://link.aps.org/doi/10.1103/PhysRevLett.79.4669}.

\bibitem{Bonamy2009}
D.~Bonamy, \enquote{Intermittency and roughening in the failure of brittle
  heterogeneous materials,} Journal of Physics D: Applied Physics \textbf{42},
  214014 (2009).
\newblock URL~: \url{http://stacks.iop.org/0022-3727/42/i=21/a=214014}.

\bibitem{SchmittbuhlRouxVilotteMaaloy1995}
J.~Schmittbuhl, S.~Roux, J.-P. Vilotte, et K.~Jorgen~M\aa{}l\o{}y,
  \enquote{Interfacial crack pinning: Effect of nonlocal interactions,} Phys.
  Rev. Lett. \textbf{74}, 1787--1790 (1995).
\newblock URL~: \url{http://link.aps.org/doi/10.1103/PhysRevLett.74.1787}.

\bibitem{RamanathanErtasFisher1997}
S.~Ramanathan, D.~Erta\ifmmode~\mbox{\c{s}}\else \c{s}\fi{}, et D.~S. Fisher,
  \enquote{Quasistatic crack propagation in heterogeneous media,} Phys. Rev.
  Lett. \textbf{79}, 873--876 (1997).
\newblock URL~: \url{http://link.aps.org/doi/10.1103/PhysRevLett.79.873}.

\bibitem{BonamySantucciPonson2008}
D.~Bonamy, S.~Santucci, et L.~Ponson, \enquote{Crackling dynamics in material
  failure as the signature of a self-organized dynamic phase transition,} Phys.
  Rev. Lett. \textbf{101}, 045501 (2008).
\newblock URL~: \url{http://link.aps.org/abstract/PRL/v101/e045501}.

\bibitem{Ponson2008}
L.~Ponson, \enquote{Depinning transition in failure of inhomogeneous brittle
  materials,} Phys. Rev. Lett. \textbf{103}, 055501 (2009).

\bibitem{SchmittbuhlMaloy1997}
J.~Schmittbuhl et K.~M\aa{}l\o{}y, \enquote{Direct observation of a self-affine
  crack propagation,} Phys. Rev. Lett. \textbf{78}, 3888--91 (1997).

\bibitem{DeleplaceSchmittbuhlMaaloy1999}
A.~Delaplace, J.~Schmittbuhl, et K.~J. M\aa{}l\o{}y, \enquote{High resolution
  description of a crack front in a heterogeneous plexiglas block,} Phys. Rev.
  E \textbf{60}, 1337--1343 (1999).
\newblock URL~: \url{http://link.aps.org/doi/10.1103/PhysRevE.60.1337}.

\bibitem{LaursonSantucciZapperi2010}
L.~Laurson, S.~Santucci, et S.~Zapperi, \enquote{Avalanches and clusters in
  planar crack front propagation,} Phys. Rev. E \textbf{81}, 046116 (2010).
\newblock URL~: \url{http://link.aps.org/doi/10.1103/PhysRevE.81.046116}.

\bibitem{SantucciGrobToussaint2010}
S.~Santucci, M.~Grob, R.~Toussaint, J.~Schmittbuhl, A.~Hansen, et K.~J. Maløy,
  \enquote{Fracture roughness scaling: A case study on planar cracks,} EPL
  (Europhysics Letters) \textbf{92}, 44001 (2010).
\newblock URL~: \url{http://stacks.iop.org/0295-5075/92/i=4/a=44001}.

\bibitem{TallakstadToussaintSantucciSchmittbuhlMaaloy2011}
K.~Tallakstad, R.~Toussaint, S.~Santucci, J.~Schmittbuhl, et K.~M\aa{}l\o{}y,
  \enquote{Local dynamics of a randomly pinned crack front during creep and
  forced propagation: An experimental study,} Phys. Rev. E \textbf{83}, 046108
  (2011).
\newblock URL~: \url{http://link.aps.org/doi/10.1103/PhysRevE.83.046108}.

\bibitem{SchafferWongZen2000}
E.~Sch\"affer et P.-z. Wong, \enquote{Contact line dynamics near the pinning
  threshold: A capillary rise and fall experiment,} Phys. Rev. E \textbf{61},
  5257--5277 (2000).
\newblock URL~: \url{http://link.aps.org/doi/10.1103/PhysRevE.61.5257}.

\bibitem{MoulinetGuthmannRolley2002}
S.~Moulinet, C.~Guthmann, et E.~Rolley, \enquote{Roughness and dynamics of a
  contact line of a viscous fluid on a disordered substrate,} Eur. Phys. J. E
  \textbf{8}, 437--443 (2002).

\bibitem{MoulinetGuthmannRolley2004}
S.~Moulinet, C.~Guthmann, et E.~Rolley, \enquote{Dissipation in the dynamics of
  a moving contact line: effect of the substrate disorder,} Eur. Phys. J. B
  \textbf{37}, 127--136 (2004).

\bibitem{LeDoussalWieseMoulinetRolley2009}
P.~{Le~Doussal}, K.~Wiese, S.~Moulinet, et E.~Rolley, \enquote{Height
  fluctuations of a contact line: {A} direct measurement of the renormalized
  disorder correlator,} EPL \textbf{87}, 56001 (2009).

\bibitem{LeDoussalWieseRaphaelGolestanian2004}
P.~L. Doussal, K.~Wiese, E.~Raphael, et R.~Golestanian, \enquote{Can non-linear
  elasticity explain contact-line roughness at depinning?} Phys. Rev. Lett.
  \textbf{96}, 015702 (2006).

\bibitem{DSFisher1998}
D.~Fisher, \enquote{Collective transport in random media: {From}
  superconductors to earthquakes,} Phys. Rep. \textbf{301}, 113--150 (1998).

\bibitem{BenZionRice1993}
Y.~Ben-Zion et J.~Rice, \enquote{{Earthquake failure sequences along a cellular
  fault zone in a three-dimensional elastic solid containing asperity and
  nonasperity regions},} Journal of Geophysical Research \textbf{98},
  14109--14131 (1993).
\newblock URL~:
  \url{http://citeseerx.ist.psu.edu/viewdoc/download?doi=10.1.1.161.5821\&amp;rep=rep1\&amp;type=pdf}.

\bibitem{BenZionRice1997}
Y.~Ben-Zion et J.~Rice, \enquote{{Dynamic simulations of slip on a smooth fault
  in an elastic solid},} Journal of Geophysical Research \textbf{102}, 17--17
  (1997).
\newblock URL~:
  \url{http://earth.usc.edu/~ybz/pubs\_recent/BZR\_JGR97/BZR\_JGR97.pdf}.

\bibitem{FisherDahmenRamanathanBenZion1997}
D.~S. Fisher, K.~Dahmen, S.~Ramanathan, et Y.~Ben-Zion, \enquote{Statistics of
  earthquakes in simple models of heterogeneous faults,} Phys. Rev. Lett.
  \textbf{78}, 4885--4888 (1997).
\newblock URL~: \url{http://link.aps.org/doi/10.1103/PhysRevLett.78.4885}.

\bibitem{Omori1894}
F.~Omori, \enquote{On the aftershocks of earthquakes,} Journal of the College
  of Science, Imperial University of Tokyo \textbf{7}, 111--200.

\bibitem{GiamarchiLeDoussalBookYoung}
T.~Giamarchi et P.~{Le~Doussal}, \enquote{Statics and dynamics of disordered
  elastic systems,} dans \enquote{Spin glasses and random fields,} , A.~Young,
  éd. (World Scientific, Singapore, 1997).

\bibitem{PlanetSantucciOrtin2009}
R.~Planet, S.~Santucci, et J.~Ort\'{\i}n, \enquote{Avalanches and non-gaussian
  fluctuations of the global velocity of imbibition fronts,} Phys. Rev. Lett.
  \textbf{102}, 094502 (2009).
\newblock URL~: \url{http://link.aps.org/doi/10.1103/PhysRevLett.102.094502}.

\bibitem{AlavaRostDube2004}
M.~{Alava}, M.~{Rost}, et M.~{Dub{\'e}}, \enquote{{Imbibition in disordered
  media},} Advances in Physics \textbf{53}, 83--175 (2004).

\bibitem{BarabisiStanleyBook}
A.-L. Barab\'asi et H.~E. Stanley, \emph{Fractal Concepts in Surface Growth}
  (Cambridge University Press, 1995).
\newblock URL~: \url{http://dx.doi.org/10.1017/CBO9780511599798}, cambridge
  Books Online.

\bibitem{HalpinHealyZhang1995}
T.~Halpin-Healy et Y.-C. Zhang, \enquote{Kinetic roughening phenomena,
  stochastic growth, directed polymers and all that,} Phys. Rep. \textbf{254},
  215--415 (1995).

\bibitem{WakitaItohMatsuyamaMatsushita}
J.~ichi Wakita, H.~Itoh, T.~Matsuyama, et M.~Matsushita, \enquote{Self-affinity
  for the growing interface of bacterial colonies,} Journal of the Physical
  Society of Japan \textbf{66}, 67--72 (1997).
\newblock URL~: \url{http://dx.doi.org/10.1143/JPSJ.66.67}.

\bibitem{HuergoPasqualeGonzalezBolzanArvia2012}
M.~A.~C. Huergo, M.~A. Pasquale, P.~H. Gonz\'alez, A.~E. Bolz\'an, et A.~J.
  Arvia, \enquote{Growth dynamics of cancer cell colonies and their comparison
  with noncancerous cells,} Phys. Rev. E \textbf{85}, 011918 (2012).
\newblock URL~: \url{http://link.aps.org/doi/10.1103/PhysRevE.85.011918}.

\bibitem{Myllys&al2001}
M.~Myllys, J.~Maunuksela, M.~Alava, T.~Ala-Nissila, J.~Merikoski, et
  J.~Timonen, \enquote{Kinetic roughening in slow combustion of paper,} Phys.
  Rev. E \textbf{64}, 036101 (2001).
\newblock URL~: \url{http://link.aps.org/doi/10.1103/PhysRevE.64.036101}.

\bibitem{TakeuchiSano2010}
K.~A. {Takeuchi} et M.~{Sano}, \enquote{{Universal Fluctuations of Growing
  Interfaces: Evidence in Turbulent Liquid Crystals},} Physical Review Letters
  \textbf{104}, 230601 (2010).

\bibitem{TakeuchiSano2011}
K.~A. {Takeuchi}, M.~{Sano}, T.~{Sasamoto}, et H.~{Spohn}, \enquote{{Growing
  interfaces uncover universal fluctuations behind scale invariance},}
  Scientific Reports \textbf{1}, 34 (2011).

\bibitem{TakeuchiSano2012}
K.~A. {Takeuchi} et M.~{Sano}, \enquote{{Evidence for Geometry-Dependent
  Universal Fluctuations of the Kardar-Parisi-Zhang Interfaces in
  Liquid-Crystal Turbulence},} Journal of Statistical Physics \textbf{147},
  853--890 (2012).

\bibitem{Takeuchi2013}
K.~A. {Takeuchi}, \enquote{{Crossover from Growing to Stationary Interfaces in
  the Kardar-Parisi-Zhang Class},} Physical Review Letters \textbf{110}, 210604
  (2013).

\bibitem{TakeuchiSano2014}
K.~A. {Takeuchi}, \enquote{{Experimental approaches to universal
  out-of-equilibrium scaling laws: turbulent liquid crystal and other
  developments},} Journal of Statistical Mechanics: Theory and Experiment
  \textbf{1}, 01006 (2014).

\bibitem{SethnaDahmenMyers2001}
J.~Sethna, K.~Dahmen, et C.~Myers, \enquote{Crackling noise,} Nature
  \textbf{410}, 242--250 (2001).

\bibitem{BakTangWiesenfeld1987}
P.~Bak, C.~Tang, et K.~Wiesenfeld, \enquote{Self-organized criticality - an
  explanation of 1/f noise,} Phys. Rev. Lett. \textbf{59}, 381--384 (1987).

\bibitem{Manna1991}
S.~Manna, \enquote{Two-state model of self-organized critical phenomena,}
  J.~Phys.~A \textbf{24}, L363--L369 (1991).

\bibitem{Dhar1999b}
D.~Dhar, \enquote{The {Abelian} sandpile and related models,} Physica A
  \textbf{263}, 4 (1999).
\newblock URL~:
  \url{http://www.citebase.org/abstract?id=oai:arXiv.org:cond-mat/9808047}.

\bibitem{PruessnerBook}
G.~Pruessner, \emph{Self-Organised Criticality: Theory, Models and
  Characterisation} (Cambridge University Press, 2012).

\bibitem{DahmenSethna1996}
K.~Dahmen et J.~Sethna, \enquote{Hysteresis, avalanches, and disorder-induced
  critical scaling: A renormalization-group approach,} Phys. Rev. B
  \textbf{53}, 14872--14905 (1996).

\bibitem{TarjusBaczykTissier2013}
G.~Tarjus, M.~Baczyk, et M.~Tissier, \enquote{Avalanches and dimensional
  reduction breakdown in the critical behavior of disordered systems,} Physical
  review letters \textbf{110}, 135703 (2013).

\bibitem{TarjusTissier2016}
G.~Tarjus et M.~Tissier, \enquote{Avalanches and perturbation theory in the
  random-field ising model,} Journal of Statistical Mechanics: Theory and
  Experiment \textbf{2016}, 023207 (2016).
\newblock URL~: \url{http://stacks.iop.org/1742-5468/2016/i=2/a=023207}.

\bibitem{LeDoussalMullerWiese2010}
P.~{Le Doussal}, M.~{M{\"u}ller}, et K.~J. {Wiese}, \enquote{{Avalanches in
  mean-field models and the Barkhausen noise in spin-glasses},} EPL
  (Europhysics Letters) \textbf{91}, 57004 (2010).

\bibitem{LeDoussalMullerWiese2012}
P.~{Le Doussal}, M.~{M{\"u}ller}, et K.~J. {Wiese}, \enquote{{Equilibrium
  avalanches in spin glasses},} Phys. Reb. B. \textbf{85}, 214402 (2012).

\bibitem{MullerWyart2015}
M.~M\"uller et M.~Wyart, \enquote{Marginal stability in structural, spin, and
  electron glasses,} Annual Review of Condensed Matter Physics \textbf{6},
  177--200 (2015).
\newblock URL~:
  \url{http://dx.doi.org/10.1146/annurev-conmatphys-031214-014614}.

\bibitem{LeDoussalWiese2014a}
P.~L. Doussal et K.~Wiese, \enquote{An exact mapping of the stochastic field
  theory for {Manna} sandpiles to interfaces in random media,} Phys. Rev. Lett.
  \textbf{114}, 110601 (2014).

\bibitem{LinLernerRossoWyart2014}
J.~Lin, E.~Lerner, A.~Rosso, et M.~Wyart, \enquote{Scaling description of the
  yielding transition in soft amorphous solids at zero temperature,}
  Proceedings of the National Academy of Sciences \textbf{111}, 14382--14387
  (2014).
\newblock URL~: \url{http://www.pnas.org/content/111/40/14382.abstract}.

\bibitem{Sinai1982}
Y.~G. Sinai, \enquote{The limiting behavior of a one-dimensional random walk in
  a random medium,} Theory of Probability \& Its Applications \textbf{27},
  256--268 (1983).
\newblock URL~: \url{http://dx.doi.org/10.1137/1127028}.

\bibitem{Kida1979}
S.~Kida, \enquote{Asymptotic properties of burgers turbulence,} Journal of
  Fluid Mechanics \textbf{93}, 337--377 (1979).
\newblock URL~: \url{http://journals.cambridge.org/article_S0022112079001932}.

\bibitem{LeDoussal2008}
P.~{Le Doussal}, \enquote{Exact results and open questions in first principle
  {functional RG},} Annals of Physics \textbf{325}, 49--150 (2009).

\bibitem{LeDoussalWiese2008a}
P.~{Le Doussal} et K.~J. Wiese, \enquote{Driven particle in a random landscape:
  disorder correlator, avalanche distribution and extreme value statistics of
  records,} Phys. Rev. E \textbf{79}, 051105 (2009).

\bibitem{ABBMTh}
B.~Alessandro, C.~Beatrice, G.~Bertotti, et A.~Montorsi, \enquote{Domain-wall
  dynamics and barkhausen effect in metallic ferromagnetic materials. i.
  theory,} Journal of Applied Physics \textbf{68} (1990).

\bibitem{ABBMEx}
B.~Alessandro, C.~Beatrice, G.~Bertotti, et A.~Montorsi, \enquote{Domain-wall
  dynamics and barkhausen effect in metallic ferromagnetic materials. ii.
  experiments,} Journal of Applied Physics \textbf{68} (1990).

\bibitem{ZapperiCizeauDurinStanley1998}
S.~Zapperi, P.~Cizeau, G.~Durin, et H.~E. Stanley, \enquote{Dynamics of a
  ferromagnetic domain wall: Avalanches, depinning transition, and the
  {Barkhausen} effect,} Phys. Rev. B \textbf{58}, 6353--6366 (1998).

\bibitem{LeDoussalWiese2012a}
P.~{Le~Doussal} et K.~J. Wiese, \enquote{Avalanche dynamics of elastic
  interfaces,} Phys. Rev. E \textbf{88}, 022106 (2013).
\newblock URL~: \url{http://link.aps.org/doi/10.1103/PhysRevE.88.022106}.

\bibitem{DobrinevskiLeDoussalWiese2011b}
A.~Dobrinevski, P.~{Le Doussal}, et K.~J. Wiese, \enquote{Non-stationary
  dynamics of the {Alessandro-Beatrice-Bertotti-Montorsi} model,} Phys. Rev. E
  \textbf{85}, 031105 (2012).

\bibitem{BertottiDurinMagni1994}
G.~Bertotti, G.~Durin, et A.~Magni, \enquote{Scaling aspects of domain wall
  dynamics and barkhausen effect in ferromagnetic materials,} Journal of
  Applied Physics \textbf{75} (1994).

\bibitem{gardiner2004handbook}
C.~W. Gardiner, \emph{Handbook of stochastic methods for physics, chemistry and
  the natural sciences}, vol.~13 de \emph{Springer Series in Synergetics}
  (Springer-Verlag, Berlin, 2004), 3è éd.

\bibitem{MiddletonLeDoussalWiese2006}
A.~A. Middleton, P.~{Le~Doussal}, et K.~J. Wiese, \enquote{Measuring functional
  renormalization group fixed-point functions for pinned manifolds,} Phys. Rev.
  Lett. \textbf{98}, 155701 (2007).

\bibitem{LeDoussalMiddletonWiese2008}
P.~{Le~Doussal}, A.~Middleton, et K.~Wiese, \enquote{Statistics of static
  avalanches in a random pinning landscape,} Phys. Rev. E \textbf{79}, 050101
  (R) (2009).
\newblock URL~:
  \url{http://www.phys.ens.fr/~wiese/abstracts/stat-avalanche-PRL.html}.

\bibitem{DelormeLeDoussalWiese2016}
M.~{Delorme}, P.~{Le Doussal}, et K.~{Wiese}, \enquote{{Distribution of joint
  local and total size and of extension for avalanches in the Brownian force
  model},} ArXiv e-prints  (2016).

\bibitem{NarayanDSFisher1993a}
O.~Narayan et D.~S. Fisher, \enquote{Threshold critical dynamics of driven
  interfaces in random media,} Phys. Rev. B \textbf{48}, 7030--42 (1993).

\bibitem{LeDoussalWiese2008c}
P.~{Le~Doussal} et K.~J. Wiese, \enquote{Size distributions of shocks and
  static avalanches from the functional renormalization group,} Phys. Rev. E
  \textbf{79}, 051106 (2009).

\bibitem{DobrinevskiLeDoussalWiese2014a}
A.~Dobrinevski, P.~{Le Doussal}, et K.~Wiese, \enquote{Avalanche shape and
  exponents beyond mean-field theory,} EPL \textbf{108}, 66002 (2014).

\bibitem{LeDoussalWiese2011b}
P.~{Le Doussal} et K.~J. Wiese, \enquote{First-principle derivation of static
  avalanche-size distribution,} Phys. Rev. E \textbf{85}, 061102 (2012).
\newblock URL~: \url{http://pre.aps.org/abstract/PRE/v85/i6/e061102}.

\bibitem{AragonKoltonDoussalWieseJagla2016}
L.~E. Arag\'on, A.~B. Kolton, P.~L. Doussal, K.~J. Wiese, et E.~A. Jagla,
  \enquote{Avalanches in tip-driven interfaces in random media,} EPL
  (Europhysics Letters) \textbf{113}, 10002 (2016).
\newblock URL~: \url{http://stacks.iop.org/0295-5075/113/i=1/a=10002}.

\bibitem{PaczuskiBoettcher1996}
M.~Paczuski et S.~Boettcher, \enquote{Universality in sandpiles, interface
  depinning, and earthquake models,} Phys. Rev. Lett. \textbf{77}, 111 (1996).
\newblock URL~: \url{http://link.aps.org/doi/10.1103/PhysRevLett.77.111}.

\bibitem{DSFisher1986}
D.~S. Fisher, \enquote{Interface fluctuations in disordered systems:
  $5-\ensuremath{\epsilon}$ expansion and failure of dimensional reduction,}
  Phys. Rev. Lett. \textbf{56} (1986).
\newblock URL~: \url{http://link.aps.org/doi/10.1103/PhysRevLett.56.1964}.

\bibitem{BalentsBouchaudMezard1996}
L.~Balents, J.~Bouchaud, et M.~M\'ezard, \enquote{The large scale energy
  landscape of randomly pinned objects,} J. Phys. I (France) \textbf{6},
  1007--20 (1996).

\bibitem{ChauveLeDoussalWiese2000a}
P.~Chauve, P.~{Le~Doussal}, et K.~Wiese, \enquote{Renormalization of pinned
  elastic systems: How does it work beyond one loop?} Phys. Rev. Lett.
  \textbf{86}, 1785--1788 (2001).

\bibitem{LeDoussalWieseChauve2003}
P.~{Le~Doussal}, K.~J. Wiese, et P.~Chauve, \enquote{Functional renormalization
  group and the field theory of disordered elastic systems,} Phys. Rev. E
  \textbf{69}, 026112 (2004).

\bibitem{SchehrDoussal2003}
G.~Schehr et P.~Le~Doussal, \enquote{Exact multilocal renormalization of the
  effective action: Application to the random sine gordon model statics and
  nonequilibrium dynamics,} Phys. Rev. E \textbf{68}, 046101 (2003).
\newblock URL~: \url{http://link.aps.org/doi/10.1103/PhysRevE.68.046101}.

\bibitem{BalentsLeDoussal2004}
L.~Balents et P.~L. Doussal, \enquote{Thermal fluctuations in pinned elastic
  systems: field theory of rare events and droplets,} Annals of Physics
  \textbf{315}, 213--303 (2005).

\bibitem{Wiese2005}
K.~Wiese, \enquote{Why one needs a functional renormalization group to survive
  in a disordered world,} Pramana \textbf{64}, 817--827 (2005).

\bibitem{WieseLeDoussal2006}
K.~Wiese et P.~L. Doussal, \enquote{Functional renormalization for disordered
  systems: Basic recipes and gourmet dishes,} Markov Processes Relat. Fields
  \textbf{13}, 777--818 (2007).

\bibitem{Delamotte2012}
B.~{Delamotte}, \enquote{{An Introduction to the Nonperturbative
  Renormalization Group},} dans \enquote{Lecture Notes in Physics, Berlin
  Springer Verlag,} , vol. 852 de \emph{Lecture Notes in Physics, Berlin
  Springer Verlag}, J.~{Polonyi} et A.~{Schwenk}, éds. (2012), vol. 852 de
  \emph{Lecture Notes in Physics, Berlin Springer Verlag}, p.~49.

\bibitem{LeDoussal2006b}
P.~{Le Doussal}, \enquote{Finite temperature {Functional RG}, droplets and
  decaying {Burgers} turbulence,} Europhys. Lett. \textbf{76}, 457--463 (2006).

\bibitem{Wetterich1993}
C.~Wetterich, \enquote{Exact evolution equation for the effective potential,}
  Physics Letters B \textbf{301}, 90 -- 94 (1993).
\newblock URL~:
  \url{http://www.sciencedirect.com/science/article/pii/037026939390726X}.

\bibitem{Morris1994}
T.~R. Morris, \enquote{The exact renormalization group and approximate
  solutions,} International Journal of Modern Physics A \textbf{9}, 2411--2449
  (1994).

\bibitem{NarayanDSFisher1992b}
O.~Narayan et D.~S. Fisher, \enquote{Critical behavior of sliding
  charge-density waves in 4-epsilon dimensions,} Phys. Rev. B \textbf{46},
  11520--49 (1992).

\bibitem{NattermannStepanowTangLeschhorn1992}
T.~Nattermann, S.~Stepanow, L.-H. Tang, et H.~Leschhorn, \enquote{Dynamics of
  interface depinning in a disordered medium,} J. Phys. II (France) \textbf{2},
  1483--8 (1992).

\bibitem{LeschhornNattermannStepanowTang1997}
H.~Leschhorn, T.~Nattermann, S.~Stepanow, et L.-H. Tang, \enquote{Driven
  interface depinning in a disordered medium,} Annalen der Physik \textbf{509},
  1--34 (1997).

\bibitem{LeDoussalWieseChauve2002}
P.~{Le~Doussal}, K.~J. Wiese, et P.~Chauve, \enquote{2-loop functional
  renormalization group analysis of the depinning transition,} Phys. Rev. B
  \textbf{66}, 174201 (2002).

\bibitem{MSR}
P.~Martin, E.~Siggia, et H.~Rose, \enquote{Statistical dynamics of classical
  systems,} Phys. Rev. \textbf{A 8}, 423--437 (1973).

\bibitem{Janssen1976}
H.-K. Janssen, \enquote{On a lagrangean for classical field dynamics and
  renormalization group calculations of dynamical critical properties,}
  Zeitschrift f{\"u}r Physik B Condensed Matter \textbf{23}, 377--380 (1976).
\newblock URL~: \url{http://dx.doi.org/10.1007/BF01316547}.

\bibitem{Janssen1992}
H.~Janssen, \enquote{On the renormalized field theory of nonlinear critical
  relaxation,} dans \enquote{From Phase Transitions to Chaos,}  (World
  Scientific, Singapore, 1992), Topics in Modern Statistical Physics, p.
  68--117.

\bibitem{TauberBook2014}
U.~T\"auber, \emph{Critical Dynamics: A Field Theory Approach to Equilibrium
  and Non-Equilibrium Scaling Behavior} (Cambridge University Press, 2014).

\bibitem{RossoLeDoussalWiese2006a}
A.~Rosso, P.~{Le~Doussal}, et K.~Wiese, \enquote{Numerical calculation of the
  functional renormalization group fixed-point functions at the depinning
  transition,} Phys. Rev. B \textbf{75}, 220201 (2007).

\bibitem{Delamotte:2002vw}
B.~Delamotte, \enquote{{A Hint of renormalization},} Am. J. Phys. \textbf{72},
  170--184 (2004).

\bibitem{LeDoussalWiese2011a}
P.~{Le Doussal} et K.~J. Wiese, \enquote{Distribution of velocities in an
  avalanche,} EPL \textbf{97}, 46004 (2012).
\newblock URL~: \url{http://www.phys.ens.fr/~wiese/pdf/epl14307-offprints.pdf}.

\bibitem{LeBlancAnghelutaDahmenGoldenfeld2013}
M.~LeBlanc, L.~Angheluta, K.~Dahmen, et N.~Goldenfeld, \enquote{Universal
  fluctuations and extreme statistics of avalanches near the depinning
  transition,} Phys. Rev. E \textbf{87}, 022126 (2013).
\newblock URL~: \url{http://link.aps.org/doi/10.1103/PhysRevE.87.022126}.

\bibitem{LeBlancAnghelutaDahmenGoldenfeld2012}
M.~LeBlanc, L.~Angheluta, K.~Dahmen, et N.~Goldenfeld, \enquote{Distribution of
  maximum velocities in avalanches near the depinning transition,} Phys. Rev.
  Lett. \textbf{109}, 105702 (2012).
\newblock URL~: \url{http://link.aps.org/doi/10.1103/PhysRevLett.109.105702}.

\bibitem{RossoLeDoussalWiese2009a}
A.~Rosso, P.~{Le~Doussal}, et K.~Wiese, \enquote{Avalanche-size distribution at
  the depinning transition: A numerical test of the theory,} Phys. Rev. B
  \textbf{80}, 144204 (2009).

\bibitem{BaldassarriColaioriCastellano2003}
A.~Baldassarri, F.~Colaiori, et C.~Castellano, \enquote{Average shape of a
  fluctuation: Universality in excursions of stochastic processes,} Phys. Rev.
  Lett. \textbf{90}, 060601 (2003).
\newblock URL~: \url{http://link.aps.org/doi/10.1103/PhysRevLett.90.060601}.

\bibitem{ColaioriZapperiDurin2004}
F.~Colaiori, S.~Zapperi, et G.~Durin, \enquote{{Shape of a Barkhausen pulse},}
  Journal of Magnetism and Magnetic Materials \textbf{272-276}, E533--E534
  (2004).
\newblock URL~:
  \url{http://linkinghub.elsevier.com/retrieve/pii/S0304885303025423}.

\bibitem{PapanikolaouBohnSommerDurinZapperiSethna2011}
S.~Papanikolaou, F.~Bohn, R.~Sommer, G.~Durin, S.~Zapperi, et J.~Sethna,
  \enquote{Universality beyond power laws and the average avalanche shape,}
  Nature Physics \textbf{7}, 316--320 (2011).
\newblock URL~: \url{http://dx.doi.org/10.1038/nphys1884}.

\bibitem{ZapperiCastellanoColaioriDurin2005}
S.~Zapperi, C.~Castellano, F.~Colaiori, et G.~Durin, \enquote{Signature of
  effective mass in crackling-noise asymmetry,} Nat Phys \textbf{1}, 46--49
  (2005).
\newblock URL~: \url{http://dx.doi.org/10.1038/nphys101}.

\bibitem{DobrinevskiLeDoussalWiese2013}
A.~Dobrinevski, P.~Le~Doussal, et K.~J. Wiese, \enquote{Statistics of
  avalanches with relaxation and barkhausen noise: A solvable model,} Phys.
  Rev. E \textbf{88}, 032106 (2013).
\newblock URL~: \url{http://link.aps.org/doi/10.1103/PhysRevE.88.032106}.

\bibitem{DurinBohnCorreaSommerDoussalWiese2016}
G.~{Durin}, F.~{Bohn}, M.~A. {Correa}, R.~L. {Sommer}, P.~{Le Doussal}, et
  K.~J. {Wiese}, \enquote{{Quantitative scaling of magnetic avalanches},} ArXiv
  e-prints  (2016).

\bibitem{BurridgeKnopoff1967}
R.~Burridge et L.~Knopoff, \enquote{Model and theoretical seismicity,} Bulletin
  of the Seismological Society of America \textbf{57}, 341--371 (1967).

\bibitem{JaglaKolton2009}
E.~Jagla et A.~Kolton, \enquote{The mechanisms of spatial and temporal
  earthquake clustering,} arXiv:\null \textbf{0901.1907} (2009).

\bibitem{JaglaLandesRosso2014}
E.~A. Jagla, F.~P. Landes, et A.~Rosso, \enquote{Viscoelastic effects in
  avalanche dynamics: A key to earthquake statistics,} Phys. Rev. Lett.
  \textbf{112}, 174301 (2014).
\newblock URL~: \url{http://link.aps.org/doi/10.1103/PhysRevLett.112.174301}.

\bibitem{Jagla2014}
E.~A. Jagla, \enquote{Aftershock production rate of driven viscoelastic
  interfaces,} Phys. Rev. E \textbf{90}, 042129 (2014).
\newblock URL~: \url{http://link.aps.org/doi/10.1103/PhysRevE.90.042129}.

\bibitem{Quastel2011}
J.~Quastel, \enquote{{Introduction to KPZ},} Current Developments in
  Mathematics  (2011).

\bibitem{SpohnLesHouches2016}
H.~{Spohn}, \enquote{{The Kardar-Parisi-Zhang equation - a statistical physics
  perspective},} ArXiv e-prints  (2016).

\bibitem{Hairer2013}
M.~{Hairer}, \enquote{{Solving the KPZ equation.}} {Ann. Math. (2)}
  \textbf{178}, 559--664 (2013).

\bibitem{ImbrieSpencer1988}
J.~Imbrie et T.~Spencer, \enquote{Diffusion of directed polymers in a random
  environment,} J. Stat. Phys. \textbf{52}, 609 (1988).

\bibitem{Bolthausen1989}
E.~Bolthausen, \enquote{A note on the diffusion of directed polymers in a
  random environment,} Comm. Math. Phys. \textbf{123}, 529--534 (1989).
\newblock URL~: \url{http://projecteuclid.org/euclid.cmp/1104178982}.

\bibitem{comets2004probabilistic}
F.~Comets, T.~Shiga, N.~Yoshida \emph{et~al.}, \enquote{Probabilistic analysis
  of directed polymers in a random environment: a review,} Advanced Studies in
  Pure Mathematics \textbf{39}, 115--142 (2004).

\bibitem{comets2003directed}
F.~Comets, T.~Shiga, et N.~Yoshida, \enquote{Directed polymers in a random
  environment: path localization and strong disorder,} Bernoulli \textbf{9},
  705--723 (2003).

\bibitem{hammersley1965first}
J.~M. Hammersley et D.~Welsh, \enquote{First-passage percolation, subadditive
  processes, stochastic networks, and generalized renewal theory,} dans
  \enquote{Bernoulli 1713 Bayes 1763 Laplace 1813,}  (Springer, 1965), p.
  61--110.

\bibitem{martin2006last}
J.~B. Martin, \enquote{Last-passage percolation with general weight
  distribution,}  URL~:
  \url{http://www.stats.ox.ac.uk/~martin/papers/lppsurvey.ps.gz}.

\bibitem{johansson2000}
K.~Johansson, \enquote{Shape fluctuations and random matrices,} Communications
  in mathematical physics \textbf{209}, 437--476 (2000).

\bibitem{KriecherbauerKrug2008}
T.~{Kriecherbauer} et J.~{Krug}, \enquote{{A pedestrian's view on interacting
  particle systems, KPZ universality, and random matrices},} ArXiv e-prints
  (2008).

\bibitem{CorwinQuastelRemenik2011}
I.~{Corwin}, J.~{Quastel}, et D.~{Remenik}, \enquote{{Renormalization fixed
  point of the KPZ universality class},} ArXiv e-prints  (2011).

\bibitem{PraehoferSpohn2001}
M.~{Praehofer} et H.~{Spohn}, \enquote{{Scale Invariance of the PNG Droplet and
  the Airy Process},} ArXiv Mathematics e-prints  (2001).

\bibitem{Sasamoto2005}
T.~Sasamoto, \enquote{Spatial correlations of the 1d kpz surface on a flat
  substrate,} Journal of Physics A: Mathematical and General \textbf{38}, L549
  (2005).
\newblock URL~: \url{http://stacks.iop.org/0305-4470/38/i=33/a=L01}.

\bibitem{QuastelRemenik2014}
J.~Quastel et D.~Remenik, \emph{Topics in Percolative and Disordered Systems}
  (Springer New York, New York, NY, 2014), chap. Airy Processes and Variational
  Problems, p. 121--171.
\newblock URL~: \url{http://dx.doi.org/10.1007/978-1-4939-0339-9_5}.

\bibitem{Dotsenko2010}
V.~Dotsenko, \enquote{Bethe ansatz derivation of the tracy-widom distribution
  for one-dimensional directed polymers,} EPL (Europhysics Letters)
  \textbf{90}, 20003 (2010).
\newblock URL~: \url{http://stacks.iop.org/0295-5075/90/i=2/a=20003}.

\bibitem{Dotsenko2016}
V.~{Dotsenko}, \enquote{{On two-time distribution functions in (1+1) random
  directed polymers},} ArXiv e-prints  (2016).

\bibitem{Johansson2015}
K.~{Johansson}, \enquote{{Two time distribution in Brownian directed
  percolation},} ArXiv e-prints  (2015).

\bibitem{FerrariSpohn2016}
P.~L. {Ferrari} et H.~{Spohn}, \enquote{{On time correlations for KPZ growth in
  one dimension},} ArXiv e-prints  (2016).

\bibitem{GueudreLeDoussal2012}
T.~{Gueudr{\'e}} et P.~{Le Doussal}, \enquote{{Directed polymer near a hard
  wall and KPZ equation in the half-space},} EPL (Europhysics Letters)
  \textbf{100}, 26006 (2012).

\bibitem{BaikDeiftJohansson1998}
J.~{Baik}, P.~{Deift}, et K.~{Johansson}, \enquote{{On the Distribution of the
  Length of the Longest Increasing Subsequence of Random Permutations},} ArXiv
  Mathematics e-prints  (1998).

\bibitem{AmirCorwinQuastel2010}
G.~{Amir}, I.~{Corwin}, et J.~{Quastel}, \enquote{{Probability Distribution of
  the Free Energy of the Continuum Directed Random Polymer in 1+1 dimensions},}
  ArXiv e-prints  (2010).

\bibitem{SasamotoSpohn2010}
T.~{Sasamoto} et H.~{Spohn}, \enquote{{Exact height distributions for the KPZ
  equation with narrow wedge initial condition},} Nuclear Physics B
  \textbf{834}, 523--542 (2010).

\bibitem{CalabreseLeDoussalRosso2010}
P.~Calabrese, P.~Le~Doussal, et A.~Rosso, \enquote{Free-energy distribution of
  the directed polymer at high temperature,} EPL (Europhysics Letters)
  \textbf{90}, 20002 (2010).

\bibitem{Johansson2003}
K.~Johansson, \enquote{Discrete polynuclear growth and determinantal
  processes,} Communications in Mathematical Physics \textbf{242}, 277--329
  (2003).
\newblock URL~: \url{http://dx.doi.org/10.1007/s00220-003-0945-y}.

\bibitem{ProlhacSpohn2011}
S.~{Prolhac} et H.~{Spohn}, \enquote{{The one-dimensional KPZ equation and the
  Airy process},} Journal of Statistical Mechanics: Theory and Experiment
  \textbf{3}, 03020 (2011).

\bibitem{BaikRains2001}
J.~Baik et E.~M. Rains, \enquote{The asymptotics of monotone subsequences of
  involutions,} Duke Math. J. \textbf{109}, 205--281 (2001).
\newblock URL~: \url{http://dx.doi.org/10.1215/S0012-7094-01-10921-6}.

\bibitem{CalabreseLeDoussal2011}
P.~{Calabrese} et P.~{Le Doussal}, \enquote{{Exact Solution for the
  Kardar-Parisi-Zhang Equation with Flat Initial Conditions},} Physical Review
  Letters \textbf{106}, 250603 (2011).

\bibitem{LeDoussalCalabrese2012}
P.~{Le Doussal} et P.~{Calabrese}, \enquote{{The KPZ equation with flat initial
  condition and the directed polymer with one free end},} Journal of
  Statistical Mechanics: Theory and Experiment \textbf{6}, 06001 (2012).

\bibitem{GuedreLeDoussalRossoHenryCalabrese2012}
T.~{Gueudr{\'e}}, P.~{Le Doussal}, A.~{Rosso}, A.~{Henry}, et P.~{Calabrese},
  \enquote{{Short-time growth of a Kardar-Parisi-Zhang interface with flat
  initial conditions},} Phys. Rev. E. \textbf{86}, 041151 (2012).

\bibitem{OrtmannQuastelRemenik2014}
J.~{Ortmann}, J.~{Quastel}, et D.~{Remenik}, \enquote{{Exact formulas for
  random growth with half-flat initial data},} ArXiv e-prints  (2014).

\bibitem{FerrariSpohn2006}
P.~L. {Ferrari} et H.~{Spohn}, \enquote{{Scaling Limit for the Space-Time
  Covariance of the Stationary Totally Asymmetric Simple Exclusion Process},}
  Communications in Mathematical Physics \textbf{265}, 1--44 (2006).

\bibitem{BaikFerrariPeche2011}
J.~{Baik}, P.~L. {Ferrari}, et S.~{P{\'e}ch{\'e}}, \enquote{{Limit process of
  stationary TASEP near the characteristic line},} ArXiv e-prints  (2009).

\bibitem{ImamuraSasamoto2012}
T.~Imamura et T.~Sasamoto, \enquote{Exact solution for the stationary
  kardar-parisi-zhang equation,} Phys. Rev. Lett. \textbf{108}, 190603 (2012).
\newblock URL~: \url{http://link.aps.org/doi/10.1103/PhysRevLett.108.190603}.

\bibitem{ImamuraSasamoto2013}
T.~{Imamura} et T.~{Sasamoto}, \enquote{{Stationary Correlations for the 1D KPZ
  Equation},} Journal of Statistical Physics \textbf{150}, 908--939 (2013).

\bibitem{BorodinCorwinFerrariVeto2015}
A.~Borodin, I.~Corwin, P.~Ferrari, et B.~Vet{\H{o}}, \enquote{Height
  fluctuations for the stationary kpz equation,} Mathematical Physics, Analysis
  and Geometry \textbf{18}, 1--95 (2015).
\newblock URL~: \url{http://dx.doi.org/10.1007/s11040-015-9189-2}.

\bibitem{HairerQuastel2015}
M.~{Hairer} et J.~{Quastel}, \enquote{{A class of growth models rescaling to
  KPZ},} ArXiv e-prints  (2015).

\bibitem{BertiniGiacomin1997}
L.~Bertini et G.~Giacomin, \enquote{Stochastic burgers and kpz equations from
  particle systems,} Comm. Math. Phys. \textbf{183}, 571--607 (1997).
\newblock URL~: \url{http://projecteuclid.org/euclid.cmp/1158328658}.

\bibitem{AlbertsKhaninQuastel2012}
T.~{Alberts}, K.~{Khanin}, et J.~{Quastel}, \enquote{{The intermediate disorder
  regime for directed polymers in dimension $1+1$},} ArXiv e-prints  (2012).

\bibitem{BustingorryLeDoussalRosso2010}
S.~{Bustingorry}, P.~{Le Doussal}, et A.~{Rosso}, \enquote{{Universal
  high-temperature regime of pinned elastic objects},} Phys. Rev. B.
  \textbf{82}, 140201 (2010).

\bibitem{OConnellYor2001}
N.~O'Connell et M.~Yor, \enquote{Brownian analogues of {B}urke's theorem,}
  Stochastic Process. Appl. \textbf{96}, 285--304 (2001).
\newblock URL~: \url{http://dx.doi.org/10.1016/S0304-4149(01)00119-3}.

\bibitem{OConnell2009}
N.~{O'Connell}, \enquote{{Directed polymers and the quantum Toda lattice},}
  ArXiv e-prints  (2009).

\bibitem{BorodinCorwinMacDo2014}
A.~Borodin et I.~Corwin, \enquote{Macdonald processes,} Probability Theory and
  Related Fields \textbf{158}, 225--400 (2014).
\newblock URL~: \url{http://dx.doi.org/10.1007/s00440-013-0482-3}.

\bibitem{AuffingerBaikCorwin2012}
A.~{Auffinger}, J.~{Baik}, et I.~{Corwin}, \enquote{{Universality for directed
  polymers in thin rectangles},} ArXiv e-prints  (2012).

\bibitem{BecKhanin2007}
J.~Bec et K.~Khanin, \enquote{Burgers turbulence,} Phys. Rep. \textbf{447},
  1--66 (2007).
\newblock URL~:
  \url{http://www.citebase.org/abstract?id=oai:arXiv.org:0704.1611}.

\bibitem{ForsterNelsonStephen1977}
D.~Forster, D.~R. Nelson, et M.~J. Stephen, \enquote{Large-distance and
  long-time properties of a randomly stirred fluid,} Phys. Rev. A \textbf{16},
  732--749 (1977).
\newblock URL~: \url{http://link.aps.org/doi/10.1103/PhysRevA.16.732}.

\bibitem{KrugSpohnGodreche1991}
J.~Krug, H.~Spohn, et C.~Godr{\`e}che, \enquote{Solids far from equilibrium,}
  Solids far from equilibrium  (1991).

\bibitem{Seppalainen2009}
T.~{Sepp{\"a}l{\"a}inen}, \enquote{{Scaling for a one-dimensional directed
  polymer with boundary conditions},} ArXiv e-prints  (2009).

\bibitem{CorwinOConnellSeppalainenZygouras2014}
I.~Corwin, N.~O’Connell, T.~Seppäläinen, et N.~Zygouras, \enquote{Tropical
  combinatorics and whittaker functions,} Duke Math. J. \textbf{163}, 513--563
  (2014).
\newblock URL~: \url{http://dx.doi.org/10.1215/00127094-2410289}.

\bibitem{burke1956}
P.~J. Burke, \enquote{The output of a queuing system,} Operations research
  \textbf{4}, 699--704 (1956).

\bibitem{schutz1997}
G.~M. Sch{\"u}tz, \enquote{Exact solution of the master equation for the
  asymmetric exclusion process,} Journal of statistical physics \textbf{88},
  427--445 (1997).

\bibitem{Kardar1987}
M.~Kardar, \enquote{Replica {Bethe} ansatz studies of two-dimensional
  interfaces with quenched random impurities,} Nucl. Phys. B \textbf{290},
  582--602 (1987).

\bibitem{LiebLiniger1963}
E.~H. Lieb et W.~Liniger, \enquote{Exact analysis of an interacting bose gas.
  i. the general solution and the ground state,} Phys. Rev. \textbf{130},
  1605--1616 (1963).
\newblock URL~: \url{http://link.aps.org/doi/10.1103/PhysRev.130.1605}.

\bibitem{franchini2011notes}
F.~Franchini, \enquote{Notes on bethe ansatz techniques,}   (2011).

\bibitem{BrunetDerrida2000a}
{\'E}.~{Brunet} et B.~{Derrida}, \enquote{{Ground state energy of a non-integer
  number of particles with {$\delta$} attractive interactions},} Physica A
  Statistical Mechanics and its Applications \textbf{279}, 398--407 (2000).

\bibitem{BrunetDerrida2000b}
{\'E}.~{Brunet} et B.~{Derrida}, \enquote{{Probability distribution of the free
  energy of a directed polymer in a random medium},} Phys. Rev. E \textbf{61},
  6789--6801 (2000).

\bibitem{ProlhacSpohn2011b}
S.~{Prolhac} et H.~{Spohn}, \enquote{{The propagator of the attractive
  delta-Bose gas in one dimension},} Journal of Mathematical Physics
  \textbf{52}, 122106--122106 (2011).

\bibitem{McGuire1964}
J.~B. {McGuire}, \enquote{{Study of Exactly Soluble One-Dimensional N-Body
  Problems},} Journal of Mathematical Physics \textbf{5}, 622--636 (1964).

\bibitem{BorodinCorwinPetrovSasamoto2015}
A.~{Borodin}, I.~{Corwin}, L.~{Petrov}, et T.~{Sasamoto}, \enquote{{Spectral
  Theory for Interacting Particle Systems Solvable by Coordinate Bethe
  Ansatz},} Communications in Mathematical Physics \textbf{339}, 1167--1245
  (2015).

\bibitem{CalabreseCaux2007}
P.~{Calabrese} et J.-S. {Caux}, \enquote{{Dynamics of the attractive 1D Bose
  gas: analytical treatment from integrability},} Journal of Statistical
  Mechanics: Theory and Experiment \textbf{8}, 08032 (2007).

\bibitem{gaudin1983}
M.~Gaudin, \emph{La fonction d'onde de Bethe}, Collection du Commissariat a
  l'Energie Atomique / Ser. scientifique (Masson, 1983).
\newblock URL~: \url{https://books.google.fr/books?id=yLLvAAAAMAAJ}.

\bibitem{Bornemann2008}
F.~{Bornemann}, \enquote{{On the Numerical Evaluation of Fredholm
  Determinants},} ArXiv e-prints  (2008).

\bibitem{BouchaudOrland1990}
J.~P. Bouchaud et H.~Orland, \enquote{On the bethe ansatz for random directed
  polymers,} Journal of Statistical Physics \textbf{61}, 877--884 (1990).
\newblock URL~: \url{http://dx.doi.org/10.1007/BF01027306}.

\bibitem{Dotsenko2points}
V.~Dotsenko, \enquote{Two-point free energy distribution function in (1+1)
  directed polymers,} Journal of Physics A: Mathematical and Theoretical
  \textbf{46}, 355001 (2013).
\newblock URL~: \url{http://stacks.iop.org/1751-8121/46/i=35/a=355001}.

\bibitem{LeDoussal2014}
P.~{Le Doussal}, \enquote{{Crossover from droplet to flat initial conditions in
  the KPZ equation from the replica Bethe ansatz},} Journal of Statistical
  Mechanics: Theory and Experiment \textbf{4}, 04018 (2014).

\bibitem{Dotsenko2013}
V.~Dotsenko, \enquote{Distribution function of the endpoint fluctuations of
  one-dimensional directed polymers in a random potential,} Journal of
  Statistical Mechanics: Theory and Experiment \textbf{2013}, P02012 (2013).
\newblock URL~: \url{http://stacks.iop.org/1742-5468/2013/i=02/a=P02012}.

\bibitem{DraiefMairesseOConnell2005}
M.~Draief, J.~Mairesse, et N.~O'Connell, \enquote{Queues, stores, and
  tableaux,} J. Appl. Probab. \textbf{42}, 1145--1167 (2005).
\newblock URL~: \url{http://dx.doi.org/10.1239/jap/1134587823}.

\bibitem{OConnel2009}
N.~{O'Connell}, \enquote{{Directed polymers and the quantum Toda lattice},}
  ArXiv e-prints  (2009).

\bibitem{BorodinCorwinRemenik2013}
A.~Borodin, I.~Corwin, et D.~Remenik, \enquote{Log-gamma polymer free energy
  fluctuations via a fredholm determinant identity,} Communications in
  Mathematical Physics \textbf{324}, 215--232 (2013).
\newblock URL~: \url{http://dx.doi.org/10.1007/s00220-013-1750-x}.

\bibitem{CorwinSeappalainenShen2015}
I.~Corwin, T.~Sepp{\"a}l{\"a}inen, et H.~Shen, \enquote{The strict-weak lattice
  polymer,} Journal of Statistical Physics \textbf{160}, 1027--1053 (2015).
\newblock URL~: \url{http://dx.doi.org/10.1007/s10955-015-1267-0}.

\bibitem{OConnellOrtmann2015}
N.~O'Connell et J.~Ortmann, \enquote{Tracy-widom asymptotics for a random
  polymer model with gamma-distributed weights,} Electron. J. Probab.
  \textbf{20}, no. 25, 1--18 (2015).
\newblock URL~: \url{http://ejp.ejpecp.org/article/view/3787}.

\bibitem{MateevPetrov2015}
K.~{Matveev} et L.~{Petrov}, \enquote{{q-randomized Robinson-Schensted-Knuth
  correspondences and random polymers},} ArXiv e-prints  (2015).

\bibitem{Povolotsky2013}
A.~M. {Povolotsky}, \enquote{{On the integrability of zero-range chipping
  models with factorized steady states},} Journal of Physics A Mathematical
  General \textbf{46}, 465205 (2013).

\bibitem{CorwinPetrov2015}
I.~Corwin et L.~Petrov, \enquote{The {\$}{\$}q{\$}{\$} q -pushasep: A new
  integrable model for traffic in {\$}{\$}1+1{\$}{\$} 1 + 1 dimension,} Journal
  of Statistical Physics \textbf{160}, 1005--1026 (2015).
\newblock URL~: \url{http://dx.doi.org/10.1007/s10955-015-1218-9}.

\bibitem{BarraquandCorwin2015}
G.~Barraquand et I.~Corwin, \enquote{Random-walk in beta-distributed random
  environment,} Probability Theory and Related Fields p. 1--60 (2016).
\newblock URL~: \url{http://dx.doi.org/10.1007/s00440-016-0699-z}.

\bibitem{TracyWidom2008}
C.~A. {Tracy} et H.~{Widom}, \enquote{{A Fredholm Determinant Representation in
  ASEP},} Journal of Statistical Physics \textbf{132}, 291--300 (2008).

\bibitem{TracyWidom2009}
C.~A. {Tracy} et H.~{Widom}, \enquote{{Asymptotics in ASEP with Step Initial
  Condition},} Communications in Mathematical Physics \textbf{290}, 129--154
  (2009).

\bibitem{RASY2013}
F.~Rassoul-Agha, T.~Seppäläinen, et A.~Yilmaz, \enquote{Quenched free energy
  and large deviations for random walks in random potentials,} Communications
  on Pure and Applied Mathematics \textbf{66}, 202--244 (2013).
\newblock URL~: \url{http://dx.doi.org/10.1002/cpa.21417}.

\bibitem{GeorgiouSeppalainen2011}
N.~{Georgiou} et T.~{Sepp{\"a}l{\"a}inen}, \enquote{{Large deviation rate
  functions for the partition function in a log-gamma distributed random
  potential},} ArXiv e-prints  (2011).

\bibitem{CometsNguyen2015}
F.~Comets et V.-L. Nguyen, \enquote{Localization in log-gamma polymers with
  boundaries,} Probability Theory and Related Fields p. 1--33 (2015).
\newblock URL~: \url{http://dx.doi.org/10.1007/s00440-015-0662-4}.

\end{thebibliography}
%%% version finale: remplacer \bibliography{} par  le contenu du bbl

\end{document}